\documentclass[twocolumn,final]{elsart3p}

\ifx\pdfoutput\undefined
  \usepackage{graphicx}
\else
  \usepackage[pdftex]{graphicx}
  \usepackage{epstopdf}
\fi

\usepackage[english]{babel}

\usepackage{amssymb}
\usepackage{amsfonts}

\input myunits.sty

\usepackage{fix2col}

\begin{document}
\selectlanguage{english}

% Title, authors, abstract, keywords, PACS
\vbox{}\vskip-1.0in

\begin{frontmatter}

%%%%%%%%%%%%%%%%%%%%%%%%%%%%%%%%%%%%%%%%%%%%%%%%%%%%%%%%%%%%%%%%%%%%%%%%%%%%
\vtop to 0pt {\center{\Large EUROPEAN ORGANIZATION FOR NUCLEAR RESEARCH}}

\vspace*{1cm}
\vbox to 0pt {\large \hfill \sc CERN--PH--EP/2007-001}
\vspace*{1.5mm}
\vbox to 0pt {\large \hfill \sc 17 January 2007}
\vspace*{-1cm}

%%%%%%%%%%%%%%%%%%%%%%%%%%%%%%%%%%%%%%%%%%%%%%%%%%%%%%%%%%%%%%%%%%%%%%%%%%%%


\title{%
%Setup and Performance of the COMPASS Experiment at CERN%
The COMPASS Experiment at CERN%
}


% \author[CERN]{The COMPASS Collaboration}
% \address[CERN]{CERN, CH-1211 Geneva 23, Switzerland}

\author[saclay]{P.~Abbon},
%% \author[protvino]{E.S.~Ageev},%  				left end 2004
\author[cern]{E.~Albrecht},
\author[dubna]{V.Yu.~Alexakhin},
\author[moscowlpi]{Yu.~Alexandrov},
\author[dubna]{G.D.~Alexeev},
\author[turin]{M.G.~Alekseev\thanksref{d}},
\author[turin]{A.~Amoroso},
\author[munichtu]{H.~Angerer},
\author[cern,dubna]{V.A.~Anosov},
\author[warsaw]{B.~Bade{\l}ek},

% ...
\address[dubna]{ Joint Institute for Nuclear Research, 141980 Dubna, Moscow region, Russia}
% ...
\address[cern]{ CERN, 1211 Geneva 23, Switzerland}
% ...
\address[moscowlpi]{Lebedev Physical Institute, 119991 Moscow, Russia}
% ...
\address[munichtu]{Technische Universit\"at M\"unchen, Physik Department, 85748 Garching, Germany}
% ...
\address[saclay]{ CEA DAPNIA/SPhN Saclay, 91191 Gif-sur-Yvette, France}
% ...
\address[turin]{ INFN Turin and University of Turin, Physics Department, 10125 Turin, Italy}
\address[warsaw]{ So{\l}tan Institute for Nuclear Studies and Warsaw University, 00-681 Warsaw, Poland}
% ...

\corauth[*]{Corresponding author. Tel: +41-22-76-76423\\
\textit{E-mail address:} Gerhard.Mallot@cern.ch}

\thanks[a]{Also at IST, Universidade T\'ecnica de Lisboa, Lisbon, Portugal}
\thanks[b]{Also at University of East Piedmont, 15100 Alessandria, Italy}
\thanks[c]{deceased}
\thanks[d]{On leave of absence from JINR Dubna}               
\thanks[e]{Also at Chubu University, Kasugai, Aichi, 487-8501 Japan}
\thanks[f]{Also at Yamagata University, Yamagata, 992-8510 Japan}
\thanks[g]{Also at KEK, Tsukuba, 305-0801 Japan}

%
% Abstract, keywords, PACS
\begin{abstract} 
\vbox{}\vskip1.0in

%  The main features and performances of the COMPASS experimental
%  setup are presented. 
  The COMPASS experiment makes use of the CERN 
  SPS high-intensity muon and hadron beams for the investigation of the
  nucleon spin structure and the spectroscopy of hadrons.  One or more
  outgoing particles are detected in coincidence with the incoming
  muon or hadron. A large polarized target inside a superconducting
  solenoid is used for the measurements with the 
  muon beam. Outgoing particles are detected by a two-stage,
  large angle 
  and large momentum range spectrometer. The setup is built using
  several types of tracking detectors, according to the expected
  incident rate, required space resolution and the solid angle to be
  covered. Particle identification is achieved using a RICH
  counter and both hadron and electromagnetic calorimeters. The setup
  has been successfully operated from 2002 onwards using a muon
  beam. Data with a hadron beam were also collected in 2004. 
  This article describes the main features and performances of the
  spectrometer 
  in 2004; 
  a short
  summary of the 2006 upgrade is also given.
\end{abstract}


\end{frontmatter}
%

% Table of contents
\tableofcontents

% Introduction 
\section{Introduction}
\label{sec:intro}
The aim of the COMPASS experiment at CERN \cite{COMPASS:96} is to study in detail how
nucleons and other hadrons are made up from quarks and gluons.
At hard scales Quantum Chromodynamics (QCD) is well established and
the agreement of experiment and theory is excellent.
However, in the non-perturbative regime,  
despite the wealth of data collected in 
the previous decades in laboratories around the world,
a fundamental understanding of hadronic structure is still missing.
 
Two main sources of information are at our disposal: 
nucleon structure functions and the hadron spectrum itself. 
While the spin-averaged structure functions and resulting parton distribution functions (PDF)
are well determined and the helicity dependent
quark PDFs have been explored during the last 15 years, 
little is known
about the polarisation of gluons in the nucleon and the 
transversity PDF. In the meson sector
the electric and magnetic polarisabilities of pions and kaons 
can shed light onto their internal dynamics.
The reported glueball states need confirmation and an extension of their 
spectrum to higher masses is mandatory. Finally,
hadrons with exotic quantum numbers and double-charmed baryons 
are ideal tools to study QCD. 

Fixed-target experiments in this field require large luminosity and 
thus high data rate capability, excellent particle identification and a wide 
angular acceptance. 
These are the main design goals for the COMPASS spectrometer 
described in this article.
The projects for the nucleon structure measurements with muon beam and 
for the spectroscopy measurements with hadron beams were originally 
launched independently in 1995 as two separate initiatives. 
The unique CERN M2 beam line, which can provide muon and hadron beams
of high quality, offered the possibility to fuse 
the two projects into a single effort and to bring together a strong 
community for QCD studies.
In the merging of the two experimental layouts many technical and
conceptual difficulties had to be overcome, in particular the 
completely different target arrangements had to be reconciled. 
This process resulted in a highly flexible and versatile setup, 
which not only can be adapted to the various planned measurements,
but also bears a large potential for future experiments.

In the following we give a brief description of the muon and hadron
programmes from which the experimental requirements were deduced.

A recent review of our present knowledge of the spin structure
of the nucleon can be found in Ref.~\cite{Bass:05}. %better \citenum{Bass:05}
The original discovery by the EMC in 1988 that the quark spins only
account for a small fraction of the nucleon spin was confirmed
with high precision during the 1990's at CERN, SLAC and later at 
DESY.
Thus the spin structure of the nucleon is not as simple as suggested
by the na\"ive quark model, and both the gluon spin and the overall parton angular momentum
are expected to contribute to the nucleon spin. 
Via the axial anomaly a very large gluon polarisation $\Delta G$
could mask the quark spin contribution and thus explain its smallness.
The gluon polarisation can be studied in deep inelastic
scattering either indirectly by the $Q^2$ evolution of the spin-dependent
structure functions or more directly via the photon--gluon
fusion process yielding a quark--antiquark pair which subsequently
fragments into hadrons. A particularly clean process is open-charm
production leading to $D$~mesons.

The detection of the decay $D^0\rightarrow K^{-}\pi^{+}$ and $\overline{D}^0\rightarrow K^{+}\pi^{-}$ (branching ratio: 3.8\%)
served as reference process for the design of the COMPASS spectrometer
for the muon programme.
%Even with only one $D$~meson observed, 
%The experiment is
%statistics limited largely due to the branching ratio of 3.8\,\%.
%Therefore 
Maximising luminosity together with large acceptance were
important goals for the design.
With present muon beam intensities only a polarised solid-state target
with a high fraction of polarisable nucleons can provide the required 
luminosity. Apart from the luminosity, the beam and target polarisations and the
dilution factor have to be taken into account to estimate the statistical
accuracy of the measured double-spin cross section asymmetries.
%Given the available muon beam intensity only a polarised solid-state
%target with a high fraction of polarisable nucleons can be used.
The measurements of quark polarisations, both longitudinal and
transverse, require a large range in momentum transfer and thus
in the muon scattering angle.
The final layout covers an opening angle of the spectrometer of
$\pm180\,\mrad$ with a luminosity of almost $5\cdot 10^{32}/\cm^2/\s$.

Due to multiple scattering in the long solid-state target the production
and decay vertices of $D$~mesons cannot be separated using a microvertex
detector, otherwise a standard
technique to improve the signal-to-noise ratio for heavy flavour production.
The $D$ identification thus has to rely entirely on the kinematic charm decay
reconstruction making excellent particle identification mandatory for
background rejection.
Over a wide kinematic range this task can only be performed using
a Ring-Imaging Cherenkov detector. Essential for the optimisation
of the signal-to-background ratio is also a good mass resolution
implying the use of high-resolution tracking devices.
A two-stage layout was adopted with a large aperture spectrometer close to
the target mainly used for the momentum range of approximately
$1-20\,\GeV/c$, followed by a small aperture spectrometer
accepting particles with higher momenta, in particular the scattered muons.
With this setup we also investigate spin structure functions, 
flavour separation, 
vector meson production, polarised $\Lambda$ physics and transverse quark distributions.

The physics aspects of the COMPASS programme with hadron beams are
reviewed in detail in Ref.~\cite{COMFUT:04}. %better \citenum{COMFUT:04}.
The observed spectrum of light hadrons shows
new states which cannot be explained within the
constituent quark model and which were interpreted as
glueballs or hybrid states.
In order to gain more insight, measurements with higher statistical
accuracy in particular in the mass range beyond $2\,\GeV/c^{2}$ have to be
performed. Different reactions are needed in order
to unravel the nature of such states.
They are either produced centrally or diffractively and thus
a good coverage for the decay products over a wide kinematic range
is required.
Some of the key decay channels involve $\eta$ or $\eta'$ with
subsequent decays into photons.
Their detection requires large-acceptance electromagnetic calorimetry.
In addition, the flavour partners of the states observed are searched
for using different beam particles ($\pi$, $K$, $p$).
Beam intensities of
up to $10^8$ particles per $5\,\s$ spill are needed imposing stringent requirements 
on the radiation hardness of the central detectors, in particular of 
the electromagnetic calorimeters.
%
%Also heavy flavour production for studies of form
%factors and long-range correlations in weak hadron decays is part of the
%COMPASS programme.
%Owing to low production cross-sections this task requires 
%high-intensity hadron beams and precision tracking for decay-point
%reconstruction. The latter is achieved by radiation-hard
%silicon detectors surrounding a thin low-mass target.

The construction of the spectrometer started after the experiment approval
by CERN in October 1998. Following a technical run in 2001 physics data
were taken during 2002--2004 \cite{Ageev:05a,Alexakhin:05a,Ageev:05b,Ageev:06a}.
Data taking resumed in 2006 after the 2005 shutdown of the CERN accelerators.
Up to now only muon data were taken,
apart from a two-week pilot run with a pion beam de\-di\-ca\-ted to the
measurement of the pion polarisability via the Primakoff reaction.
Hadron beam experiments are scheduled to start in 2007.
Depending on the beam availability the present COMPASS physics
programme will be completed around 2010.
Future plans involving measurements of generalised parton distribution
functions, detailed measurements of transversity and an extension of
the spectroscopy studies are presently being discussed.

The following sections describe in more detail the general layout
and the choice of technologies
%(Section~2),
(Sec.~\ref{sec:layout}),
the beam line, the targets, tracking and particle identification,
the triggers, readout electronics, data acquisition and detector control
%(Sections~3--9).
(Sec.~\ref{sec:beam}--\ref{sec:det_ctrl_mon}).
%Section~10 
Section~\ref{sec:performance} 
represents a snapshot of the data processing procedures
and of the spectrometer performance. The substantial detector
upgrades implemented during the 2005 shutdown are described in
%Section~11.
Sec.~\ref{sec:upgrade}.
The article ends with a summary and an outlook.

Throughout this paper the following kinematic variables will be used:
$E$ ($E'$) is the energy of the incoming (scattered) muon, $M$
the nucleon mass, $m$ the muon mass,
$\theta$ the muon scattering angle in the laboratory system,
$Q^2 $  the negative squared four-momentum
of the virtual photon, $\nu=E-E'$ its energy
in the laboratory system, i.e. the energy loss of the
muon, $y=\nu/E$ the fractional energy loss and
$x_{\rm Bj}=Q^2/2M \nu$ the Bjorken scaling variable.

% Layout
\section{Layout of the spectrometer} 
\label{sec:layout} 
    
\subsection{General overview} 
\label{sec:layout.overview} 

The COMPASS physics programme imposes specific requirements to the 
experimental setup, as illustrated in the introduction to this article. 
They are: large angle and momentum acceptance, including the 
request to track particles scattered at extremely small angles, 
precise kinematic reconstruction of the events together with efficient
particle identification and good mass  
resolution. Operation at high luminosity imposes capabilities of high beam
intensity and counting rates, high trigger rates and huge data flows.  
 
%These needs are well illustrated by one of the most challenging 
%process to be reconstructed with the COMPASS spectrometer: the 
%Photon-Gluon Fusion in which a D$^0$ meson is produced in the muon 
%scattering off the polarised nuclear solid state target, a technique 
%which prevents the employment of microvertex detectors and, thus, the 
%direct detection of the charm decay.  D$^0$ is detected via its 
%hadronic decay in charged particles and the favourite channel is the 
%decay into a kaon and a pion pair. In Fig. 
%\ref{fig:layout.angle_vs_momentum} the kaon and pion momenta and polar 
%angles from Monte Carlo simulation are shown for a muon beam with 
%energy of $160\,\GeV$, illustrating the wide ranges of the angle and 
%momentum of the decay products. The expected detection rate is as low 
%as ?(??), making the measurement affordable only with high luminosity 
%operation. The combinatorial background can completely mask the 
%signal: precise determination of the kinematic parameters and hadron 
%identification are mandatory. 
 
\begin{figure*}[tbp] 
  \begin{center} 
    \includegraphics[width=\textwidth]{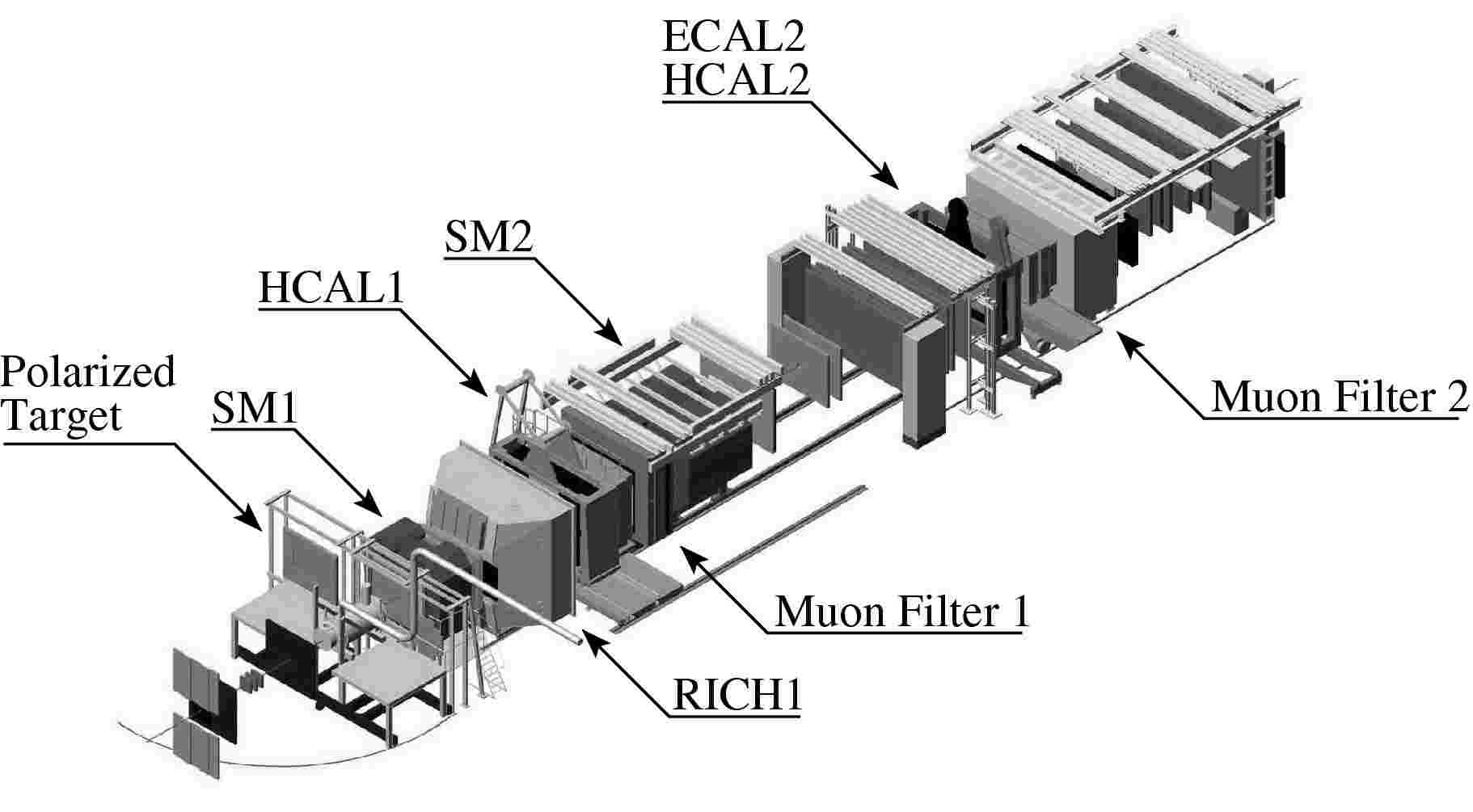} 
    \includegraphics[width=\textwidth]{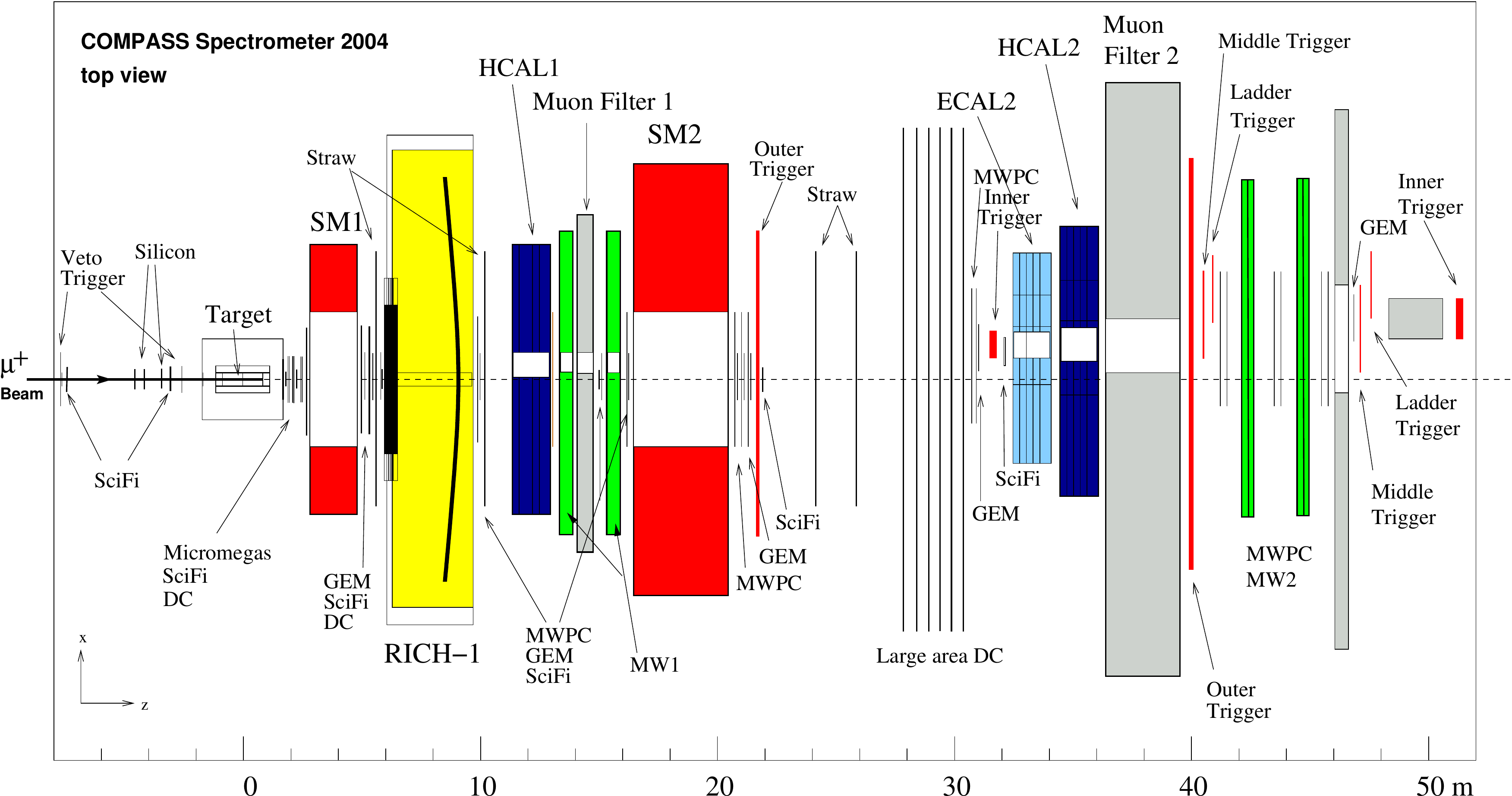} 
  \end{center} 
  \caption{Compass 2004 muon setup (top) artistic view, (bottom) top 
    view (for detector names, see text).}  
  \label{fig:layout.setup} 
\end{figure*} 
 
%The features outlined above determine the main constraints on the 
%setup.  
The basic layout of the COMPASS spectrometer, as it was used 
in 2004, is shown in Fig.~\ref{fig:layout.setup}. Three parts can 
be distinguished.  The first part includes the detectors upstream of 
the target, which measures the incoming beam particles.  The second 
and the third part of the setup are located downstream of the 
target, and extend over a total length of $50\,\m$. These are the large 
angle spectrometer and the small angle spectrometer,  
respectively.  The use of two spectrometers for the outgoing particles 
is a consequence of the large momentum range and the large angular 
acceptance requirements.  Each of the two spectrometers is built 
around an analysing magnet, preceded and followed by telescopes of 
trackers and completed by a hadron calorimeter and by a muon filter 
station for high energy muon identification. A RICH detector for 
hadron identification is part of the large angle spectrometer. The
small angle spectrometer includes an  
electromagnetic calorimeter. 
 
The flexibility required by the broad spectrum of the  
COMPASS physics programme has been implemented  
by mounting  
%a large set of  
huge setup elements 
on rails, allowing them to be positioned   
at variable distances from the experimental target:  
the RICH, the first hadron calorimeter, the first muon filter,  
the second analysing magnet and the trackers fixed to it can move  
longitudinally on rails. 
 
Tables~\ref{tab:layout.detectors} and \ref{tab:layout.hodoscopes} 
provide an overview of the  
different detectors used in COMPASS and their main parameters. The 
detectors are grouped according to their positions and functions in 
the spectrometer. 
\begin{table*} 
  \centering 
  \caption{Overview of detectors used in COMPASS, together with 
    their respective main parameters, grouped according to their 
    geometrical positions along the beam line (stations) and
    functions in the 
    spectrometer. The first column 
    shows the naming convention for the respective stations. 
    The second column gives the number of detectors making up these stations,
    while the third column specifies the coordinates  
    measured by the detectors. Here, e.g.\ $XY$ means that both 
    projections are measured by each detector, while $X/Y$ means that 
    only one of two coordinates $X$ or $Y$ is measured by one of the 
    detectors.  
    Typical values for 
    resolutions of one detector at standard COMPASS muon beam conditions are
    given, where  
    appropriate, in the sixth column. 
    These numbers correspond to an average over all detectors of this kind in
    the experiment, and hence may include contributions from pile-up, magnetic fringe
    fields, or reconstruction inefficiencies.  
    Here, $\sigma_\mathrm{s}$ denotes the 
    r.m.s.\  spatial resolution along one coordinate, $\sigma_t$ the r.m.s.\  time  
    resolution, $\sigma_\mathrm{ph}$ 
    the single photon resolution, $\sigma_\mathrm{ring}$
    the ring resolution. More detailed information on
    the respective detectors can be found in the section of the present paper
    specified in the last column.} 
  \label{tab:layout.detectors} 
  \begin{minipage}{\textwidth}
    \renewcommand{\thefootnote}{\thempfootnote}
  \scriptsize 
  \centering
%%  \begin{tabular}[htbp]{|c|c|c|c|c|c|c|c|} \hline 
  \begin{tabular*}{\textwidth}[htbp]{@{\extracolsep{\fill}}cccccccc} \hline 
    Station & \# of & Planes & \# of ch. & Active area & 
    \multicolumn{2}{c}{Resolution} & Sec.\\ 
    & dets. & per det. & per det. & $X\times Y\,(\cm^2)$ & 
    \multicolumn{2}{c}{} &\\ \hline 
    \multicolumn{8}{c}{Beam detectors} \\ \hline 
    BM01-04        & 4 & $Y$   & 64  & $6-12\times 9-23$  & 
    \multicolumn{2}{c}{$\sigma_\mathrm{s}=1.3-2.5\,\mm$, 
    $\sigma_t=0.3\,\ns$} & \ref{sec:beam.muon.bms}\\ 
%    BMS 2      & 1 & $Y$   & 64  & $12\times 9$ & 
%    \multicolumn{2}{c}{$\sigma_\mathrm{s}=1.3-2.5\,\mm$, 
%    $\sigma_t=0.3\,\ns$} & \ref{sec:} \\ 
%    BMS 3      & 1 & $Y$   & 64  & $12\times 10$ & 
%    \multicolumn{2}{c}{$\sigma_\mathrm{s}=1.3-2.5\,\mm$, 
%    $\sigma_t=0.3\,\ns$} & \ref{sec:} \\ 
%    BMS 4      & 1 & $Y$   & 64  & $6\times 23$ & 
%    \multicolumn{2}{c}{$\sigma_\mathrm{s}=1.3-2.5\,\mm$, 
%    $\sigma_t=0.3\,\ns$} & \ref{sec:} \\ 
    BM05         & 2 & $Y$   & 64      & $12\times 16$ & 
    \multicolumn{2}{c}{$\sigma_\mathrm{s}=0.7\,\mm$, 
    $\sigma_t=0.5\,\ns$} & \ref{sec:beam.muon.bms} \\ 
    BM06         & 2 & $Y$   & 128     & $12\times 16$ & 
    \multicolumn{2}{c}{$\sigma_\mathrm{s}=0.4\,\mm$, 
    $\sigma_t=0.5\,\ns$} & \ref{sec:beam.muon.bms} \\ 
%    BMS 6      & 1 & $Y$   & 128  & $12\times 16$ & 
%    \multicolumn{2}{c}{$\sigma_\mathrm{s}=1.3-2.5\,\mm$, 
%    $\sigma_t=0.3\,\ns$} & \ref{sec:} \\ 
    SciFi 1,2 & 2 & $XY$ & 192 & $3.9\times 3.9$ & 
    \multicolumn{2}{c}{$\sigma_\mathrm{s}=130\,\mum$, $\sigma_t=0.4\,\ns$}
    & \ref{sec:tracking.vsat.scifi} \\ 
    Silicon & 2 & $XYUV$ & 2304 & $5\times 7$ & 
    \multicolumn{2}{c}{$\sigma_\mathrm{s}=8-11\,\mum$, $\sigma_t=2.5\,\ns$}
    & \ref{sec:tracking.vsat.silicon} \\  
    \hline 
    \multicolumn{8}{c}{Large angle spectrometer} \\ \hline 
    SciFi 3,4 & 2 & $XYU$ & 384 & $5.3\times 5.3$ & 
    \multicolumn{2}{c}{$\sigma_\mathrm{s}=130\,\mum$, $\sigma_t=0.4\,\ns$}
    & \ref{sec:tracking.vsat.scifi} \\  
    Micromegas & 12 & $X/Y/U/V$ & 1024 & $40\times 40$ &  
    \multicolumn{2}{c}{$\sigma_\mathrm{s}=90\,\mum$, $\sigma_t=9\,\ns$}&
    \ref{sec:tracking.sat.micromegas} \\  
    DC & 3 & $XYUV$& 1408 & $180\times 127$ &  
    \multicolumn{2}{c}{$\sigma_\mathrm{s}=190\,\mum$}&
    \ref{sec:tracking.lat.dc} \\  
    Straw & 9 & $X/Y/U/V$ & 892 & $323\times 280$ & 
    \multicolumn{2}{c}{$\sigma_\mathrm{s}=190\,\mum$ \protect\footnote{Resolution
    measured for $6\,\mm$ straw tubes only, corresponding to an active area of
    $110\times 350\,\cm^2$}}&
    \ref{sec:tracking.lat.straw} \\  
    GEM 1-4 & 8 & $XY/UV$ & 1536 & $31\times 31$ & 
    \multicolumn{2}{c}{$\sigma_\mathrm{s}=70\,\mum$, $\sigma_t=12\,\ns$} &
    \ref{sec:tracking.sat.gem} \\  
    SciFi 5 &1& $XY$ & 320 & $8.4\times 8.4$ & 
    \multicolumn{2}{c}{$\sigma_\mathrm{s}=170\,\mum$, $\sigma_t=0.4\,\ns$}
    & \ref{sec:tracking.vsat.scifi} \\  
    RICH-1 &8&1 (pads)&10368&$60\times 
    120$&\multicolumn{2}{c}{$\sigma_\mathrm{ph}=1.2\,\mrad$} &
    \ref{sec:pid.rich} \\  
       &&&&&\multicolumn{2}{c}{$\sigma_\mathrm{ring}=0.55\,\mrad$ (for
    $\beta=1$)} & \\  
    MWPC A$^*$ &1&$XUVY$&2768&$178\times 120$& 
    \multicolumn{2}{c}{$\sigma_\mathrm{s}=1.6\,\mm$} &
    \ref{sec:tracking.lat.mwpc} \\  
    HCAL1 & 1 & 1 & 480 & $420\times 300$ & 
    \multicolumn{2}{c}{$\Delta E/E=0.59/\sqrt{E/\GeV}\oplus 
      0.08$} & \ref{sec:pid.calo.hcal1} \\ 
    MW1 &8&$X/Y$&$1184/928$&$473\times 405$& 
    \multicolumn{2}{c}{$\sigma_\mathrm{s}=3\,\mm$} & \ref{sec:pid.muon.mw1}
    \\ \hline  
    \multicolumn{8}{c}{Small angle spectrometer} \\ \hline 
    GEM 5-11 & 14 & $XY/UV$ & 1536 & $31\times 31$ & 
    \multicolumn{2}{c}{$\sigma_\mathrm{s}=70\,\mum$, $\sigma_t=12\,\ns$} &
    \ref{sec:tracking.sat.gem} \\  
    MWPC A &7&$XUV$&2256&$178\times 120$& 
    \multicolumn{2}{c}{$\sigma_\mathrm{s}=1.6\,\mm$} &
    \ref{sec:tracking.lat.mwpc} \\  
    SciFi 6  &1& $XYU$ & 462 & $10\times 10$& 
    \multicolumn{2}{c}{$\sigma_\mathrm{s}=210\,\mum$, $\sigma_t=0.4\,\ns$}
    & \ref{sec:tracking.vsat.scifi} \\  
    SciFi 7  &1& $XY$ & 286 & $10\times 10$& 
    \multicolumn{2}{c}{$\sigma_\mathrm{s}=210\,\mum$, $\sigma_t=0.4\,\ns$}
    & \ref{sec:tracking.vsat.scifi} \\  
    SciFi 8  &1& $XY$ & 352 & $12.3\times 12.3$& 
    \multicolumn{2}{c}{$\sigma_\mathrm{s}=210\,\mum$, $\sigma_t=0.4\,\ns$}
    & \ref{sec:tracking.vsat.scifi} \\  
    Straw & 6 & $X/Y/U/V$ & 892 & $323\times 280$ & 
    \multicolumn{2}{c}{$\sigma_\mathrm{s}=190\,\mum$ \protect\footnotemark[\value{mpfootnote}]}&
    \ref{sec:tracking.lat.straw} \\  
    Large area DC &6&$XY/XU/XV$&500&$500\times 250$& 
    \multicolumn{2}{c}{$\sigma_\mathrm{s}=0.5\,\mm$} &
    \ref{sec:tracking.lat.w45} \\  
    ECAL2 & 1 & 1 & 2972 & $245\times 184$ & 
    \multicolumn{2}{c}{$\Delta E/E=0.06/\sqrt{E/\GeV}\oplus 
      0.02$} & \ref{sec:pid.calo.ecal} \\ 
    HCAL2 & 1 & 1 & 216 & $440\times 200$ & 
    \multicolumn{2}{c}{$\Delta E/E=0.66/\sqrt{E/\GeV}\oplus 
      0.05$} & \ref{sec:pid.calo.hcal2} \\ 
    MWPC B &6&$XU/XV$&1504&$178\times 90$& 
    \multicolumn{2}{c}{$\sigma_\mathrm{s}=1.6\,\mm$} &
    \ref{sec:tracking.lat.mwpc} \\  
    MW2 &2&$XYV$&840&$447\times 
    202$&\multicolumn{2}{c}{$\sigma_\mathrm{s}=0.6-0.9\,\mm$} &
    \ref{sec:pid.muon.mw2} \\ \hline
  \end{tabular*} 
  \normalsize 
\end{minipage}
\end{table*} 
\begin{table} 
  \centering 
  \caption{Overview of trigger detectors used in COMPASS, together with 
    their respective main parameters, grouped according to their 
    functions in the spectrometer.  
    The second column gives the number of detectors at this particular 
    position, while the third column specifies the coordinates 
    measured by these detectors. Here, $XY$ means that both 
    projections are measured by each detector.} 
  \label{tab:layout.hodoscopes} 
  \scriptsize 
%  \begin{tabular}[htbp]{|c|r|c|c|c|} \hline 
  \begin{tabular*}{\columnwidth}[htbp]{@{\extracolsep{\fill}}ccccc} \hline  
    Det. name & \# of & Planes & \# of ch. & Active area \\ 
    & dets. & per det. & per det. & $X\times Y\,(\cm^2)$ \\ \hline 
    \multicolumn{5}{c}{Trigger hodoscopes} \\ \hline 
    Inner&1&$X$&64&$17.3\times 32$\\ 
         &1&$X$&64&$35.3\times 51$\\ 
    Ladder&1&$X$&32&$128.2\times 40$ \\ 
          &1&$X$&32&$168.2\times 47.5$ \\ 
    Middle&1&$XY$&40/32&$120\times 102$ \\ 
          &1&$XY$&40/32&$150\times 120$ \\ 
    Outer&1&$Y$&16&$200\times 100$ \\ 
         &1&$Y$&32&$480\times 225$ \\ \hline 
    \multicolumn{5}{c}{Veto detectors} \\ \hline 
    Veto 1&1& &34&$250\times 320$ \\ 
    Veto 2&1& &4&$30\times 30$ \\ 
    Veto BL&1& &4&$50\times 50$ \\ \hline
  \end{tabular*} 
  \normalsize 
\end{table} 
 
An introductory overview of the experimental apparatus is provided in 
the following: the beam spectrometer in
Sec.~\ref{sec:layout.beam_spectrometer}, the large angle  
spectrometer in Sec.~\ref{sec:layout.las}, the small angle  
spectrometer in Sec.~\ref{sec:layout.sas}, 
the trackers in Sec.~\ref{sec:layout.trackers} and 
the muon filters in Sec.~\ref{sec:layout.muon_filters}.  Setup 
elements specific to the physics programme with muon beam and to the 
first measurement performed with a hadron beam are considered in 
Sec.~\ref{sec:layout.muon_setup}   
and Sec.~\ref{sec:layout.hadron_setup}, respectively. 
 
%\begin{figure}[tbp] 
%  \begin{center} 
%    \includegraphics[width=0.5\columnwidth,angle=270]{./figures/anglemomentum_pi} 
%    \includegraphics[width=0.5\columnwidth]{./figures/anglemomentum_pi} 
%    \includegraphics[width=0.5\columnwidth,angle=270]{./figures/anglemomentum_K} 
%    \includegraphics[width=0.5\columnwidth]{./figures/anglemomentum_K} 
%  \end{center} 
%  \caption{\small Angles of outgoing particles as a function of the momentum: 
%    a) pions and b) kaons} 
%  \label{fig:layout.angle_vs_momentum} 
%\end{figure} 
 
\subsection{Beam telescope and beam spectrometer} 
\label{sec:layout.beam_spectrometer} 
The first part of the setup includes the Beam Momentum Station 
(BMS), located along the beam line about $100\,\m$ upstream of the 
experimental  
hall.  
This beam 
spectrometer measures the momentum of the incoming muon on an 
event  
by event base; it includes an analysing magnet and two telescopes of 
tracking stations formed by scintillator hodoscopes and 
scintillating fibre (SciFi) detectors. 
 
A precise track 
reconstruction of the incident particle is provided 
by fast trackers located 
upstream of the target. There are two stations of 
scintillating fibres and three stations of silicon microstrip detectors. 
Scintillator veto counters define  
the beam spot size and separate the beam from the beam halo. 
 
\subsection{Large angle spectrometer} 
\label{sec:layout.las} 
The second part, i.e.\ the Large Angle Spectrometer (LAS), has been 
designed to ensure $180\,\mrad$ polar acceptance.  It is built around 
the SM1 magnet, which is preceded and followed by telescopes of 
trackers. 
 
SM1 is a dipole magnet located $4\,\m$ downstream 
of the  
target centre. It is $110\,\cm$ long, has a horizontal gap of 
$229\,\cm$ and a   
vertical gap of $152\,\cm$ in the middle. The pole tips of the magnet 
are   
wedge-shaped with the apex of the edge facing the target, so that the 
tracks pointing to the target are orthogonal to the field lines.  The 
SM1 vertical size matches the required  
angular acceptance of $\pm 180\,\mrad$.  The main component of the 
field  
goes from top to bottom. Its field integral was measured 
\cite{Kurek:02} to be $1.0\,\T\m$ and corresponds to a deflection of 
$300\,\mrad$ for particles with a momentum of $1\,\GeV/c$. Due to the
bending power of SM1, the LAS detectors located downstream  
of SM1
need to have an angular acceptance of $\pm 250\,\mrad$ in the 
horizontal plane. 
 
The SM1 magnet is followed by a RICH detector with large 
transverse dimensions to match the LAS acceptance requirement, which 
is used to identify charged hadrons with momenta ranging from a few 
$\GeV/c$ to $43\,\GeV/c$. The LAS is completed by a large hadron
calorimeter (HCAL1) with a central hole matching the second spectrometer 
acceptance.  
The calorimeter detects outgoing hadrons and is used 
in the trigger formation. The LAS is completed by a muon filter. 
 
\subsection{Small angle spectrometer} 
\label{sec:layout.sas} 
The third part of the COMPASS setup, the Small Angle Spectrometer 
(SAS),  
detects particles at small angles ($\pm 30\,\mrad$) and large momenta 
of  
$5\,\GeV/c$ and higher. Its central element is the $4\,\m$ long SM2 magnet, 
located $18\,\m$ downstream of the target centre and preceded and followed by 
telescopes of trackers. 
 
SM2 is a rectangular shape dipole magnet with a gap of $2\times 
1\,\m^2$  
and a total field integral of $4.4\,\T\m$ for its nominal current of 
$4000\,\A$.  
As for SM1, its main field component is in the vertical direction. 
The SM2 magnet was used in several experiments prior to COMPASS; its 
magnetic field is known from previous measurements \cite{Adams:97a}.   
The 
downstream part of the SAS part includes electromagnetic and hadron 
calorimeters and a muon filter.  Each of these elements has a hole
matching the acceptance of the quasi-real photon trigger.
The electromagnetic calorimeter (ECAL2) 
is used to detect gammas and 
neutral pions. The SAS hadron calorimeter (HCAL2), as well as HCAL1,  
is used in the  
trigger formation. A second muon filter is positioned at the  
downstream end of the spectrometer. 
 
\subsection{Tracking detectors} 
\label{sec:layout.trackers} 
The particle flux per unit transverse surface varies by more than five 
orders of magnitude in the different regions included in the overall  
spectrometer acceptance. Along the beam, or close to the 
target, the detectors must combine a high particle rate capability (up 
to a few MHz/channel) with an excellent space resolution ($100\,\mum$ 
and better). The amount of material along the beam path has to remain 
at a minimum in order to minimise multiple scattering and secondary 
interactions.   
These requests are particularly severe upstream of the SM1 magnet 
where the incident flux is further increased because of the large 
number of low energy secondary particles coming from the target region. Far 
from the beam, the resolution constraint can be relaxed, but larger areas need
to be covered. Different 
tracking techniques, including novel ones, are employed in regions at 
different distance from the beam axis, in order to match the requirements
concerning rate capability, space and time resolution as well as the size of the
surface to be instrumented. 
Different 
varieties of large gaseous detectors based on wire amplification are used for
the regions further away from the beam, with their central regions deactivated
in order not to exceed their rate capability. The near-beam and beam regions
are covered by 
fast scintillating, gaseous and silicon tracking
detectors, respectively, with active areas overlapping the dead zones of the
larger detectors to guarantee efficient track reconstruction and good relative
alignment. 

The tracking detectors are grouped as (see also
Table~\ref{tab:layout.detectors}): 
\begin{itemize} 
\item Very Small Area Trackers (VSAT) - These detectors, small in
  size, must combine high flux capabilities and excellent space or
  time resolutions. The area in and around the beam    
  is covered by eight scintillating fibres stations, and, upstream of the
  target, by three stations of double-sided silicon microstrip
  detectors. Their lateral 
  sizes vary from $4\,\cm$ to $12\,\cm$, to take into account the beam
  divergence depending   
  on the position along the beam axis.
\item Small Area Trackers (SAT) - For distances from the beam larger 
  than $2.5\,\cm$ medium size detectors, featuring high space
  resolution and minimum material budget are required. We use three
  Micromegas  
  (Micromesh Gaseous Structure) stations, and 11 GEM (Gas Electron
  Multiplier) 
  stations. Each Micromegas station is composed of four planes and
  has an active  
  area of $40\times 40\,\cm^2$. All three stations are located between 
  the target and the SM1 magnet. 
  Each GEM station consists of two detectors with an active area of
  $31\times 31\,\cm^2$, each measuring two coordinates.  
  The 11 GEM stations cover the region from the 
  downstream side of SM1 to the far end of the COMPASS setup. Both 
  Micromegas and GEM detectors have central dead zones with $5\,\cm$ 
  diameter. 
\item Large Area Trackers (LAT) - At large angles the trackers provide
  good spatial resolution and cover the large areas defined by the
  experimental setup acceptance. In the LAS, particles emerging at
  large 
  angles are 
  tracked by three 
  Drift Chambers (DC), one located   
  upstream of SM1 and two immediately downstream of it. All DC have an active
  area of $1.8\times 1.3\,\m^2$  
  with a central dead zone of $30\,\cm$ diameter. They are followed by
  three   
  stations of straw drift tubes, two upstream and one downstream of
  the RICH counter. Each straw station consists of two planes of size 
  $323\times 280\,\cm^2$ and one plane of size $325\times
  243\,\cm^2$, all of which have   
  a central dead zone of 
  $20\times 20\,\cm^2$. 
  From downstream of the RICH counter to the far 
  end of the setup the particles scattered at relatively small angles 
  are detected by 14 multi-wire proportional chamber (MWPC) stations 
  with active areas of $1.8\times 0.9-1.2\,\m^2$ and the  
  diameters of their 
  insensitive central zones increasing along the beam line 
  from $16$ to $22\,\cm$. 
  The outer region downstream of SM2 is covered by 
  two additional straw 
  stations of the same sizes as above, and by  
  six large area drift chambers of $5.0\times 2.5\,\m^2$ active surface and  
  $50\,\cm$ or $100\,\cm$ diameter central dead zone. 
\end{itemize} 
 
\subsection{Muon filters} 
\label{sec:layout.muon_filters} 
Identification of the scattered muons is performed by two dedicated 
muon filters. The design principle of a muon filter includes an 
absorber layer, preceded and followed by tracker stations (Muon Walls) with 
moderate space resolution. The absorber is thick enough to stop 
incoming hadrons.  Muons are positively identified when a track can be 
reconstructed in both sets of trackers placed upstream and 
downstream of the absorber. 
 
The first Muon Wall (MW1) is located at the downstream end of LAS, in front 
of SM2. It consists of 
two stations of squared drift tubes, each with an 
active area of $4.8\times 4.1\,\m^2$ and a central hole of 
$1.4\times 0.9\,\m^2$. An iron wall, $60\,\cm$ thick, is placed between the 
two stations. 
 
The second Muon Wall (MW2) is installed at the very end of the SAS. The 
absorber is a $2.4\,\m$ thick concrete wall. The portion of the trajectory 
upstream of the concrete wall is reconstructed by the SAS trackers, 
while downstream of it there are two dedicated stations of 
steel drift tubes with an active surface of $4.5\times 2.0\,\m^2$ each. 
 
\subsection{Setup for muon beam programme} 
\label{sec:layout.muon_setup} 
While the large majority of the spectrometer components were designed 
to match the needs of the entire COMPASS physics programme, some 
elements are specific to the measurements with muon beam, as shown in
Fig.~\ref{fig:layout.setup}.  
 
\begin{figure}[tbp] 
  \begin{center} 
    \includegraphics[width=\columnwidth]{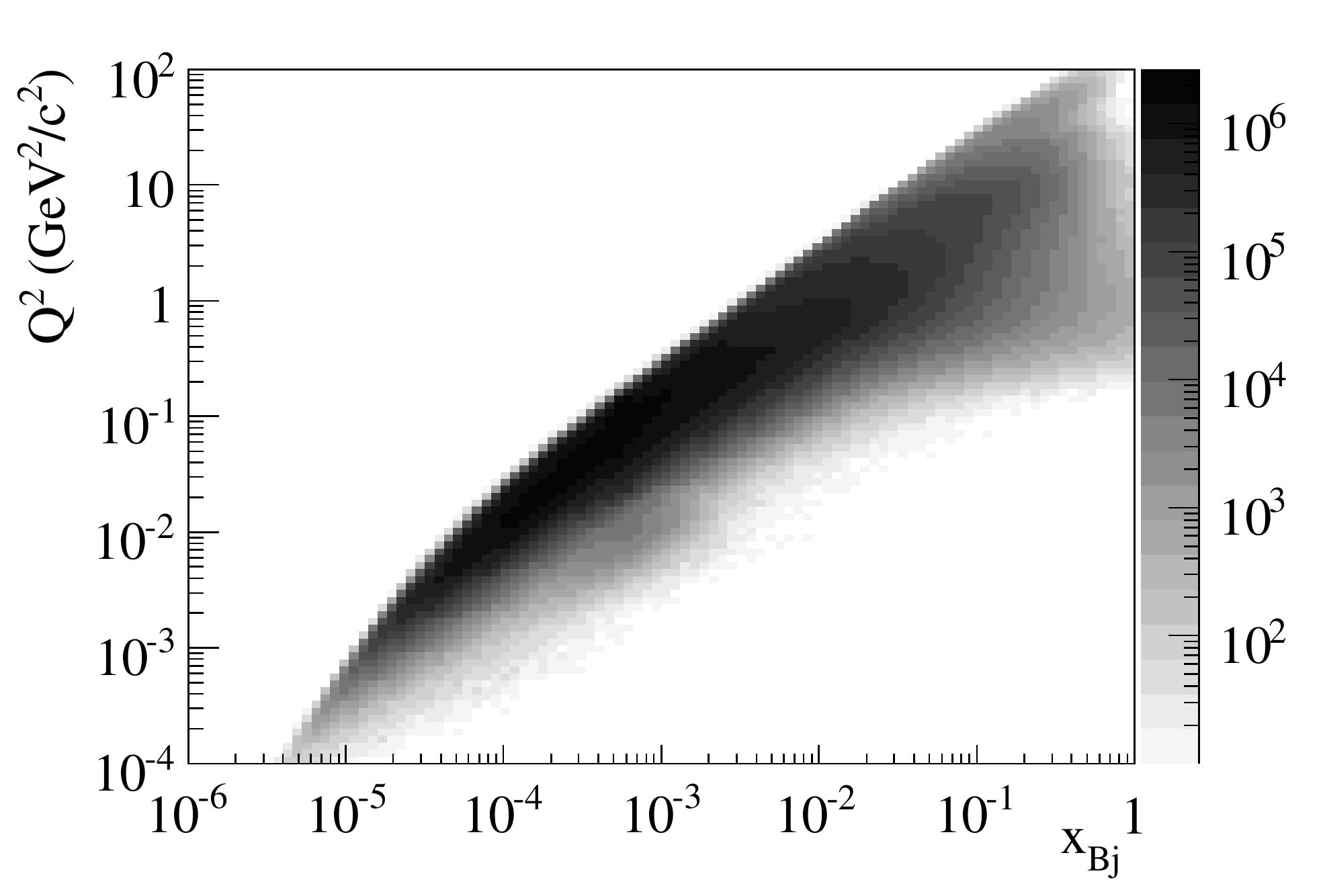} 
  \end{center} 
  \caption{\small $x_\mathrm{Bj}-Q^{2}$ region covered by the COMPASS spectrometer
for the
    $160\,\GeV/c$ muon beam.} 
  \label{fig:layout.kine} 
\end{figure}

For the measurements with the muon beam the kinematic phase space covered  
by the spectrometer described above is expressed in terms of 
$Q^2$ and $x_\mathrm{Bj}$. Taking into account the geometrical 
acceptance of the setup,
the kinematics domain covered by COMPASS for incident energies of 
$160\,\GeV$ extends to values of $Q^2$ up to
$100\,(\GeV/c)^2$ and to values of  
$x_\mathrm{Bj}$ down to $\EE*{-5}$ as shown in Fig.~\ref{fig:layout.kine}. 
 
Specific to the measurements with muon beam is the solid state 
polarised target. The target  
material is contained in two oppositely polarised target cells. The 
two cells are $60\,\cm$ long with $3\,\cm$ diameter, separated by a $10\,\cm$ 
interval. A highly homogeneous magnetic field is required 
to establish and preserve the target polarisation.
In the years 2002-2004, the superconducting solenoid 
magnet previously used by the Spin Muon Collaboration (SMC) was in operation.
This magnet was originally designed for inclusive deep 
inelastic scattering experiments only; its angular aperture 
of $\pm 70\,\mrad$ does not cover  
the whole phase space as required by the 
COMPASS physics programme. In 2006 it was replaced by a new, 
dedicated solenoid with an angular aperture of $\pm$180~mrad. 
 
The trigger system is designed to provide a minimum bias selection of 
inelastic scattering events, namely to trigger on scattered muons. 
This is obtained by correlating the information of two stations of 
trigger hodoscopes formed by fast scintillator counters. In order to 
cope with the muon counting rates, strongly depending on the distance 
from the beam axis, the trigger system is formed by four subsystems, 
which make use of hodoscopes of different size and granularity.  In 
addition to these hodoscopes, the hadron calorimeters are also used in 
order to select events with a minimum hadron energy in the final 
state. This criterion is particularly important to trigger on events 
with muons scattered at very small angles. Finally the trigger signal 
is completed by the beam veto counters. 
A stand-alone calorimetric trigger is added to cover the high $Q^2$
range
where the scattered muon does not reach the trigger hodoscopes.

\subsection{Setup for measurements with hadron beam in 2004} 
\label{sec:layout.hadron_setup} 
A first measurement with a pion beam was performed in the last weeks 
of the 2004 running period. The COMPASS setup was then used for a 
measurement of the pion electric and magnetic polarisabilities via 
Primakoff scattering.  In this measurement the incident pion beam is 
scattered off a thin solid target and the scattered pion is detected 
in coincidence with an outgoing photon. 
 
Several modifications were applied in order to adapt the setup to 
this hadron beam measurement. The large polarised target system was removed and replaced by a solid target holder, surrounded by a barrel 
shaped detector designed to measure low energy target fragments. This detector  
consists of scintillating counters inside an electromagnetic 
calorimeter.  Two sandwiches of scintillating counters and lead foils 
were used to veto on photons and charged particles emitted at large 
angles. Two silicon microstrip telescopes were installed, with two and 
three stations upstream and downstream of the target, respectively, 
providing high angular resolution. The multiplicity information 
from the second telescope was used online at the event filter level. 
Scattering off materials along the beam path was minimised by removing 
the BMS and one of the scintillating fibre stations upstream of the target. 
For the same reason, three out of the six scintillating fibre 
stations downstream of the target were also removed. 
The size of the electromagnetic calorimeter central hole was 
reduced in order to fit the hadron beam size. Finally, the trigger on
scattered pions was provided by  
a dedicated scintillator hodoscope.
% 

%%% Local Variables: 
%%% mode: latex
%%% TeX-master: t
%%% End: 

% Beam
\section{Beam line}
\label{sec:beam}
The CERN SPS beam line M2 can be tuned for either
high-intensity positive muon beams up to $190\,\GeV/c$ or
high-intensity hadron (mainly proton or mainly pion, positive or
negative) beams up to $280\,\GeV/c$. Negative muon beams are also
available, although with lower intensities. On request a low-energy,
low-intensity tertiary electron beam can be used for test and
calibration purposes. The changes between the various beam modes are
fast and fully controlled from a computer terminal.

\subsection{The guiding principles and optics of the muon beam} 
\label{sec:beam.muon.optics}
The muon beam is
derived from a very intense primary proton beam, extracted from the
CERN SPS at $400\,\GeV/c$ momentum, that impinges on a Beryllium target
with $500\,\mm$ thickness (T6).  Thinner targets can be selected for
lower flux,  
if required. The nominal proton flux available for COMPASS is
$1.2\EE{13}$ protons during $4.8\,\s$ long spills, within a $16.8\,\s$
long SPS 
cycle. A section of six acceptance quadrupoles and a set of three dipoles
selects a high pion flux within a momentum band of up to $\pm 10\%$
around a nominal momentum up to $225\,\GeV/c$ and within a geometrical 
acceptance of about $3\pi\,\mu\sr$. 
At the production target the pion flux has a kaon contamination of about $3.6\%$. 
The pions are
transported along a $600\,\m$ long channel, consisting
of regularly spaced alternately focusing and defocusing (FODO) quadrupoles
with a phase advance of $60^{\circ}$ per cell. Along this channel a
fraction of the pions decay into a muon and a neutrino. Both pions and
a large fraction of the muons produced in the decays are transported
until the muons are focused on and the hadrons are stopped in a hadron
absorber made of 9 motorised modules of Beryllium, $1.1\,\m$ long
each. 

The hadron absorber is located inside the aperture of a series of 3
dipole magnets, providing an upward deflection of $4.8\,\mrad$ each.
These dipoles are followed by a fourth magnet, providing an additional
deflection of $9.6\,\mrad$, resulting in a total deflection of
$24\,\mrad$ for a good momentum separation. The dipole section is
followed by a series of acceptance quadrupoles for the muons. The
accepted muon beam is subsequently cleaned and momentum selected by
two horizontal and three vertical magnetic collimators. All the five
collimators are toroids whose gap can be adjusted to match the profile
of the useful beam. The muons are transported to the surface level by
a second $250\,\m$ long FODO channel. Finally the muons are bent back
onto a horizontal axis by three 5 metres long dipole magnets,
surrounded by 4 hodoscopes and 2 scintillating fibres planes for
momentum measurement, and focused onto the polarised target. The
nominal momentum of the muon section of the beam is lower than the one
of the hadron section, with a maximum of $190\,\GeV/c$ with a momentum spread
usually between $\pm 3\%$ and $\pm 5\%$ RMS. Typically the muon
momentum is chosen to be around $90\%-94\%$ of the central hadron
momentum in order to provide the best compromise between muon flux and
polarisation. The final section of the beam comprises several
additional bending and quadrupole magnets that fine-steer the beam on
the target and, during transverse polarisation data taking, compensate
for the horizontal deflection induced by the $0.5\,\T$ transverse
dipole field of the polarised target.

\subsection{Muon beam parameters and performance}
\label{sec:beam.muon.parameters}

The principles of the muon beam, as optimised for the experiments
prior to COMPASS are described in more details in
Ref.\cite{Doble:94}. In order to meet the COMPASS requirements for a
high intensity muon beam, the proton intensity on the Beryllium target
was increased by about a factor of 2.5. Further increase was obtained
by re-aligning the beam section after the production target and by
retuning the openings of several collimators and scrapers. In addition,
the SPS flat top (extraction time) was increased from $2\,\s$ to
$4.8\,\s$, at the expense of a slight decrease of the maximum proton
energy. Due to these modifications, the overall beam intensity
(muons/spill) was increased by a factor of 5, and the beam duty cycle
improved by more than a factor of 2.  
 
\begin{table*}[tbp]
  \centering
  \caption{Parameters and performance of the $160\,\GeV/c$ muon beam.}
  \begin{tabular*}{\textwidth}{@{\extracolsep{\fill}}lc} \hline
    {\bf Beam parameters} & {\bf Measured} \\ \hline
    Beam momentum ($p_\mu$)/($p_\pi$) & ($160\,\GeV/c$)/($172\,\GeV/c$) \\ 
    Proton flux on T6 per SPS cycle & $1.2\EE{13}$ \\ 
    Focussed muon flux per SPS cycle & $2\EE{8}$ \\ 
    Beam polarisation & $(-80\pm 4)\%$ \\ 
    Spot size at COMPASS target ($\sigma_x \times \sigma_y$) &
    $8\times 8\,\mm^2$ \\ 
    Divergence at COMPASS target ($\sigma_x \times \sigma_y$) &
    $0.4\times 0.8\,\mrad$ \\
    Muon halo within $15\,\cm$ from beam axis & $16\%$ \\ 
    Halo in experiment ($3.2\times 2.5\,\m^2$) at
    $\left|x,y\right|>15\,\cm$ & $7\%$ \\ \hline
  \end{tabular*}
  \label{tab:beam.muon.parameters}
\end{table*}

The nominal parameters of the positive muon beam are listed in
Table~\ref{tab:beam.muon.parameters}. The 
muon momentum can be chosen between $60$ and $190\,\GeV/c$. The maximum 
authorised muon flux is $2\EE{8}$ muons per SPS cycle, the limitation
being imposed by 
radio-protection guidelines. This flux can be obtained at
the nominal COMPASS setting of $160\,\GeV/c$  
and below, but is out of reach at higher momenta and for negative
muons. 

When arriving in the experimental hall, the muon beam is accompanied
by a large halo, primarily composed of muons that could not be
significantly deflected or absorbed. The muon halo is defined as the
number of incident particles measured outside the area crossed by the
nominal muon beam. The outer part of the halo is  measured in the
first large veto counter with a surface of $2.50\times 3.20\,\m^{2}$
and a $30\times 30\,\cm^{2}$ hole in the middle. It amounts to about
$7\,\%$ of the nominal muon beam. The inner part of the halo, which
also includes the tails of the beam distribution, is detected by the
inner veto counters whose dimensions are $30\times 30\,\cm^{2}$ with a
hole of $4\,\cm$ diameter; it represents about $16\,\%$ of the muon
beam.  

\begin{figure}[tbp]
  \centering
  \includegraphics[width=\columnwidth]{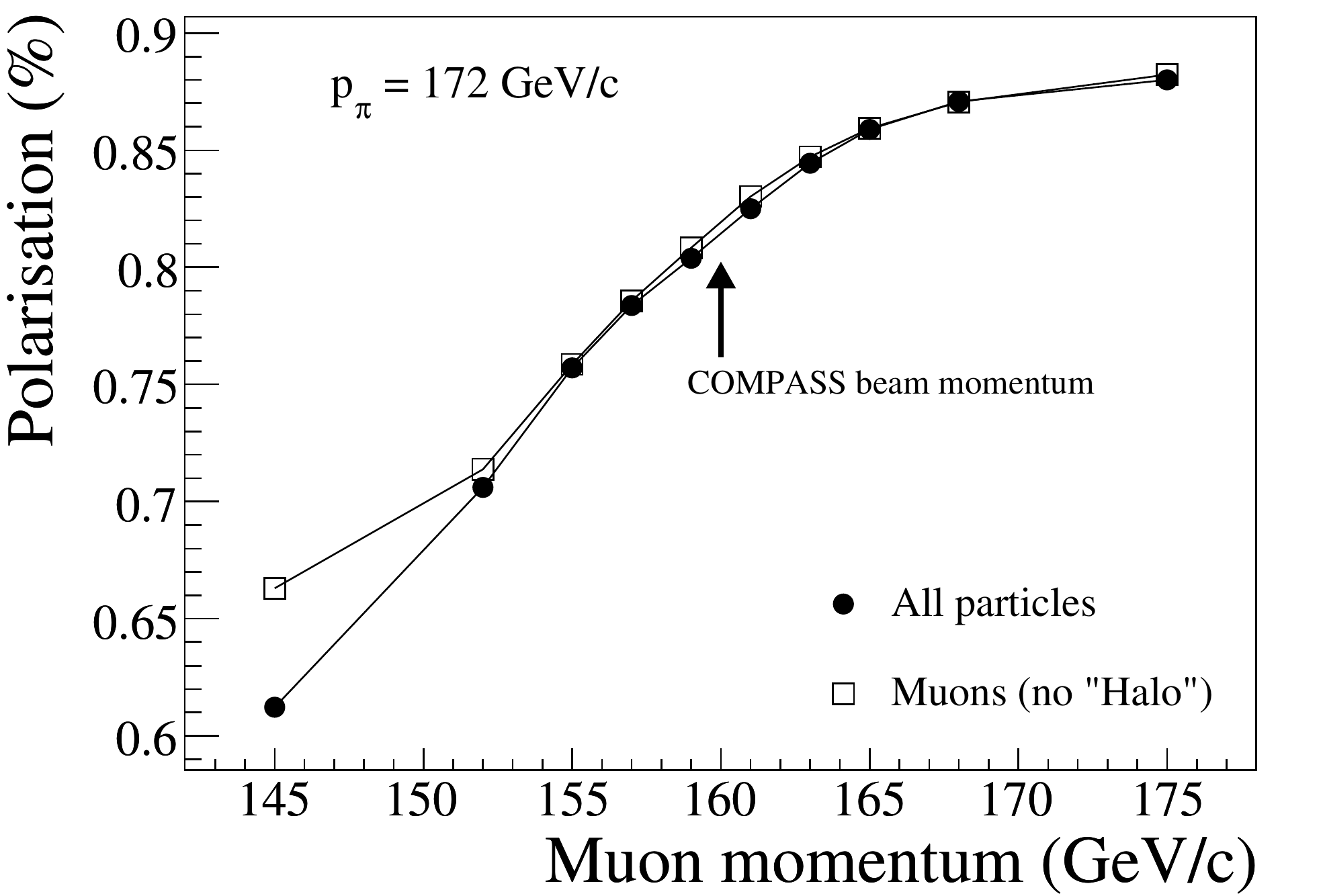}
  \caption{The muon beam polarisation (absolute value) as a function
    of the central muon 
    momentum, assuming a central hadron momentum of $172\,\GeV/c$.} 
  \label{fig:beam.muon.polarisation}
\end{figure}

Due to the parity violating nature of the pion decay, the COMPASS muon
beam is naturally polarised. The average beam polarisation results
from the integration of all individual muon helicities over the phase
space defined by the beam optics. It strongly depends on the ratio
between muon and pion momenta. This is illustrated in
Fig.~\ref{fig:beam.muon.polarisation}, where the muon polarisation is
shown as a function of the muon momentum, assuming a fixed pion momentum
of $172\,\GeV/c$. The final muon polarisation value of $(-80\pm 4)\%$ 
in the 2004 run also includes a tiny correction due to the kaon 
component of the pion beam. 

\begin{figure}[tbp]
  \centering
    \includegraphics[width=\columnwidth]{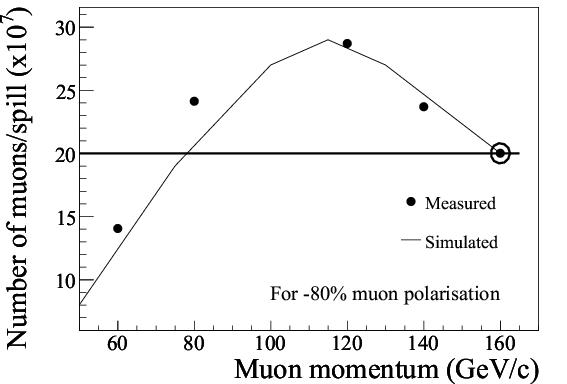}
    \caption{The maximum muon flux per SPS cycle as a function of the muon
    momentum, assuming a $p_\mu/p_\pi$ ratio corresponding to $-80\%$
    positive muon polarisation. The points are measurements at various beam energies. 
The solid curve is a result from a simulation of the beam optics.} 
  \label{fig:beam.muon.flux}
\end{figure}

The statistical factor of merit of the COMPASS experiment is
proportional to the beam intensity and to the square of the muon
polarisation. The factor of merit is optimised for a muon polarisation
of $-80\%$; the maximum allowed flux of $2\EE{8}$ muons per SPS cycle
is then achieved for all momenta between $80$ and $160\,\GeV/c$. This
is visible in Fig.~\ref{fig:beam.muon.flux} where the measured
intensities are compared to a prediction from the beam simulation
software. Higher polarisation values could also be reached, but at the
expense of less intense muon fluxes. For standard COMPASS data taking,
a beam momentum of $160\,\GeV/c$ is selected.  

\subsection{Muon beam momentum measurement}
\label{sec:beam.muon.bms}

In order to make maximum use of the incident flux, the momentum spread
of the beam as defined by the beam optics is large and can reach
$5\%$. An accurate determination of the  kinematical parameters
therefore requires a measurement of the momentum of each individual
muon. This is done by the Beam Momentum Station (BMS).  

\begin{figure*}[tbp]
  \centering
  \includegraphics[width=\textwidth]{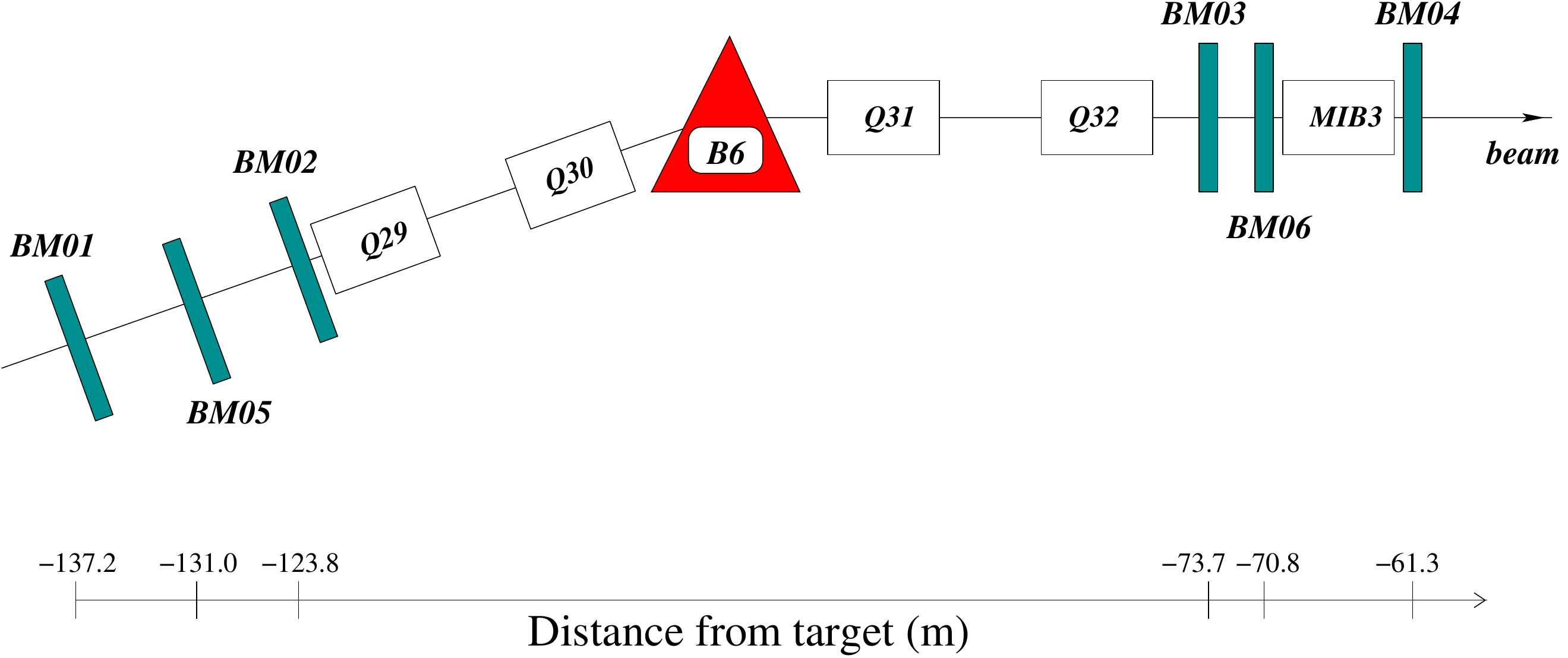}
  \caption{Layout of the Beam Momentum Station for the COMPASS muon beam.}
  \label{fig:beam.muon.bms}
\end{figure*}

Fig.~\ref{fig:beam.muon.bms} shows the detectors composing the BMS.
Three consecutive dipole magnets (B6) compose the last large vertical
bend ($30\,\mrad$) that brings the muon beam close to the horizontal
direction before entering the experimental hall.  The B6 dipoles are
surrounded by a system of four quadrupoles and six beam detectors. Four of these
(BM01--BM04)) are scintillator hodoscopes with horizontal scintillator
strips already used in previous experiments \cite{Allkofer:81}.  Each
hodoscope is made up of 64 elements, $5\,\mm$ high in the dispersive
plane, with an overlap of a few tenths of a $\mm$ in order to avoid
efficiency losses.  A thickness of $20\,\mm$ along the beam ensures a
large output signal.  In the central regions, the scintillator strips
are horizontally divided into several elements such that the particle
flux per element does not exceed $1\EE{7}\,\s^{-1}$, even for the
highest beam intensity.  Since the beam cross section varies from one
plane to another, the element length varies from $10\,\mm$ to
$60\,\mm$.  The readout is done using fast photomultiplier
tubes (PMT).  The time resolution achieved is $0.3\,\ns$.

In order to cope with the high beam current and multiple-hit
environment of the COMPASS experiment, two scintillating fibre
hodoscopes (BM05, BM06) were added, one in between each of the
existing hodoscope pairs. These two planes provide additional
redundancy in the track matching between the beam momentum station
and the detectors located in front of the target, thus increasing the
overall beam detection efficiency.

The design of these hodoscopes is similar to the scintillating fibre
detectors used in the spectrometer (see
Sec.~\ref{sec:tracking.vsat.scifi}). Each plane has a size of
$12\,\cm$ horizontally and $16\,\cm$ vertically and is made of stacks
of $2\,\mm$ round scintillating fibres. Each stack has 4 fibres along
the beam direction.  The overlap of two adjacent stacks is chosen such
that if the traversed path in a single fibre falls below $1.4\,\mm$,
the path in the adjacent fibre exceeds $1.4\,\mm$. The minimum path
through the scintillator material registered in a single channel is
therefore $5.6\,\mm$.  The BM05 plane consists of 64 channels of two
adjacent stacks each.  The BM06 plane has 128 channels all made out of
single stacks, in order to achieve the desired resolution in the
dispersive plane. The design was chosen, such that the maximum rate
per channel does not exceed $3\EE{6}\,\s^{-1}$.

Simulated beam tracks have been used to parameterize the momentum
dependence of the track coordinates in these six detectors. This
parameterization is used to calculate the momentum of each muon track
to a precision of $\leq 1\%$. The reconstruction efficiency is
$\approx93 \%$.

During event reconstruction the efficiency and the purity of the beam
momentum station are further improved by using the information
obtained by the tracking detectors located in front of the target. The
incident tracks corresponding to the good events are reconstructed and
back--propagated from the target region to the beam momentum station.
%%%
The spatial correlation between the extrapolated track and the actual BMS
hits is used to select among ambiguous beam candidates. If
there are not enough hits in the BMS to reconstruct the momentum, a rescue
algorithm is used. This algorithm relies on the determination of track
angles at the first stage of BMS and at the target region. Track parts
are then combined using time correlation.
%The spatial correlation between extrapolated tracks the track angles 
%at the second stage of
%the beam momentum station and the zone in front of the target 
%is then independently determined 
%for both the vertical and the horizontal angles.  
 
\subsection{Beam optics for the hadron beam}
\label{sec:beam.hadron.optics}
A high-intensity secondary hadron beam is obtained by moving the nine
motorised hadron absorber modules out of the beam and loading settings
corresponding to a single momentum all along the beam line. Up to
$225\,\GeV/c$, the front end of the beam line is operated with the same
optics as for the muon mode of the beam. At higher momenta a different
optics is used in the acceptance quadrupoles, giving access to $280\,\GeV/c$. The beam optics is optimised for
momentum resolution. The beam is composed of a (momentum dependent)
mixture of pions, protons and kaons. For the tagging of individual
beam particles, a pair of differential Cherenkov counters
\cite{Bovet:80} (CEDAR) is foreseen in the final section of the beam.
The beam optics is optimised to provide a wide and parallel beam as
required for the CEDAR counters, while delivering a relatively small
beam spot at the COMPASS target in the experimental hall.

\begin{table*}[tbp]
  \centering
  \caption{Parameters and performance of the $190\,\GeV/c$ negative
    hadron beam.} 
  \begin{tabular*}{\textwidth}{@{\extracolsep{\fill}}lc} \hline
%%  \begin{tabular}{|l|c|} \hline
    {\bf Beam parameters} & {\bf Measured} \\ \hline
    Beam momentum & $190\,\GeV/c$ \\ 
    Hadron flux at COMPASS per SPS cycle & $\leq \EE*{8}$ \\ 
    Proportion of negative pions & $95 \%$ \\
    Proportion of negative kaons & $4.5 \%$ \\
    Other components (mainly antiprotons)& $0.5 \%$ \\ 
    Typical spot size at COMPASS target ($\sigma_x \times \sigma_y$) &
    $3\times 3\,\mm^2$ \\ \hline
  \end{tabular*}
    \label{tab:beam.hadron.parameters}
\end{table*}

The parameters of the negative hadron beam for a momentum of
$190\,\GeV/c$ are listed in
Table~\ref{tab:beam.hadron.parameters}. For positive beams the proportions of the various
particles change: at $190\,\GeV/c$ the positive beam consists of
$71.5\%$ protons, $25.5\%$ pions and $3.0\%$ kaons. The maximum
allowed hadron flux is $\EE*{8}$ particles per SPS cycle, limited by
radiation safety rules assuming less than $20\%$ interaction length
material along the beam path.
 
\subsection{Electron beam}
\label{sec:beam.electron}
On request a $40\,\GeV/c$ tertiary electron beam can be provided by
selecting a $100\,\GeV/c$ negative secondary beam, which impinges on a
$5\,\mm$ thick lead converter, located about $50\,\m$ upstream of the
hadron absorbers, which are moved out of the beam for this purpose. The
downstream part of the beam line is set to $40\,\GeV/c$ negative particles, 
so that only
the electrons that have lost $60\,\GeV$ due to Bremsstrahlung in the
converter 
are transported to the experiment. The
electron flux is typically small, of a few thousands per SPS cycle. In
COMPASS the electron beam is used for an absolute calibration of the
electromagnetic calorimeters.

%%% Local Variables: 
%%% mode: latex
%%% TeX-master: "compass_spec"
%%% End: 

% Targets
\section{Targets}
\label{sec:targets}

\subsection{Polarised target}
\label{sec:targets.pt}

The COMPASS muon programme aims to measure cross section asymmetries
$\Delta \sigma /(2 {\overline \sigma})$ where $\Delta \sigma$ is the
difference between the cross sections of a given process for two
different spin configurations and ${\overline \sigma}$ the spin
averaged cross section.  The corresponding observable counting rate
asymmetry is $A_{\mathrm{obs}} = ( P_{\mu} P_\mathrm{T} f) (\Delta \sigma / 2
{\overline \sigma})$, where $P_{\mu}$ and $P_\mathrm{T}$ are the beam and
target polarisations, respectively, and $f$ the fraction of polarisable
material inside the target. The use of a polarised target is thus
mandatory and, in addition, the factors $P_\mathrm{T}$ and $f$ must be
made as 
large as possible in order to optimise the statistical significance of
the results. Furthermore, due to the limited muon flux,
% ($1.2\EE{7}\,\s^{-1}$), a target thickness of about $60\,\g/\cm^2$ is
% needed to reach the luminosity of a high precision experiment
%($\approx 4\EE{32}\,\cm^{-2}\s^{-1}$).
a solid state polarised target, much 
thicker than those commonly used in electron beams, is required.

While electron spins can be aligned in a magnetic field and give rise
to a large polarisation at equilibrium for a low enough temperature,
only a negligible nuclear spin polarisation can be reached.
Therefore, solid state polarised targets rely on dynamic nuclear
polarisation (DNP) which transfers the electron polarisation to the
nuclear spins by means of a microwave field \cite{Abragam:89a}.  This
process requires a material containing some amount of paramagnetic
centres, e.g.\ created by irradiation, a temperature below $1\,\K$
and a strong and homogeneous 
magnetic 
field.

Deuterated lithium ($^6$LiD) has been chosen as isoscalar target.
This material  allows to reach a high degree of deuteron
polarisation ($>40\%$) and has a very favourable composition
\cite{Ball:04a,Goertz:95a,Bultmann:99a}. Indeed, since $^6$Li can be
considered to a good approximation as a spin-0 $^4$He nucleus and a
deuteron, the fraction of polarisable material $f$ is of the order of
0.35, taking into account also the He content in the target region. 
The irradiated ammonia (NH$_3$), which will be used as polarised
proton target, has a less favourable composition ($f \approx 0.15$) but
can be polarised to a higher degree ($>80\%$).  Spin asymmetries are
measured using a target divided in two cells, which are exposed to the
same beam flux but polarised in opposite directions. In order to
cancel acceptance effects which could mask the physics asymmetries,
the spin directions must be frequently inverted by rotating the
solenoid field.  During this process, the polarisation must be
maintained by a transverse field which is also needed for data taking
in so-called ``transverse mode'', i.e. with orthogonal directions of the
beam and target polarisations.
In addition, the sign of the polarisation in
the target cells is inverted two or three times per year by rebuilding the
polarisations with opposite microwave frequencies.
% For review on polarisation using paramagnetic centres and on spin
% temperature theory see Refs. \cite{Goldman:70a,Goertz:04a}.  The
% Spin Muon Collaboration (SMC) has published many papers on the
% polarised target, see citations in Ref. \cite{Adams:99a}.  For
% recent review in this field the proceedings of Polarised Solid
% Targets and Techniques are a good reference \cite{PSTT:03} or review
% papers Refs. \cite{Goertz:02a,Kisselev:00a,Crabb:97a}.  For low
% temperature techniques there are good text books
% \cite{Lounasmaa:74a,Pobell:96a}.  There is a historical review on
% $^6$LiD target material in Ref. \cite{Ball:04a}. The work done by
% Bochum group to prepare a large amount of the irradiated $^6$LiD
% target material is described in Refs.
% \cite{Goertz:95a,Bultmann:99a,Meier:01a}.  The $^6$LiD has large
% fraction of polarisable nucleons leading to a figure of merit about
% two times higher than that of the deuterated butanol used in the SMC
% target \cite{Meier:01a,Takabayashi:02a,Doshita:05a}.
%

\begin{figure*}[tbp]
  \begin{center}
    \includegraphics[width=\textwidth]{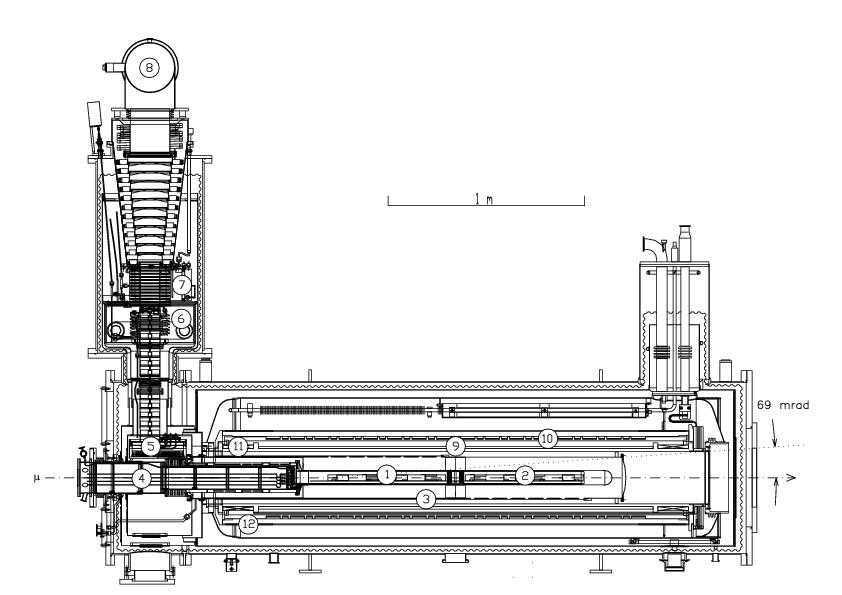}
  \end{center}
  \caption{\small Side view of the COMPASS polarised target: (1)
    upstream target cell and (2) downstream target cell inside mixing
    chamber, (3) microwave cavity, (4) target holder, (5) still ($^3$He
    evaporator), (6) $^4$He evaporator, (7) $^4$He liquid/gas phase
    separator, (8) $^3$He pumping port, (9) solenoid coil, (10)
    correction coils, (11) end compensation coil, (12) dipole coil.
    The muon beam enters from the left. The two halves of the
    microwave cavity are separated by a thin microwave stopper.  }
  \label{fig:targets.pt.smcDR}
\end{figure*}

The COMPASS polarised target (see Fig.~\ref{fig:targets.pt.smcDR}) has
been designed to meet these requirements. It incorporates several
elements previously used by the SMC experiment \cite{Adams:99a}.
% Up to 2004, the SMC magnet has been used, limiting the acceptance to 
% $\pm\ 70$ mrad. 
% The large aperture solenoid designed to match the full
% COMPASS spectrometer acceptance ($\pm$ 180 mrad) has been put in operation
% in 2006 (see section 11.1).

The superconducting solenoid (see
Fig.~\ref{fig:targets.pt.smcDR},(9)) produces a $2.5\,\T$ magnetic field
along the beam direction. Sixteen corrections coils
(Fig.~\ref{fig:targets.pt.smcDR},(10)) are
used to obtain an axial homogeneity better than
$20\,\mathrm{ppm}$  
in a volume $1500\,\mm$ long, and $50\,\mm$ in diameter \cite{Dael:92a}.
The transverse holding field of $0.42\,\T$ 
is produced by a dipole coil (see Fig.~\ref{fig:targets.pt.smcDR},(12))
and deviates at most by 10\% from its nominal value inside the
target volume.

The $^3$He/$^4$He dilution refrigerator is filled with liquid helium from the
gas/liquid phase separator (see Fig.~\ref{fig:targets.pt.smcDR},(7)).
The cold gas from the separator cools down the outer and inner
vertical and horizontal thermal screens around the dilution refrigerator 
at nominal temperatures of $80\,\K$ and $4\,\K$, respectively.  The incoming
$^3$He gas is also cooled with cold gas from the separator. 
Needle valve controlled lines are used to
fill the $^4$He evaporator (see Fig.~\ref{fig:targets.pt.smcDR},(6))
with liquid helium and to cool the microwave cavity (see
Fig.~\ref{fig:targets.pt.smcDR},(3)).  The nominal operation
temperatures of the cavity and the $^4$He evaporator are $3\,\K$ and
$1.5\,\K$, respectively.

A microwave cavity (Fig.~\ref{fig:targets.pt.smcDR},(3)) similar to the one previously used by SMC \cite{Adams:99a} was built. The amount of unpolarised material along the beam was minimised by reducing the thickness of the microwave stopper, and by modifying the downstream end window \cite{Ball:03a}. The $60\,\cm$ long target cells
(see Fig.~\ref{fig:targets.pt.smcDR},(1),(2)) have a diameter of $3\,\cm$ and are 
separated by $10\,\cm$. The cells are
made of a polyamide mesh in order to improve the heat exchange between the crystals
and the liquid helium. They are fixed in the centre of an aramid fibre epoxy tube, which itself
is fixed to the target holder isolation vacuum tube (see Fig. \ref{fig:targets.pt.smcDR},(4)).
The target cells are filled with $^6$LiD crystals of $2-4\,\mm$ size 
\cite{Meier:01a}; the volume between the target material crystals
is filled with a mixture of liquid $^3$He/$^4$He.
The $^6$LiD mass in each target cell is $170-180\,\g$
\cite{Neliba:04a}, and depends on the packing factor (between 0.49 and 0.54) achieved 
during the filling. 
% added sentence FH
The isotopic
dilution of deuterons with 0.5\% of protons and $^6$Li with 4.2\% of
$^7$Li was determined by NMR measurements from the polarised target
material \cite{Kondo:04a}.
Each cell contains five NMR coils used for the
local monitoring of the polarisation.

\begin{figure*}[tbp]
  \begin{center}
    \includegraphics[width=\textwidth]{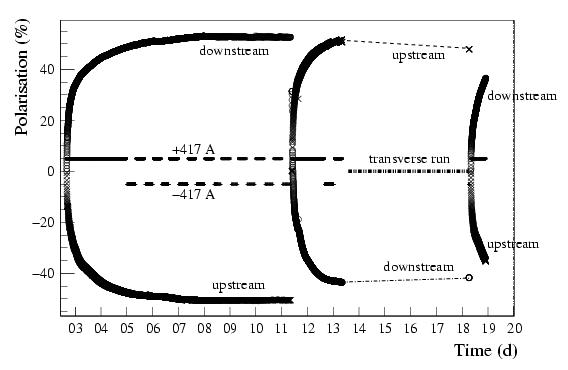}
  \end{center}
  \caption{\small Typical average polarisations in the upstream and
    downstream target cells during 20 days of the 2004 run.  After day
    11, the polarisations in the target cells are reversed by changing
    the microwave frequencies.  Data are taken in transverse mode from
    day 13 to day 18 and a new field reversal by microwaves is performed at
    the end of the period. The current of $\pm 417\,\A$ corresponds to
    an axial field  
    of $2.5\,\T$.  }
  \label{fig:targets.pt.polaris}
\end{figure*}

The target material is polarised via dynamic nuclear polarisation,
obtained by irradiating the paramagnetic centres with
microwaves at frequencies of $70.2-70.3\,\GHz$ at 2.5\,T and at temperature
of about $200\,\mK$.
The microwave radiation is generated with two extended
interaction oscillator tubes (EIO) \cite{Adams:99a}. The density of the
paramagnetic centres is of the order of 10$^{-4}-$10$^{-3}$ per
nucleus \cite{Goertz:95a,Meier:01a}.  An additional modulation of the
microwave frequency of about $5\,\MHz$ \cite{Ball:03a,Gautheron:04b} 
helps enhancing the polarisation.  
% The polarisation process produces
% heat which warms up the mixing chamber to $200-300\,\mK$ \cite{Doshita:04a}.
A deuteron
polarisation $|P| > 40$\% is reached within 24 hours in a $2.5\,\T$ field
with a $^3$He flow of $80-120\,\mathrm{mmol}/\s$ in the dilution
refrigerator.  The 
maximum polarisation difference between the upstream and downstream
cells $|P_{\mathrm{up}} - P_{\mathrm{down}}| > 100$\% is reached in five days
\cite{Koivuniemi:04b}, see Fig.~\ref{fig:targets.pt.polaris}.

At least once during a data taking period, 
the mixing chamber is filled with only $^4$He to perform the thermal
 equilibrium (TE) calibration of the polarisation
\cite{Kondo:04a} in a $2.5\,\T$ solenoid field at a
constant temperature in the range $1.0-1.6\,\K$ \cite{Kondo:04a}.
The deuteron in the $^6$LiD material has a single
$16.38\,\MHz$ NMR line about $3\,\kHz$ wide for a $2.5\,\T$ field \cite{Koivuniemi:04a}.
The spin magnetisation
of $^6$LiD reaches good thermal equilibrium at this temperature in
about $15\,\Hour$.  The polarisation of the target
material is calculated from the helium temperature measured by the 
$^3$He vapour pressure \cite{Lounasmaa:74a,Pobell:96a}. The intensity
of the measured TE NMR signal is used to calibrate the polarisation
measured during the dynamic nuclear polarisation process, when the spin system is not
anymore in thermal equilibrium with the helium.

During data taking in transverse mode, the target material is kept in frozen spin
mode below $90\,\mK$
and the spin direction is
maintained by the $0.42\,\T$ transverse dipole field. The polarisation is reversed
by exchanging the microwave frequencies of the two cells.
The polarisation is
measured in the longitudinal $2.5\,\T$ field at the end of each
transverse data taking period of about 6 days. The relaxation rate in
frozen spin mode is $(0.4-1.0)\,\%/\mathrm{d}$ in the $0.42\,\T$ 
field and $(0.05-0.10)\,\%/\mathrm{d}$ in the $2.5\,\T$ field.

% In the remaining part of this section, we present the main elements of
% the COMPASS polarised target and briefly describe some critical aspects
% of its mode of operation.

\subsection{Targets for hadron beams}
\label{sec:targets.hadron}
During the 2004 hadron run COMPASS collected data for the measurement
of the pion 
polarisabilities via Primakoff scattering and for the diffractive meson production
studies in parallel. As the two measurements require targets with
different characteristics, few solid state targets have been prepared
and exchanged during the data taking. This section briefly describes
the choice of target materials and geometries and their main physics
motivations.

The majority of the 2004 hadron data has been collected with a target
optimised for Primakoff scattering. Since this process is enhanced
over the diffractive background when targets with large atomic numbers
are used \cite{Antipov:82}, lead was chosen as material. To study
systematic effects, additional measurements of the $Z^2$-dependence of
the Primakoff cross section using Copper and Carbon were performed (see
Table~\ref{tab:sec:targets.hadron:targets}).  All targets consisted of
simple discs with a diameter of~$3\,\cm$, corresponding to more than
3~sigma of the beam width, while the thickness was determined by the
required resolution to properly separate the electromagnetic
scattering from the diffractive background. For that, the squared 
four-momentum transferred to the target nucleus, $t$, should be
measured with a precision better than $5\EE{-4}\,\GeV^2/c^2$. The
largest contribution to the uncertainty 
on $t$ comes from the multiple scattering in the target, allowing for
a maximum thickness of about 0.5 radiation lengths.

\begin{table}
  \begin{center}
    \caption{List of available target materials for the Primakoff and
      diffractive measurements. Here, $X_0$ denotes the radiation length
      and $\lambda_\mathrm{I}$ the nuclear interaction length.}
    \label{tab:sec:targets.hadron:targets}
    % \begin{tabular}{|c|c|c|c|c|} \hline 
    \begin{tabular*}{\columnwidth}{@{\extracolsep{\fill}}lccc} \hline 
      Material & Thickness $x$ & $x/X_0$ & $x/\lambda_\mathrm{I}$ \\ \hline  
      Lead & $3\,\mm$ ($2\,\mm
      +1\,\mm$ segmented)
      & 0.53 & 0.029 \\ 
      Copper & $3.5\,\mm$ & 0.24 & 0.037 \\
      Carbon & $23\,\mm$ & 0.12 & 0.086 \\ \hline
    \end{tabular*}
  \end{center}
\end{table}

The study of diffractively produced hybrid mesons requires targets
with low atomic numbers, such as liquid hydrogen or paraffin, to
minimise the multiple scattering. On the other hand, the requirement
of running the polarisability and hybrid meson programmes in parallel
excluded the use of hydrogen targets. As a compromise the Carbon
target has been also used for the diffractive scattering studies.

Both measurement imply a small energy transfer.  In order to reject
the hard scattering events in the offline analysis, the targets were
inserted into a barrel-shaped veto system, called Recoil Veto, that
measured the recoil energy of the target fragments produced in the
reaction.  The Recoil Veto consists of an inner cylindrical layer of
12 scintillator strips, with a diameter of $\sim 10\,\cm$, surrounded by
an outer layer of 96 lead glass blocks. The recoil energy is measured
from the combined information of the energy loss in the scintillator
strips and the Cherenkov light produced in the lead glass blocks.  The
target material is placed at the centre of the Recoil Veto with a
lightweight support made of foam.

%%% Local Variables: 
%%% mode: latex
%%% TeX-master: "compass_spec"
%%% End: 

% Tracking
\section{Tracking detectors}
\label{sec:tracking}
The tracking system of COMPASS comprises many tracking
stations, distributed over the entire length of the spectrometer. 
Each tracking station consists of a set of detectors of the same type,
located at approximately the same $z$-coordinate along the beam. In a
station, the  
trajectory of a charged particle is measured in several projections 
transverse to the beam direction in order to reduce ambiguities. In
the following we use the terms $X$- and $Y$-plane  
to designate the 
group of channels within a station 
measuring the
horizontal and vertical 
coordinates, respectively, of 
the particle penetration point. Similarly, the terms $U$- and $V$-plane
describe 
all 
channels measuring projections onto axes rotated clockwise and
anticlockwise, 
respectively, with respect to the $x$-axis. Note that the dipole
magnets bend the particle trajectories in the horizontal plane. 
Many
different detector technologies of varying rate capability, resolution, and
active area are in use, dictated by the increasing particle rates closer
to the beam axis, and by the spectrometer acceptance.   

Section~\ref{sec:tracking.vsat} describes the Very Small Area Trackers
(VSAT), which cover the beam region up to a radial distance of $2.5$ -
$3\,\cm$. The very high rate of beam particles 
in this area (up to about
$\EE*{5}\,\s^{-1}\mm^{-2}$ in the centre of the muon beam) requires  
excellent time or position resolution of the corresponding detectors
in order to 
identify hits belonging to the same track.  
Scintillating fibres (see Sec.~\ref{sec:tracking.vsat.scifi}) and silicon
microstrip detectors (see Sec.~\ref{sec:tracking.vsat.silicon})
fulfil this task.
 
The intermediate region at a radial
distance of $2.5\,\cm$ to $30$ - $40\,\cm$ is
covered by the Small Area Trackers (SAT, see Sec.~\ref{sec:tracking.sat}),
and is the domain of micropattern gas detectors. Here, two novel
devices -- Micromegas (see Sec.~\ref{sec:tracking.sat.micromegas}) and GEM
detectors (see Sec.~\ref{sec:tracking.sat.gem})-- are employed successfully 
for the first time in a large-scale 
particle physics experiment. These detectors combine high
rate capability (up to about $\EE*{4}\,\s^{-1}\mm^{-2}$) and good spatial
resolution (better than $100\,\mum$) with low material budget over fairly large sizes. 

The reduced flux in the outermost regions,
covered by the Large Area Tracker (LAT,
see Sec.~\ref{sec:tracking.lat}), allows the use of 
drift chambers (\ref{sec:tracking.lat.dc},
see Sec.~\ref{sec:tracking.lat.w45}), 
straw tube chambers (see Sec.~\ref{sec:tracking.lat.straw}), 
and multiwire proportional counters (see Sec.~\ref{sec:tracking.lat.mwpc}). 

\subsection{Very small area trackers}
\label{sec:tracking.vsat}
\subsubsection{Scintillating fibre detectors}
\label{sec:tracking.vsat.scifi}
The purpose of scintillating fibre (SciFi) detectors in the COMPASS
experiment is to provide tracking of incoming and scattered
beam particles as well as of all other charged reaction products in
and very near the centre of the primary beam. 

As the hit rate can reach $3\EE{6}\,\s^{-1}$ per
fibre in the centre of the muon beam, hits can be assigned to the
corresponding track by time correlation only,
whereas spatial correlation would be far too
ambiguous. Time correlation is also used to link the incoming muon
with the scattered muon track, as well as with the trigger and the
information from the beam momentum station. 

For the muon program, a total of eight SciFi detector
stations are used.  Two pairs of stations are placed upstream (no.\
1, 2) 
and 
downstream (no.\ 3, 4) of the target, two more pairs upstream (no.\ 5,
6) and
downstream (no.\ 7, 8) of the second spectrometer magnet (SM2). The
main 
parameters of the different stations are given in
Table~\ref{tab:trk.vsat.scifi.parameters}. 
\begin{table*}[tbp]
  \centering
  \caption{Parameters of SciFi stations in COMPASS. Column~3 specifies
    the number of fibre layers per projection, columns~4 and 7 give
    the size of the square active area and the
    number of channels for each projection, respectively. Column~8 lists
    the thickness of the respective station in units of radiation lengths
    ($X_0$).} 
  \label{tab:trk.vsat.scifi.parameters}
  \scriptsize
  % \begin{tabular}{|c|l|l|l|l|l|l|l|} \hline
  \begin{tabular*}{\textwidth}{@{\extracolsep{\fill}}clrllllr} \hline
    No. & Proj. & \# of & Size & Fibre \o & Pitch & \#
    of ch. & Thickness\\ 
    &       & layers & ($\cm^2$) & ($\mm$) & ($\mm$) & & ($X_0$)
    \\ \hline
    1,2 & $X,Y$ & 14 & $3.9^2,3.9^2$ & 0.5 & 0.41 & $96,96$ &
    $1.64\%$ \\ 
    3,4 & $X,Y,U$ & 14 & $5.3^2,5.3^2,5.3^2$ & 0.5 & 0.41 &
    $128,128,128$ & $2.46\%$ \\ 
    5 & $X,Y$ & 12 & $8.4^2,8.4^2$ & 0.75 & 0.52 & $160,160$ &
    $2.1\%$ \\ 
    6 & $X,Y,U$ & 8 & $10.0^2,10.0^2,12.3^2$ & 1.0 & 0.70 &
    $143,143,176$ & $2.79\%$ \\
    7 & $X,Y$ & 8 & $10.0^2,10.0^2$ & 1.0 & 0.70 & $143,143$ &
    $1.86\%$ \\ 
    8 & $X,Y$ & 8 & $12.3^2,12.3^2$ & 1.0 & 0.70 & $176,176$ &
    $1.86\%$ \\ \hline
  \end{tabular*}
  \normalsize
\end{table*}
In total the
eight SciFi stations make use more than 2500 PMT channels including about
8000 fibres. Details may be found in
\cite{Bisplinghoff:02,Horikawa:04}. 
Each station consists of at least two projections, one vertically
($Y$) and one 
horizontally ($X$) 
sensitive. Three stations (no.\ 3, 4, 6) comprise an additional inclined
($\sim 45^\circ$) 
projection ($U$).  

\begin{figure}[tbp]
  \centering
  \includegraphics[width=\columnwidth]{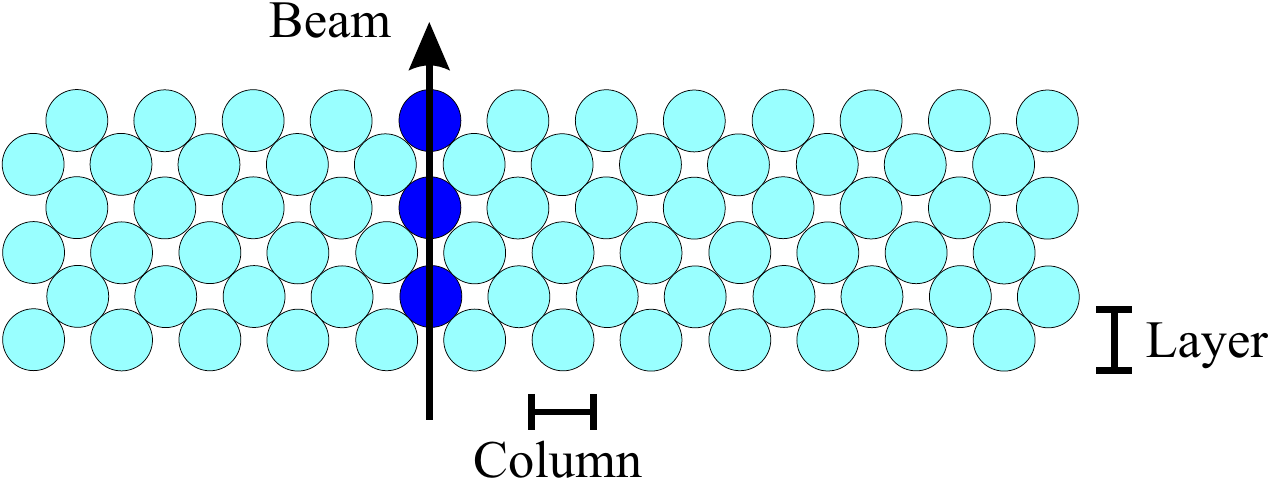}
  \caption{Fibre configuration of a SciFi plane (the actual
    number of fibre layers per plane is 8, 12 or 14, depending on the
    station).}  
  \label{fig:trk.vsat.scifi.configuration}
\end{figure}

In order to provide a sufficient amount of photoelectrons (at least
about 20 per minimum ionising particle), several layers of
fibres are stacked for each projection, the fibre axes of one layer
being shifted with respect to the ones of the next layer (see
Fig.~\ref{fig:trk.vsat.scifi.configuration}). 
The overlap of fibres
%%from two adjacent layers 
is chosen sufficiently large in order to avoid relying on the detection
of tracks with only grazing incidence of the particles. 
The light output of a group of fibres lined up in 
beam direction (labelled ``column'' in
Fig.~\ref{fig:trk.vsat.scifi.configuration}) is collected on one
photon detector channel. The number of fibres in one column is 
seven for stations~1--4, six for station~5, and four
for stations~6--8, and is chosen to achieve the required time
resolution and at the same time minimise the amount of material in the
beam. 

As fibre material we chose Kuraray SCSF-78MJ \cite{Kuraray} for
all SciFi stations.
The scintillation light is guided by clear (not
scintillating) fibres of lengths between $0.5\,\m$ 
and $3\,\m$. It is then detected by 16-channel multi-anode
PMTs (Hamamatsu H6568  
\cite{Hamamatsu}) followed by
fast leading edge discriminators \cite{Gorin:00} and pipelined
TDCs (see Sec.~\ref{sec:daq.digital.f1}).

Stations 1--4 have an r.m.s. spatial resolution of $130\,\mum$, 
station 5 of $170\,\mum$ and stations 6--8 of $210\,\mum$,
with local variations which are consistent with
fluctuations of the order of $10\%$ of the fibre diameter.  
The intrinsic detection efficiency of the SciFi stations was measured
to be $\geq99\%$. Due to occupancy in the readout in the high
intensity region the efficiency is slightly lower,  
varying between $96\%$ and $99\%$ for the various stations.

\begin{figure}[tbp]
  \centering
%  \includegraphics[width=\columnwidth]{./figures/ScifiFig_2_9_11_05}
%  \includegraphics[width=\columnwidth,viewport=100 460 460
%  710]{./figures/scifi_neu}
  \includegraphics[width=\columnwidth]{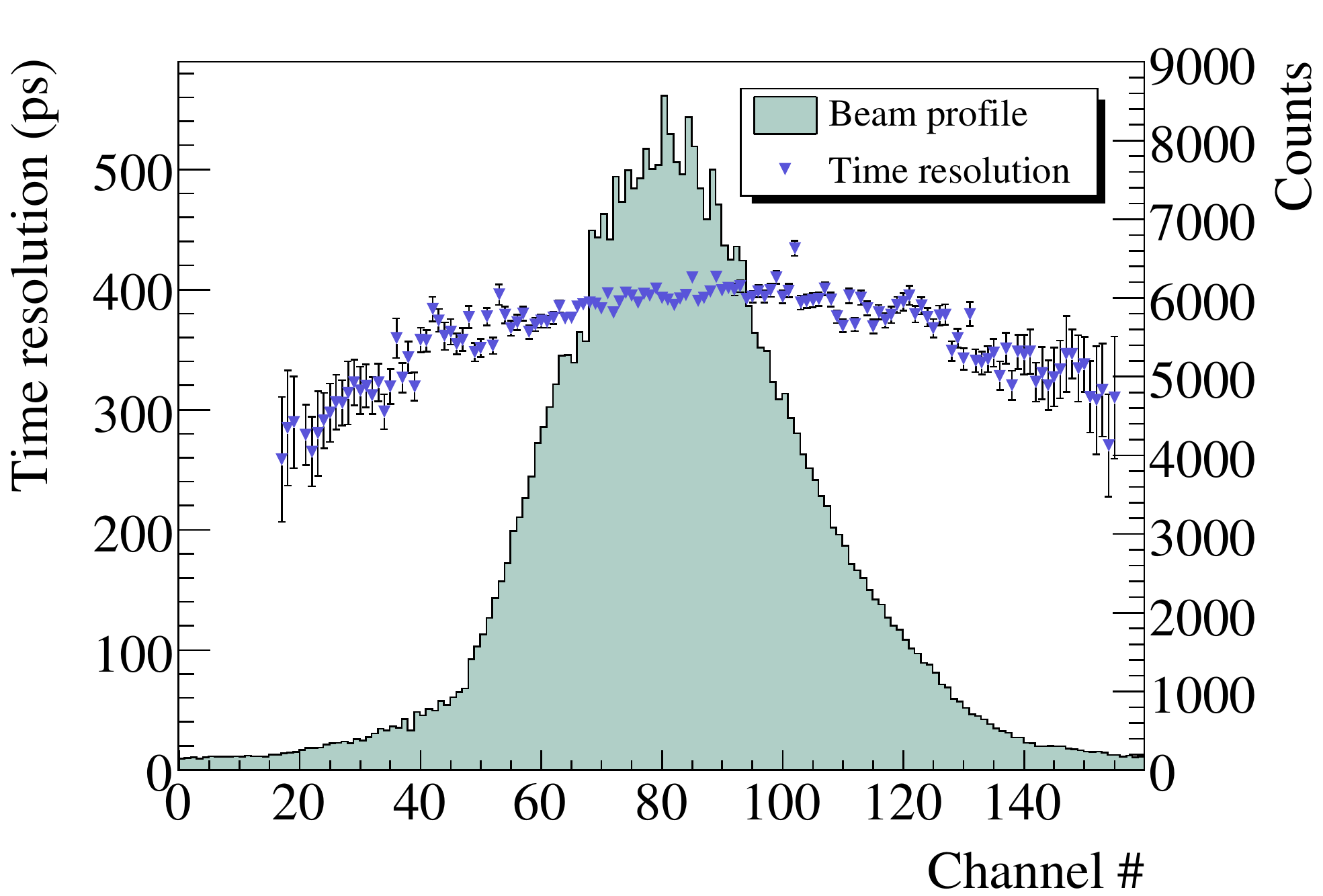}
  \caption{Time resolution (r.m.s.) 
    of a single SciFi $Y$ plane   
    across the beam region (triangles) together 
    with the beam profile (histogram).}  
  \label{fig:trk.vsat.scifi.timeres}
\end{figure}

The obtained time resolution using one plane is 
nearly constant for all channels. R.m.s.\ values between $350\,\ps$ and
$450\,\ps$ were obtained for the central regions of the
various 
planes. 
This can be seen from 
Fig.~\ref{fig:trk.vsat.scifi.timeres}, where the
profile of the intensity distribution in the beam region measured by
one SciFi plane is shown together with the obtained time resolution.
The time resolution across the plane shows a smooth curve with slightly
better values in the outer region where the intensity is low.

Within the three years of operation all SciFi-detector showed a very
stable operation, and there is no indication of ageing or radiation
damage.  

For the 2006 run an additional SciFi station was added
to the 
setup in order to increase reconstruction efficiency of scattered
muon tracks near the beam. It consists of $U$ and $V$ planes and 
was positioned about $110\,\cm$ upstream of SciFi station 6.
Each plane of the new station has an active area of $12.3\times
6.3\,\cm^2$, and is read out by 96 channels. The other parameters
are equal to those of the stations 7 and 8.
%

%%% Local Variables: 
%%% mode: latex
%%% TeX-master: t
%%% End: 

\subsubsection{Silicon microstrip detectors}
\label{sec:tracking.vsat.silicon}
The COMPASS silicon microstrip detectors are used for 
the detection of the incoming muon beam track,
and, for the hadron program, for
vertex and track 
reconstruction downstream of the target. 
The high beam intensity in
COMPASS requires a 
radiation hard detector design and an excellent spatial and time
resolution.  

The silicon wafer, optimised for high fluences, was originally
designed and developed for the HERA-B experiment \cite{Abt:00a}. 
The $300\,\mum$
thick n-type wafer has an active area of $5 \times 7 \,\cm^2$.
The 1280 readout strips on the n-side ($54.6\,\mum$ pitch) are
perpendicular to the 1024 readout strips on the p-side ($51.7\,\mum$
pitch), so that with one wafer two-dimensional position information
can be obtained. 
This reduces the material budget by a factor of
two as compared to a single-sided readout.

\begin{figure}[tbp]
  \centerline{\includegraphics[width=\columnwidth,
    clip=on]{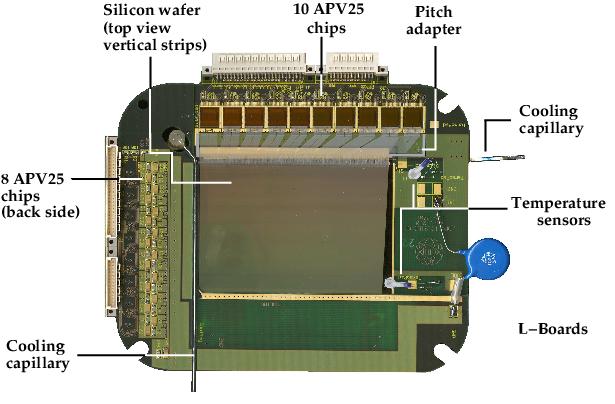}} 
  \caption{Front view of a COMPASS silicon
    detector.}\label{fig:tracking.vsat.silicon.module} 
\end{figure}

The silicon wafer is glued with silicone glue onto a frame consisting
of two L-shaped
Printed Circuit Boards (L-board) forming a detector. The readout
strips, which are tilted by $2.5^{\circ}$ with respect to the wafer
edge, are connected via $25\,\mum$ Aluminium wire bonds and a glass pitch
adapter to the  
front-end chips. Along two wafer edges a capillary is soldered onto the
back side of the L-board and is electrically insulated by a connector
of epoxy material. The capillary is flushed with gaseous nitrogen
($400\,\liter/\Hour$) to cool the front-end chips. The setup of a
COMPASS silicon  
detector is shown in Fig.~\ref{fig:tracking.vsat.silicon.module}.

The analogue signals induced on the microstrips are read out using the
APV25 front-end chip, a 128-channel preamplifier/shaper ASIC with
analogue 
pipeline, originally developed for the CMS silicon microstrip tracker
\cite{French:01a}. Each channel of the APV25 consists of an
inverter stage with unit amplification to allow signals of both
polarity to be processed, and a CR-RC type shaping amplifier with a
time constant of $50\,\ns$.  The amplifier output amplitudes are
sampled at a frequency of $38.88\,\MHz$, using the reference clock of
the trigger control system (TCS) of the experiment, and stored in a
192 cell analogue pipeline. Upon arrival of an external trigger at the
chip, the cells corresponding to the known trigger latency (up to
$4\,\mus$) are flagged for readout.  The analogue levels of the flagged
cells for 128 channels are then multiplexed at $20\,\MHz$ onto a
single differential output. In order to obtain time information from
the signal shape, not only the sample corresponding to the peak of an
in-time signal is transferred, but in addition two samples on the
rising edge of such a signal are read out.  While the sampling in the
APV25 as well as the trigger signal are synchronised to the reference
clock of the TCS, the passage of a particle is not. The resulting
shift between the TCS clock phase and the actual time of particle
passage, being randomly distributed in the $25.7\,\ns$ window for each
event, is corrected during the reconstruction. %%\cite{Angerer:05a}.  
It
is determined by the difference of the rising edge of the TCS clock
and the trigger time, and is measured via a TDC event by event.
The multiplexed analogue data stream from each APV25 chip is
digitised by a $10$ bit flash ADC, described in
Sec.~\ref{sec:daq.analog.sgadc}.

Two detectors make up one silicon station. They are mounted
back-to-back on a fibre-glass frame such that  
% and housed in a cryostat.
one detector measures the horizontal ($X$) and vertical ($Y$) 
coordinates of a particle trajectory, while the other 
is rotated around the
beam axis by 5$^\circ$,  
providing two additional projections ($U$, $V$). 
\cite{Angerer:03a}.  
The wafers are oriented such that the $X$ and 
$U$ planes constitute the n-side, and the $Y$
and $V$ planes the p-side of the wafer, respectively.  

%The residuals were calculated for a hadron run using silicon detectors
%only, so that the track error could be easily deconvoluted. In
%Fig.~\ref{fig:tracking.vsat.silicon.residual} a standard residual distribution
%for a 
%$VY$ projection is shown, where we obtain $\sigma=8.1\,\mum$.  The
%spatial resolution depends strongly on the cluster size (number of
%neighbouring strips combined to one hit) since for more than one hit
%strip the spatial information can be refined by calculating the mean
%of strip coordinates of one cluster, weighted by the corresponding
%amplitude. The ratio of cluster size~1 and cluster size~2 is mainly
%given by the silicon wafer's design, because only on the p-side
%(VY-projections) the wafer is equipped with intermediate strips, which
%improve the charge devision. For cluster size~1 the achieved spatial
%resolution is in the interval $\sigma = 10 - 14\,\mum$ and for
%cluster size~2 $\sigma = 4 - 6 \,\mum$ respectively.

\begin{figure}[tbp]
  \centerline{\includegraphics[width=\columnwidth, clip=on]
    {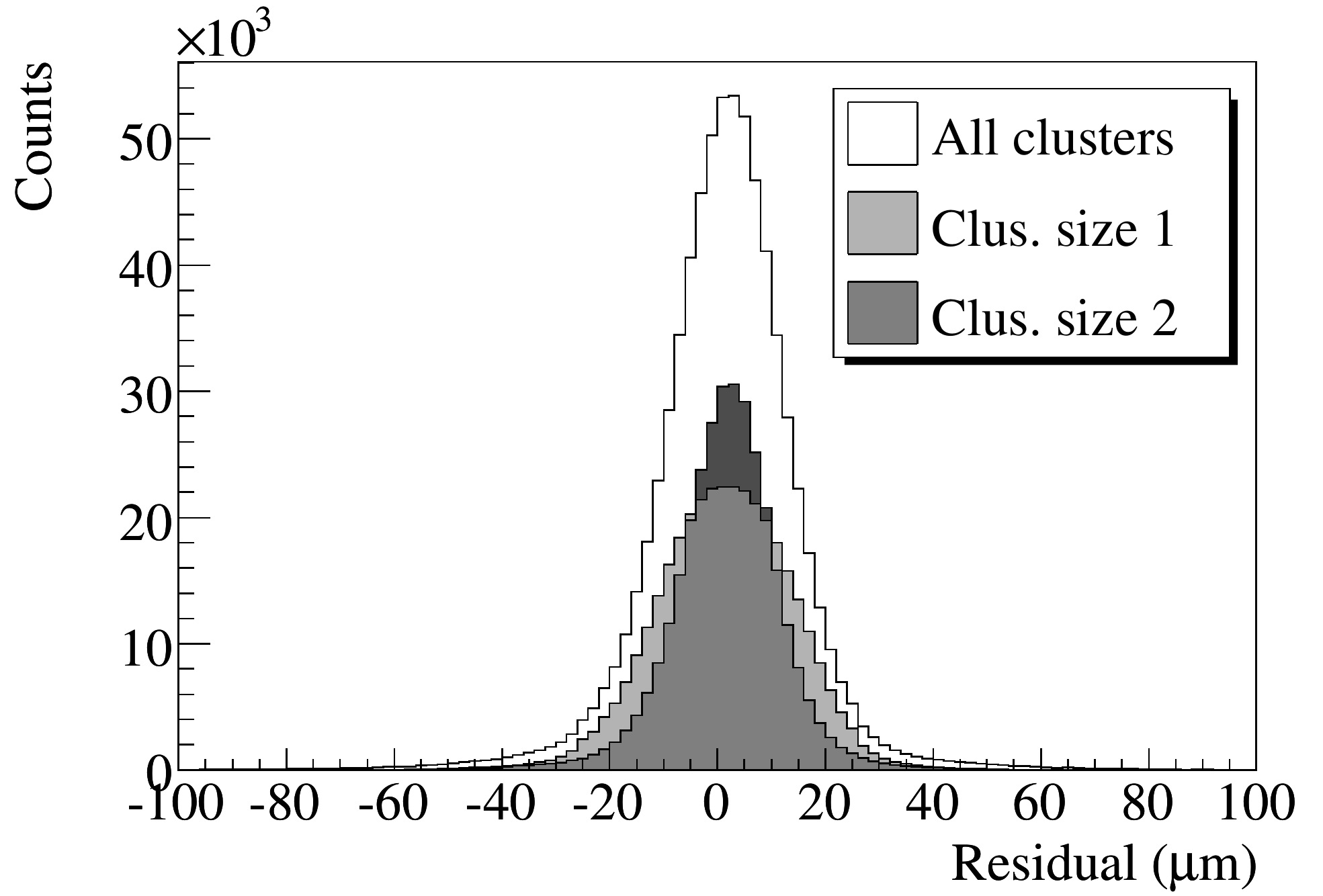}} 
  \caption{Residual distribution for one silicon $Y$-plane.  
    In light grey the distribution for cluster size~1 and in dark grey
    for cluster size~2 are shown. 
    % The $\sigma$ for is $9.0\,\mum$, deconvoluted with the track
    % error $\sigma$ is $8.8\,\mum$
  }
  \label{fig:tracking.vsat.silicon.residualY}
\end{figure}

\begin{figure}[tbp]
  \centerline{\includegraphics[width=\columnwidth, clip=on]
    {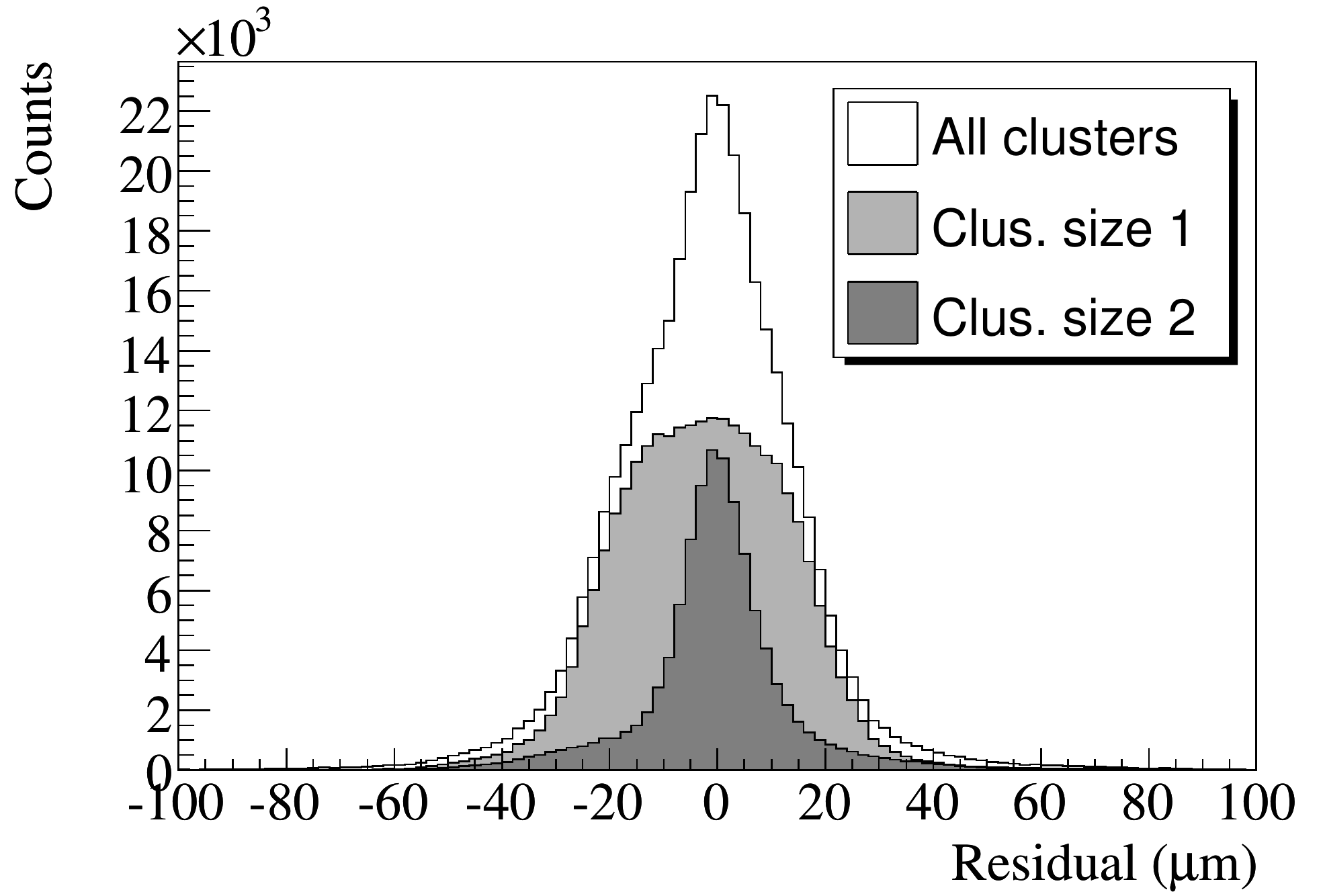}} 
  \caption{Residual distribution for one silicon $X$-plane.  
    In light grey the distribution for cluster size~1 and in dark grey
    for cluster size~2 are shown.}
  \label{fig:tracking.vsat.silicon.residualX}
\end{figure}

The residuals were calculated for standard muon run conditions using 
silicon detectors only, so that the track error ($<3\,\mum$) could be 
easily deconvoluted. The
spatial resolution strongly depends on the cluster size (number of
neighbouring hit strips combined to one cluster) since for more than one hit
strip the spatial information can be refined by calculating the mean
of strip coordinates of one cluster, weighted by the corresponding
amplitude 
and taking into account the strip response function.  
%The charge cloud is not shared linearly by the contributing strips.
%This was corrected in order to calculate the precise cluster position.
The ratio of hits with cluster size~2 to hits with
cluster size~1  
is 1.0 for the p-side ($Y$-, $V$-planes) and 0.4 for the n-side ($X$-,
$U$-planes), and  
is mainly
given by the design of the wafer. 
The charge sharing is improved on the p-side by additional capacitive
coupling due to 
intermediate strips. 
%because only on the p-side
%($VY$-projections) the wafer is equipped with intermediate strips, which
%improve the charge devision. 
This results in an improved spatial resolution for the $Y$- and
$V$-planes compared to the $X$- and $U$-planes, as can be seen from
the residual distributions shown in  
Fig.~\ref{fig:tracking.vsat.silicon.residualY} for a $Y$-plane, and 
Fig.~\ref{fig:tracking.vsat.silicon.residualX} for an $X$-plane,
respectively.  
%% For cluster size~1 the obtained spatial
%% resolution (r.m.s.) is 
%in the interval $\sigma = 10 - 14\,\mum$ and for
%cluster size~2 $\sigma = 4 - 6 \,\mum$, respectively. 
%% $10\,\mum$ and $14\,\mum$ for the p- and n-side, respectively, for cluster
%% size~2 it is $6\,\mum$ and $4\,\mum$. 
The average spatial resolution of the COMPASS silicon detectors is
$8\,\mum$ for the p-side, and 
$11\,\mum$ for the n-side. 

\begin{figure}[tbp]
  \centerline{\includegraphics[width=\columnwidth, clip=on]
    {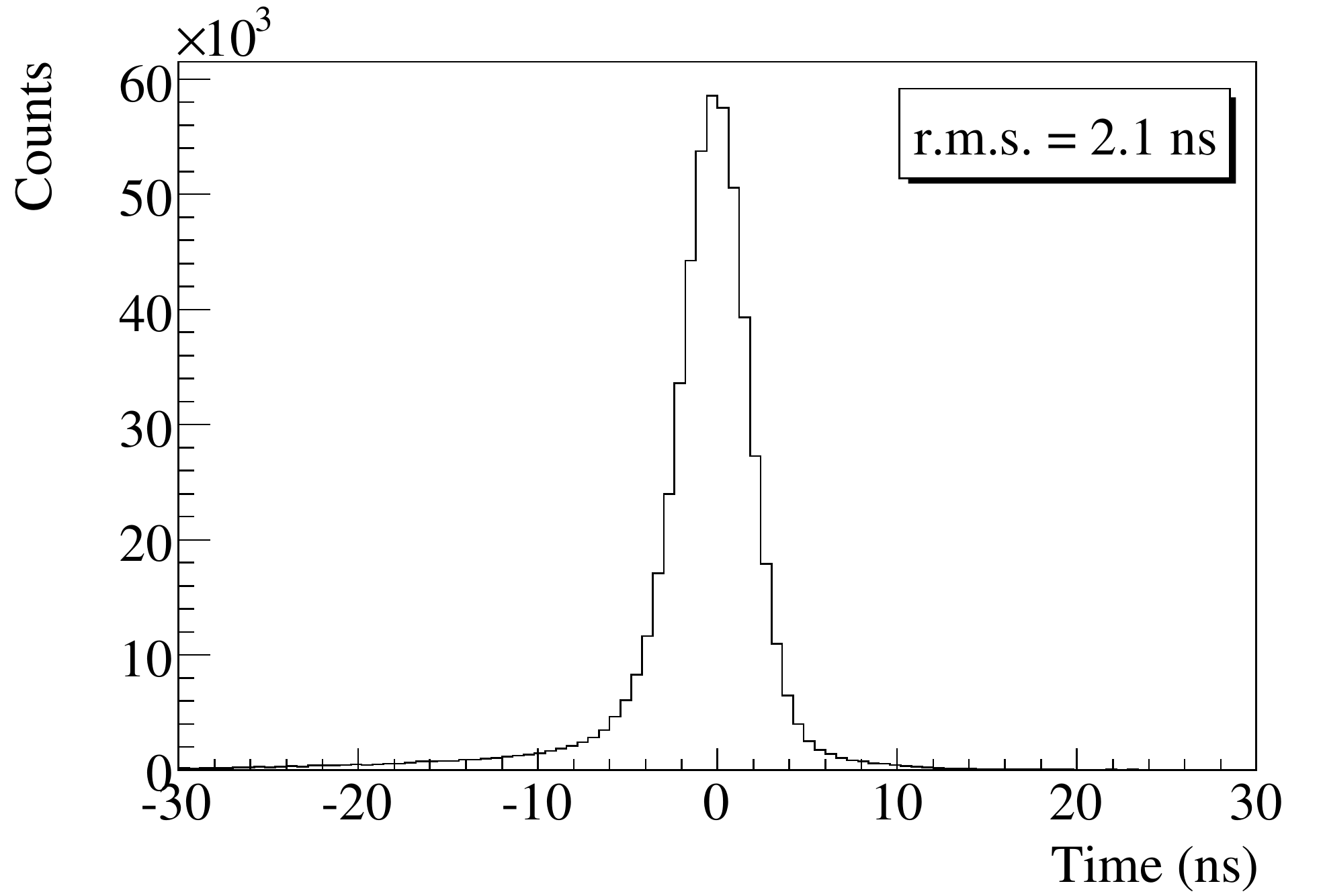}} 
  \caption{Signal time distribution of a
    single projection of a silicon detector}
  \label{fig:tracking.vsat.silicon.time_track}
\end{figure}

Figure~\ref{fig:tracking.vsat.silicon.time_track} shows the signal
time 
distribution for one plane of a silicon detector. The average time
resolution 
was found to be 
$\left < \sigma_t \right > = 2.5\,\ns$.

For the COMPASS muon data taking period 2002 two silicon stations
were used as beam telescope, while for 2003 and 2004 three silicon
stations were employed for this purpose. 
In each muon data
taking period each  
silicon detector was exposed to a fluence of
about $\EE*{13}\,\mathrm{muons}/\cm^2$ in the central region. 
%A drop of the charge collection
%efficiency (CCE) 
An increase of noise combined with a decrease of signal amplitude
due to radiation damage was observed for the central  
region of the silicon detectors. These effects could be
compensated by increasing the 
depletion voltage by about $15\,\V$ on average for each data taking period.

For  
the COMPASS hadron pilot run in 2004, two silicon stations were
installed upstream of the target for the detection of beam tracks 
and three downstream of the target for
vertex and track reconstruction. During this 
period the central regions of the detectors were irradiated by
$8\EE{11}\,\mathrm{pions}/\cm^2$. 
For the future COMPASS data taking periods with high intensity hadron
beams a fluence 
of about $\EE*{13}\,\mathrm{hadrons}/\cm^2$ will be reached in the
central beam area, requiring advanced methods to increase the
radiation hardness of silicon detectors. 
In COMPASS this will be achieved by 
exploiting the Lazarus effect \cite{Palmieri:98}, which results in a
%partial 
recovery of the 
charge collection efficiency (CCE) for irradiated detectors when
operated at cryogenic temperatures. It has been   
shown experimentally, that the CCE recovery is greatest for operation 
temperatures around $130\,\K$. 
The silicon detector system used at present has already been designed
for such 
cryogenic operation \cite{Angerer:03a}. To this end the
silicon detectors are housed in vacuum tight cryostats with
low mass and light tight detector windows. 
%% \cite{Angerer:05a}. 
The detectors will be cooled by 
flushing the capillary along the wafer edge (see Fig.
\ref{fig:tracking.vsat.silicon.module}) with liquid nitrogen
instead of gaseous nitrogen. The liquid 
nitrogen distribution system is currently being developed. 

%%% Local Variables: 
%%% mode: latex
%%% TeX-master: t
%%% TeX-master: t
%%% End: 

% 
\subsection{Small area trackers}
\label{sec:tracking.sat}
\subsubsection{Micromegas detectors}
\label{sec:tracking.sat.micromegas}

COMPASS is the first high energy experiment using Micromegas
(Micromesh Gaseous Structure) 
detectors \cite{Thers:01,Kunne:03,Bernet:05b}.
Twelve detectors, with 1024 strips
each, assembled in 3 stations of 4 planes each ($X$, $Y$, $U$, $V$),
track particles in the 
$1\,\m$ long region between the polarised target solenoid and the first
dipole magnet. 

\begin{figure}[tbp]
  \begin{center}
    \includegraphics[width=\columnwidth]{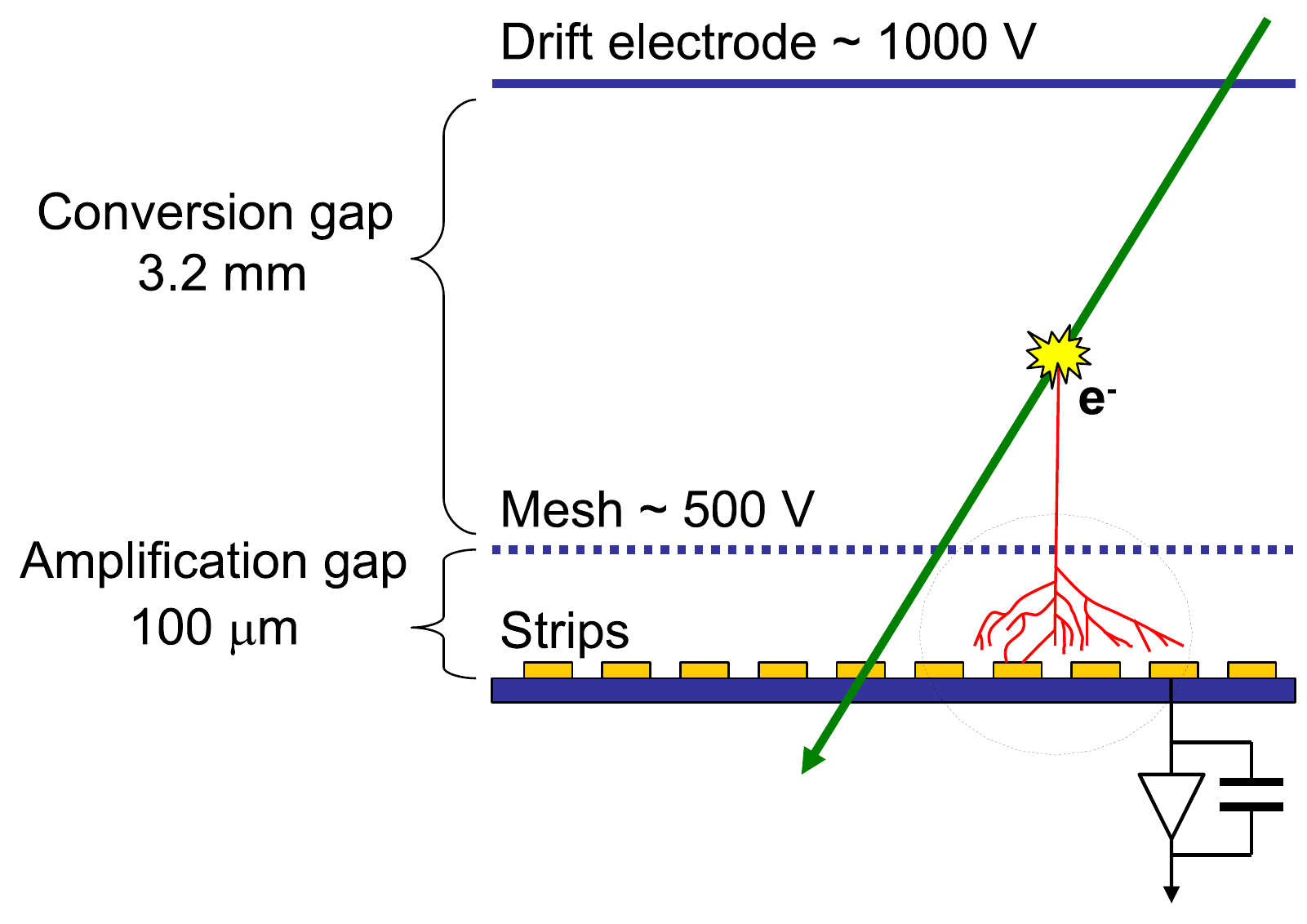}
  \end{center}
  \caption{Principle of a Micromegas detector.}
  \label{fig:tracking.sat.mm.principle}
\end{figure}

The Micromegas detector is based on a parallel plate electrode
structure and a set of parallel microstrips for readout. The special
feature of 
this detector is the presence of a metallic micromesh which separates
the gaseous volume into two regions: a conversion gap where the
ionisation takes place and the resulting primary electrons drift in a
moderate field (here about $1\,\kV/\cm$ over $3.2\,\mm$), and an
amplification gap where a higher field (here $50\,\kV/\cm$ over
$100\,\mum$) 
produces an avalanche which results in a large number of electron/ion
pairs (see Fig.~\ref{fig:tracking.sat.mm.principle}). 
The field
configuration near the mesh is such that most of the ions from the
avalanche are captured by the mesh and do not drift back into the
conversion gap. Consequently the ions drift over a maximum distance of
$100\,\mum$ and the width of the signal induced by the ions cannot
exceed the drift time over that distance, that is about $100\,\ns$.
The 
fast evacuation of positive ions combined with the reduced transverse
diffusion of the electrons and the high granularity of the detector
result in a high rate capability.

The gas mixture used is Ne/C$_2$H$_6$/CF$_4$ (80/10/10), 
optimised for good time resolution. In addition, it minimises  
the discharge rate to $0.03$ discharges per detector and per beam spill
\cite{Bernet:05b}.

\begin{figure}[tbp]
  \begin{center}
    \includegraphics[width=\columnwidth]{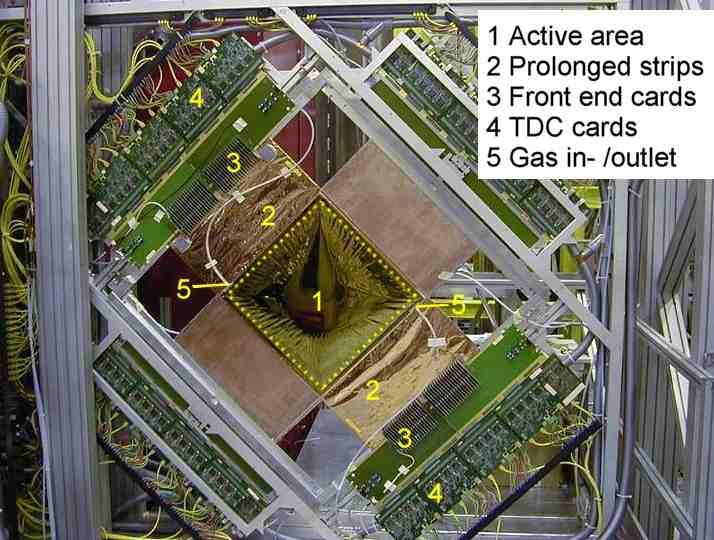}
  \end{center}
  \caption{A Micromegas doublet ($U$ and $V$ projections) in the
    COMPASS 
    experiment.  The active zone is the $40\times 40\,\cm^2$ internal
    square (1). Strips are extended (2) in order to keep the front-end
    electronics (3) outside the acceptance of the spectrometer.}
  \label{fig:tracking.sat.mm.uv}
\end{figure}

The detector has an active area of $40\times 40\,\cm^2$ and a central
dead 
zone of $5\,\cm$ in diameter. The strip pitch is $360\,\mum$ for the
central part of the detector ($512\,\mathrm{strips}$), and $420\,\mum$ 
for the 
outer part ($2\times 256\,\mathrm{strips}$). In 
order to minimise the amount of material inside the acceptance of the
spectrometer, the readout PC boards are positioned $35\,\cm$ further
away by extending the readout strips outside the active area (see
Fig.~\ref{fig:tracking.sat.mm.uv}). The
thickness of one detector plane in the active area is about $0.3\%$
of a radiation length. 

The Micromegas are assembled in doublets of two identical detectors
mounted back to back, and rotated by $90^\circ$ with respect to one
another, so that a doublet measures two orthogonal coordinates.
Fig.~\ref{fig:tracking.sat.mm.uv} shows a $UV$ doublet (strips 
at $+45^\circ$ in $U$-plane, and at $-45^\circ$ in $V$-plane).

A digital readout based on the custom-made SFE16 chip
\cite{Delagnes:00} is used. The chip, a 16-channel low-noise
(ENC $900\,e^-$ at $68\,\pF$) charge
preamplifier-filter-discriminator, was designed in order to stand
high counting rates (up to $200\,\kHz/\mathrm{channel}$). The 
time window of the chip is $220\,\ns$ for typical experimental
conditions. Its peaking time of $85\,\ns$ is matched to the signal
rise time 
for a $100\,\mum$ amplification gap with the present gas
mixture. 
The SFE16 chips are connected via LVDS
links to F1-TDC chips in multi-hit mode (see
Sec.~\ref{sec:daq.digital.f1}). Both the 
leading and trailing edge times of the analogue signal are recorded. 
On the one hand, the weighted average of these two measurements
yields an improved determination of the mean time by correcting for the
walk. 
The signal amplitude, on the other hand, can be determined indirectly
from the time over threshold, i.e. from the time difference of
the two measurements. 

\begin{figure}[tbp]
  \begin{center}
    \includegraphics[width=\columnwidth, angle=0]
%    {./figures/mm_timeres_MM02V_nim2.pdf}
    {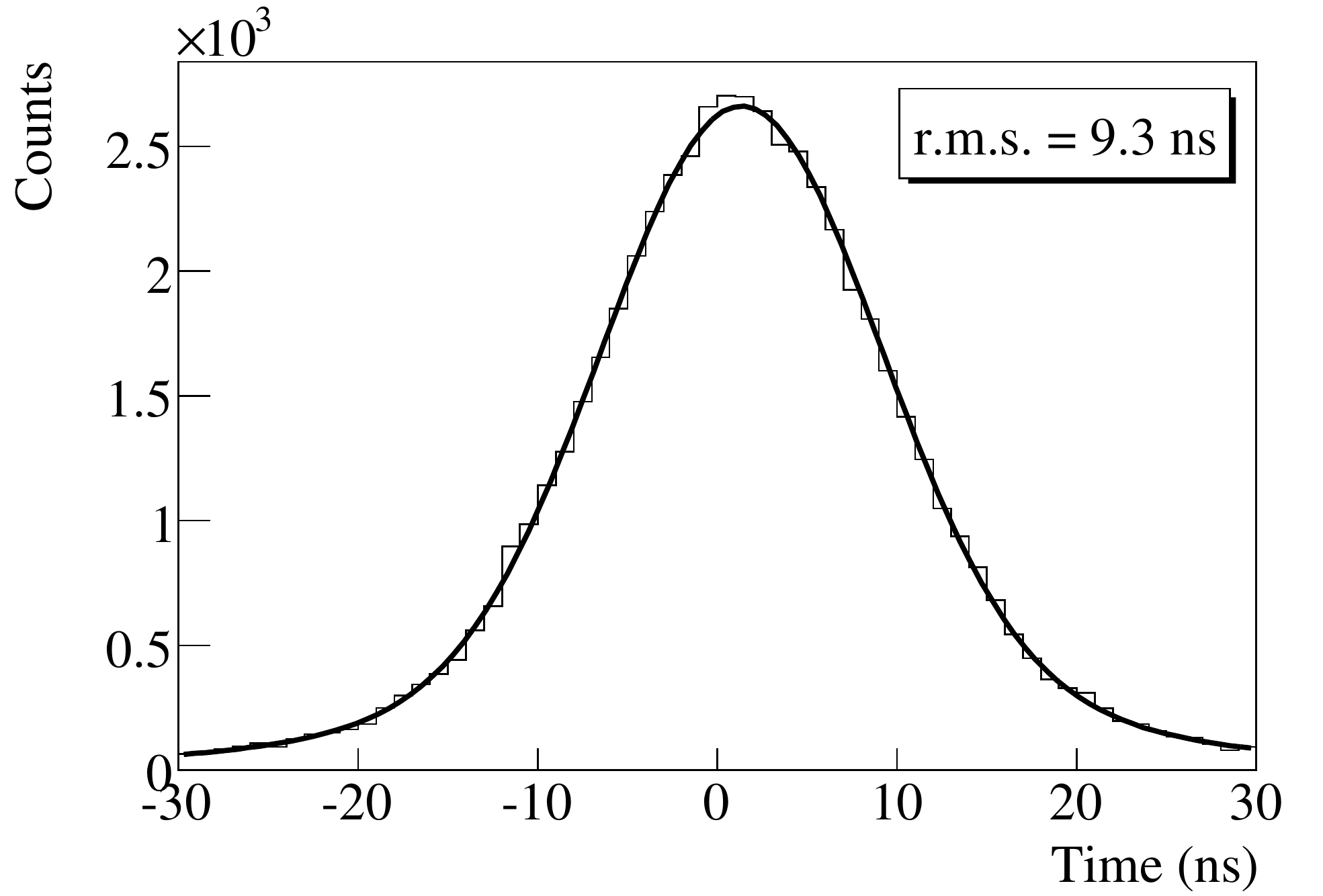}
  \end{center}
  \caption{Distribution of the mean time measured by a Micromegas with
    respect to the track time for
    nominal intensity data. The r.m.s.\ width is
    $9.3\,\ns$.} 
  \label{fig:tracking.sat.mm.meantime}
\end{figure}

In COMPASS, the Micromegas see an integrated flux of $30\,\MHz$,
reaching 
$450\,\kHz/\cm^2$ close to the dead zone.  The time resolution, the
efficiency and the position resolution have been measured 
%both at low intensity and 
in COMPASS nominal data taking conditions of  
$4\EE{7}\,\mu/\mathrm{s}$ scattered on the one radiation length
target, i.e.\ 
$100-200\,\kHz$ per strip, in the fringe fields of the target
solenoid and the first dipole. The obtained mean time resolution is  
$9.3\,\ns$, as shown in
Fig.~\ref{fig:tracking.sat.mm.meantime}. 
%% From NIM paper
%% C. Bernet thesis: 10.0 mus
Only signals within a
time window of $\pm 50\,\ns$ are used to combine adjacent hits into
clusters. The average cluster size is $2.6$ for the strips with
$360\,\mum$ pitch.

The average efficiency of all Micromegas detectors was determined using
charged particle tracks reconstructed in at least 20 planes of the 
spectrometer. It reaches $97\%$ at nominal beam intensity.
%% C. Bernet thesis: 96.5%
% \begin{figure}[t]
%   \begin{center}
%     \includegraphics[width=\columnwidth,
%     angle=0]{./figures/mm_eff2d_MM02V_nim2.pdf}
%   \end{center}
%   \caption{Two-dim.\ efficiency of V plane for nominal intensity data.
%     The hole in the canter corresponds to the central dead zone.}
%   \label{fig:tracking.sat.mm.effic}
% \end{figure}

\begin{figure}[tbp]
  \begin{center}
    \includegraphics[width=\columnwidth]
%    {./figures/mm_2v_residuals.pdf}
    {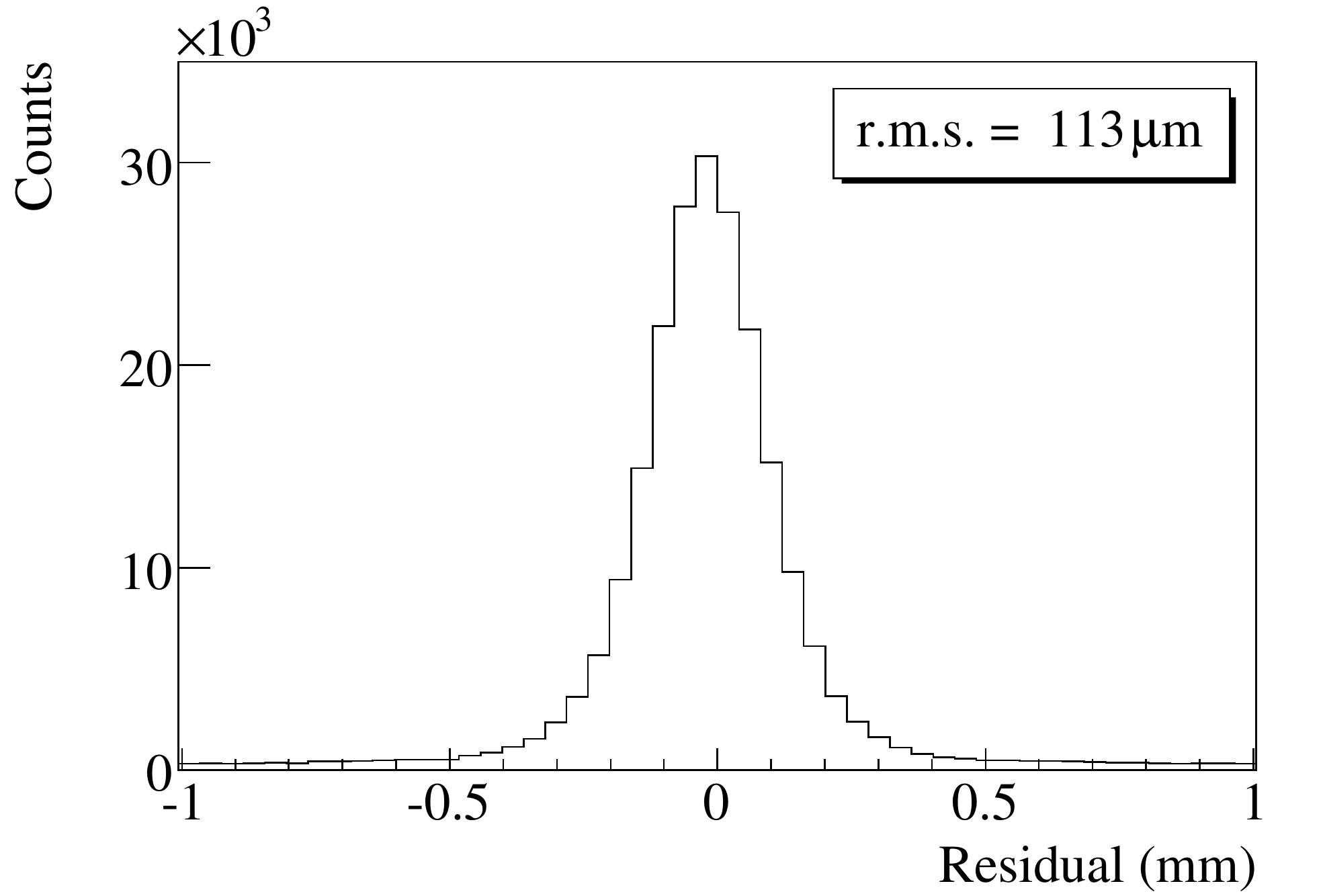}
  \end{center}
  \caption{Distribution of residuals of a Micromegas
    detector at high intensity. The r.m.s.\ width is
    $113\,\mum$, to which the detector contributes $92\,\mum$.}
  \label{fig:tracking.sat.mm.residuals}
\end{figure}

To evaluate the spatial resolution, incident tracks are reconstructed
using the hits in 11 Micromegas, and the residuals in the 12th one are
calculated. Fig.~\ref{fig:tracking.sat.mm.residuals} shows the
distribution of residuals for the full active area of one Micromegas detector. 
% for the central zone of the detector where the pitch is $360\,\mum$. 
Deconvoluting the precision of the track, 
we obtain a spatial 
% resolution of $70\,\mum$ at low intensity. 
resolution of $90\,\mum$, averaged over all Micromegas detectors at
nominal beam intensity. 
%% From NIM paper
%% C. Bernet thesis: 93mum
Their position within the spectrometer between the target solenoid and the
first spectrometer dipole implies that 
they operate in the
fringe field of both magnets, which
%reaches $0.3\,\T$ for some detectors. 
exerts a Lorentz force on the drifting electrons. Tracks detected in the
Micromegas cover angles up to $70\,\mrad$.

%The twelve Micromegas sense the fringe field of the first spectrometer
% dipole, which can reach 0.3 T at some detector location. For these
% detectors, efficiency and timing resolution measured with field ON
% were found to be similar to those measured with field OFF.

During the COMPASS data taking period 2002 -- 2004 a total charge of 
$1\,\mC/\mm^2$ was accumulated in the 
sensitive region closest to the beam. 
The mean amplitude of the signals was
continuously monitored for all detectors. No
variation of amplitude (and thus of gain) was observed between the
beginning and the end of the period. We conclude that no ageing has
been observed and that the detector is robust and stable.

%%% Local Variables: 
%%% mode: latex
%%% TeX-master: t
%%% End: 

\subsubsection{GEM detectors}
\label{sec:tracking.sat.gem}
\begin{figure}[tbp]
  \centering
  \includegraphics[width=\columnwidth]{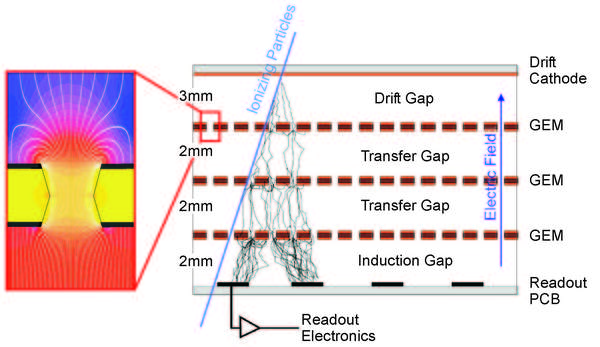}
  \caption{Schematic cross section of a triple GEM
    detector. The insert shows the electric field configuration for
    typical GEM voltages.}
  \label{fig:tracking.sat.gem.principle}
\end{figure}

COMPASS is the first high-luminosity particle physics experiment to
employ gaseous micropattern detectors with amplification in Gas
Electron Multiplier (GEM) 
\cite{Sauli:97} foils only. The GEM consists of a $50\,\mum$ thin
Polyimide foil (APICAL$^{\mathrm{\circledR}}$ AV \cite{Apical}) with Cu
cladding on 
both sides, into which a large number of micro-holes (about $\EE*{4}/\cm^2$,
diameter $70\,\mum$) 
has been chemically etched using
photolithographic techniques. Upon application of a potential
difference of several $100\,\V$ across the foil, avalanche
multiplication of primary electrons drifting into the holes is
achieved when the foil is inserted between parallel plate electrodes
of a gas-filled chamber. Suitable electric fields extract the electrons
from the holes on the other side of the foil and guide them to the next
amplification stage or to the 
readout anode. The insert in Fig.~\ref{fig:tracking.sat.gem.principle}
depicts the electric field lines 
in the vicinity of a GEM hole for typical voltage settings. 

As shown in Fig.~\ref{fig:tracking.sat.gem.principle}, the COMPASS GEM 
detectors consist of three GEM amplification stages, stacked on top of
each other, and separated by thin spacer grids of $2\,\mm$ height
\cite{Altunbas:02a}.  
This scheme, developed for COMPASS together with a number of
additional features as segmented GEM foils and asymmetric gain
sharing between the three foils, guarantees a safe and stable
operation without electrical 
discharges in a high-intensity particle beam
\cite{Ketzer:01b,Bachmann:01e,Ketzer:02a}, 
and has been
adopted by various other experiments \cite{Bencivenni:02a,Bozzo:04}. 
The detectors are operated in an Ar/CO$_2$ (70/30) gas mixture, chosen
for its convenient features such as large drift velocity, low
diffusion, non-flammability, and non-polymerising properties.  

\begin{figure}[tbp]
  \centering
  \includegraphics[width=\columnwidth]{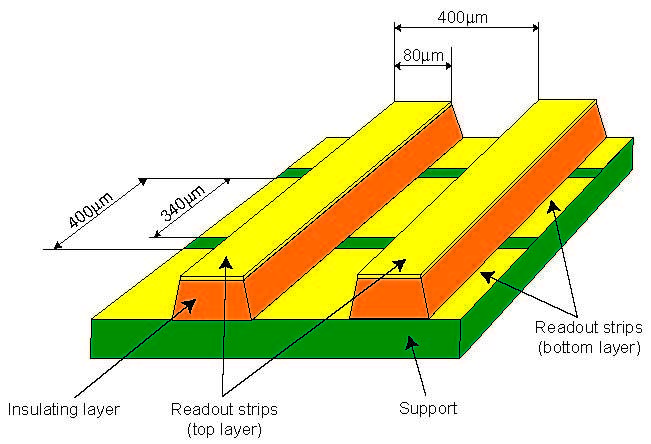}
  \caption{Schematic view of the two-dimensional
    readout structure of the COMPASS GEM detectors.}
  \label{fig:tracking.sat.gem.2dstrip.schematics}
\end{figure}

The electron cloud emerging from the last GEM induces a fast signal on
the readout anode, which is segmented in two sets of 768 strips with a
pitch of $400\,\mum$ each,
perpendicular to each other and separated by a thin insulating layer,
as shown in Fig.~\ref{fig:tracking.sat.gem.2dstrip.schematics}. 
% \begin{figure}[tbp]
%   \centering
%   \includegraphics[width=\columnwidth]{./figures/2d-stripPCB_microscope}
%   \caption{Microscope photograph of the two-dimensional readout
%     structure.}
%   \label{fig:tracking.sat.gem.2dstrip.photo}
% \end{figure}
For each particle trajectory one detector consequently records two
projections of the track with highly
correlated amplitudes, a feature which significantly reduces ambiguities in
multi-hit events \cite{Ketzer:04a}. 

The active area of each GEM detector is $31\times 31\,\cm^2$. 
The central region with a diameter of $5\,\cm$ is deactivated during normal
high-intensity physics runs by lowering the potential difference
across the last 
GEM foil in order to avoid too high occupancies on the central strips. At 
beam intensities below \mbox{$2\EE{7}/\mathrm{spill}$} this area can be
activated remotely 
to allow the detector to be aligned using beam tracks (see
Sec.~\ref{sec:performance.alignment}).  
%a feature which is of great importance for alignment of
%the spectrometer with beam tracks. 

In order not to spoil the mass resolution of the spectrometer due to
multiple scattering in the detector, great care has been devoted to
minimise the amount of material in the active area of the detector. 
To this end, the gas volume is enclosed on both sides by 
a light-weight honeycomb structure, into which a circular hole of
$5\,\cm$ diameter has been cut where the centre of the beam passes
through. The material budget of one detector, corresponding to the
measurement of two projections of a particle trajectory, amounts to
$0.4\%$ of a radiation length in the centre, and to $0.7\%$ in the
periphery. 

The signals on the strips are read out using the 
APV25 front-end chip, in the same way as described in
Sec.~\ref{sec:tracking.vsat.silicon} for the silicon microstrip
detectors. 
The readout strips are wire-bonded to the front-end PCB housing three
front-end chips.   
Since this chip lacks proper protection against
overcurrents from potential gas discharges, an external protection
network consisting of a double-diode clamp (BAV99) and a $220\,\pF$
coupling capacitor was added in front of each input channel. A glass
pitch adapter with aluminium strips is used to bring the strip pitch
down to $44\,\mum$ to match the input pitch of the APV25. 

Two GEM detectors are mounted back-to-back, 
%onto a large area tracker, 
%covering its dead zone close to the beam. 
forming one GEM station. 
One detector is rotated by $45^\circ$ with respect to the other,
resulting 
in the measurement of a charged particle trajectory in four
projections (labelled $XY$ and $UV$). Partial overlap with a large
area tracker located at the same position along the beam guarantees
complete track reconstruction and alignment. 

%\paragraph{Performance}
%\label{sec:tracking.sat.gem.performance}
The intrinsic properties of the triple GEM detectors have been
extensively studied in test beams and in COMPASS using low-intensity
beams without magnetic fields \cite{Ketzer:02a}. It was found that a
total effective gain of $8000$ is required in order to efficiently
detect minimum ionising particles in both projections. 
% \begin{figure}[tbp]
%   \centering
%   \includegraphics[width=\columnwidth]{./figures/PlateausXY_new}
%   \caption{Efficiency plateaus.}
%   \label{fig:tracking.sat.gem.eff_vs_voltage}
% \end{figure}

% Figure~\ref{fig:tracking.sat.gem.eff_map} shows a map of the
% background-corrected 2d-efficiency, i.e.\ the efficiency to detect 
% both projections at the same time, for one particular GEM detector.  
% \begin{figure}[tbp]
%   \centering
%   \includegraphics[width=\columnwidth]{./figures/GM06UV_2dEfficiency_linear_grey}
%   \caption{Map of the 2D-efficiency, determined at standard physics
%     conditions.} 
%   \label{fig:tracking.sat.gem.eff_map}
% \end{figure}
% Apart from local inefficiencies due to spacer grids, which account for a
% loss of less than $2\%$, the distribution is found to be very uniform
% across the detector surface at an average value over all detectors of
% $\langle\varepsilon_\mathrm{2D}\rangle = 95.6\%$. 
At nominal muon beam conditions 
the efficiency   
to detect a particle trajectory in at least one of the two 
projections, averaged over all GEM detectors in COMPASS, was determined to be  
$97.2\%$ \cite{Ketzer:04a}, with variations between different detectors on the
per cent level.  
Apart from local inefficiencies due to the spacer grids, which account for 
a loss of efficiency of less than $2\%$, the distribution is found to be very
uniform  
across each detector surface. 

\begin{figure}[tbp]
  \centering
  \includegraphics[width=\columnwidth]{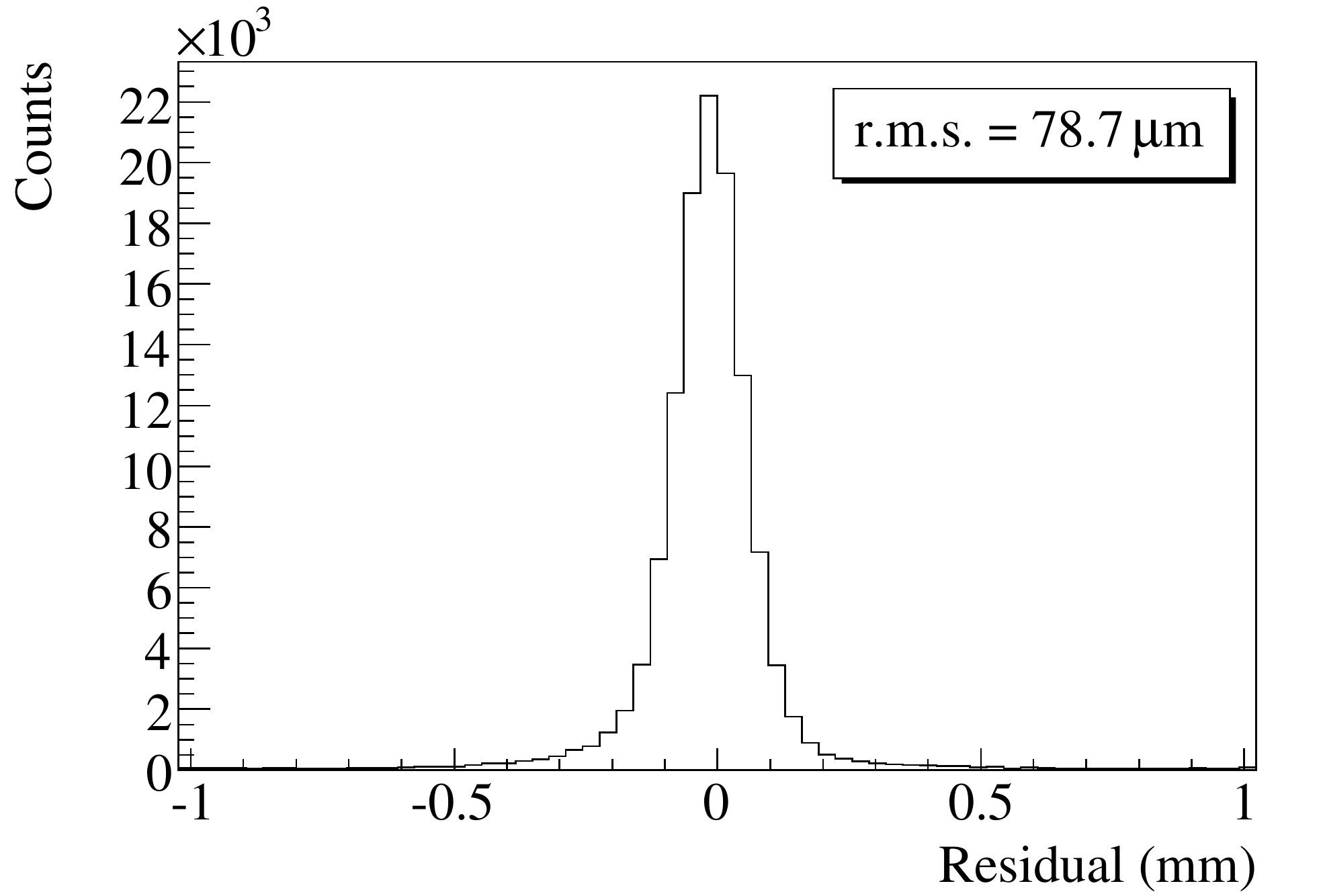}  
  \caption{Residual distribution for one GEM projection in 
    standard high-intensity muon beam conditions \cite{Ketzer:04a}. For this particular
    plane, the r.m.s.\ width of
    the distribution is $78.7\,\mum$, to which the detector contributes
    $66.4\,\mum$.} 
  \label{fig:tracking.sat.gem.spaceres}
\end{figure}

An offline clustering algorithm combines 
hits from adjacent strips to yield an improved value for
the position of a particle trajectory. The average number of strips
per cluster is $3.1$ for the top layer of strips, and $3.6$ for the bottom
layer \cite{Ketzer:02a}, consistent with the lateral diffusion of the
charge cloud in the GEM stack.  
Since the GEM detectors are the most precise tracking devices in
COMPASS downstream of the first dipole magnet, their spatial
resolutions at standard high-intensity muon beam 
conditions were measured using other GEM detectors only, so
that the track error could easily be deconvoluted. 
Figure~\ref{fig:tracking.sat.gem.spaceres} shows the distribution of 
residuals, i.\ e.\ the difference along one coordinate of expected
track and measured cluster position, plotted for all hits on one projection of
a GEM detector \cite{Ketzer:04a}.
Deconvoluting the track error, the resolutions for all GEM planes in the
spectrometer are 
found to be distributed around an average value of 
$70\,\mum$. 
This value includes a contribution of overlapping clusters due to pile-up 
in high intensity conditions of about $20\,\mum$. Variations for single
detectors are due to the effect of the fringe field of the first spectrometer
magnet, and the influence of multiple scattering in the material preceding the
respective detector.  
For all detectors the distortion of charge clouds in the 
immediate vicinity of the spacer grids deteriorates the average
resolution by about 
$4\,\mum$. 
% Discarding tracks
% with angles to the beam axis larger than $3.5\,\mrad$, an average
% spatial resolution over all GEM planes of $69.6\,\mum$ is
% obtained. 

\begin{figure}[tbp]
  \centering
  \includegraphics[width=\columnwidth]{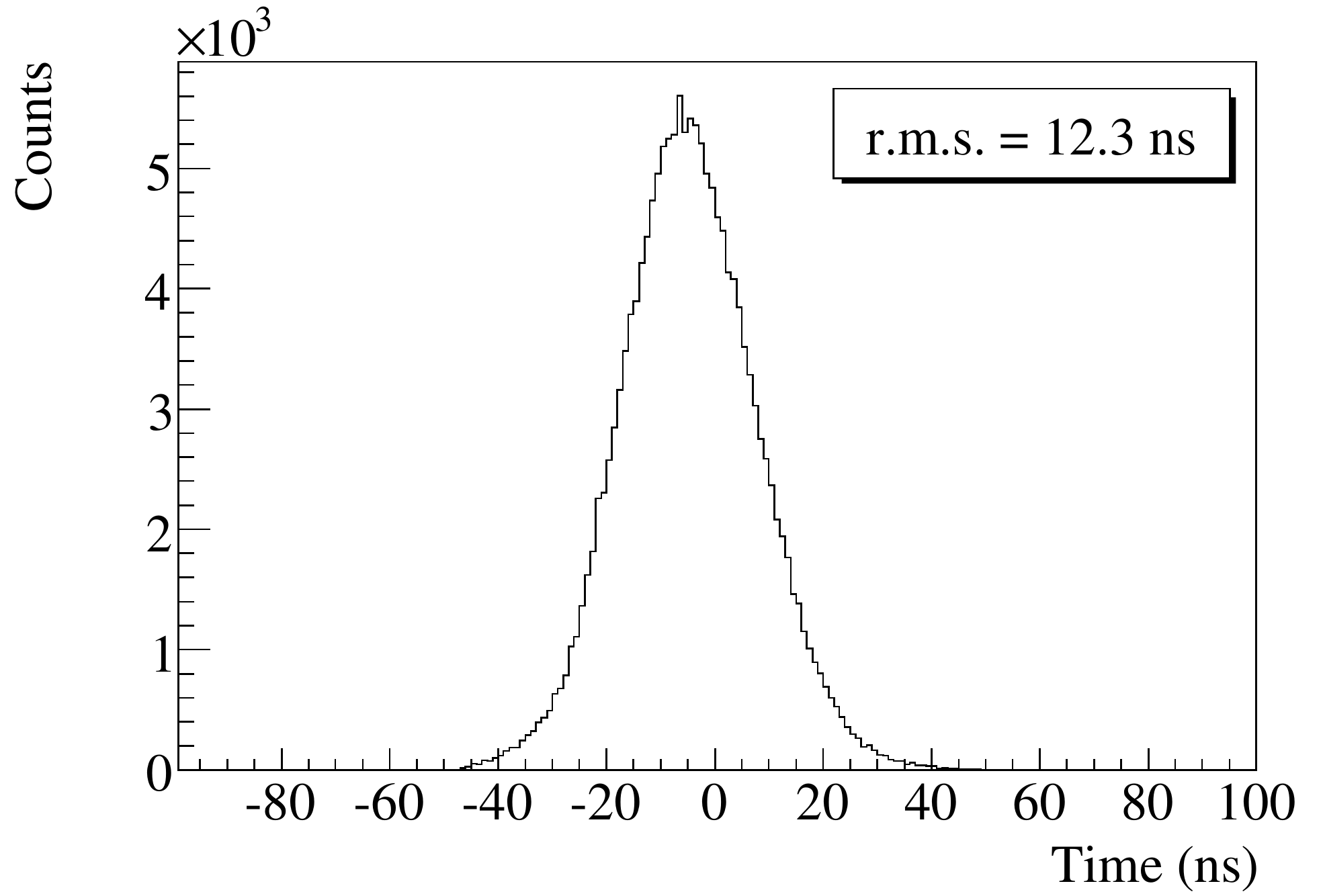}
  \caption{Distribution of cluster time measured by a single GEM plane by
    sampling the 
    analogue signal at $40\,\MHz$, with respect to the track time
    \cite{Ketzer:04a}.} 
  \label{fig:tracking.sat.gem.timeres}
\end{figure}

In addition to an improved spatial resolution the analogue readout
method also allows to extract time information by sampling the signal
at three consecutive points in time. Knowing the detector response to
a minimum ionising particle, the hit time can be determined from
ratios of the three measured amplitudes. With this method, an average
time
resolution of $12\,\ns$ was found for
the GEM detectors in the
high intensity muon beam \cite{Ketzer:04a}, as can be seen from
Fig.~\ref{fig:tracking.sat.gem.timeres}. 

In total, 11 GEM detector stations, i.e.\ 22 detectors, are installed
in COMPASS. Out of these, seven were operational from 2001  
on, three stations were added in 2002, and one additional station was
added for the 2004 run. Depending on the position in the spectrometer,
particle rates as high as $25\,\kHz/\mm^2$ are observed close to the
central inactive area, equivalent to a total collected charge since 2002 of
more than $2\,\mC/\mm^2$. Despite of this high-radiation environment,
no degradation of performance has been observed. Laboratory
tests, in which a total charge of $7\,\mC/\mm^2$ was collected using
Cu X-rays without loss of gain \cite{Altunbas:03a}, 
show that the  
GEM detectors will operate reliably well beyond the second phase of
COMPASS, which started in 2006. 
% \begin{figure}[tbp]
%   \centering
%   \includegraphics[width=\columnwidth]{./figures/GM06UV_Rate_22379}
%   \caption{Rate of charged particles.}
%   \label{fig:tracking.sat.gem.rate}
% \end{figure}

% \begin{figure}[tbp]
%   \centering
%   \includegraphics[width=\columnwidth]{./figures/pulseheight_vs_charge_c}
%   \caption{Variation of the corrected gain as a function of
%     accumulated charge. For reference measurements, a single wire
%     proportional counter (SWPC) was
%     installed in the same gas line in front of the GEM detector.}
%   \label{fig:tracking.sat.gem.gain_vs_charge}
% \end{figure}
%

%\clearpage

%%% Local Variables: 
%%% mode: latex
%%% TeX-master: "../compass_spec.tex"
%%% End: 

%
\subsection{Large area trackers}
\label{sec:tracking.lat}
\subsubsection{Drift chambers}
\label{sec:tracking.lat.dc}

% Three Drift Chambers (DC) were developed and built in order to ensure
% reliable tracking in the vicinity of the SM1 magnet. The chambers
% fulfil the severe criteria imposed by the experimental conditions in
% this region and for the required kinematics: large active area (more
% than $1\,\m^2$) combined with good spatial resolution (better than
% $200\,\mum$) and minimised material budget. Even more demanding for
% drift detectors, they are be able to stand high incident rates
% ($300\,\kHz$/channel and higher) with minimal loss in local
% efficiency.
%
Three identical drift chambers (DC) are 
installed in COMPASS. Their design was 
optimised for operation upstream of the first dipole magnet
(SM1), where the total particle flux through the chamber is higher by
almost a factor of three compared to the downstream side due to the 
low-energy background which is bent away by the magnet. One DC is
installed 
upstream, and two DCs  
downstream of the SM1 magnet. 
All three DCs have an active area of
$180\times 127\,\cm^2$, fully covering the acceptance of the 
SMC target magnet upstream as well as downstream of SM1. 

Each DC consists of eight layers of wires with four different
inclinations:  
vertical ($X$), horizontal ($Y$) and tilted by $20^\circ$($U$) and
$-20^\circ$($V$) with respect to the vertical direction. The tilt
angle 
and the ordering of planes ($XYUV$ along the beam) was chosen in
order to 
minimise the number of fake 
tracks during the reconstruction. 

Each layer of wires consists of 176 sensitive wires of $20\,\mum$
diameter, 
alternated with a total of 177 potential wires with $100\,\mum$
diameter, and is  
enclosed by two Mylar$^{\mathrm{\circledR}}$ \cite{Dupont} cathode foils of $25\,\mum$
thickness, coated with about $10\,\mum$ of graphite, 
defining a gas 
gap of $8\,\mm$ extent. 
Two consecutive layers of the same
inclination are staggered by $3.5\,\mm$ (half a drift cell) in order to
solve left-right ambiguities.   
During operation of the chamber the cathode foils, the
sensitive wires and the potential wires are kept at around
$-1700\,\V$, 
$0\,\V$ and $-1700\,\V$, respectively. 
The total material budget of each detector (8 layers) along the beam
path, including the gas mixture, is $0.32\%$ of a radiation length. 

\begin{figure}[tbp]
  \begin{center}
    \includegraphics[width=\columnwidth]{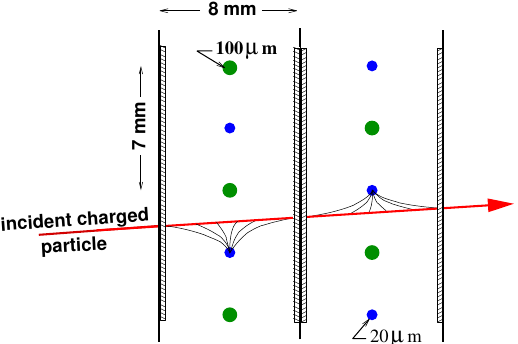}
  \end{center}
  \caption{\small Drift cell geometry of the COMPASS drift chambers.}
  \label{fig:tracking.lat.dc.cell}
\end{figure}

Drift cell boundaries (Fig.~\ref{fig:tracking.lat.dc.cell}) are
defined by the 
cathode foils, normal to the beam direction, and by two potential
wires separated by $7\,\mm$. 
The choice of a small drift cell size
($8\times 7\,\mm^2$) was triggered by counting rate considerations.
Smaller drift cells decrease the incident flux per cell and reduce the
electron drift time. The reduced drift time has an additional
advantage: it allows the use of a shorter time window and consequently
minimises 
the number of uncorrelated particles. 

\begin{figure}[tbp]
  \begin{center}
    \includegraphics[width=\columnwidth]{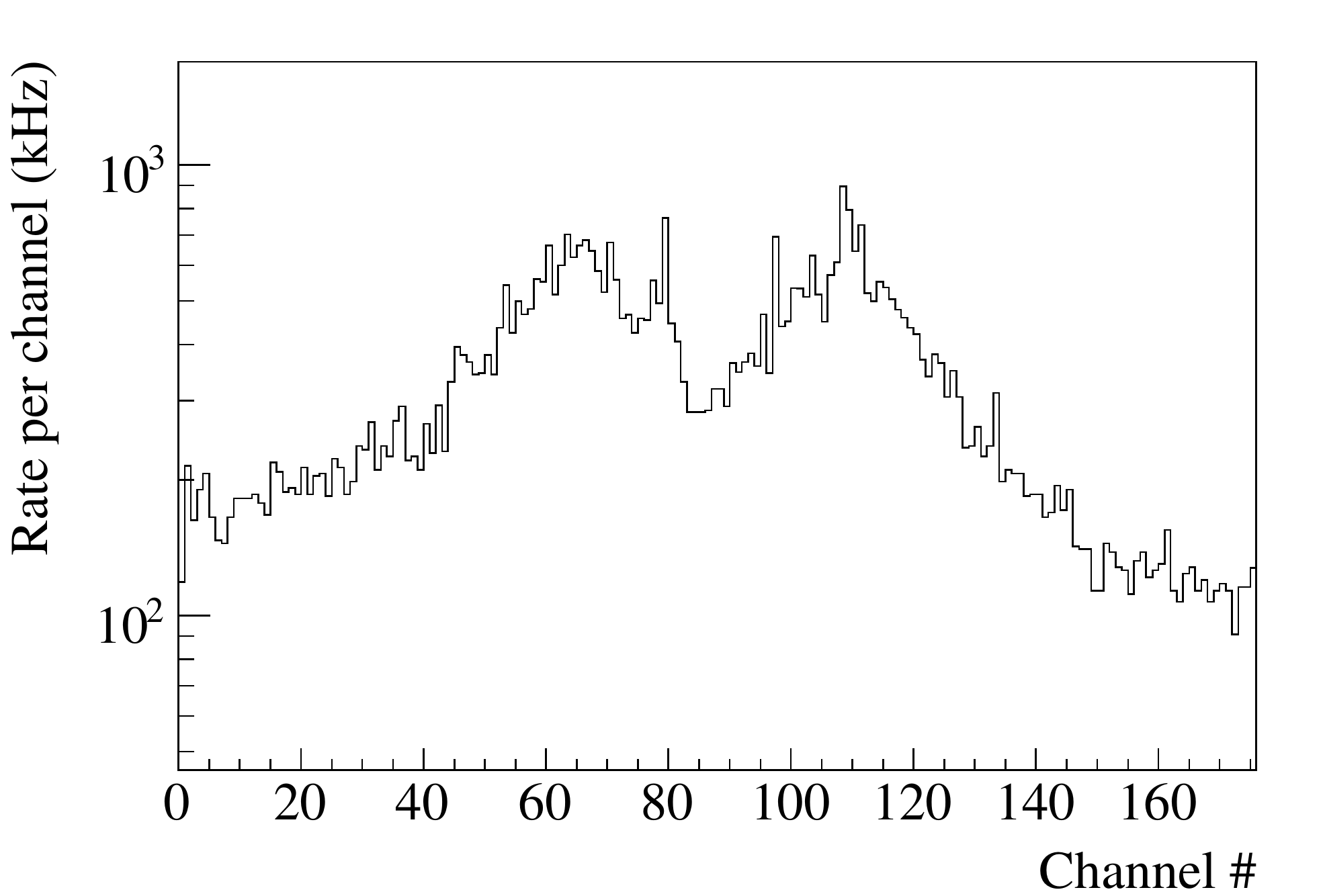}
  \end{center}
  \caption{\small Hit rate versus wire number for one DC wire layer
    upstream of SM1 at nominal beam intensity with the 
    central zone 
    deactivated.} 
  \label{fig:tracking.lat.dc.rate}
\end{figure}

Fig.~\ref{fig:tracking.lat.dc.rate} shows that the hit rate per wire
at a
distance of $15\,\cm$ from the beam reaches
$800\,\kHz/\mathrm{wire}$ upstream of SM1. Downstream of SM1, the
maximum hit rate per wire is reduced to $300\,\kHz$.  
In order to avoid even higher rates 
near the beam a central
dead zone of $30\,\cm$ diameter 
was implemented for all layers. 
As a segmented part of the whole
cathode foil, this dead zone has an independent high voltage supply.
During standard data taking the central zone is deactivated by keeping
the high voltage (HV) supply at low enough voltage, so that the local
efficiency 
vanishes. For alignment purposes at low beam intensity, the dead
zone is activated by setting the HV to the nominal value of
the potential wires ($-1700\,\V$).

The gas mixture was chosen in order to fulfil several constraints.
First it should ensure good spatial resolution. Second it should
feature a nearly linear time-vs-distance dependence (RT relation).  
Third it has to be fast enough so that the occupancy time is
minimised. Finally it should provide good efficiency and a large
HV plateau. The best compromise was obtained by a
mixture of Ar/C$_2$H$_6$/CF$_4$ (45/45/10). The Argon
component ensures high primary electron rate (about $100$ per MIP), the
C$_2$H$_6$ component serves as a quencher and the 
CF$_4$ is used to increase the drift velocity, to 
$77\,\mm/\ns$ ($13\,\ns/\mm$). 
With this mixture the
gain was measured to be $2\EE{4}$ for a HV of $1800\,\V$, 
while 
full efficiency is obtained at a voltage slightly below $1600\,\V$.  
In order to limit cathode currents below $15\,\muA$, DC1 was operated
at $1650\,\V$, DC2/3 at $1750\,\V$. 

The detector output signals are read by the ASD8-B chip \cite{Newcomer:93},
an eight-channel preamplifier / amplifier / discriminator with $3\,\ns$
rise time. Eight ASD8 chips are mounted on a single analogue board
of 64 channels. 
A discriminator threshold of about $25000\,\mathrm{e}^-$ is set to
each pair of wire layers (352 wires); it can be independently adjusted
on each ASD8 chip.   
The discriminated
signals are sent to the F1-TDC digitisation board. The
64-channel F1-TDC board used for the drift chambers is identical to the
one used for the Micromegas
detectors (see Sec.~\ref{sec:tracking.sat.micromegas}). 

Particular care was taken in minimising the electronic noise of the
whole system. In both analogue and digital boards the signal wire 
layers are enclosed in between two grounded copper planes. Multiple
connections 
are made between the board ground and the detector structure;
this common ground is also shared with the
low-voltage power supplies. As a result the measured
noise figure at nominal
threshold is in the range 
$10-100\,\Hz/\mathrm{channel}$. 
   
\begin{figure}[tbp]
  \begin{center}
    \includegraphics[width=\columnwidth]{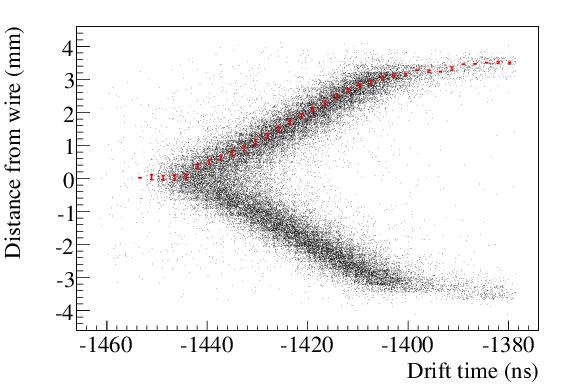}
  \end{center}
  \caption{Measured RT relation for one layer of a DC.}  
  \label{fig:tracking.lat.dc.rt}
\end{figure}

The performances of the DCs were studied at several beam
conditions, the incident flux being roughly proportional to the beam
intensity. 
Figure~\ref{fig:tracking.lat.dc.rt} shows the RT relation 
for one layer of a DC, measured in a low intensity beam. 

At nominal COMPASS beam 
conditions the mean layer efficiency is $95\%$ or higher, the
efficiency being higher for the DC detectors located downstream of
SM1. Nearly all inefficiency is due to pile-up at high
rates. 
% Fig.~\ref{fig:trk_dc.eff2D} shows a bi-dimensional plot of a
% horizontal wire layer obtained in nominal conditions. 
% \begin{figure}[tbp]
%   \begin{center}
%     \includegraphics[width=\columnwidth]{./figures/trk_dc_eff2D}
%   \end{center}
%   \caption{\small Two dimensional efficiency of a DC layer at nominal
%     beam conditions.}
%   \label{fig:trk_dc.eff2D}
% \end{figure}
% The deactivated
% zone in the middle is clearly seen as well as the electrical
% connection to it, along which the inefficiency is maximum. We note
% that this local inefficiency is compensated by the redundancy of the
% whole system, so that the effect on the global track reconstruction is
% negligible.

\begin{figure}[tbp]
  \begin{center}
    \includegraphics[width=\columnwidth]{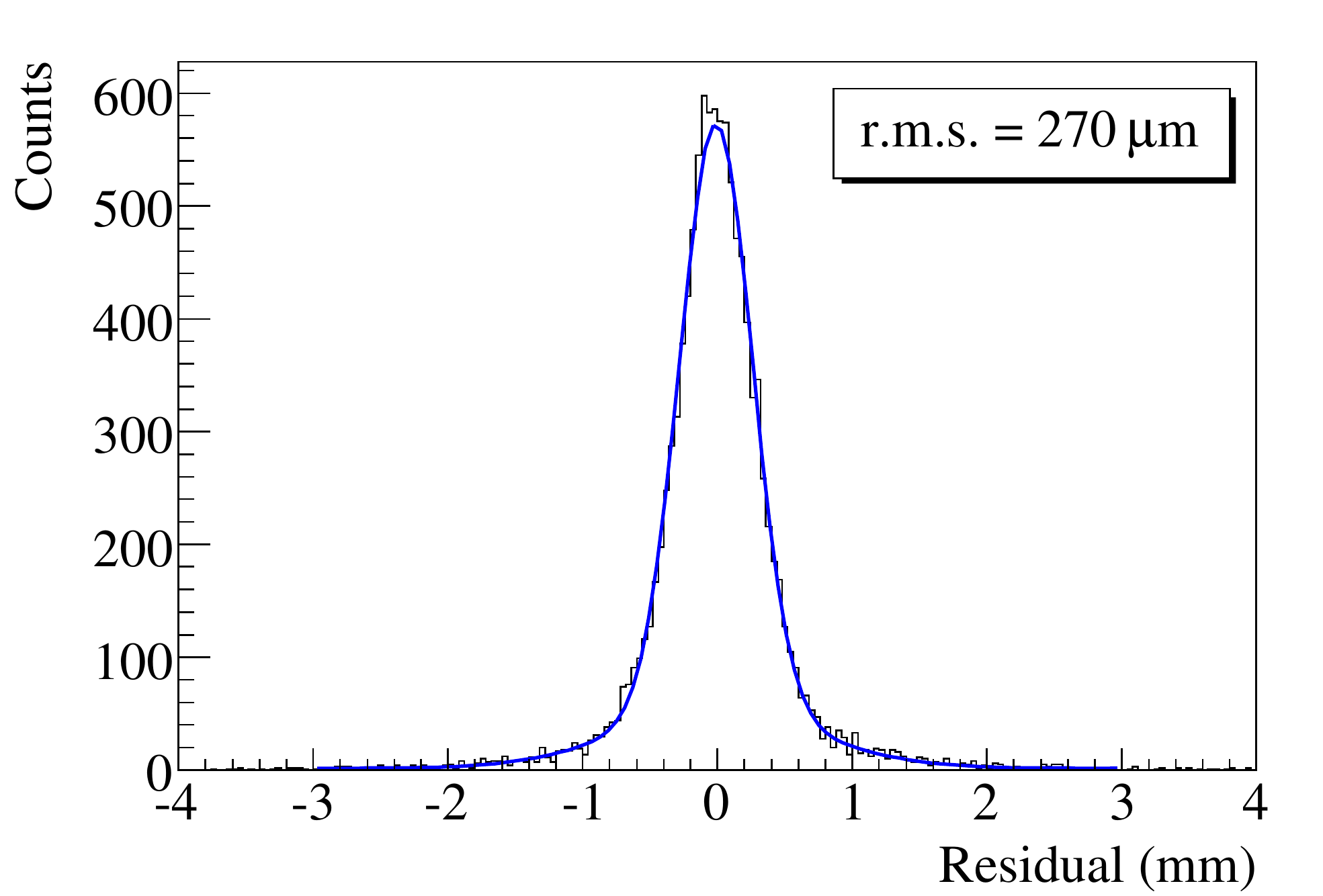}
  \end{center}
  \caption{\small Single DC layer residual distribution in the high
    intensity 
    beam (here: DC1 $X$).} 
  \label{fig:tracking.lat.dc.resol}
\end{figure}

The spatial resolution of the DC detectors was evaluated using the
residuals of the fitted tracks for each wire layer. Combining the
residuals 
from two wire layers with the same orientation allows us to separate
the 
intrinsic layer resolution from the uncertainty due to the track
fitting. 
At nominal muon beam intensity a mean value for the resolution of a
single DC wire 
layer of $270\,\mum$ was measured, averaged over all layers and
over the full active surface, with maximum deviations from this value
of $\pm 20\,\mum$ (see Fig.~\ref{fig:tracking.lat.dc.resol}). 
Compared to the expected intrinsic resolution, this value includes a
deterioration due to two effects:
(i) the large  
halo coming along with the high intensity beam, which contributes
about $30\,\mum$, and (ii)   
the large fringe field of the SM1 magnet present in all DCs, which 
exerts a Lorentz force on the drifting electrons, 
and contributes about $20\,\mum$ for the $X$, $U$ and $V$ layers.  
For the $Y$ layers the electrons drift mainly parallel to the magnetic
field, limiting the Lorentz effect to a few $\mum$ only. 
%Part of this effect is accounted for in the corresponding RT
%relation. 
The resulting resolution of one DC (8 wire layers) for the
horizontal coordinate (bending direction of the dipole) is 
$110\,\mum$, while it is $170\,\mum$ for the vertical coordinate.

%%% Local Variables: 
%%% mode: latex
%%% TeX-master: t
%%% End: 

\subsubsection{Straw tube chambers}
\label{sec:tracking.lat.straw}
Straw tube drift chambers \cite{Bychkov:05} are used for the tracking
%and momentum determination 
of charged particles produced at large scattering angles
($15-200\,\mrad$) in the Large Area Tracking section (LAT) of
COMPASS downstream of the first spectrometer magnet
(SM1). 

The straw tubes are made of two layers of thin plastic
films. The 
inner layer consists of a carbon loaded 
Kapton$^{\mathrm{\circledR}}$ 160 XC 370 \cite{Dupont} foil with a
thickness of $40\,\mum$. It is glued onto the second layer, an
aluminised Kapton$^{\mathrm{\circledR}}$  
foil of $12\,\mum$ thickness.
The anode wires are made of gold-plated tungsten with $30\,\mum$
diameter. They are centred in the straw tubes by two 
end-plugs and four small plastic spacers, which are positioned
at intervals of about $60\,\cm$ along each tube. The counting gas is
supplied through the end-plugs and a
gas-manifold, which is integrated into the aluminium frame construction.

\begin{figure}[tbp]
  \begin{center}
    \includegraphics[width=\columnwidth]{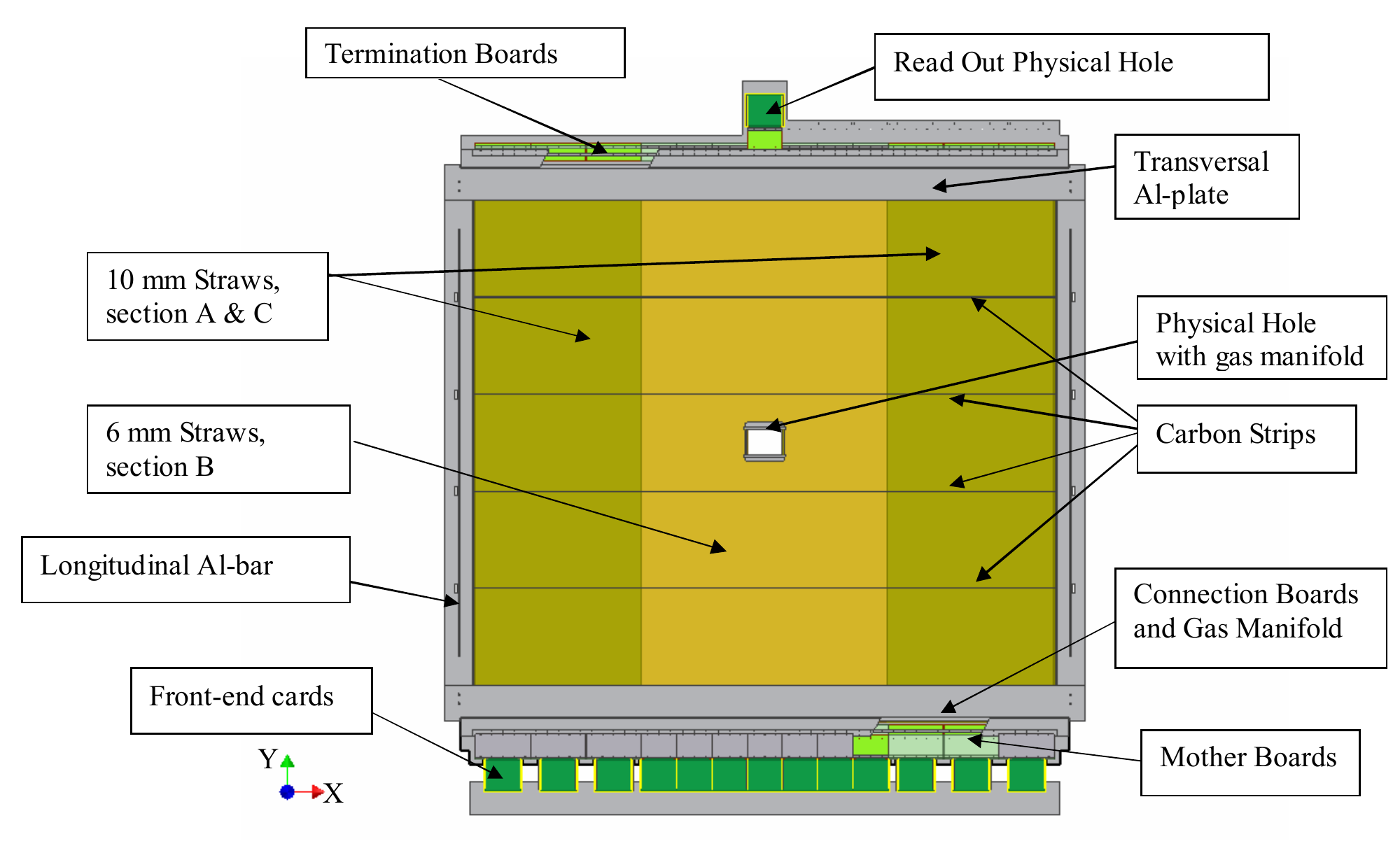}
  \end{center}
  \caption{Schematic view of a COMPASS straw detector (type X)
    \cite{Bychkov:05}.} 
  \label{fig:tracking.lat.straw.detector}
\end{figure}

In total 12440 straw tubes are assembled into 15 detectors. To resolve
left-right ambiguities of a particle 
trajectory, each
detector consists of two staggered layers of 
straws, which are glued together and mounted onto an aluminium
frame for 
mechanical stability. 
Each detector has an active area of about $9\,\m^2$ and 
is divided into three sections. 
Fig.~\ref{fig:tracking.lat.straw.detector} shows a schematic view of a
detector (type X). 
The central part (section B, see
Fig.~\ref{fig:tracking.lat.straw.detector}), being closer to the beam
axis, is exposed to higher rates. This part is made of 190 long and 64
short straws per layer, all with an outer diameter of 
$6.14\,\mm$, forming a central dead zone of about $20\times
20\,\cm^{2}$.  
It has a rectangular hole without material of about $20
\times 10\,\cm^{2}$ for the beam.
The outer two parts (section A and C) each 
have 96 straws with $9.65\,\mm$ outer diameter. The chosen
diameters are a compromise between minimising the number of channels
and production cost, and keeping the occupancy in each tube below
$2\%$ at maximum beam rates. As a fast counting gas a mixture of
Ar/CO$_{2}$/CF$_{4}$ (74/6/20) is used. The straw tubes are operated
at a high voltage of $1950\,\V$, corresponding to a gain of
$6\EE{4}$. 

In order to measure three projections of a particle trajectory, one
station 
consists of three detectors, one with vertical,  
one with horizontal and one with inclined straw tubes. 
The detectors with 
inclined straws are rotated by $10^\circ$ with respect to the
vertical ones. The detectors with vertical and inclined straws are of
the same type (called type X), while the ones with horizontal
straws have a slightly different geometry (type Y). The exact
dimensions of both types are given in
table~\ref{tab:tracking.lat.straw.geometry}.
\begin{table*}[tbp]
  \caption{Geometrical properties of $X$ and $Y$ types of straw
    detectors.  
  }
  \label{tab:tracking.lat.straw.geometry}
  \begin{center}
    \begin{footnotesize}
%      \begin{tabular}{@{\extracolsep{\fill}}|c|p{2.3cm}|p{1.5cm}|p{1.5cm}|p{1.5cm}|p{1.7cm}|p{2cm}|}   
      \begin{tabular*}{\textwidth}{@{\extracolsep{\fill}}cp{2.3cm}p{1.5cm}p{1.5cm}p{1.5cm}p{1.7cm}p{2cm}}   
        \hline 
        Type&\mbox{Sensitive area} $X\times Y$ ($\mm^2$)& Length \mbox{of straws}
        ($\mm$)&\multicolumn{2}{p{3.0cm}}{Number of straws with
          outer diameter of}&Number \mbox{of readout} channels&
        Overall dimensions $X\times Y$ ($\mm^2$) \\  
        &         &  & $6.14\,\mm$  & $9.65\,\mm$ &   & \\ \hline 
        &                        & 3202 & 380 &    384             &
        &                       \\  
        \raisebox{1.5ex}[0cm][0cm] {X}  &  \raisebox{1.5ex}[0cm][0cm]
        {$3232\times 2802$} & 1523 &  128 &         &
        \raisebox{1.5ex}[0cm][0cm] {892} &  \raisebox{1.5ex}[0cm][0cm]
        {$3570\times 4117$}\\ \hline 
        &                     & 3652 & 320 &   256      &        &
        \\   
        \raisebox{1.5ex}[0cm][0cm] {Y} &  \raisebox{1.5ex}[0cm][0cm]
        {$3254\times 2427$} &1752 & 128 &  &  \raisebox{1.5ex}[0cm][0cm]
        {704} &  \raisebox{1.5ex}[0cm][0cm] {$4567\times 3160$} \\ \hline      
      \end{tabular*}   
    \end{footnotesize}
  \end{center}
\end{table*}

The thickness of one detector along the beam direction is 
$40\,\mm$. The amount of material in the active part of the detector
was minimised with respect to scattering and secondary
interactions, and corresponds to $0.2\%$ of a radiation length for one
detector, not taking into account the detector gas.  

The length of a COMPASS straw tube increases with increasing
humidity. The relative elongation was measured to be  
about $3\EE{-5}$ for a humidity change of $1\%$ \cite{Bychkov:05},
resulting in increased tension on the frame or, even worse, in a
bending of straws. 
In order to keep the humidity constant, each straw
station is surrounded by a protective gas volume of N$_2$ gas,
enclosed by $12\,\mum$ thin Mylar$^{\mathrm{\circledR}}$ \cite{Dupont}
foils which are aluminised on both 
sides. 
%This 
%adds $\%$ of a radiation length to the total material budget of a
%straw detector. 

The mechanical 
precision of wires averaged over all detectors was determined to be
$170\,\mum$ (r.m.s.) by measuring the coordinates of the wires at all
spacer and end-plug positions with a triple stereo imaging X-ray
scanner \cite{Platzer:05}. The deviation of the wire with respect to
its nominal position is then used in the offline data analysis as a correction
to the measured coordinate, reducing the uncertainty of the mechanical wire
position to $60\,\mum$.  

The electrical connections to the anode wires are fed through the gas
tight volume by connecting boards. The straws are read out on the
bottom side only, while they are terminated on the top side by
termination 
boards. 
The signals of the straws are amplified and
discriminated by the ASD8-B chip, already described in
Sec.~\ref{sec:tracking.lat.dc}, and digitised by the F1-TDC chip, 
described in 
Sec.~\ref{sec:daq.digital.f1}.
The discriminator
thresholds are 
typically set to $6\,\fC$, resulting in a noise level of
$5\,\kHz/\mathrm{channel}$.  
Eight ASD8 and F1-TDC chips, corresponding to sixty-four channels, are
assembled on one front-end card. Each front-end card is plugged to a
motherboard, which also contains the high voltage distribution
circuit and a test pulse line for calibration. 
Motherboards and front-end cards are supported by the
chamber frame.   

\begin{figure}[tbp]
  \begin{center}
    \includegraphics[width=\columnwidth]{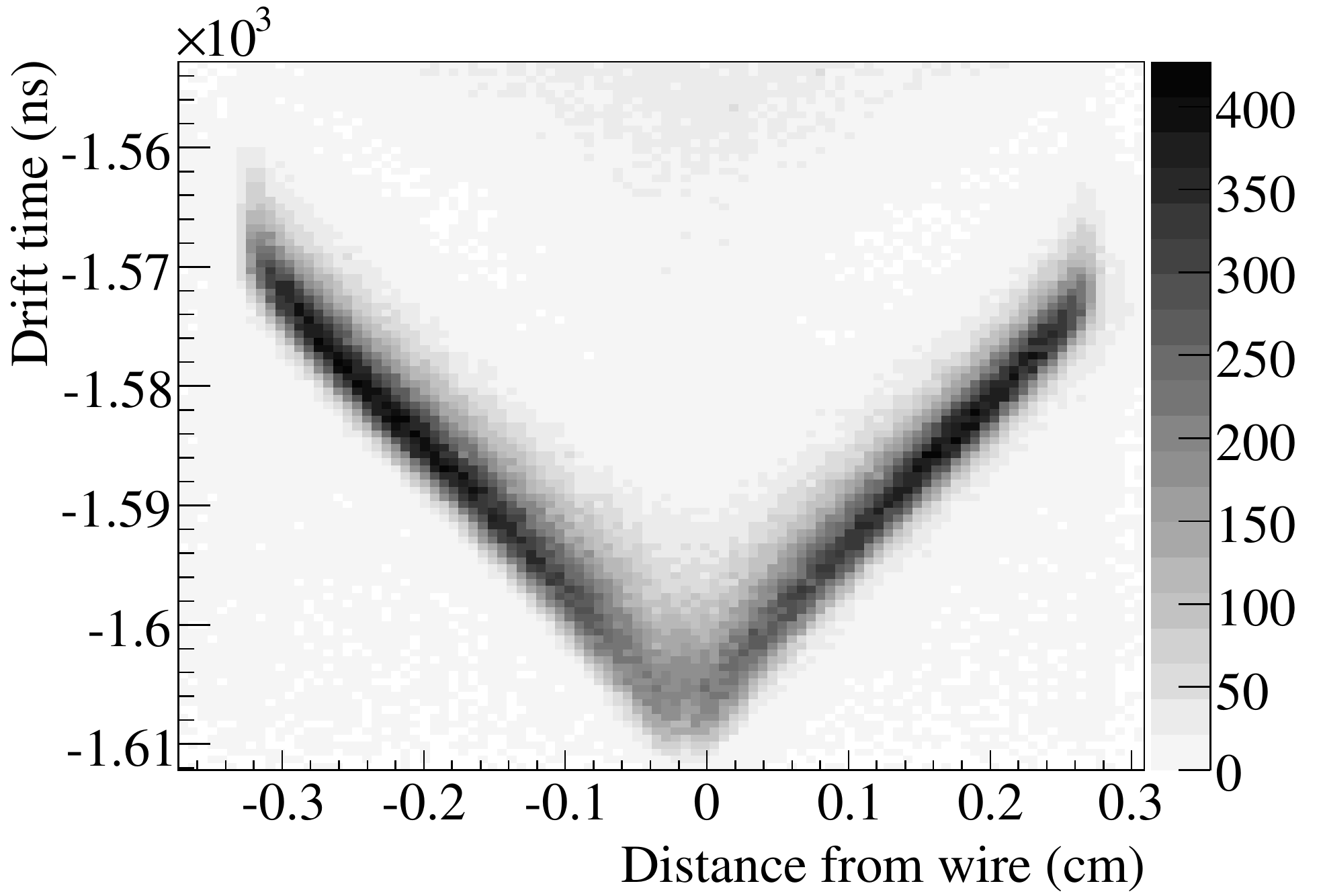}
  \end{center}
  \caption{RT relation for a single $6\,\mm$ straw tube for particle
    tracks measured with the 
    $160\,\GeV/c$ muon beam~\cite{Bychkov:05}.} 
%    assuming 
%    7 points, between which it is interpolated linearly. The picture
%    is taken  
%    from reference~\cite{Bychkov:05}.}
  \label{fig:tracking.lat.straw.rt}
\end{figure}

The performance of the straw tube chambers has been evaluated in both
COMPASS low and high intensity beams.  
At nominal muon beam intensity 
tracks 
reconstructed by the COMPASS offline analysis software (CORAL, 
see Sec.~\ref{sec:performance.reco}) 
were 
used to determine, for each wire separately, the relation between 
the measured drift times
and the 
distances of tracks to the anode wire. 
Figure~\ref{fig:tracking.lat.straw.rt} shows this RT relation for one single
$6\,\mm$ straw tube. 
%By fitting the V-distribution, one
%obtains the RT relation, the coordinate $x_V$-fit of the wire and the
%time offset t0 of the TDC-channel. After these free parameters have
%been fixed, 

\begin{figure}[tbp]
  \begin{center}
    \includegraphics[width=\columnwidth]{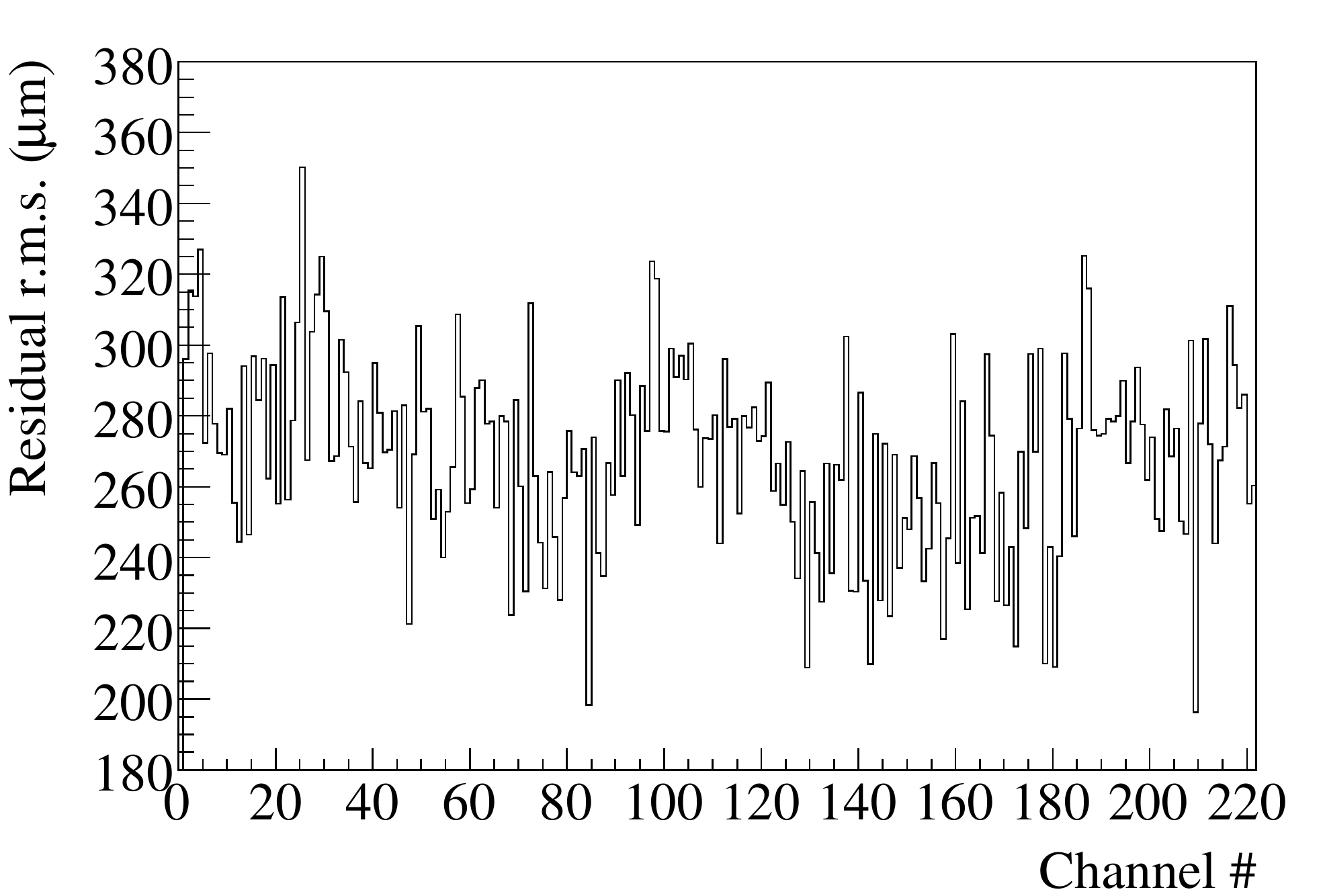}
  \end{center}
  \caption{R.m.s.\ widths of single-channel residual distributions
    versus channels number 
    for all $6\,\mm$ straw
    tubes of one layer.} 
  \label{fig:tracking.lat.straw.residual_vs_ch}
\end{figure}

Once this relation has been established,  
the resolution of a given straw tube can
be determined from the r.m.s.\ width of the distribution of track residuals,
taking into account the corrections from the X-ray scan of wire positions.  
Fig.~\ref{fig:tracking.lat.straw.residual_vs_ch} shows the r.m.s.\
widths of single-channel  
residual distributions versus channel number 
for the central part ($6\,\mm$ tubes) of one straw tube 
layer, where the number of good tracks from other detectors is still
sufficiently high.    
Averaging over several layers of $6\,\mm$ straws, a mean value for the
resolution 
of $270\,\mum$ for one straw layer under nominal COMPASS muon beam conditions
is obtained. For one straw detector (two layers), the average resolution is
then $190\,\mum$. 
Intrinsic resolutions without the effects of temperature variations
and of the overall tracking are given in Ref.~\cite{Bychkov:05}. 
Dedicated beam tests already showed earlier that resolutions are the same for
$6\,\mm$ and $10\,\mm$ straws \cite{Bychkov:05}.  
% Ilgner: 243mum (6mm), 271mum (10mm) single layer resolution. 
In these tests it was also shown that 
the mean efficiency of a straw detector, i.e.\ the probability for
either of the two layers to detect a hit, is higher than $99\%$, with the 
inefficiencies being 
concentrated along the mechanical edges of the detector.  

%%% Local Variables: 
%%% mode: latex
%%% TeX-master: t
%%% End: 

\subsubsection{Multiwire proportional chambers}
\label{sec:tracking.lat.mwpc}
The tracking of particles at large radial distances to the beam in the
SAS is mainly based on a system of multiwire proportional
chambers (MWPC). A total of 34 wire layers, corresponding to about $25000$
detector channels, is installed and operated since the year 2001. All
layers are characterised by a wire length of about $1\,\m$, a wire
diameter of $20\,\mum$, a wire pitch of $2\,\mm$ and an anode/cathode
gap of $8\,\mm$. 

In COMPASS three different types of MWPC are used, named
type-A, type-A$^{\star}$ and type-B. Type-A detectors have three anode
wire 
layers, one vertical ($X$) and two tilted by $\pm 10.14^{\circ}$
with respect to the 
vertical axis ($U$, $V$), and an active area of $178 \times
120\,\cm^2$. 
Type-A$^{\star}$ detectors are similar to type-A, with an additional 
horizontal wire layer ($Y$). Type-B detectors have a smaller active
area 
($178 \times  
80\,\cm^2$) and only two wire layers, one vertical ($X$) and one 
tilted by $10.14^{\circ}$ ($U$ or $V$). Type-B stations are composed of
two detectors with inclined wire layers with opposite orientations,
fixed together; only three layers, one vertical and two tilted, are read
out. 
All wire layers are enclosed from both sides by $10\,\mum$ thick 
graphite coated Mylar$^{\mathrm{\circledR}}$ \cite{Dupont} cathode
foils, to provide field symmetry and to 
enclose the detector gas. 
A central dead zone of $16-22\,\mm$ diameter, depending
on the location of the chamber along the beam axis, was realized by
removing the graphite coating from the foils. The total thickness
corresponds to $0.2\%$, $0.3\%$, and $0.2\%$ of a 
radiation length for A-, A$^\star$-, and B-type chambers,
respectively. 
The detector characteristics are summarised in
Table~\ref{tab:tracking.lat.mwpc.summary}.
\begin{table*}[tbp]
  \caption{Characteristics of the COMPASS MWPC detectors.}
  \label{tab:tracking.lat.mwpc.summary}
  \begin{center}
%    \begin{tabular}{|l|c|c|c|} \hline
    \begin{tabular*}{\textwidth}{@{\extracolsep{\fill}}lccc} \hline
      % & \multicolumn{3}{c|}{\large Omega} \\ \hline
      & A-type & A$^{\ast}$-type & B-type \\ \hline
      \# of chambers   &   7  &  1  &  6              \\
%      External dimensions   &   $202\times 146\,\cm^2$  & $202\times
%      146\,\cm^2$ 
%      & $202\times 96\,\cm^2$                 \\
      Active area          &   $178\times 120\,\cm^2$    & $178\times
      120\,\cm^2$  
      &   $178\times 80\,\cm^2$                             \\
      \# of layers/chamber  &   3  &  4   &   2                           \\
      Planes &   $X$, $U$, $V$ &  $X$, $U$, $V$, $Y$ & $X$, $U/V$         \\
      % &   \multicolumn{3}{c|}{(t=10.14$^{\circ}$ with respect to v)} \\
      Dead zone $\oslash$ & $16 - 20\,\mm$ & $16\,\mm$ &
      $22\,\mm$\\ 
      Wire pitch  &  $2\,\mm$ & $2\,\mm$ & $2\,\mm$  \\
      Anode/cathode gap & $8\,\mm$ & $8\,\mm$ & $8\,\mm$ \\
      \# of wires/plane & 752 ($X$, $U$, $V$), 512 ($Y$) & 752 ($X$,
      $U$, $V$), 512 ($Y$) & 752 ($X$, $U$, $V$), 512 ($Y$)  \\
      % HV:                   &   \multicolumn{3}{c|}{?.? KV} \\
      % readout:              &   \multicolumn{3}{c|}{chip MADIV} \\
      % gas                   &   \multicolumn{3}{c|}{?} \\
      % provenienza:          & \multicolumn{3}{c|}{Spettrometro OMEGA}\\
      % totale fili:          &   \multicolumn{3}{c|}{~27000}  \\
      \hline
      % \multicolumn{4}{l}{\small{(1): v = vertical, t = 10.14$^{\circ}$
      %     w.r.t. v, h = horizontal}} 
    \end{tabular*}
  \end{center}
\end{table*}

The MWPC are operated with a gas mixture of Ar/CO$_2$/CF$_4$ in
proportions 74/6/20. The addition of a fast gas 
as CF$_4$ is a crucial requirement to operate MWPC detectors in high
rate environments, without introducing
excessive detector dead time. The typical time jitter of the wire
signal with the chosen gas mixture is about $80\,\ns$, 
compatible with the typical electron drift velocity in the
gas mixture used. At nominal high voltage of $4250\,\V$ the gain is between
$3.5\EE{4}$ and $4\EE{4}$. 

The read-out electronics is distributed on two printed circuit boards:
the mother board, 
fixed to the chamber frame, and the
front-end board, where the signals of the chamber are discriminated and
digitised. The data is transferred through a fast serial link to the
CATCH modules.
%\begin{figure}
%  \centering
%  \includegraphics[width=0.9\columnwidth]{figures/tracking_mwpc_readout}
%  \caption{Scheme of the front-end card housing the
%    preamplifier/discriminators, digitising electronics, and
%    interfaces.}
%  \label{fig:tracking.lat.mwpc.readout}
%\end{figure}
Figure~\ref{fig:compass.daq.mwpc.readout} (see
Sec.~\ref{sec:daq.catch.cmc}) shows the 
functional block 
diagram of the read-out electronics.  

The front-end electronic is organised in triplets of cards that can
read 64 channels per board. Each card houses the MAD4
preamplifier/discriminator chips \cite{Gonella:01} 
with a peaking time of $5\,\ns$, 
%\cite{pegoraro,ieee99},
%originally developed for the readout of the muon 
%detectors of the CMS barrel, 
%the digitising chips F1 
the F1-TDC chips 
(see Sec.~\ref{sec:daq.digital.f1}), 
and the threshold DACs. The three cards are connected together by a
fast LVDS bus, which distributes the reference clock,
trigger and 
synchronisation signals. Production cost has been reduced by housing
both analogue and digital parts on the same PCB. Moreover central and
side boards share a common design, except for the bus 
%arbiter 
controller 
and
the parallel-to-serial converter chips for fast
data transmission to the CATCH module (see
Sec.~\ref{sec:daq.catch}), which are 
housed only on the central board. For noise prevention all the
voltages required by the card components are regulated on--board, and
separate grounds for analogue and digital parts are used. 
A test system has been developed, capable of injecting a charge
pulse similar to the signal of a minimum ionising 
particle crossing the chamber into each
wire. The system is composed of a VME control
board and pulser boxes fixed to the frame of each MWPC detector. It
allows tests and calibration of the front-end cards and of the
read-out system, including the input connectors.

\begin{figure}[tbp]
  \centering
  \includegraphics[width=\columnwidth]{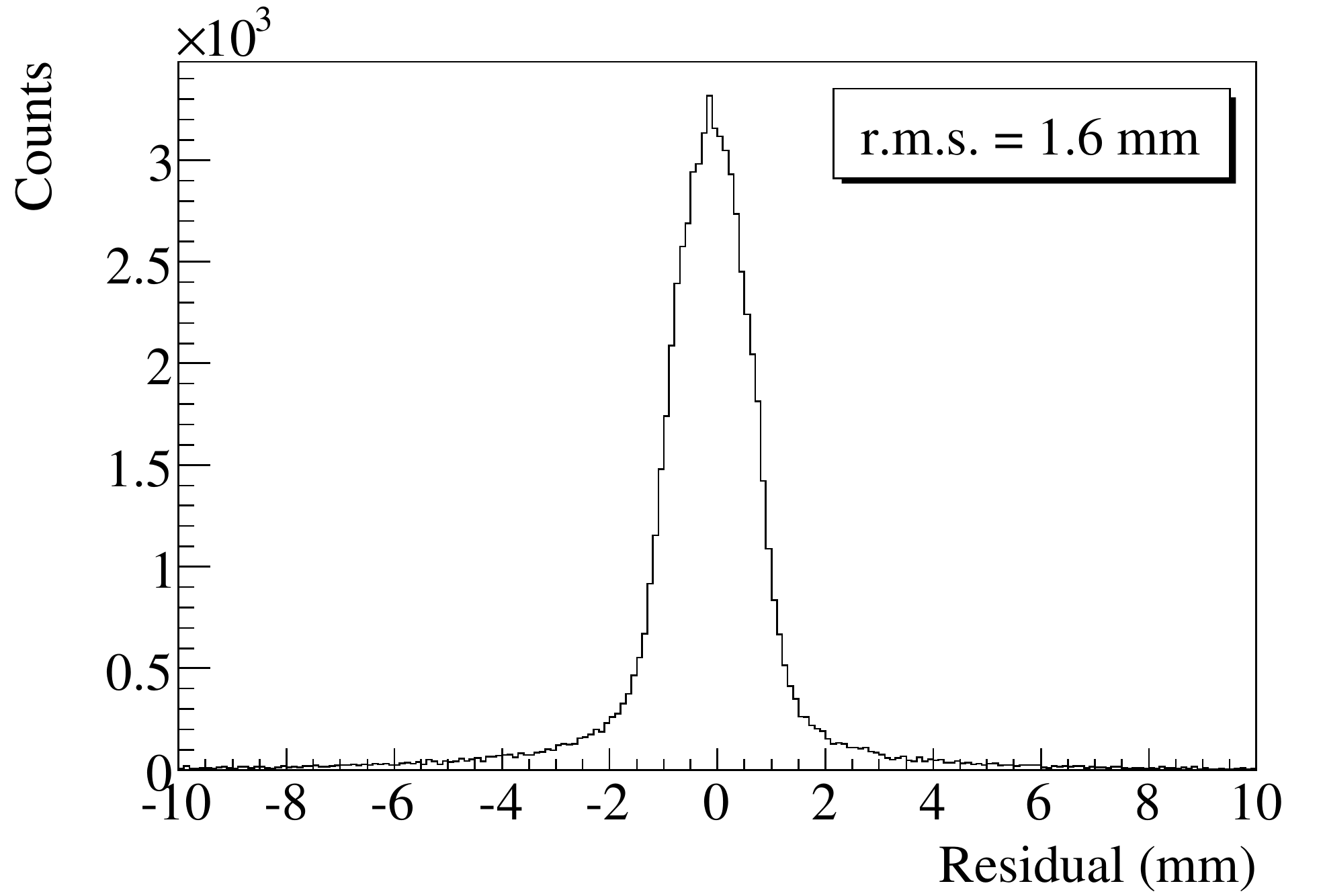}
  \caption{Typical MWPC residual distribution for standard physics
    conditions, showing the difference 
    between wire cluster position and extrapolated track position
    along the axis perpendicular to the wire layer.} 
  \label{fig:tracking.lat.mwpc.resi1}
\end{figure}

\begin{figure}[tbp]
  \centering
  \includegraphics[width=\columnwidth]{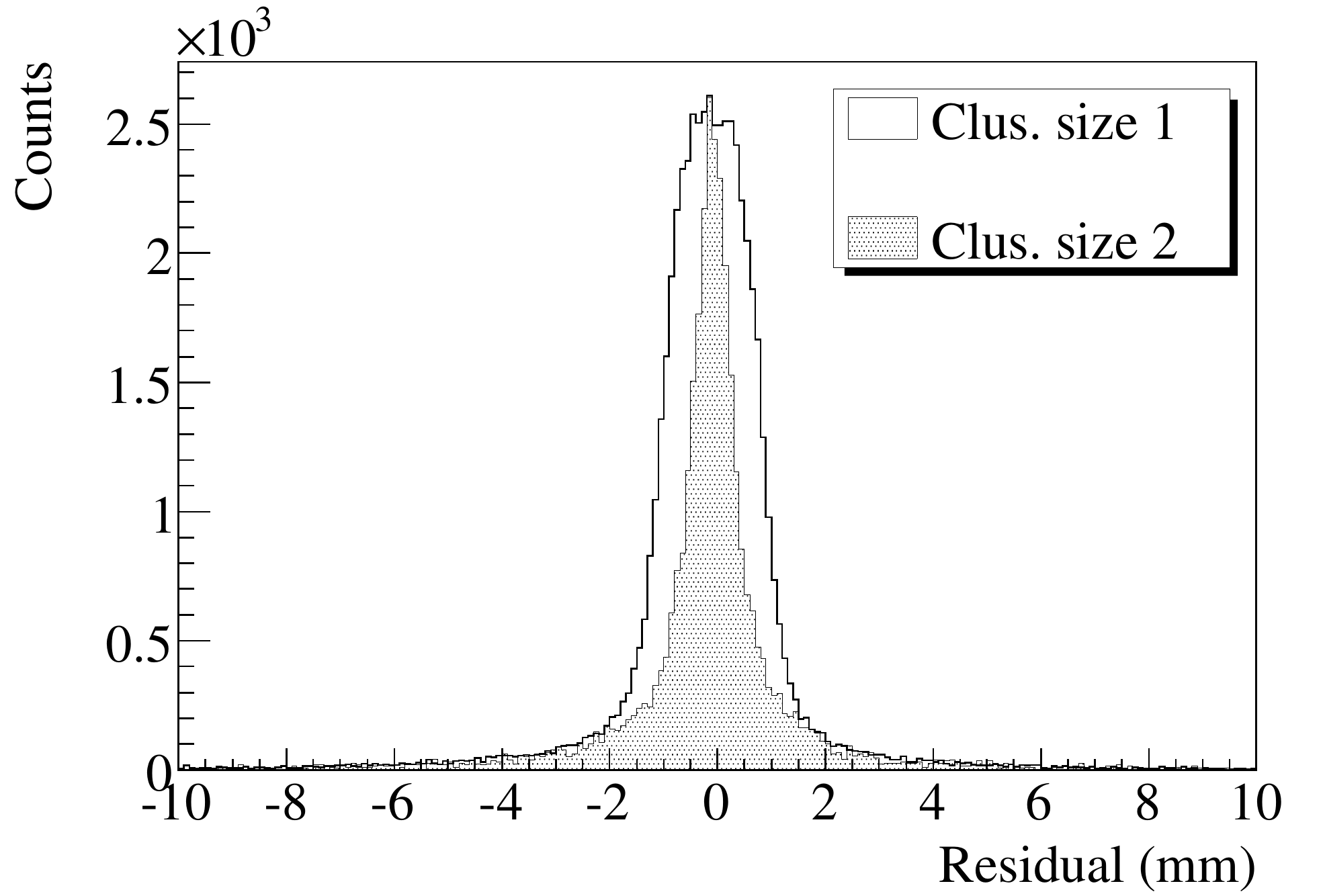}
  \caption{Residual distributions of an MWPC layer for 
    cluster sizes 1 (transparent) and 2 (grey).} 
  \label{fig:tracking.lat.mwpc.resi2}
\end{figure}

In a typical MWPC an ionising particle traversing the detector may
induce a signal on several neighbouring wires. The position of a hit
can then be calculated from the weighted mean of coordinates of
adjacent wires with signals (clusters). 
For the COMPASS MWPC, the cluster size distribution has an
approximately exponentially decreasing shape with a probability of
about $30\%$ for 
a cluster size larger than unity.  
%The typical cluster 
%size distribution of the MWPC is shown in
%Fig.~\ref{fig:tracking.lat.mwpc.cls}; the probability of having more
%than one channel fired is about 30\%. Hit clusterisation has to be
%applied to achieve the best spatial resolution.  
%\begin{figure}
%  \begin{center}
%    % \includegraphics[width=\textwidth]
%    % {compass_figures/2002/mwpc_anal/high_i/PA04X1_cls}
%  \includegraphics[width=0.9\columnwidth]{figures/tracking_mwpc_cluster_size}
%  \end{center}
%  \caption{Cluster size distribution for one MWPC layer. The
%    probability of cluster size $> 1$ is around 30\%.}
%  \label{fig:tracking.lat.mwpc.cls}
%\end{figure}
The spatial resolution of each single MWPC layer was measured in a 
standard high-intensity muon beam using
a sub-sample of tracks reconstructed in the GEM 
stations coupled to the MWPCs. The MWPC layers have been excluded from the 
tracking. The residual distribution has an r.m.s.\ width 
of $1.6\,\mm$ (see Fig.~\ref{fig:tracking.lat.mwpc.resi1}). 
%as expected from 
%a proportional counter with $2\,\mm$ wire spacing. 
The beneficial effect of wire clusterisation on the measured spatial
resolution is  
evident 
from a separate analysis of clusters with different sizes, as shown in 
Fig.~\ref{fig:tracking.lat.mwpc.resi2} for cluster size one and two. 
%The grey histograms represent the measured hit residual after clusterisation, 
%while the white histograms show the residual distribution of each hit wire 
%separately. The improvement given by the clusterisation procedure is clearly 
%visible.

\begin{figure}[tbp]
  \centering
  \includegraphics[width=\columnwidth]{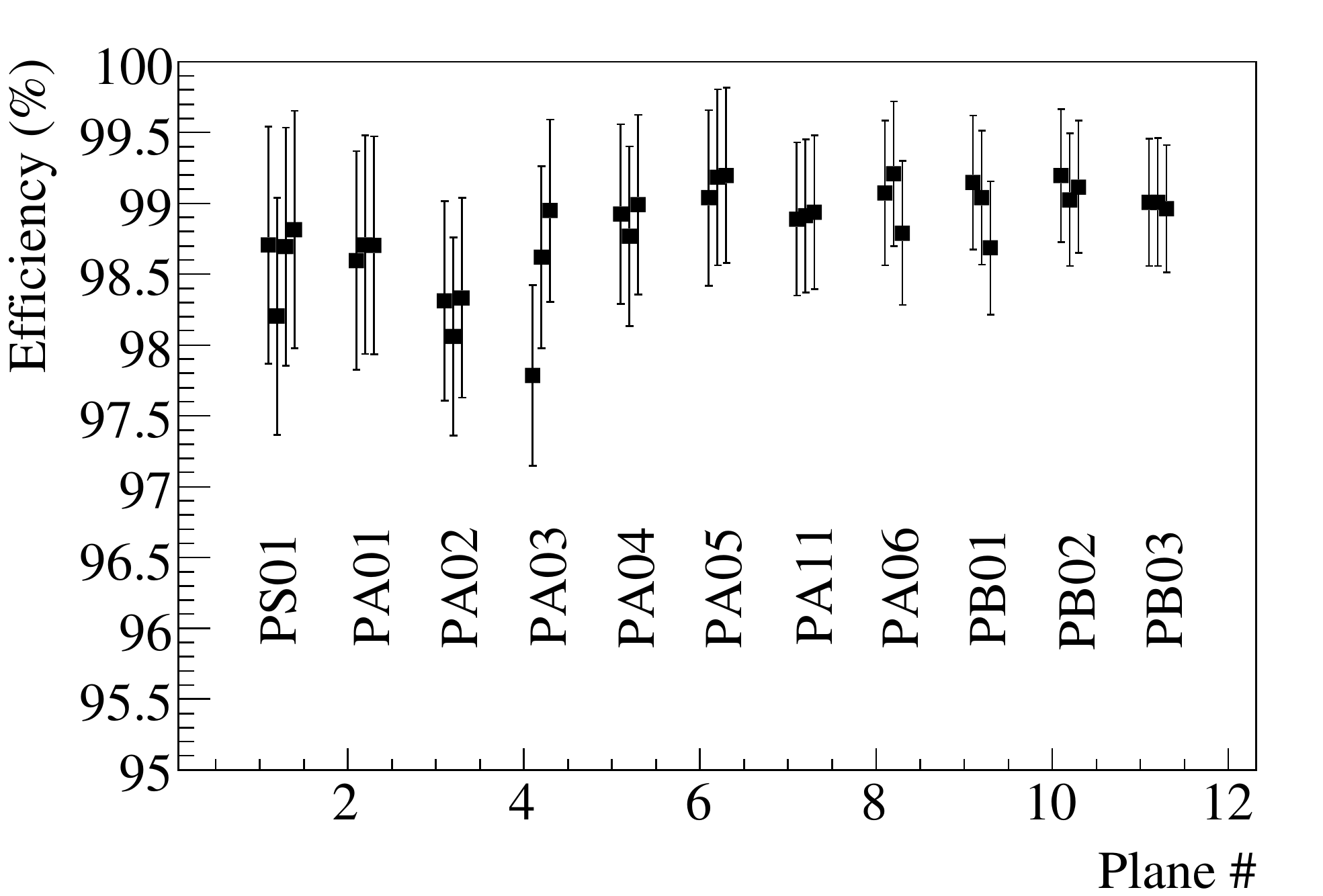}
  \caption{Efficiencies of all MWPC stations during the 2004 data
    taking period. The average efficiency is $>98\%$.} 
  \label{fig:tracking.lat.mwpc.eff}
\end{figure}

The MWPC detectors have been operated in stable conditions at nominal
high voltage since the year 2001. Fig.~\ref{fig:tracking.lat.mwpc.eff}
shows the average efficiencies of all MWPC stations during the 2004
data taking period. 

%%% Local Variables: 
%%% mode: latex
%%% TeX-master: "~/tex/compass/paper/spectro2005/compass_spec"
%%% End: 

\subsubsection{Large area drift chambers}
\label{sec:tracking.lat.w45}
To provide tracking for charged particles deflected by a large 
angle in the COMPASS SAS a system of six large area drift 
chambers is used. A detailed description of the prototype chamber can
be found in \cite{Brasse:76}. The basic detector characteristics are
summarised 
in Table~\ref{tab:tracking.lat.w45.summary}. 
\begin{table*}[tbp]
  \caption{Basic characteristics of the COMPASS large area drift chambers.}
  \label{tab:tracking.lat.w45.summary}
  \begin{center}
%    \begin{tabular}{|l|c|c|c|c|c|} \hline
    \begin{tabular*}{\textwidth}{@{\extracolsep{\fill}}lccccc} \hline
      &  $XY$-type  & $XV$-type  & $XU$-type& $YV$-type &
      $YU$-type  \\ \hline
      \# of chambers  &   2       &  1  &  1 & 1 & 1    \\
      Active area      &$500\times 250\,\cm^2$&$500\times 250\,\cm^2$ &
      $500\times 250\,\cm^2$ & $500\times 250\,\cm^2$ & $500\times
      250\,\cm^2$ \\ 
      \# of layers/chamber   &    4      &  4  &  4 & 4  & 4     \\
      Planes & $X$, $Y$ &  $X$, $V$   & $X$, $U$ & $Y$, $V$ & $Y$, $U$
      \\
      Dead zone $\oslash$ & $500\,\mm$ & $1000\,\mm$ & $1000\,\mm$ &
      $1000\,\mm$ & $1000\,\mm$ \\
      Anode wire pitch  & $4\,\cm$ & $4\,\cm$ & $4\,\cm$ & $4\,\cm$ & $4\,\cm$  \\
      Anode/cathode gap   & $10\,\mm$ & $10\,\mm$ & $10\,\mm$ &
      $10\,\mm$ & $10\,\mm$  \\
      \# of wires/plane & 260 ($X$), 130 ($Y$) & 260 ($X$), 288 ($V$)
      & 
      260 ($X$), 288 ($U$) & 130 ($Y$), 288 ($V$) & 130 ($Y$), 288 
      ($U$) \\
%      HV                  &   \multicolumn{5}{c}{$1925\,\V$} \\
%      Readout             &   \multicolumn{5}{c}{MAD4 + F1-TDC} \\
%      Gas mixture         & \multicolumn{5}{c}{Ar/CF$_4$/CO$_2$ (85/10/5)}\\
%      Space resolution/plane &   \multicolumn{5}{c}{ $450\,\mum$}  \\
      \hline
      % \multicolumn{4}{l}{\small{(1): v = vertical, t+ = 30$^{\circ}$
      %     w.r.t. v, h = horizontal, t- = -30$^{\circ}$ w.r.t. v }} 
    \end{tabular*}
  \end{center}
\end{table*}

Each chamber has an active area of $5
\times 2.5\,\m^2$, and consists of 4 sensitive anode wire layers with a
wire pitch of $4\,\cm$, separated by layers of cathode wires with a
pitch of $2\,\mm$. 
The
total number of  
readout channels is 
$2750$. 
All chambers have two planes, each plane consisting of two wire
layers. 
Four of the chambers have $X$ layers, coupled with $Y$ ($XY$-type), $V$  
($+30^{\circ}$ with respect to the $X$ layer, $XV$-type), or $U$
($-30^{\circ}$, $XU$-type) layers. The two other chambers are of
$YV$-type and $YU$-type.   
%All chambers have two $X$ wire layers, and either two $Y$, $V$
%($+30^{\circ}$ with respect to the $X$ layer), or $U$ ($-30^{\circ}$) layers. 
%Two chambers are 
%composed of two $X$ (vertical) and two $Y$ (horizontal) layers, two chambers
%of two $X$ and two $V$ (inclined by $+30^{\circ}$ with
%respect to the $X$
%layer), and two 
%chambers with two $X$ and two $U$ (inclined by $-30^{\circ}$ with
%respect to the $X$ layer)  
%layers.
% The
% orientations of the planes are $X$, $Y$, $V$ ($+30^{\circ}$) and $U$
% ($-30^{\circ}$ with respect to the $X$ layer).   
The two wire layers with the same orientation within one chamber are 
shifted by half the wire pitch. 
  
The diameter of
the anode wire is $20\,\mum$, while the cathode is made of $100\,\mum$
wires. The anode/cathode gap is $10\,\mm$. The cathode 
wires are inclined by $5^{\circ}$ with respect to vertical direction to 
provide 
better field homogeneity. The signal wires are separated 
with field wires of $200\,\mum$ diameter. 

\begin{figure}[tbp]
  \centering
  \includegraphics[width=\columnwidth]{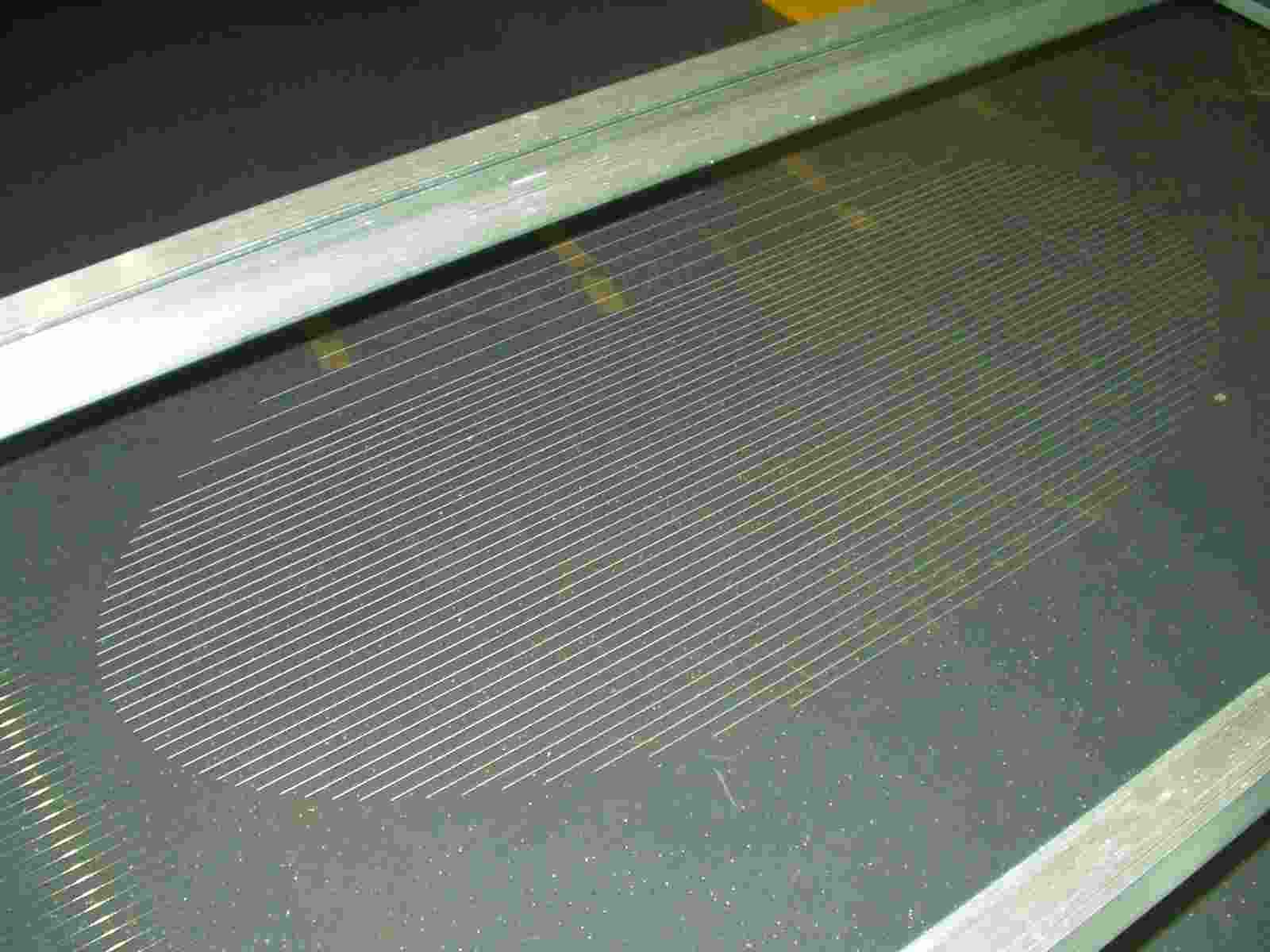}
  \caption{The dead region of a large area drift chamber. In the central 
  part of the chamber (diameter 1 meter) the sensitive wire thickness 
  is 5 times larger than in the active part of the chamber.} 
  \label{fig:tracking.lat.w45.dead_r}
\end{figure}

To guarantee a reliable operation in a high particle rate  
environment, a dead region with a diameter of $0.5\,\m$ and of $1\,\m$
was    
implemented in the centre of each layer 
of $XY$-type,  
and of $XV$, $XU$, $YV$ and $YU$-type chambers, respectively. 
To this end a galvanisation technique was developed in order to enlarge
the radius of the sense wires up to $100\,\mum$, thus preventing 
gas amplification in this region (see
Fig.~\ref{fig:tracking.lat.w45.dead_r}).

A fast CF$_4$-based gas mixture,  
Ar/CF$_4$/CO$_2$ (85/10/5), 
is used to increase the drift velocity, which is important for 
efficient track  
reconstruction. The signal wires are operated at a high voltage of
$1925\,\V$, the field wire potential is kept at $-800\,\V$.
%%% corresponding to a gain of $xxx$.  

Similarly to the MWPCs (see Sec.~\ref{sec:tracking.lat.mwpc}), 
the readout electronics is based on MAD4 amplifiers/discriminators 
and F1-TDC chips. 
% In contrast to the MWPCs, the 
% $F1$ chips are used in TDC mode (drift time measurements), not in 
% LATCH mode as for the MWPCs (fired channel mode). 
%the thresholds were set 
%up by a separate VME-CPU driven line. 
A threshold value of $5\,\fC$ is 
used. 

\begin{figure}[tbp]
  \centering
  \includegraphics[width=\columnwidth]{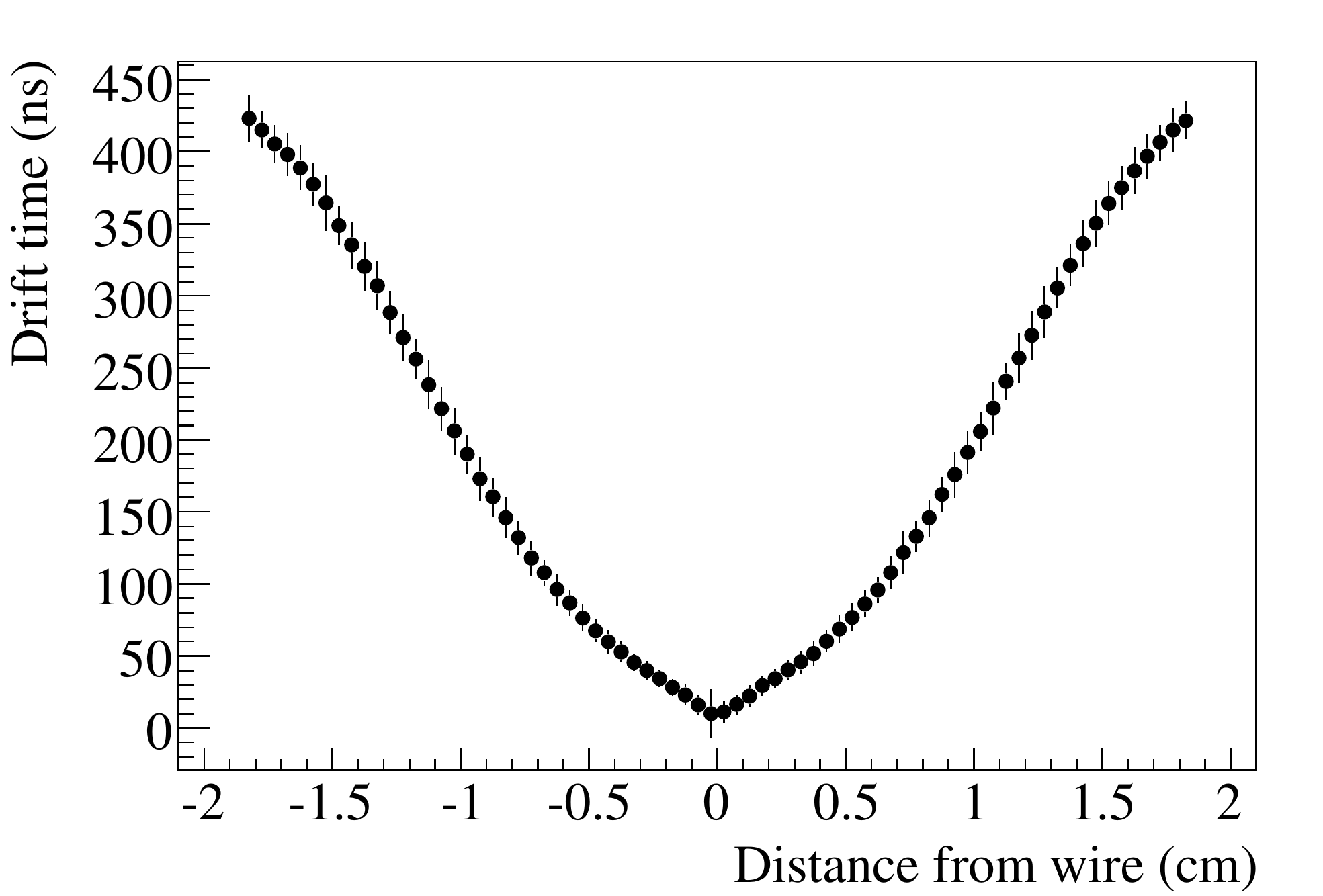}
  \caption{RT relation for one large area drift chamber.}
  \label{fig:tracking.lat.w45.rt}
\end{figure}

The average layer efficiency was measured to be $93\%$. 
The gas mixture and high voltage values were optimised in order to
homogenise   
the drift velocity in the drift gap. 
The resulting RT relation plot of one large area drift chamber 
is shown in Fig. ~\ref{fig:tracking.lat.w45.rt}.
A mean spatial resolution of $0.5\,\mm$ was achieved in the 2004 run.

%

%%% Local Variables: 
%%% mode: latex
%%% TeX-master: "compass_spec"
%%% End: 

% Particle Id

\section{Particle identification}

\label{sec:pid}

The COMPASS Large Angle and Small Angle Spectrometers (LAS and SAS) include 
several particle identification detectors. A RICH counter located in the 
first spectrometer (RICH-1, see Sec.~\ref{sec:pid.rich}) separates 
outgoing hadrons into pions, kaons and protons, up to momenta as 
large as $43\,\GeV/c$. Two hadron calorimeters 
(HCAL1 and HCAL2, see Sec.~\ref{sec:pid.calo.hcal1} and 
\ref{sec:pid.calo.hcal2}) measure the energy of hadrons and provide 
a complementary trigger signal. An electromagnetic calorimeter 
(ECAL2, see Sec.~\ref{sec:pid.calo.ecal}) determines the energies of the 
photons and electrons emitted at small angles. Finally, in both LAS and SAS, scattered muons are 
detected in two muon wall systems (MW1 and MW2, see 
Sec.~\ref{sec:pid.muon.mw1} and \ref{sec:pid.muon.mw2}) both consisting 
of medium resolution tracking detectors combined with a hadron absorber.

\subsection{RICH-1 detector}

\label{sec:pid.rich}

%%%%%%%%%%%%%%%%%%%%%%%%%%%%%%%%%%%%%%%%%%%%%
%\subsection{RICH-1 detector} 
%\label{sec:pid.rich} 
%%%%%%%%%%%%%%%%%%%%%%%%%%%%%%%%%%%%%% 
%% Written by S. Dalla Torre, edited by S. Platchkov 
%%%%%%%%%%%%%%%%%%%%%%%%%%%%%%%%%%%%%% 
 
\begin{figure*}[tbp]  
  %%%%% 2 columns !  
  \begin{center}  
    \includegraphics[width=\textwidth]{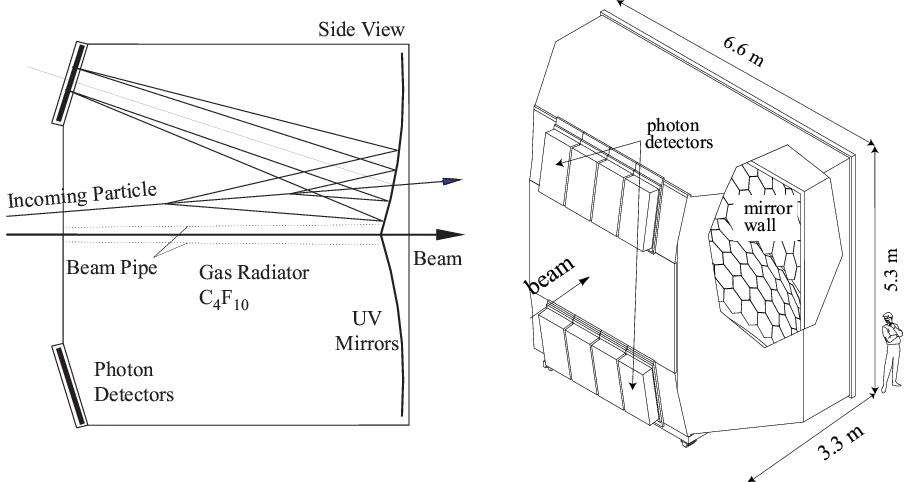}  
  \end{center}  
  \caption{\small COMPASS RICH-1: principle and artistic view.}  
  %%%%%%% from NIMA 433, 207  
  \label{fig:pid.rich.principle}  
\end{figure*}

The COMPASS RICH-1 \cite{Albrecht:05} is a large-size Ring Imaging   
Cherenkov detector which performs hadron identification in the domain  
between $5\,\GeV/c$ and $43\,\GeV/c$. It has large transverse  
dimensions (it covers the whole angular acceptance of the COMPASS LAS,  
i.e. $\pm250\,\mrad$ in the horizontal plane and $\pm180\,\mrad$ in  
the vertical plane), high-rate capability and introduces minimum  
material in the region of the spectrometer acceptance. Its  
large-volume vessel (see Fig.~\ref{fig:pid.rich.principle}) is filled with  
C$_4$F$_{10}$ radiator gas. Cherenkov photons emitted in the gas are  
reflected by two spherical mirror surfaces. The photons are converted  
to electrons by the CsI photocathodes of eight MWPCs, which amplify
the single photoelectrons   
and detect them.  
  
Hadron identification in the multi-$10\,\GeV/c$ momentum domain imposes  
C$_4$F$_{10}$ as a radiator gas, thanks to its low chromaticity, in  
spite of its high refractive index. The quest for sufficient number of  
Cherenkov photons for this gas determines the overall length of the  
radiator vessel (see Fig.~\ref{fig:pid.rich.principle}) to be of about  
$3\,\m$. Requirements for low background, minimum material in the  
spectrometer acceptance and the need to operate the MWPC-CsI  
detectors in an environment of a reduced particle flux, dictate the  
final RICH-1 geometry: the photon detectors are placed far from the beam  
line and outside the spectrometer acceptance. The corresponding mirror  
system consists of two reflecting surfaces located off beam axis, such  
that the Cherenkov ring images are focused outside the LAS acceptance.  
  
The RICH-1 geometry results in a photon detector surface of $5.6\,\m^2$.  
The surface is covered with eight proportional chambers (MWPCs),  
equipped with CsI photon converter layers. CsI photon converters have  
good quantum efficiency for wavelengths below $200\,\nm$ only; this  
property implies compatibility in the very UV (VUV) domain of both the mirror  
system and the gas radiator. Quartz windows separate the radiator gas  
from the photon detector; the quartz optical properties impose the  
lower limit of the useful wavelength at $\approx 165\,\nm$. A dedicated  
radiator gas system establishes continuous gas circulation in a closed  
loop and ensures both optimum VUV transparency and constant relative pressure  
in the vessel.  
  
\subsubsection{RICH-1 gas and gas system}  
\label{sec:pid.rich.gas}  
  
The RICH-1 gas radiator vessel has a length of $3\,\m$ and a volume of  
about $80\,\m^3$.  The refractive index of the C$_4$F$_{10}$ gas  
($n-1=0.0015$ for $7\,\eV$ photons) and its low chromaticity  
($\mathrm{d}\!\!\; n/\mathrm{d}\!\!\; E\sim 5\EE{-5}\,\eV^{-1}$ at $7\,\eV$) 
make it adequate for hadron  
identification above $10\,\GeV/c$.  The gas is handled by a dedicated  
gas system \cite{Albrecht:03c}, which controls the pressure inside the  
vessel, keeps the radiator gas transparent in the VUV domain, fills  
the vessel with the radiator gas and recovers it in the storage tank.  
  
The gas in the vessel must be kept in well controlled relative  
pressure conditions in order to avoid damages to the two thin vessel  
walls in the acceptance region, and to avoid mechanical deformation of  
the vessel itself, which can be transfered to the mirror wall.  The  
vessel pressure is kept constant within $1\,\Pa$ over months of  
operation, while the maximum allowed variations are an order of  
magnitude larger.  
  
The radiator transparency in the light wavelength region between 160  
and $200\,\nm$ is essential for RICH-1 operation, as it influences  
directly the number of photons observed per ring. The commercially  
available C$_4$F$_{10}$ material is fully opaque in the VUV domain, as  
it contains VUV absorbing impurities \cite{Albrecht:04e}; a cleaning  
procedure is then mandatory before the insertion of the gas in the  
RICH vessel. Cleaning has been performed in liquid phase up to year  
2001 and in gas phase later, with typical material losses related to  
the cleaning procedure of around 20\%. The gas is also constantly  
filtered from water vapour and oxygen contaminations during operation.  
This is done by continuously circulating the gas in a closed loop  
circuit through a Cu catalyst at $\approx 40^\circ$C to remove oxygen  
and through a 5A molecular sieve at $\approx 15^\circ$C to trap the  
water vapour traces. As a result, water vapour traces below 1~ppm and  
O$_2$ below 3~ppm are routinely obtained.  
  
\begin{figure}[tbp]  
  \begin{center}  
    \includegraphics[width=\columnwidth]{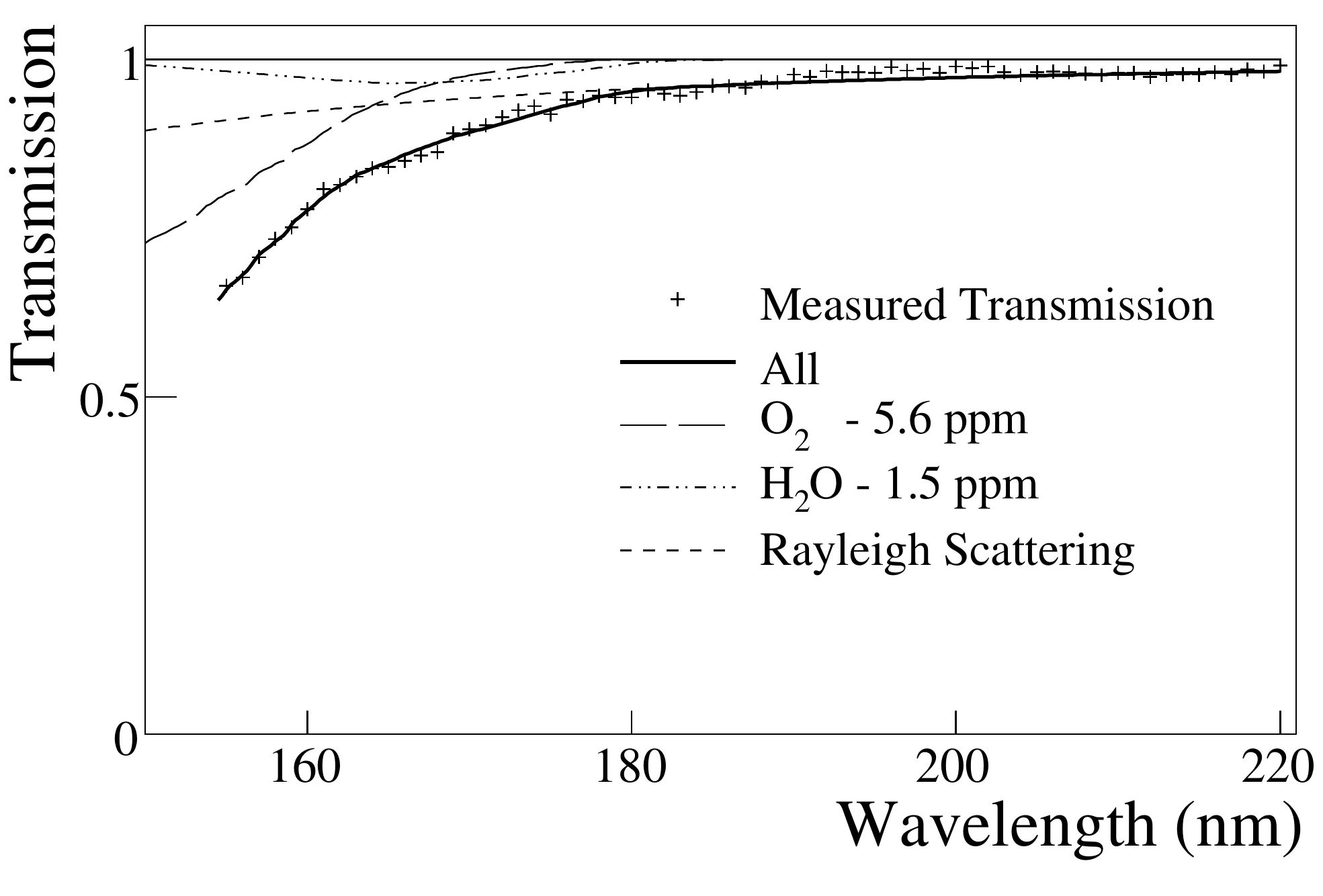}  
  \end{center}  
  \caption{\small Typical UV light transmission through 1.87 m of  
    C$_4$F$_{10}$, as measured online during data taking (crosses).  
    The solid curve is a fit to the data. The main contributions 
    to the UV light absorption are also shown.}  
  \label{fig:pid.rich.gas_trasnparency}  
\end{figure}

The VUV light transmission is measured online using two complementary  
systems. A dedicated setup performs an integral measurement over the range  
from $160\,\nm$ to $210\,\nm$ using a UV lamp and a solar-blind  
photomultiplier. A system based  
on the use of an UV lamp and a monochromator 
measures the transmission as a  
function of the light wavelength in the VUV range of interest. 
Good transparency of the  
gas radiator, stable over months, was achieved from 2003. 
Figure~\ref{fig:pid.rich.gas_trasnparency} shows a typical transmission curve  
in the VUV domain, with the main contributions to the VUV light  
absorption: Rayleigh scattering, O$_2$ and H$_2$O. 
  
\subsubsection{RICH-1 mirror system}  
\label{sec:pid.rich.mirrors}  
  
The RICH-1 optical system \cite{Albrecht:03b} consists of two VUV  
reflecting spherical surfaces of total area larger than $21\,\m^{2}$ and a  
radius of curvature of $6600\,\mm$. It was designed to focus the  
images outside the spectrometer acceptance on the photon detectors.  
The surface of the plane photon detectors is a rough approximation of  
the spherical focal surface. The two mirror surfaces are a mosaic type  
composition of 116 spherical mirror units: 68 of them are regular  
hexagons with a side length of $261\,\mm$, the other 48 are pentagons  
with six different sizes. The clearance left between adjacent mirrors  
results in a 4\% loss of reflecting surface. This optical arrangement,  
coupled to the $3\,\m$ long radiator, results in a geometrical  
aberration of $0.32\,\mrad$ for images produced by particles incident at  
angles of a few $\mrad$, and increasing for particles incident at
larger angles.   
  
\begin{figure}[tbp]  
  \begin{center}  
    \includegraphics[width=\columnwidth]  
    {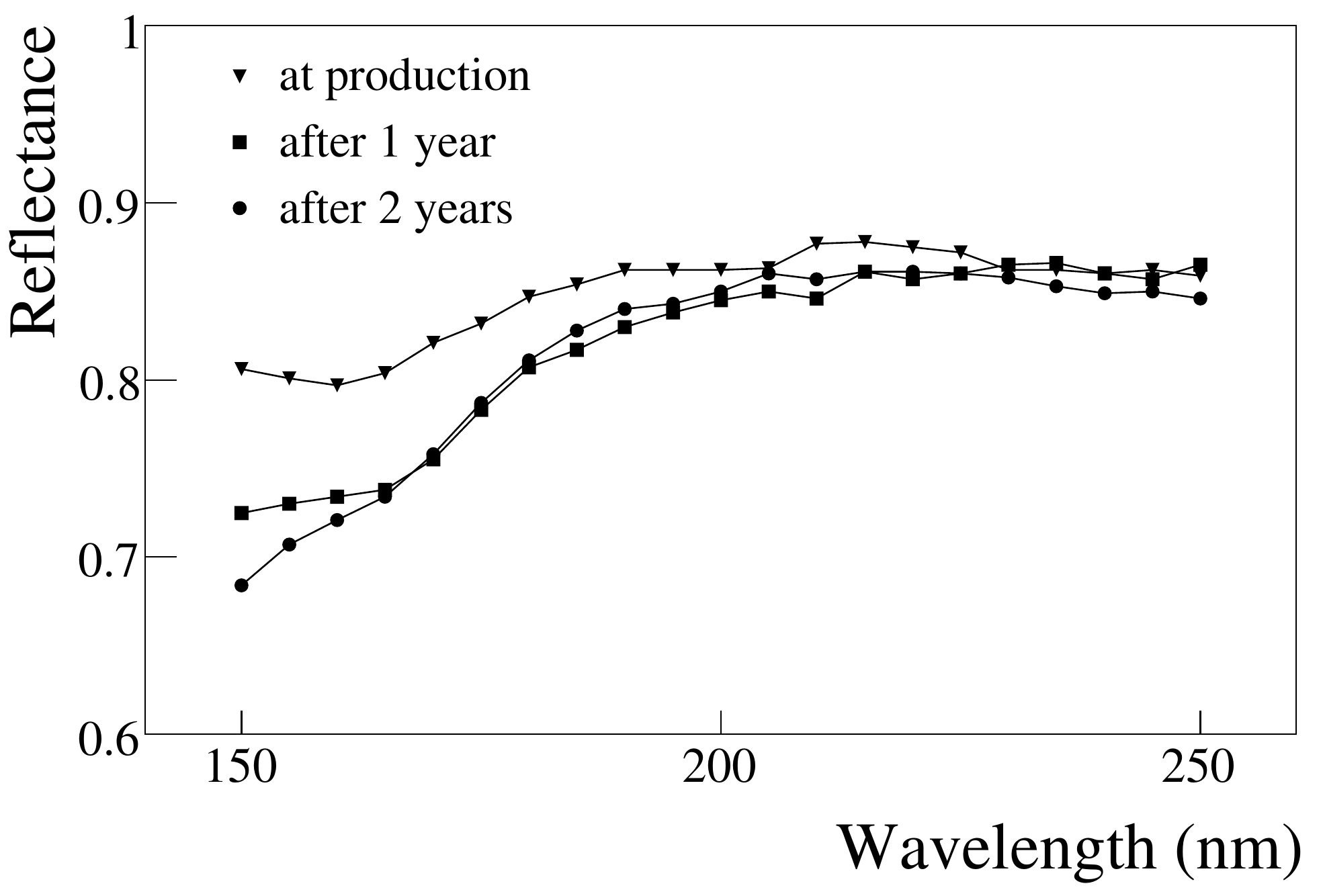}  
  \end{center}  
  \caption{\small Measured reflectance for a typical mirror piece. The  
    measurements have been performed shortly after production, 1 year  
    and 2 years later.}  
  \label{fig:pid.rich.mirror_reflectance}  
\end{figure}

The mirror substrate is a borosilicate glass, $7\,\mm$ thick,  
corresponding to about 5.5\% of a radiation length.  All individual  
substrate pieces have been characterised by measuring the radius of  
curvature and the diameter of the image of a point-like source.  The  
roughness of the polished surfaces has been measured by sampling.  
Good reflectance in the VUV  
region is obtained by applying a reflective layer (Al, $80\,\nm$)  
covered by a protective layer (MgF$_2$, $30\,\nm$ ).  The reflectance,  
as measured shortly after the production in the useful wavelength  
interval ($165-200\,\nm$), was found to be in the range $83-87$\%. The  
measurements were repeated after 1 and 2 years of operation. Except a  
short term degradation at small wavelengths, stable reflectance  
values above $165\,\nm$ (see Fig. \ref{fig:pid.rich.mirror_reflectance})  
have been observed.  
  
The mechanical structure supporting the mirror wall has a net-like  
configuration designed to minimise the amount of material in the  
spectrometer acceptance; the mirrors are suspended to the nodal points, 
which lay on a  
sphere; as a consequence, only angular adjustment of the mirror units  
is needed.  Each mirror is equipped with a joint which allows it to  
rotate around two orthogonal axes with angular resolution better than  
$0.1\,\mrad$.  The overall amount of material used for the mechanical  
supports is equivalent to 2.5\% of a radiation length.  
  
Alignment of the RICH-1 mirrors is performed inside the vessel. Since  
the loci of the spherical surfaces lay outside the vessel volume, an  
original alignment procedure was adopted. The
coordinates of the two sphere centres are known in the vessel  
reference frame. The coordinates of a theodolite are measured in the  
same frame. The theodolite axis is oriented along the straight line  
joining its centre and the centre of the sphere (reference line). If  
the mirror which is just in front of the theodolite is perfectly  
aligned, the normal to the mirror surface at the intersection point  
with the reference line will also lie along this line. In the alignment  
procedure, the mirror is rotated until the two lines  
coincide. The accepted residual misalignment angle is $<0.1\,\mrad$  
corresponding to the precision with which the reference line is  
defined.

\subsubsection{RICH-1 photon detectors}  
\label{sec:pid.rich.PD}  
  
\begin{figure}[tbp]  
  \begin{center}  
    \includegraphics[width=\columnwidth]{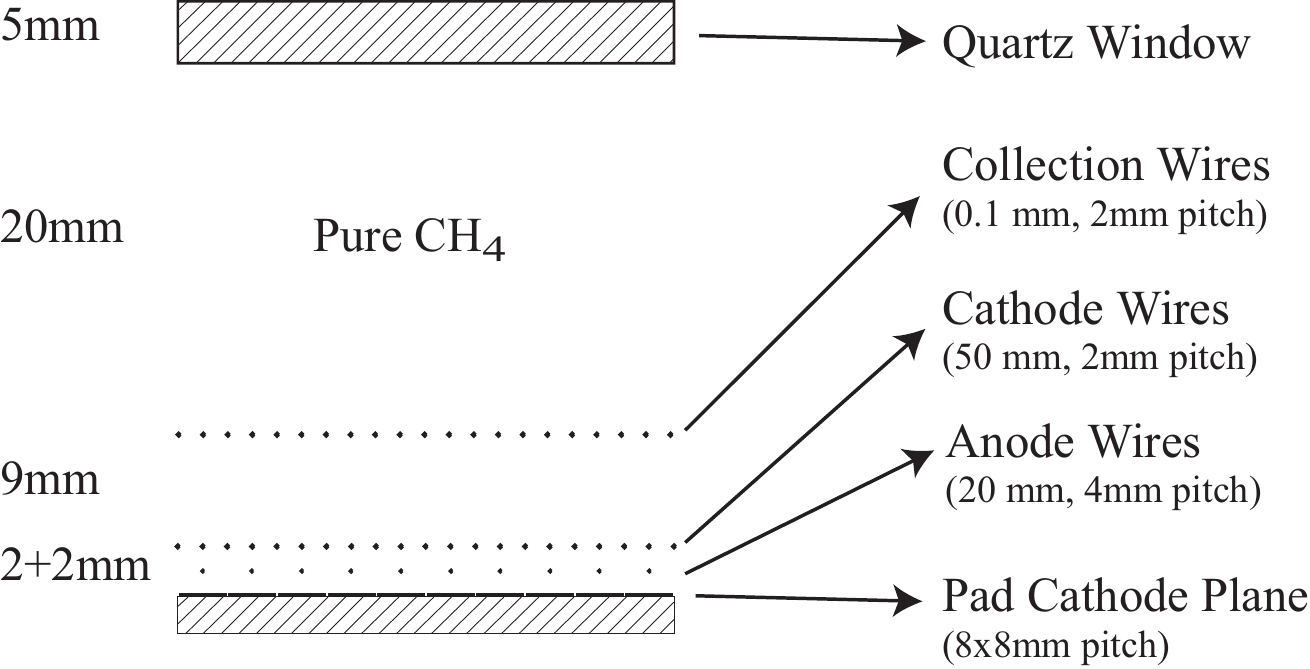}  
  \end{center}  
  \caption{\small Cross section of the RICH-1 photon detector.}  
  \label{fig:pid.rich.det-structure}  
\end{figure}

The photon sensitive surface area of RICH-1 is equipped with eight 
large-size MWPCs  
($576\times 1152\,\mm^2$) CsI photocathodes  
\cite{Albrecht:03a}: this is the largest photon detection system of  
this kind in operation so far.  The main parameters of the MWPCs are  
those optimised during the RD26 collaboration studies  
\cite{RD26Collaboration:93,RD26Collaboration:94,RD26Collaboration:96}:
$20\,\mum$ diameter wires, $4\,\mm$ wire   
pitch, $2\,\mm$ anode-cathode gap and a photocathode surface segmented in  
$8\times8\,\mm^2$ pads. The eight photon detectors have 82944 pad  
channels in total. A quartz window separates the detector from the  
vessel. The MWPC chambers are filled with pure methane, as it is  
transparent to photons in the useful energy range. The overall  
detector structure is illustrated in Fig.~\ref{fig:pid.rich.det-structure}.  
  
The CsI coating of the photocathodes has been performed at CERN following  
a procedure established by the RD26 collaboration and later optimised  
for ALICE HMPID \cite{ALICEcollaboration:98} and COMPASS applications.  
Severe constrains have been imposed on the tolerances of the MWPC  
mechanical parameters: the required precision on anode-cathode gap is  
$50\,\mum$ , the anode wire mechanical tension is within 5\% of the  
nominal values.  
  
The large photocathode elements are never exposed to air after  
coating with the CsI layer, as impurities and, in particular, water  
vapour can degrade the quantum efficiency.  The possible contamination  
is monitored by measuring the oxygen level of the atmosphere to which  
the photocathodes are exposed.  For this goal, the photocathode  
units are always handled in dedicated glove boxes.  A dedicated  
photocathode transportation system has been designed and built; it  
allows the transportation of up to four photocathodes, each protected  
with a gas-tight cover and under a continuous flux of nitrogen  
filtered in a closed loop. 
 
\begin{figure*}[tbp]  
  %%%%% 2 columns !  
  \begin{center}  
    \includegraphics[angle=-90,width=\textwidth]{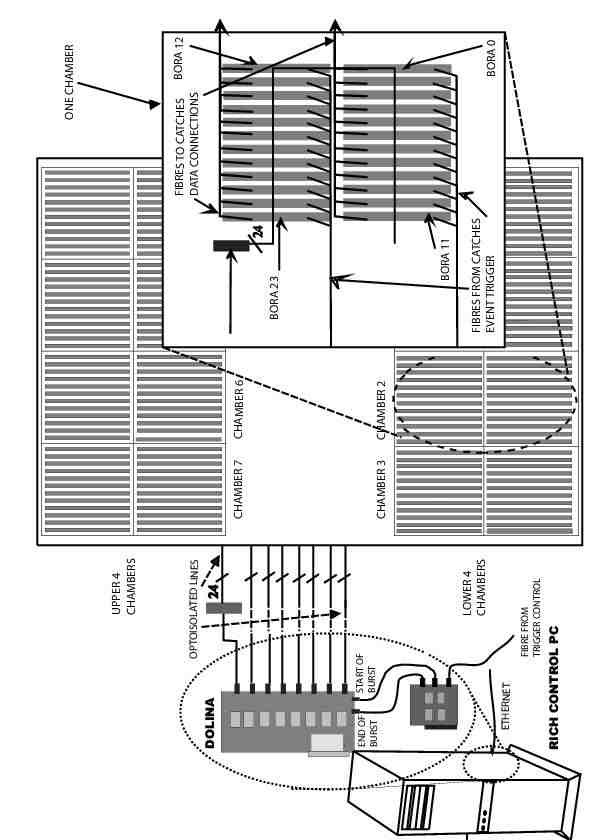}  
  \end{center}  
  \caption{\small The figure shows a top-level view of the front-end  
    data acquisition of RICH-1. Each of the 192 BORA boards is  
    directly connected through a fibre optical link to the COMPASS DAQ  
    system for data transfer and through a slower network to the  
    controlling PC computer.}  
  %%%%%%% from NIMA 433, 426  
  \label{fig:pid.rich.ro_scheme}  
\end{figure*}    
  
MWPCs with CsI photocathodes must be operated at moderate gain due to  
the presence of the photosensitive CsI layer. Because of the  
radioactive environment the photon detectors may occasionally exhibit  
electrical instabilities with long ($\sim$1~day) recovery time,  
clearly related to the combined effect of the level of the applied  
voltage and the ionising particle flux \cite{Albrecht:04}. Tiny local  
imperfections of the detectors favour the instabilities  
\cite{Albrecht:03d}. Great care in the mechanical precision of the  
MWPCs as well as in the uniformity of the anode wire tension and  
validation in the high-flux Gamma Irradiation Facility at CERN have  
allowed us to obtain stable operation at gains up to  
$5\EE{4}$. The electrical stability of the MWPCs with CsI has been  
improved in time. In 2001, 25\% of the MWPCs area was stable; in 2002,  
75\%; in 2003 and 2004, 97\%.  
  
\subsubsection{RICH-1 readout system}  
\label{sec:pid.rich.RO}  
  
Signals generated in MWPCs by single electrons, as it is the case for  
the photoelectrons due to the conversion of Cherenkov photons, have  
amplitude spectra described by Polya functions. For small detector  
gain, the Polya function has its maximum at zero. It is therefore  
mandatory to have a readout system with reduced electronics noise and  
precise control of the effective threshold setting; in COMPASS RICH-1  
this is obtained by using a modified version of the GASSIPLEX chip  
\cite{Santiard:94} and adjusting the threshold for each  
readout channel independently.  GASSIPLEX has optimum matching with
the detector   
signal, even if it has some intrinsic limitations: an effective dead  
time, resulting from the time required for baseline restoration after  
the release of the track-and-hold signal.  The COMPASS-GASSIPLEX chip  
has 16 input channels, 2 parallel output channels, a reduced readout  
time ($<500\,\ns$), $3\,\mus$ baseline restoration time, a noise figure  
of $600\,\e^{-}+10\,\e^{-}/\pF$ ENC and a gain of $6\,\mV/\fC$.  
  
\begin{figure}[tbp]  
  \begin{center}  
    \includegraphics[width=\columnwidth]{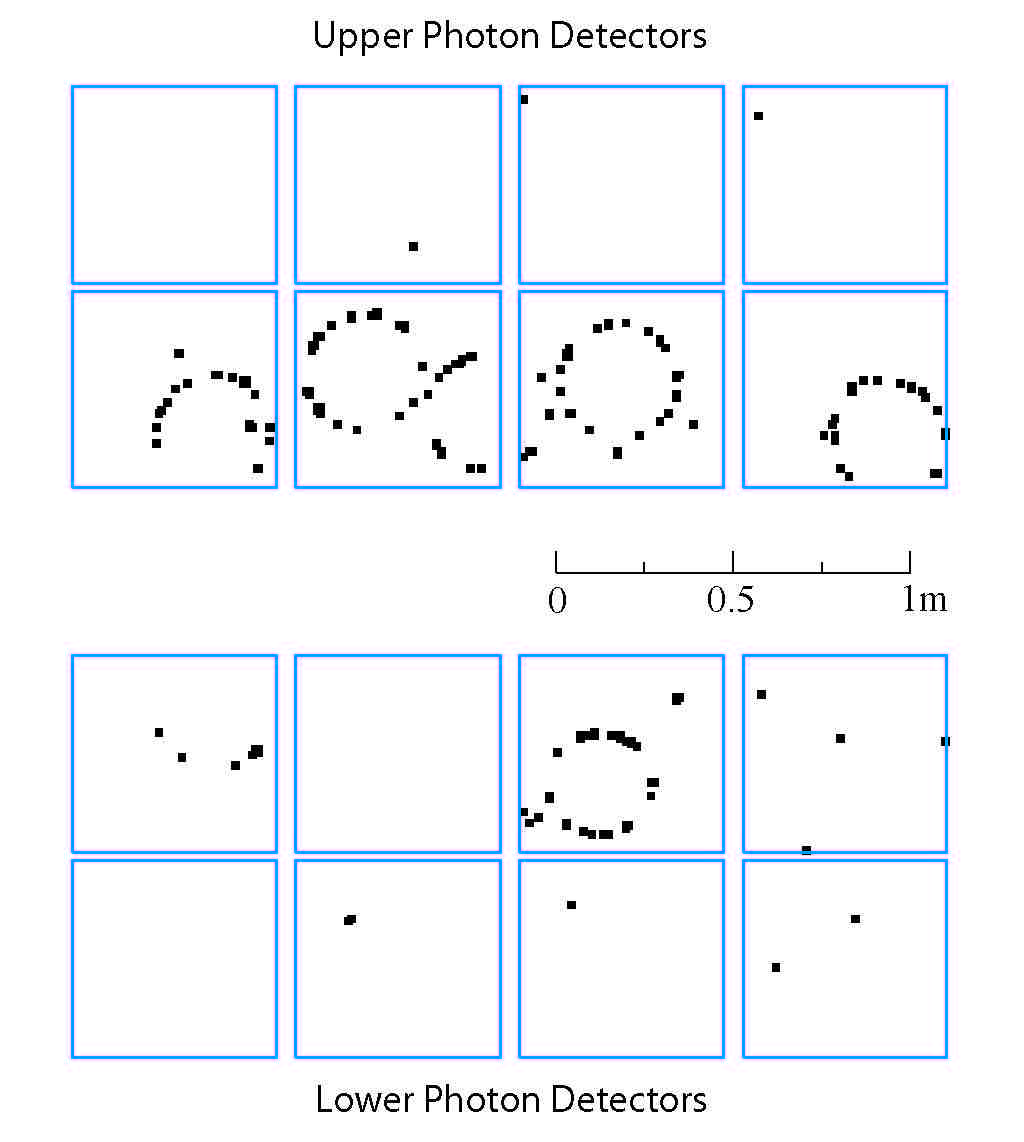}  
    \vspace{-2ex}  
    \caption{An event from the online event display of COMPASS RICH-1;  
      the squares represent the hits with signal amplitudes larger  
     than a threshold, individually set for each channel.}  
    \label{fig:pid.rich.event_display}  
  \end{center}  
\end{figure}  
  
The COMPASS-GASSIPLEX chips are coupled to an analogue readout system  
of 82944 parallel channels in total.  The chips are housed in 192  
identical large front-end boards (BORA board), with 432 input channels  
each. The BORA boards also house DSP and FPGA processors allowing  
local data treatment.  An average noise level of 1100 electrons r.m.s.  
is obtained during operation of the detector.  The BORA board acquires  
the data, digitises it in 10 bits, applies threshold and zero  
suppression and transmits event data.  Digitisation is automatically  
performed after each trigger and data are stored in a FIFO array with  
up to 128 event capability; this design results in data sparsification  
downstream of the digitisation stage. The subsequent on-board data  
reduction is performed in $13\,\mus$ .  
  
The data are transmitted through an optical fibre to the global  
acquisition system at a data transmission rate of
$40\,\mathrm{MB}/\s$. Figure~\ref{fig:pid.rich.ro_scheme} shows the
general    
architecture   
of the readout system. The main feature of the readout architecture  
is its distributed computing power, resulting in an easily  
reconfigurable system. This feature allows a characterisation of all  
analogue channels by locally measuring and calculating noise and  
pedestal.  
  
Eight parallel DSP networks allow the BORA control and, in particular,  
reconfiguring and reprogramming the DSPs and FPGAs as well as  
acquiring sample events to monitor the overall operation of the  
system.  These networks are handled by DOLINA, a PC resident  
multiprocessor board housing eight DSPs.  A high-level control  
application software runs on the PC, allowing for a 
reconfiguration of the BORA control. All BORAs are optoisolated from DOLINA  
through eight specific optoisolating boards, one for each DSP  
network, in order to avoid grounding interference between the PC and  
the detector.  
  
The readout system has been tested up to trigger rates of $75\,\kHz$  
with maximum pixel occupancy of 20\%. 
   
\subsubsection{RICH-1 characterisation in COMPASS environment}  
\label{sec:pid.rich.characterization}  

\begin{figure}[tbp]
  \centerline{\includegraphics[width=\columnwidth]
    {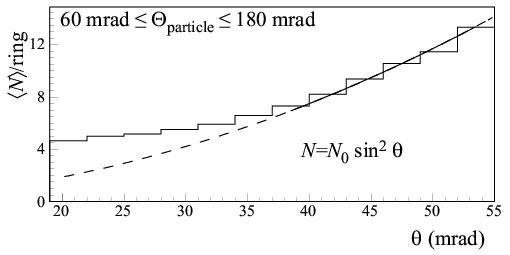}} 
  \caption{Mean number $N$ of detected photons per particle for the
    reconstructed ring images, as a function of the measured Cherenkov
    angle $\theta$. The line is the best fit of the   
    function $N=N_0\sin^2\theta$ in the range above  
    $36\,\mrad$.}
  \label{fig:pid.rich.Number_Photons}
\end{figure}  

\begin{figure}[tbp]
  \centerline{\includegraphics[width=\columnwidth]
    {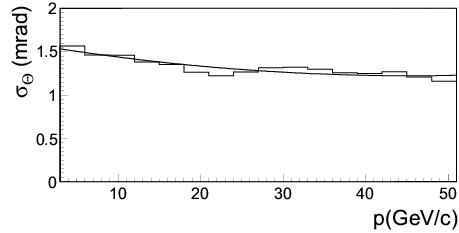}} 
  \caption{Resolution of the Cherenkov angle for the reconstructed
    ring images,  
   provided by each single photon, versus the particle momentum for a
   sample of  
   identified pions.}
  \label{fig:pid.rich.Photon_Resolution}
\end{figure}

\begin{figure}[tbp]
  \centerline{\includegraphics[width=\columnwidth]
    {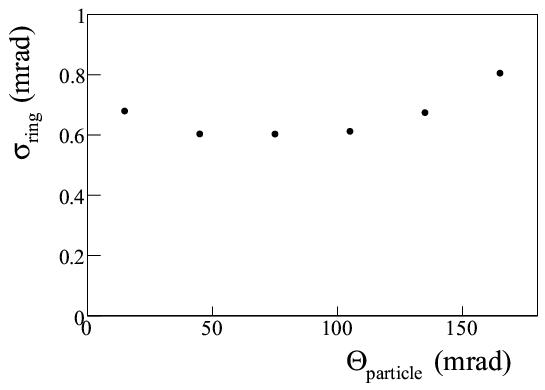}} 
  \caption{Cherenkov angle resolution for the reconstructed ring
    images, as obtained from all detected photons,  
    versus the polar angle of the particle track.}
  \label{fig:pid.rich.Ring_Resolution}
\end{figure}

Figure~\ref{fig:pid.rich.event_display} shows an example of a RICH-1 event. 
The pattern recognition allows a determination of some of the most relevant 
detector parameters: number of photons per event, Cherenkov angle and Cherenkov angle resolution.  
Figure~\ref{fig:pid.rich.Number_Photons} shows the mean number  
$N$ of detected photons per particle versus the measured Cherenkov angle  
$\theta$.  The line is the best fit of the function  
$N=N_0\sin^2\theta$ fitted in the range above $36\,\mrad$.  
For smaller Cherenkov angles, the mean number of detected photons per  
particle is small, and part of the images are not reconstructed, as a  
minimum number of 4 detected photons per ring is required. The fit clearly  
indicates that, at saturation, namely at the maximum value of the Cherenkov  
angle corresponding to $\theta=55\,\mrad$, a mean value of $\sim 14$  
detected photons is obtained. The plot in
Fig.~\ref{fig:pid.rich.Photon_Resolution} gives,   
for a sample of pions, the resolution of the Cherenkov angle  
measurement provided by each single photon as a function of the particle  
momentum. At saturation, the resulting resolution is  
$\sim 1.2\,\mrad$. The resolution on the Cherenkov  
angle obtained from all the detected photons versus the polar angle of
the particle track is presented in
Fig.~\ref{fig:pid.rich.Ring_Resolution}. Its value depends on the   
resolution for single photon measurement and on the number of detected  
photons, as well as on the signal dilution due to the background. The  
minimum value is around $0.5\,\mrad$.  The achieved resolutions allows  
pions and kaons separation at $2.5\,\sigma$ level up to $43\,\GeV/c$.  

Additional details about the RICH-1 performances are given in
Sec.~\ref{sec:recon.rich}.

%%%%%%%%%%%%%%%%%%%%%%%%%%%%%%%%%%%%%%%%%%%%%  
%%% Local Variables: 
%%% mode: latex
%%% TeX-master: t
%%% End: 

%%%%%%%%%%%%%%%%%%%%%%%%%%%%%%%%%%%%%%

\subsection{Muon identification}

\label{sec:pid.muon}

The muon identification is performed by two
detector systems, one in the LAS and one in the SAS part.
Both systems are made of a set of tracking stations and a hadron absorber
followed by a second set of tracking stations. 
Such a structure permits to distinguish muons from track segments 
induced by the typical
backgrounds like hadronic punch through from the hadron calorimeters.
The muon filtering system in the LAS consists of two stations 
MW1, separated by a $60\,\cm$ thick iron absorber 
(Muon Filter 1). In the SAS the tracking behind SM2 is used in
combination with a $2.4\,\m$ thick concrete absorber (Muon Filter 2) 
followed by two stations of MW2 and three MWPC
stations  
(see Fig.~\ref{fig:layout.setup}).

\subsubsection{Muon wall 1}

\label{sec:pid.muon.mw1}

The basic element of the MW1 system is the gaseous wire detector
called Mini Drift Tube (MDT). The geometry is based on plastic
streamer tubes (see reviews in \cite{Iarocci:83,Busza:88}).  A
modified version of the MDTs with fully metallic cathodes was produced
for the COMPASS experiment. The MDT is working in proportional mode.
This allows the detector to withstand the high-rate background conditions
of the COMPASS experiment.

\begin{figure}[tbp]
  \begin{center}
    \includegraphics[width=\columnwidth,viewport=120 600 470
    780,clip=true]{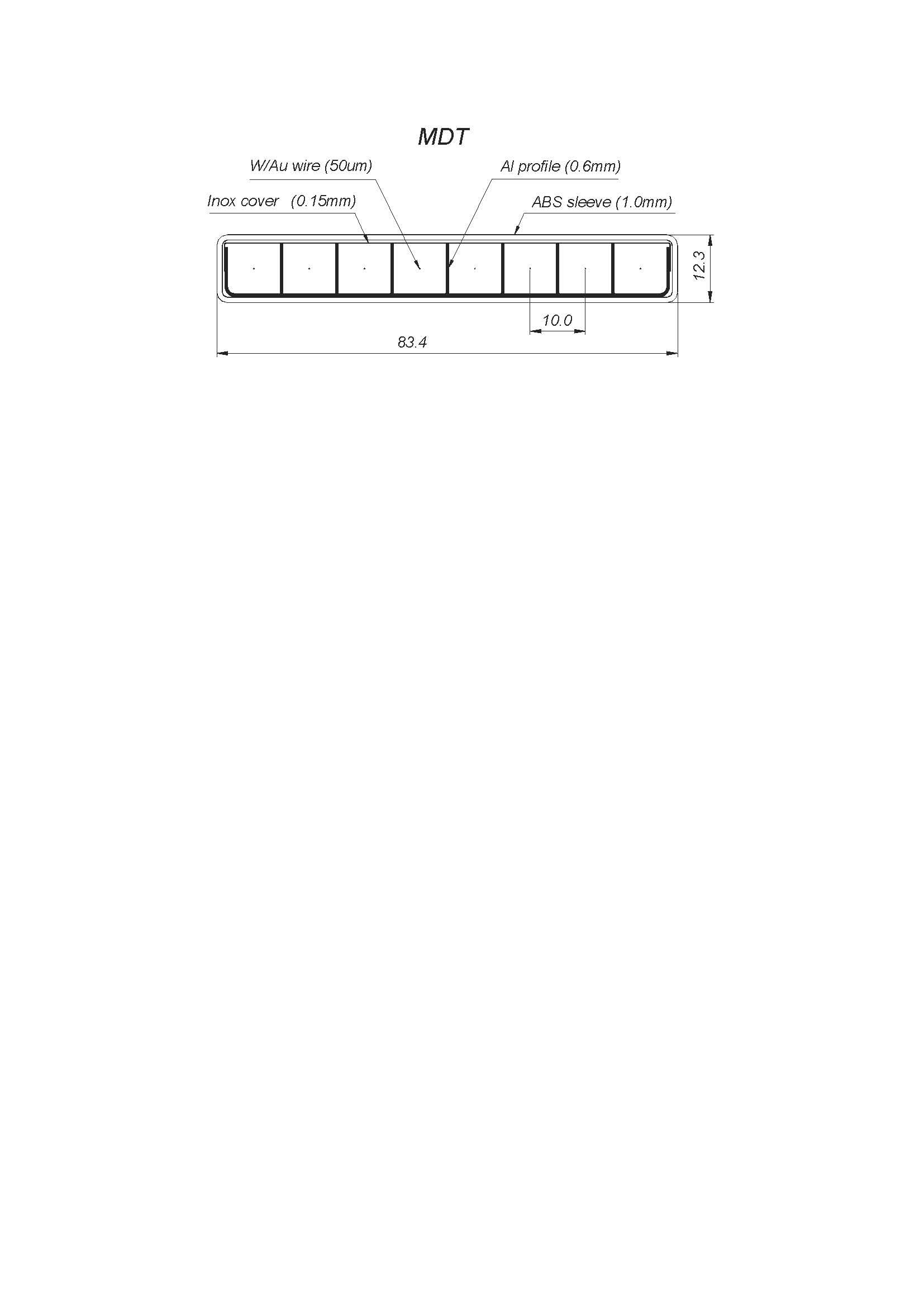}  
  \end{center}
  \caption{\small Cross section of a MDT module; all dimensions
    (except for the wires) are given in millimetres.}
  \label{fig:compass.mw1_mdt}
\end{figure}

An MDT module consists of an eight cell aluminium comb extrusion with a
wall thickness of $0.6\,\mm$ covered on top by a $0.15\,\mm$ thick
stainless steel foil, gold plated tungsten wires ($50\,\mum$ in
diameter) strung in the centre of the cells and an ABS plastic
envelope (sleeve) with thickness around $1\,\mm$ to house the interior
of the module. The wires are thermally glued to polyethylene plastic
spacers (not shown on Fig.~\ref{fig:compass.mw1_mdt}) at equal
distances of less than $1\,\m$ along the length of the MDT to provide
electrostatic stability.  The wire pitch is $10\,\mm$. The two
end plugs are thermally welded at the ends of the MDT, thus forming
together with the envelope the gas-tight volume of the module.
% Both endplugs have gas inlet/outlet connector and one of them has 8
% signal connectors (one per wire) and one high voltage connector.
The plastic envelope also serves for the HV insulation as the negative
voltage is applied to the metallic cathode. The anode wires are
grounded through the input of amplifiers.  Details of the
manufacturing and the testing are given in~\cite{Abazov:03,Abazov:04}.

\begin{figure}[tbp]
  \begin{center}
%    \includegraphics[width=\columnwidth,bb= 100 550 500
%    780,clip=true]{./figures/mw1_mw1}
    \includegraphics[width=\columnwidth,viewport=220 550 640 780,
    clip=true]{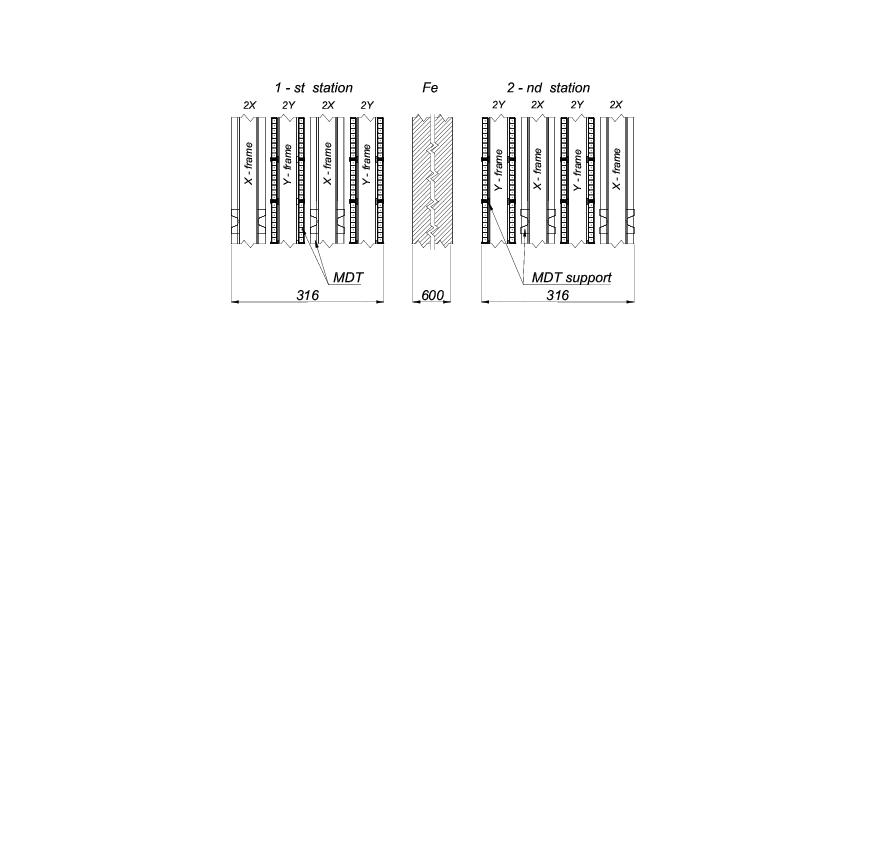}
  \end{center}
  \caption{\small Schematic cross--sectional side view of MW1; all
    dimensions are given in millimetres. Vertically only part
    ($255\,\mm$) of the stations are shown.}
  \label{fig:compass.mw1_mw1}
\end{figure}

The schematic view of the MW1 system is shown in
Fig.~\ref{fig:compass.mw1_mw1}. The system consists of two
stations separated by a $60\,\cm$ thick iron absorber. Each station
has four detectors with two planes of MDTs on both sides. Vertical and
horizontal tubes provide the $X$ and $Y$ coordinates, respectively.  The
outer surface of each station is covered with thin ($1\,\mm$ thick)
aluminium sheets for mechanical, electrostatic and noise protection.

All frames have a rectangular shape in the $XY$ plane and a hole in
their centre matching the acceptance of SM2.  The active areas are
$4845 \times 4050\,\mm^2$ (hole $1445 \times 880\,\mm^2$) and $4730 \times
4165\,\mm^2$ (hole $1475 \times 765\,\mm^2$) for the $X$ and the $Y$ planes,
respectively. The final planarity of the detector surface was measured
to be within $\pm0.5\,\mm$.  The total number of MDT modules in the
system is 1056 (8448 wires).
% They have four types in length to fit the four types of the
% sub-frames (two for X and two for Y frames). The active length of
% MDTs is slightly shorter then mechanical one due to the size of the
% endplugs (57.5 mm). The mechanical parameters of MDTs are given in
% Table~\ref{table:compass.mdts}.
The MW1 system provides a measurement of up to eight points per track
in each projection with the coordinate accuracy of $10/\sqrt{12}\,\mm$
typical for the $10\,\mm$ wire pitch.
% This is generally enough keeping in mind the multiple scattering in
% HCal 1 and iron filter.

% Table 1. Types and numbers of MDTs
% __________________________________ type total length, mm active
% length, mm quantity ---- ---------------- ----------------- --------
% X big 4165 4050 320 X small 1700 1585 272 Y big 4845 4730 320 Y
% small 1742.5 1627.5 144

% \begin{table}[hbtp]
%%\begin{center}
%  \caption{\small Types and numbers of MDTs.}
%  \begin{tabular}{| l | c | c | c|}
%\hline %\hline
%\ Type &        Total length, mm &      Active length, mm &     Quantity \\
%\hline
%X big &         4165 &                  4050 &                  320 \\
%X small &       1700 &                  1585 &                  272 \\
%Y big &         4845 &                  4730 &                  320 \\
%Y small &       1742.5 &                1627.5 &                144 \\
%\hline %\hline
%\end{tabular}
%%\caption{\small Types and numbers of MDTs.}
%\label{table:compass.mdts}
%% \end{center}
% \end{table}

Both MW1 stations have an individual suspension system supported by
Muon Filter 1 and the SM2 magnet.
% Each module is hanging freely on two ball joints and is also
% connected to the filter at its bottom; the connection is made
% adjustable to fix the verticality of the modules within $\pm$0.5 mm.
It also provides electrical insulation between the iron filter and the
stations and serves as a 'clean ground' for the front-end electronics
and its low voltage power supply.  The negative HV is applied to the
metallic cathode of the MDTs through individual $1\,\MOhm$ resistors.

The gas mixture used is Ar/CO$_2$ (70/30). It is selected because no
ageing effects are observed up to $1\,\C/\cm$, while it is fast enough
(drift time is below $150\,\ns$), non-flammable and cheap.
The total gas volume of MW1 is $3.2\,\m^3$.

The front-end electronics of MW1 consists of an analogue and a digital
part.  The latter one is located in 6U-crates on the iron absorber at a
maximum distance of $7\,\m$ from the analogue cards.  The analogue part
consists of 264 amplifier-discriminator boards with 32
input/output channels each. The cards use ASIC
amplifier~\cite{Alexeev:01a} and discriminator~\cite{Alexeev:99}
chips; their
design and parameters may be found in Ref.~\cite{Alexeev:01b}.
% The negative current pulse from each MDT wire is fed to ADB input by
% individual coaxial cable, the screen is grounded to the frame right
% near MDT detector and also at the ADB card input connector.  The
% signal is terminated at one end of the MDT detector (near
% electronics) by the total impedance about 280 Ohm which is close to
% the wave impedance of the MDT cell; this value is comprised by
% serial connection of 220 Ohm (resistor embedded in the MDT
% end-plug), 10 Ohm at the ADB card and 46 Ohm of amplifier input. The
% common threshold is applied to all 32 channels and is driven by the
% digital part. The ADB cards are connected to the digital cards by
% high density (80 wires) twisted-pair cables.
The digital part~\cite{Alexeev:05} consists of 44 digital cards (192
channels each) which are housed in five crates. The digital MW1 card
% converts the logical differential signals (LVDS) generated by the
% ADB cards into MW1 coordinate points. Each DMW1 card
includes six F1-TDC chips (see Sec.~\ref{sec:daq.digital.f1}) which
are used as 32-channel input registers (latch mode).
% Even working in latch mode the F1 chip permits to measure the drift
% time of MDT (one time per four consequetive channels/wires, 4.6 nsec
% resolution). This feature may be in principle used to further
% improve the coordinate accuracy of detector, which requires
% additional evaluation.
%
% During initialization phase DMW1 cards are loaded through a 10 Mbaud
% serial link with the control information such as: threshold level,
% duration of the strobe for input signals, trigger latency, trigger
% window and so on. The data from DMW1 cards are transmitted via a
% high rate serial hotlink (40 MB/sec) to the standard COMPASS CATCH
% cards of the DAQ system.

The whole MW1 system showed very stable performance during nominal data taking
conditions. All modules were operated with the same high voltage and
threshold settings. 
Typical values are $2050\,\V$ and $1\,\muA$ for the
high voltage and threshold, respectively; the signal amplitudes
vary between 5 and $7\,\mu \A$.

\begin{figure*}[tbp]
  \begin{center}
    \includegraphics[width=\textwidth]{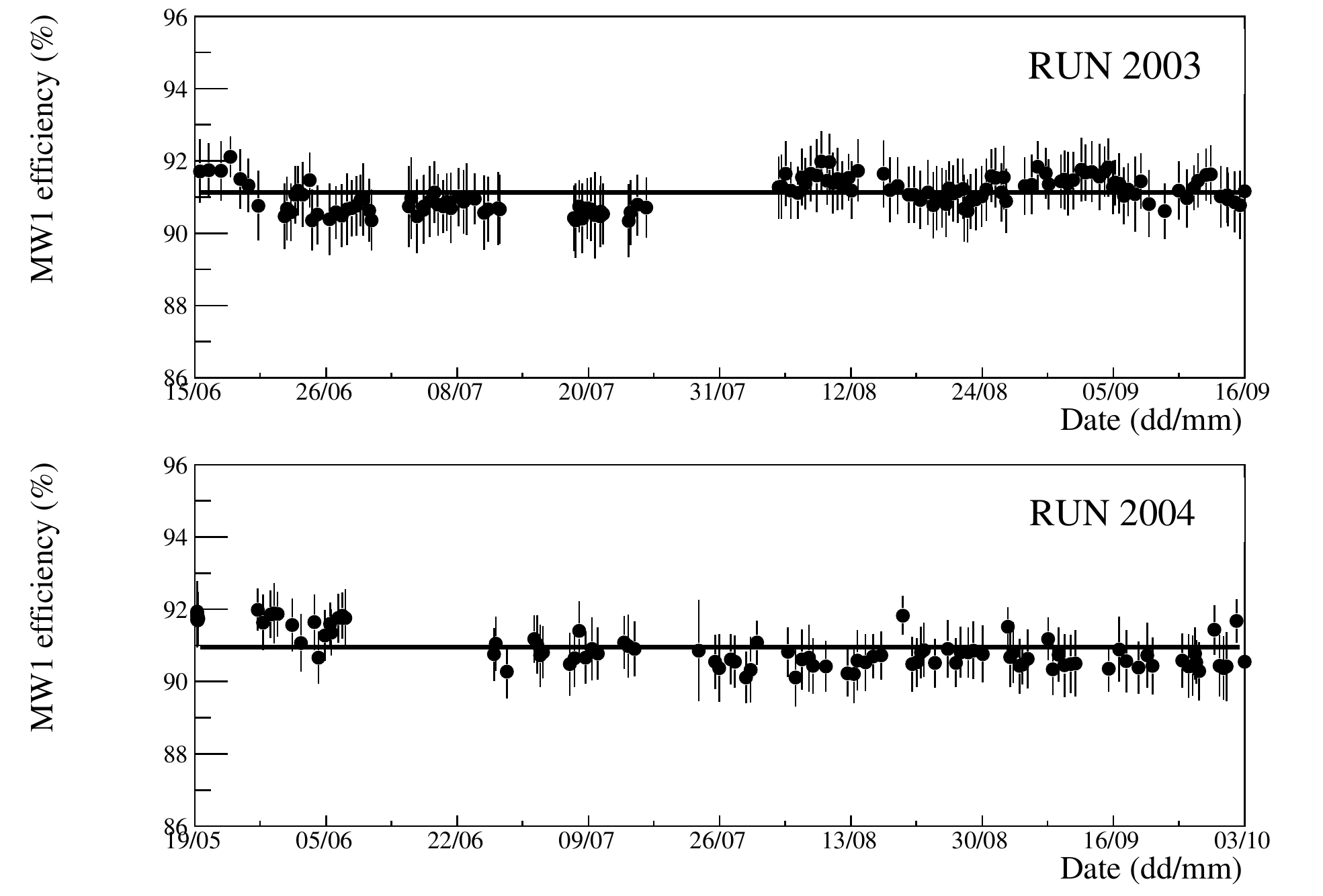}
  \end{center}
  \caption{\small The MW1 averaged tracking efficiency per plane as
    a function of time (years 2003 and 2004).}
  \label{fig:compass.mw1_eff}
\end{figure*}

% Stability of the MW1 tracking efficiency is the most relevant
% parameter which defines LAS muon ID thus, together with other
% COMPASS systems like tracking and calorimetry, minimizing the
% systematic errors for the measurements with muons.
The stability was monitored for all 16 planes of MW1.
Figure~\ref{fig:compass.mw1_eff} illustrates this for the years 2003
and 2004.  Each point in Fig.~\ref{fig:compass.mw1_eff} represents the
tracking efficiency averaged over the all 16 planes of MW1.  The
monitoring was made under the following conditions: only good data
taking physical runs were used and the selected events were in fact
halo muons with trajectories quasi-perpendicular to the detector
planes. The value of the obtained efficiency is typically 91\%, close
to the geometrical one for the MDT detectors plane of 88.5\% (just few
percent higher due to the small divergence of the halo muons).  The
variation of the single plane efficiencies of about 1\% diminishes to
0.2\% when using the whole MW1 system with at least five (out of
eight) points for a
track.
% on the percent level is not important as a muon in each projection
% is measured by the few points per track. For example, taking 5
% points as a track diminishes a 1\% variation to only 0.2\% for the
% system tracking efficiency.

%%% Local Variables: 
%%% mode: latex
%%% TeX-master: "compass_spec"
%%% End: 

\subsubsection{Muon wall 2}

\label{sec:pid.muon.mw2}

%%%%   To be written by Y. KHokhlov.

Muon wall 2 (MW2) consists of two identical stations of layers of
drift tubes.  Each of the two stations consists of 6 layers with an
active area of $447\times 202 \,\cm^2$ grouped into double
layers, each mounted to a separate steel frame. The three double
layers have vertical, horizontal and inclined (at $-15^\circ$ w.r.t. to
the vertical) tubes, respectively.

\begin{figure}[tbp]
  %%%%% 2 columns !
  \begin{center}
%    \includegraphics[width=\columnwidth,bb= 50 80 550
%    380,clip=true]{./figures/MW2_tub_lay}
    \includegraphics[width=\columnwidth,clip=true]{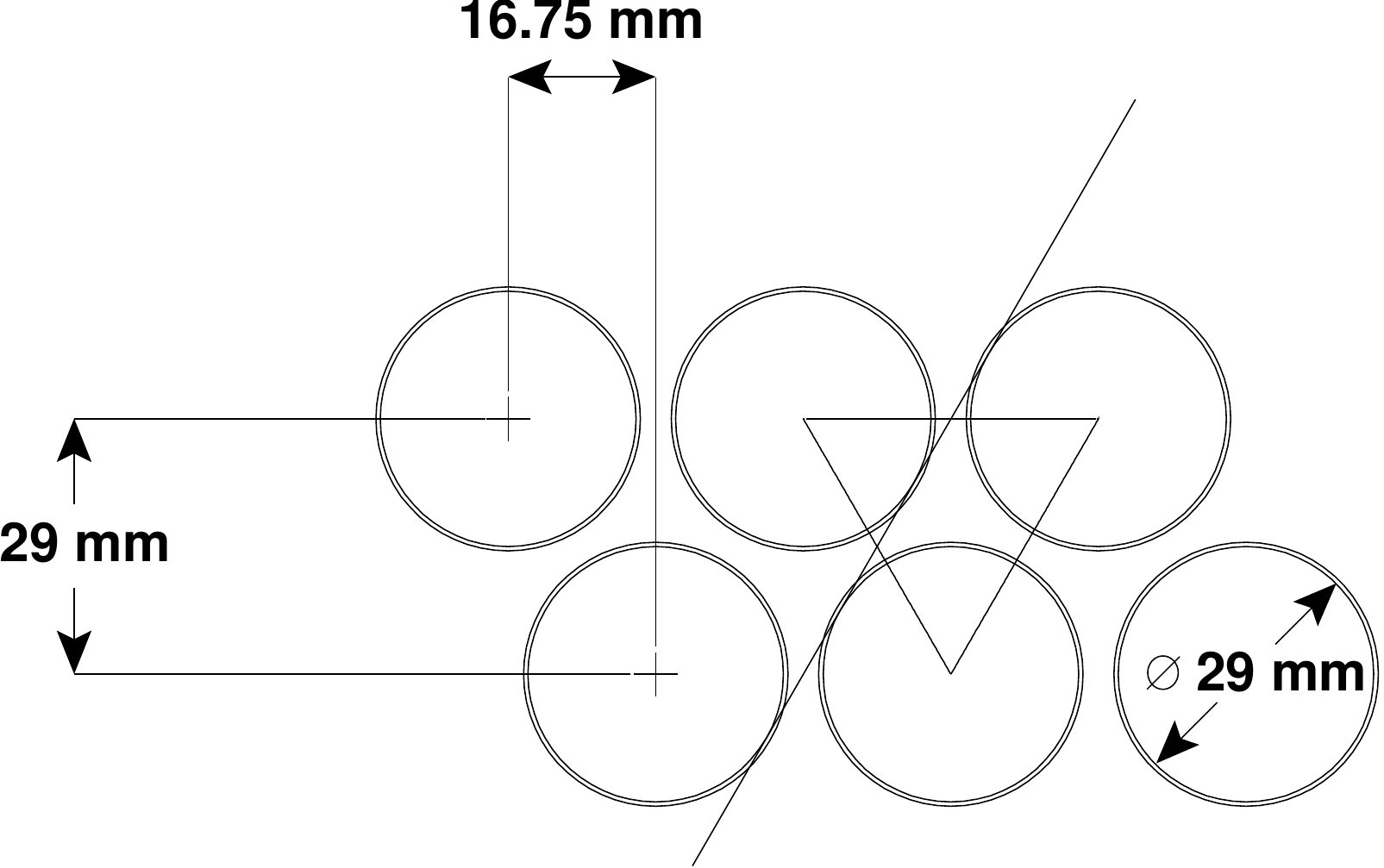}
  \end{center}
  \caption{\small  The layout of the tubes in a double layer of MW2
     planes. Dimensions in mm.
An imaginary equilateral triangle formed by 
centres of three adjacent tubes and a common tangent to 
their interior  are shown.
% of 29 mm in diameter 
}
\label{fig:pid.muon.mw2.tub_lay}
\end{figure}

The stainless steel drift tubes with an inner diameter of $29\,\mm$
and a wall thickness $0.5\,\mm$ were originally designed for the muon
system of the D\O\ experiment \cite{Brown:89}.
% They are implemented to the MW2 with some minor modifications,
% particularly of the mounting of the amplifiers.
The layout of the tubes in a double layer is shown in
Fig.~\ref{fig:pid.muon.mw2.tub_lay}.  The pitch of the wires in a
layer is $33.5\,\mm$.  Each tube is inserted into precise guide holes
in the frame and fixed with a clamp. The removable fixation allows a
replacement of damaged or malfunctioning tubes. The total numbers of
tubes in the two stations is 1680.

A rectangular hole in each plane of the detector with a size of
$1\times 0.8\,\m^2$ around the beam is realised using properly
shortened tubes.  The hole is covered by the MWPC-B (see
Sec.~\ref{sec:tracking.lat.mwpc}) stations, which partly overlap with the
sensitive area of MW2.

One end of each tube hosts a high voltage filter with a $1\,\MOhm$
resistor connected with the gold-plated tungsten anode wire of
$50\,\mum$ in diameter. The high voltage is delivered to a group of
filters through a distribution board mounted on the layer frame.
The other end of the tube is equipped with a separate capacitively
coupled amplifier-discriminator (AD) connected directly to the signal
(anode) and grounding pins of the tube, the latter being soldered to
the tube body. The effective threshold of the AD is set to $9\,\fC$.
The LVDS outputs of a group of up to 64 ADs are collected with a short
(up to $0.5\,\m$) flat cable on a printed circuit which is fixed to
the plane frame. This PCB hosts the F1-TDC board (see
Sec.~\ref{sec:daq.digital.f1}) operating in the ``standard
resolution'' mode, and also supplies the low voltage including
threshold offset to the ADs.

% The inner volumes of tubes in one layer are connected in series by
% means of short flexible pipes,
%% with a silicon rubber tubes pulled up the PET tubes,
% forming a single gas volume of $\sim 300-400\,{\rm l}$. The detector
% operates at atmospheric pressure with a bubbling gas outlet. The gas
% is supplied through six copper lines, one line per layer. The
% typical flow is $16\,{\rm l}/{\rm h}$ per line and can be increased
% roughly by factor 5 for a fast flush. It is planned to split the
% existing six lines into twelve to decrease the pressure drop along
% the tube-like volume, to facilitate the search for a leak and to
% quicken the start up filling of the detector.

The gas mixture used, Ar/CH$_4$ (75/25), is known for a saturated and
rather fast drift of 
electrons, a wide working plateau and stable performance against
radiation ageing.  The full drift time of $\sim 240 \,\ns$ at the
working high voltage of $+3\,\kV$ determines the two-track resolution.
% with a mean inverse drift velocity of $\sim 16 ns/mm$.

% \begin{figure*}[tbp]
%%%%%   2 columns !
%   \begin{center}
%     \includegraphics[width=\textwidth]{./figures/MW2_RT_scatnfit}
%   \end{center}
%   \caption{\small Distance vs. time scatter plot for a single MW2
%     plane (a) and the corresponding fit of R(T) relation (b).}
%%%%%%%%%   from NIMA 433, 207
%   \label{fig:pid.muon.mw2.RT}
% \end{figure*}

\begin{figure}[tbp]
  %%%%% 2 columns !
  \begin{center}
    \includegraphics[width=\columnwidth]{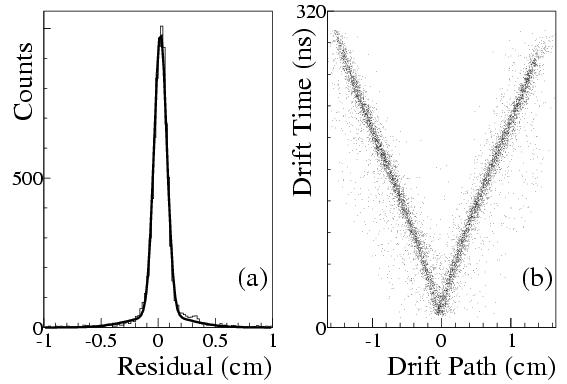}
  \end{center}
  \caption{\small (a) Histogram (thin line) of coordinate residuals
    for a single MW2 plane with the two Gaussian fit superimposed
    (thick line). (b) Reconstructed track position with respect to the
    nominal wire position vs. time for a single plane.}
  %%%%%%%% from NIMA 433, 207
  \label{fig:pid.muon.mw2.res_RT}
\end{figure}

The space resolution of the tracking through MW2 includes the
intrinsic resolution of the detector itself and the track fitting
error, so it is worse for the $Y$ planes, as the corresponding
projection measurement is generally less redundant, particularly in
this zone of the spectrometer.
% The resolution is characterized by the overall distribution
%% (Fig.\ref{fig:pid.muon.mw2.res}(a)
% (Fig.\ref{fig:pid.muon.mw2.res_RT}(a)) of the residuals between the
% coordinate of the crossing of a reconstructed track with the wires
% plane and the nominal position of the closest wire.
Fig.~\ref{fig:pid.muon.mw2.res_RT}(a) shows the residuals for a single
MW2 layer. The shape of the distribution is approximated with two
Gaussians with a common mean: a narrow ``core''
% from the ``true'' hits associated with the well measured tracks
with a $\sigma$ of 0.53 up to $0.94\, \mm$ (for $Y$ planes) which
amounts to 79\% to 91\% of the entries, and the wide ``halo'' with a
width between 2.0 and $4.8\,\mm$. The spread is mostly due to the
irregular spacing of the tubes. 

Figure~\ref{fig:pid.muon.mw2.res_RT}(b) shows the two dimensional
distribution of the measured time of the hit with respect to the trigger time
%% after a constant offset subtraction
versus the distance of the reconstructed track to the wire.
%% (projected or in plane ?)
The correlation is fitted with a linear $R(t)$ relation giving an
effective drift velocity of between 5.8 and 6.2$\,\cm/\mus$.
% , excluding the region $\pm$ $0.3\,\cm$ near the wire where some
% non-linearity is visible.

The tracking efficiency for a single plane averaged both over all
wires and the wires length
% is shown in Fig.\ref{fig:pid.muon.mw2.res}(b) as a function of the
% measured track coordinate, that is perpendicular to the direction of
% the wires.
is varying between 81\% and 84\%.  It is determined as the ratio of
the number of reconstructed tracks having a (registered) hit in the
layer not farther than $5\,\mm$ apart from the track crossing, to the
total number of reconstructed tracks passing the plane within its
acceptance.  So, accounting for the ratio of the inner diameter
($29\,\mm$) of the tube to the pitch ($33.5 \,\mm$), a value of 83 \%
corresponds to 96\% drift tube efficiency.
% MW2 included to track reconstruction ? with what R(T) ?
 
% \begin{figure*}[tbp]
%%%%%   2 columns !
%   \begin{center}
%     \includegraphics[width=\textwidth]{./figures/MW2_res_eff}
%   \end{center}
%   \caption{\small (a) Histogram (thin line) of residuals for a
%     single MW2 layer with the two Gaussian fit superimposed (thick
%     line).  (b) The efficiency of the layer averaged over the wires
%     length vs.  coordinate of the track}
%   \label{fig:pid.muon.mw2.res}
% \end{figure*}

% \begin{table}
%\caption{ Characteristics of MW2 planes during  run 34940 in y.2004. The ``u'' and ``d'' stand for ``upstream'' and ``downstream'' planes in a double layer w.r.t. the beam direction.} 
%\begin{tabular}{|l|l|l|l|l|l|l|l|l|l|l|l|l|}
%\hline
%  & \tablehead{1}{r}{b}{Single\\outlet}
%
% \small Plane & \tiny X1u &  \tiny X1d & \tiny Y1u & \tiny Y1d & \tiny V1u & \tiny V1d & \tiny X2u & \tiny X2d & \tiny Y2u & \tiny Y2d & \tiny V2u &\tiny V2d\cr
% %\hline
% %\small Eff.,\,\% & 83 & 83 & 83  & 71$^{(1)}$  & 84  & 83  & 81  & 82  & 82  & 83  & 82  &82\cr
% %{\small Drift vel.,} & & & & & & & & & & & & \cr
% %{\small $\cm/\mus$} & 5.89 & 5.93 & 6.14 & 6.21 & 5.78 & 5.83 & 5.90 & 5.82 & 6.09 & 6.07 & 5.76 &5.77\cr
% %{\small Res.,$\,\mm$} & 0.56 & 0.54 &  0.94 & 0.93 & 0.57 & 0.60 & 0.64 & 0.62 & 0.71 & 0.79 & 0.64 &0.63\cr
%
%\hline \multicolumn{13}{l}{\small{(1) The lower efficiency for Y1d
%    plane is due to 2 ``dead'' channels in the vicinity of its
%    center}}
% \end{tabular}
% \label{tab:mw2.char}
% \end{table}

%%%%%%%%%%%%%%%%%%%%%%%%%%%%%%%%%%%%%%%%%%%%%%%%

\subsection{Calorimetry}

\label{sec:pid.calo}

Two hadron calorimeters  and one electromagnetic calorimeter are
used in COMPASS setup (see
Fig.~\ref{fig:layout.setup}). Both hadron calorimeters are sampling
calorimeters using stacks of iron and scintillator plates. They 
are located before
the muon filters. Both hadron calorimeters serve a double purpose in
the experiment. They measure the energy of hadrons produced in the
target and participate in triggering on inelastic muon scattering
events
(see Sec.~\ref{sec:trigger.muon}).
ECAL2 is a homogeneous calorimeter
consisting of lead glass blocks located before 
HCAL2. It serves to measure the energy of
electromagnetic showers. In addition it was used in the trigger for
the 
measurement of Primakoff reactions with the pion beam 
(see Sec.~\ref{sec:trigger.hadron}).

\subsubsection{Hadron calorimeter 1}

\label{sec:pid.calo.hcal1}

\begin{figure}[tbp]
  \begin{center}
    \includegraphics[width=\columnwidth]{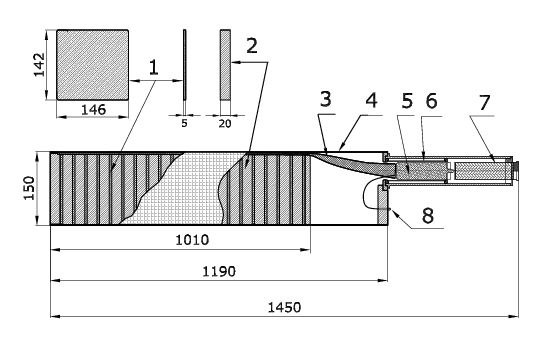}
  \end{center}
  \caption { \small{Structure of the HCAL1 module: $1$-scintillators,
      $2$-iron plates, $3$-light guide, $4$-container,
      $5$-PMT, $6$-PMT magnetic shielding,
      $7$-Cockcroft-Walton divider, $8$-optical connector for LED
      control. Dimensions are in $\mm$.  }}
  \label{fig:pid.module}
\end{figure}

The hadron calorimeter 1 (HCAL1) is placed (see Fig.
\ref{fig:layout.setup}) before MW1.  HCAL1 
has a modular structure, each module consisting of $40$ layers of
iron and scintillator plates, $20\,\mm$ and $5\,\mm$ thick,
respectively, amounting to 4.8 nuclear interaction lengths.  The
structure of a calorimeter module and its basic dimensions are shown
in Fig.~\ref{fig:pid.module} \cite{Gavr:04}.  Monte Carlo simulations
for hadrons and electrons were performed in the 10 -- $100\,\GeV$
energy range. These particles are almost fully absorbed in such a
calorimeter.

The 480 calorimeter modules were assembled and framed in a matrix of
$28\,\mbox{(horizontal)}\,\times 20\,\mbox{(vertical)} $ with 12
modules removed from each corner.  There is a rectangular window of
$8\times4$ modules at the centre of the matrix for the passage of the
beam and scattered muons. The outside dimensions of the HCAL1 are
$4.2\times3\,\m^2$ and useful surface is $10.8\,\m^2$.  The
calorimeter and its frame are mounted on a platform
% (fig.\ref{fig:compass.HCAL1_photo})
which can be moved across the beam axis.
% \begin{figure}[ptb]
%   \begin{center}
%     \includegraphics[width=\columnwidth,clip]{figures/hcal1}
%   \end{center}
%   \caption { \small{HCAL1 photo}}
%   \label{fig:compass.HCAL1_photo}
% \end{figure}
 
The resolution of a sampling calorimeter depends on the qualities of
scintillators, light guides and PMTs. The scintillators of HCAL1 have
been produced by molding under pressure from granulated polystyrene
PSM-115 mixed with P-terphenyl ($1.5\,\%$) and POPOP (1,4-Bis-[2-(5-Phenyloxazolyl)]-Benzene) ($0.04\,\%$). The
light from the scintillators is collected by a single flat light guide
placed on the open sides of the scintillators with an air gap of $0.6$
-- $0.8\,\mm$. The wavelength shifting light guides are fabricated
from organic glass (CO-95, CO-120) and painted by cumarine K-30
solved in alcohol. The amount of light emitted and collected from a
single scintillator traversed by a minimum ionising particle is enough
to produce $4$ -- $6$ photoelectrons at the PMT photocathode
($12$-cascade FEU-84-3 PMT).  They have multi-sodium photocathodes with
quantum efficiencies between $0.18$ and $0.26$ at the wavelength of
$460\,\nm$, a large dynamic range, currents up to $5\,\mA$, a typical
pulse rise time of $15$ -- $18\,\ns$ and a full pulse width of $40 $
-- $50\,\ns$ at the level of $0.1$ of its amplitude. A
Cockcroft-Walton voltage multiplication scheme is used to supply the
PMT dynodes with high voltages from the basic $100\,\V$.

The signals from the PMTs are sent via $50\,\Ohm$ cables of about
$140\,\m$ length to fast analogue-to-digital converters (see
Sec.~\ref{sec:daq.analog.caladc}). Small fractions of the signals are
fed into the fast summation system for trigger purposes (see
Sec.~\ref{sec:trigger.muon.calo}).

The main characteristics of the calorimeter -- linearity of the
response versus energy, $e/\pi$ ratio, energy and space resolutions --
were determined using the negative hadron and lepton beams at the CERN
X5 beam line with energies between $10$ and $100\,\GeV$.  The energy
resolution of HCAL1 as a function of the energy for pions can be
parameterized by
$\sigma(E)/E=(59.4\pm2.9)\,\%/\sqrt{E}\bigoplus(7.6\pm0.4)\,\%$, with
the energy $E$ in units of $\GeV$. A similar dependence is expected from MC
simulations. The average value of the $e/\pi$ ratio, calculated from
the positions of the electron and pion ADC spectra at the same energy,
is $1.2\pm0.1$. The spatial resolution $\sigma_{\rm x,y}=14\pm2\,\mm$
was measured by scanning the beam over the central module of a
$5\times5$ matrix and determining of the shower centre of gravity.

The stability during data taking is continuously monitored using a LED
system based on a single light emitting diode (LED) of the MARL 110106
type. The light of the LED is distributed and delivered to all $480$
modules by optical fibres about $3\,\m$ long and $1\,\mm$ in diameter.
A stable photodiode controls the intensity of the LED. The relative
stability of HCAL1 was controlled to 2\%.  The information is also
used to introduce corrections to the detected energy, if necessary.

At the beginning of each COMPASS data taking period, the calibration
of the HCAL1 modules is checked by using halo muons.
% The amplitude spectra are measured as well as relative delays of
% signals. The spectra were approximated by Landau distributions and
% normalisation coefficients determined from the positions of the
% maxima which correspond to the energy of $1.8\,\GeV$ deposited in
% the modules by the halo muons. The maxima were determined with a
% precision of about $0.1\,\GeV$.
By measuring the amplitude spectra the most probable energy loss of
halo muons corresponding to $1.8\,\GeV$ is determined with a precision
of $0.1\,\GeV$.  In addition the relative signal delays are equalised.

% The hadron energy release in HCAL1 is reconstructed offline summing
% up energies deposited in clusters of adjacent modules.The size of a
% cluster ($SC$) is determined as a number of adjacent modules, in
% which $\ge90\%$ of the particle energy is deposited. Clustersize
% $SC>3$ is the optimal value for hadron identification in HCAL1.

\begin{figure}[tbp]
  \begin{center}
    \includegraphics[width=\columnwidth,clip]{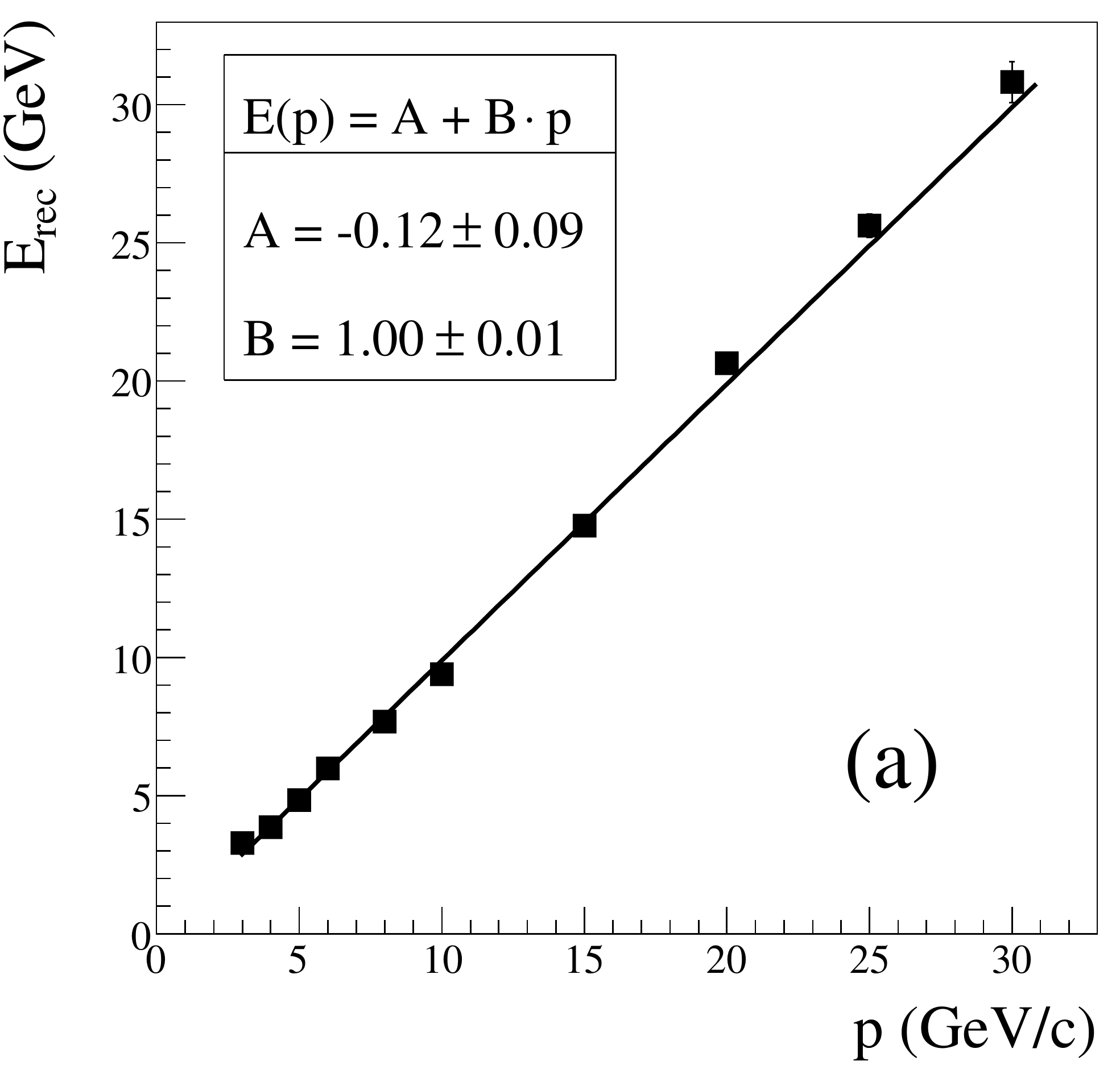}
    \includegraphics[width=\columnwidth,clip]{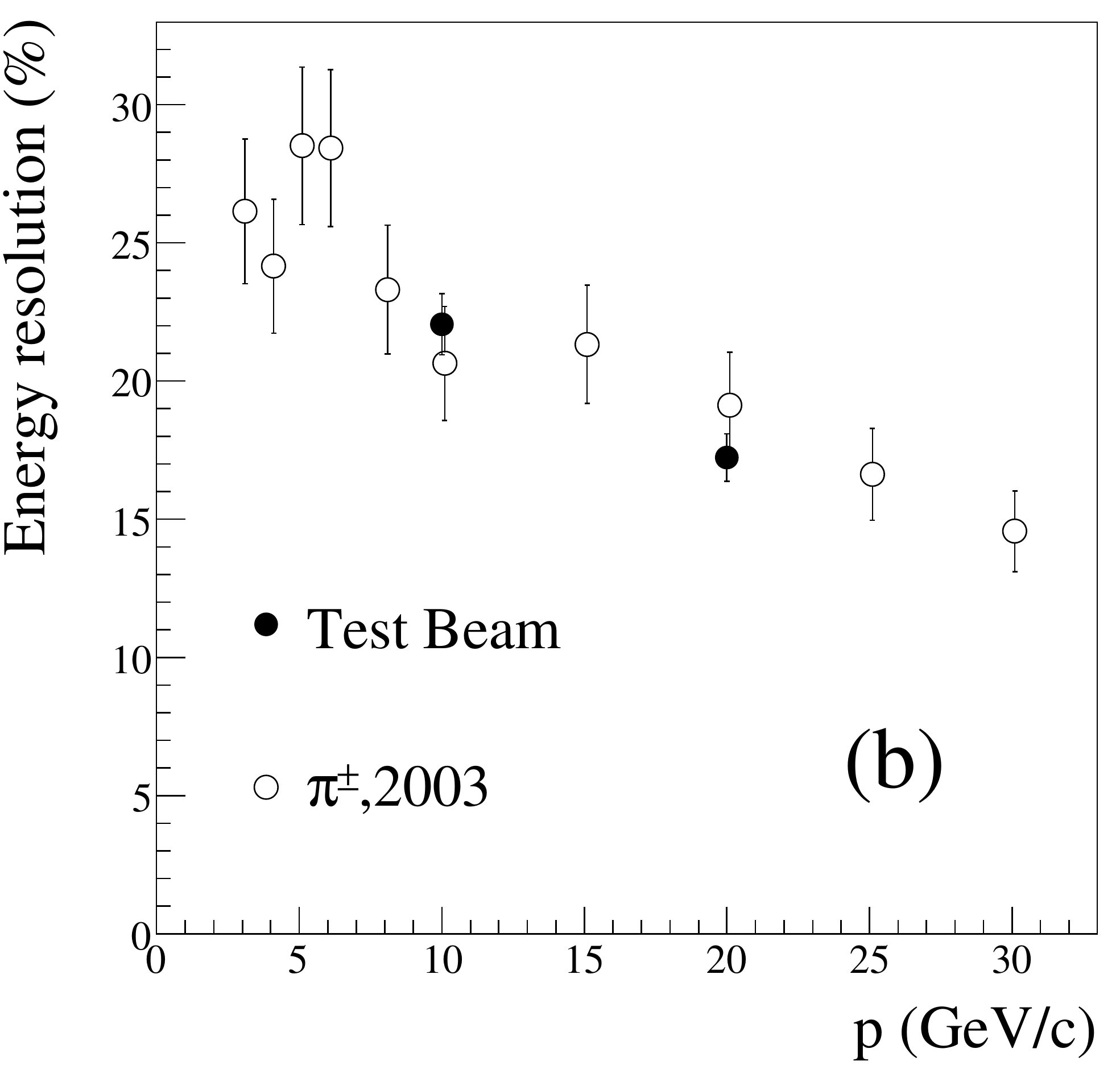}
  \end{center}
  \caption { \small{HCAL1 energy reconstruction (a) and energy
      resolution (b) as function of momentum for identified hadrons
      (the full symbols denote test beam results).  }}
  \label{fig:compass.HCAL1_resol}
\end{figure}

The energy deposited in the calorimeter is compared to the energy
(momentum) of the associated particle reconstructed by the COMPASS
reconstruction program CORAL.  For this purpose the associated
particles are combined in ten momentum bins of $0.25\,\GeV/c$ width
around the central values $3,5,...30\, \GeV/c$ and the central parts
of the corresponding HCAL1 energy spectra are fitted by Gaussian
distributions. The positions of the peaks and the corresponding
resolution $\sigma$ as a function of the track momentum are shown in
Fig.~\ref{fig:compass.HCAL1_resol}. A linear fit $E(p)=A+B\cdot p$ confirms
the linearity of the calorimeter energy response and the correctness
of the calibration. The constant term $A$ is close to zero within
errors and the slope $B$ is equal to $1$ with an uncertainty of about
$1\%$.

The efficiency of HCAL1 includes also the efficiency of the cluster
search and the energy reconstruction. Thus, the total HCAL1 efficiency
depends in principle on the particle energy. For particles with
momenta above $5\,\GeV/c$ the efficiency is almost constant and close
to $100\%$.

%%%%%%%%%%%%%%%%%%%%%%%%%%%%%%%%%%%%%%%%%%%%%%%%

\subsubsection{Hadron calorimeter 2}

\label{sec:pid.calo.hcal2}

The hadron calorimeter 2 (HCAL2) takes the form of a matrix of $22\times 10$ modules, 
arranged on a mobile platform. The basic modules are sandwich counters
 with $20\times 20\, \cm ^2$ transverse dimensions. The calorimeter
has a hole with the dimensions of $2\times 2$ modules to pass the high 
intensity beam. Two types of modules are used in the detector. Most
of them consist of thirty-six  $25\, \mm$ thick steel plates, 
interleaved with $5\, \mm$ thick  scintillator sheets.
The modules were previously used in the NA12
experiment~\cite{Binon:87} but the method of light collection from the
scintillators and the HV bases for the PMs were modified. The overall 
thickness of the counters is five nuclear interaction 
lengths for pions and seven for protons. The 
central $8\times6$~cells are filled with thicker modules consisting of forty 
layers.
 
\begin{figure}[tbp]
  \begin{center} 
    \includegraphics[width=\columnwidth,viewport= 100 200 500 580]{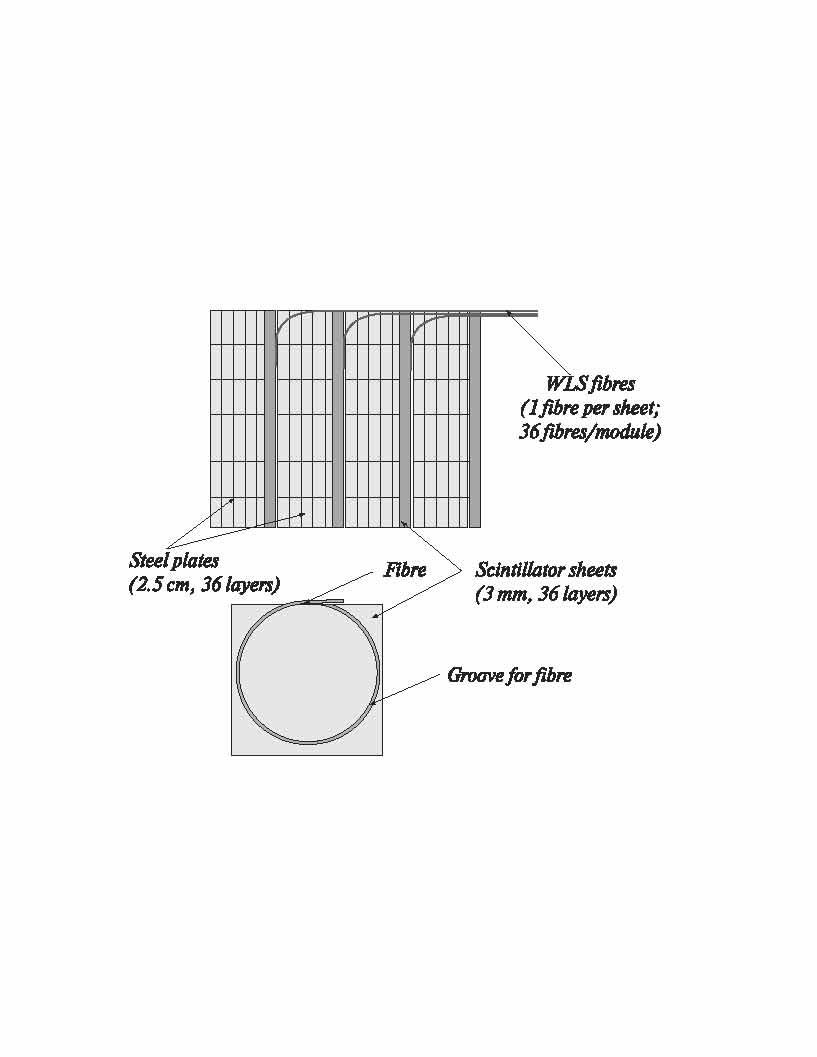}
  \end{center}
  \caption{\small Principle of the fibre light readout of HCAL2,
top: side view of part of a module, bottom: front view of a
  scintillator plate with the fibre readout.}
  \label{fig:compass.HCAL2_lightcoll}
\end{figure}

The principle of the light collection from the scintillator sheets is
presented in  Fig.~\ref{fig:compass.HCAL2_lightcoll}.  The readout of 
the scintillation light is done by wavelength shifting 
fibres of $1\,\mm$ diameter placed in a circular groove in each 
scintillator sheet. 
The bundle of fibres from all sheets collects the light onto the S-20
type photocathode 
of a FEU-84-3 PMT. All scintillator fibre pairs were characterised 
with a radioactive source and sorted to equalise the module light
output.
%optimize the overall detector resolution.  
%The photomultiplier and its divider are placed in a housing
%at the end of module. 
A silicon compound provides the optical contact between the bundle of
fibres and the PMT. 
The computer controlled HV divider is of Cockcroft-Walton type. The PMT signals are 
measured with the specially designed 12-bit fast analogue-to-digital
converters (see Sec.~\ref{sec:daq.analog.caladc}).  
Small fractions of the signals are fed into the fast summation system for trigger purposes. 
The timing spread of the signals from different PMTs is compensated with pieces of 
coaxial cables. 

The calorimeter has a light monitoring 
system common for all modules using a light pulse from a group of light
emitting diodes fed to the PMs through fibre 
light guides. The monitoring system is used for amplification adjustment and control.

\begin{figure}[tbp]
  \begin{center}
    \includegraphics[width=\columnwidth]{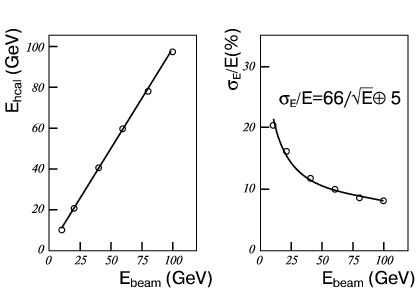}
  \end{center}
  \caption{\small Energy deposition in HCAL2 (left) and
  energy resolution (right) as a function of the beam energy.}
  \label{fig:compass.HCAL2_energy_resolution}
\end{figure}

As for HCAL1 the characteristics of the HCAL2 modules were determined
 at the X5 test beam  
of the CERN SPS using a matrix of $5\times5$ modules. 
HCAL2 has a good linearity in the range from $10$ to $100\,\GeV$, the energy resolution is
 $\sigma (E)/E = (66/\sqrt{E} \bigoplus  5)\,\%$, with the energy $E$
in units of $\GeV$ (see Fig.~\ref{fig:compass.HCAL2_energy_resolution}).  
%In Fig.~\ref{fig:compass.HCAl2_uniformity} the calorimter response is 
%presented, when the beam was scanned across the surface of 
%the test setup. The uniformity is better than $\pm 2$\%.
The uniformity of the calorimeter response was checked by scanning
the
test beam across the surface of the test setup. It was found to be
better than 2\%.
 
The final calibrations of HCAL2 are carried in the
halo of the muon beam of $160\,\GeV$ 
irradiating the whole calorimeter in the same way as for HCAL1.
The efficiency for hadrons is close to 100\% for hadrons with energies
above $10\,\GeV$. 
%The 
%high voltages on all modules have been adjusted to get the same signal
%response for the muon energy deposition.
%Then the final calibration was done. 
%Muon spectra for a few channels are presented in
%fig.\ref{fig:compass.HCAL2_muon}. 
%The muon energy deposition 
%corresponds to $\approx 2\, \GeV$ (More precise?).
%\begin{figure}[tbp]
%  \begin{center}
%    \includegraphics[width=\columnwidth]{./figures/HCAL2_muon}
%  \end{center}
%  \caption{\small HCAL2 muon spectra.}
%  \label{fig:compass.HCAL2_muon}
%\end{figure}

\subsubsection{Electromagnetic calorimeter}

\label{sec:pid.calo.ecal}

% \subsubsection{Electromagnetic Calorimeter}
% \label{sec:pid.calo.ecal}
% \bibitem{Binon:86} F. Binon et al., Nucl. Instr. and Meth. A248
%   (1986) 86.
% \bibitem{Alde:86} D. Alde et al., Nucl. Phys. B269 (1986) 485.  To
%   be written by V.Polyakov

The electromagnetic calorimeter ECAL2 in the SAS part of the COMPASS 
spectrometer consists of
2972 (a matrix of $64\times 48$) lead glass modules 
with $38\times 38\times 450\, \mm ^3$ 
dimensions amounting to 16 radiation lengths 
that were previously used in the
GAMS-4000 spectrometer~\cite{Alde:86,Binon:86}.
A high energy gamma ray (or electron) incident on ECAL2 develops
an electromagnetic shower inside the lead glass. The electrons and
positrons from a shower emit Cherenkov light on their way through the
glass.  The amount of Cherenkov light is proportional to the energy
deposited in each counter. Each lead glass block is viewed at one
end by a PMT which measures the intensity of the
light emitted at that counter.
%The
%mechanical support structure of GAMS-4000, mainly the cassette, was
%modified for COMPASS requirements. 
%The layout of the detector and the schematic drawing of
%the cassette structure are shown in
%fig.\ref{fig:pid.calo.ecal.gams4000}.  

The modules are
installed inside a frame, which can be moved vertically and
horizontally by $2.5\,\m$ for calibration and maintenance.
%nside the ECAL2 chariot and horizontally by 2.5 m together with the
%hariot along platform rails. 
The ECAL2 platform can be moved on
rails along beam axis. 
%The distance between the
%target and the ECAL2 is equal to 32 m. 
A hole of $10\times 10$ modules in
the centre allows passage of the beam particles.
Two types of lead glasses, TF1 (SF2 class) and TF101
(radiation-hardened by adding 0.2\% of cerium), are used in the
detector. The radiation-hardened modules (approximately 800) are
placed around the central hole. They can tolerate a dose of a few 
krad~\cite{Kobayashi:94}. 
%\begin{figure}[tbp]
%  \begin{center}
%    \includegraphics[width=\columnwidth]{./figures/ecal2_view}
%  \end{center}
%  \begin{center}
%    \includegraphics[width=0.6\columnwidth]{./figures/cassette_mu}
%  \end{center}
%  \caption{\small The layout and the cassette structure of the second
%    electromagnetic calorimeter.}
%  \label{fig:pid.calo.ecal.gams4000}
%\end{figure}

\begin{figure}[tbp]
  \begin{center}
    \includegraphics[width=\columnwidth,viewport= 10 230 560 450,clip=true]
   {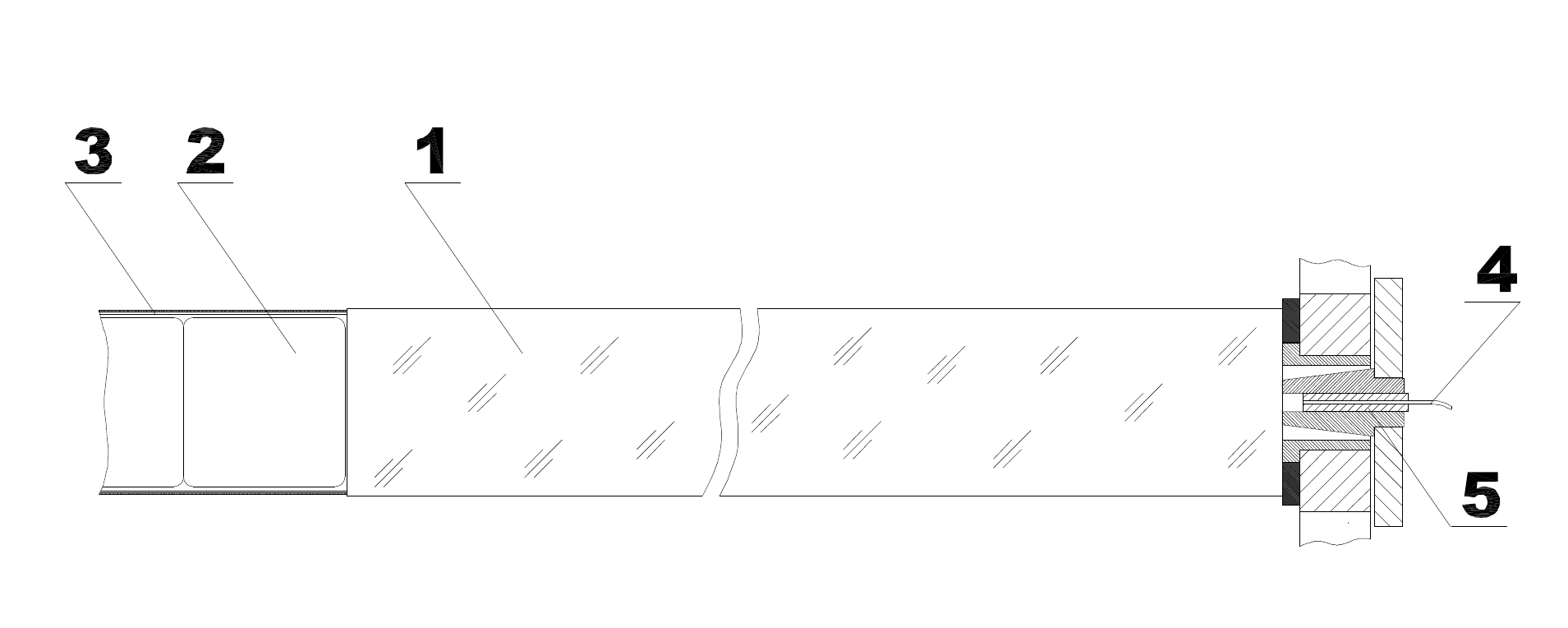}
  \end{center}
  \caption{\small A counter of the ECAL2 calorimeter: 1) TF1-000 lead
%    glass radiator 2) FEU-84-3 PMT 3) silicone compound 4)
%    plastis cup 5) permalloy magnetic screen 6) porous rubber 7)
%    cassette front panel 8) plastic flange 9) back plane of the
%    monitoring system box 10) guartz fiber to distribute light pulse
%    of the monitoring system 11) light guide connector 12) plastic
%    washer.}
    glass radiator, 2) FEU-84-3 PMT, 3) permalloy 
    magnetic screen, 4) quartz fibre to distribute the light pulse
    of the monitoring system, 5) light guide connector.}
  \label{fig:pid.calo.ecal.LG_blocks}
\end{figure}

A lead glass module is shown in
Fig.~\ref{fig:pid.calo.ecal.LG_blocks}. The accuracy of the transverse 
dimension is $0.05\,\mm$. A plastic cylinder enclosing the PMT is glued with epoxy to
one end of the glass
prism. The PMT is wrapped into three layers of
permalloy of $0.1\,\mm$ thickness. It is pressed to the glass by an elastic
plate.  The optical contact with the radiator is made through silicon
grease.  The lead glass radiator is wrapped in aluminised mylar. The
aluminised side of the mylar is protected with special varnish, which
prevents destruction of the reflecting layer and diffusion of
aluminium atoms into the glass.  The FEU-84-3 PMT with a trialkaline
photocathode of S-20 type, $34\,\mm$ in diameter, has been chosen for the
counters. About 1000 photoelectrons per $\GeV$ energy deposit are obtained.

Special high voltage dividers have been designed in order to obtain a maximum
range of linearity.  The divider and the electronics of the computer
controlled high voltage supply are described in \cite{Binon:83}.
For gain monitoring purposes, light pulses produced by a group of LEDs
are delivered by a 2 m quartz fibre to the front of each counter.
The anode signals of all counters are measured by two types of
analogue-to-digital converters, FIADC and SADC 
(see Sec.~\ref{sec:daq.analog.caladc}).

The calibration of ECAL2 is carried out at the COMPASS experiment
with the $40\, \GeV$ 
electron beam and repeated for each data taking period.
The whole calorimeter is moved to expose all modules to the
beam and a few thousand events for each module are collected.

The resolutions and linearity were measured previously
\cite{Alde:86,Binon:86,Binon:81}.
For the  energy and space resolution
 $\sigma (E)/E = 5.5\%/\sqrt{E} \bigoplus  1.5\%$ and $\sigma (x) =
 6\,\mm/\sqrt{E} \bigoplus  0.5\,\mm$ were obtained, respectively,
 with the energy $E$ in units of $\GeV$.  
The linearity was better
 than 1\% in the energy range from $0.5$ to $40\,\GeV$.

% Trigger
\section{Trigger}
\label{sec:trigger}
The trigger system has to serve several purposes: to select event
candidates
in a high rate environment with a decision time below $500\,\ns$ and
minimum dead time, to provide an event time reference 
and generate strobes for gating some of the analog-to-digital
converters, and to trigger the readout of detectors and front-end
electronics.

The trigger system is based on fast hodoscope signals, energy deposits
in calorimeters and a veto system. Depending on the incident beam
--- muons or hadrons --- and on the kinematics of the reactions different
elements are combined to form the trigger signal.

\subsection{Muon beam}
\label{sec:trigger.muon}
 
The COMPASS setup for the
muon beam is designed for an as large kinematical acceptance in $Q^2$ as
possible
ranging from $Q^2 \approx 0$ to the maximum allowed by kinematics.
Simultaneously a large range in the energy loss $\nu$ is required.

Events with $Q^2>0.5\,(\GeV/c)^2$ are mainly triggered by using the
scattered muon information only, as it was done in previous muon experiments
\cite{Allkofer:81}.
The muons are measured in two horizontal scintillator hodoscopes in
order to determine
the projection of the muon scattering angle $\theta$ in the
non-bending plane and to check its compatibility with the target
position (vertical target pointing). To suppress events due
to halo
muons, a veto system is added to the trigger system.

\begin{figure}[tbp]
\begin{center}
\includegraphics[width=\columnwidth]{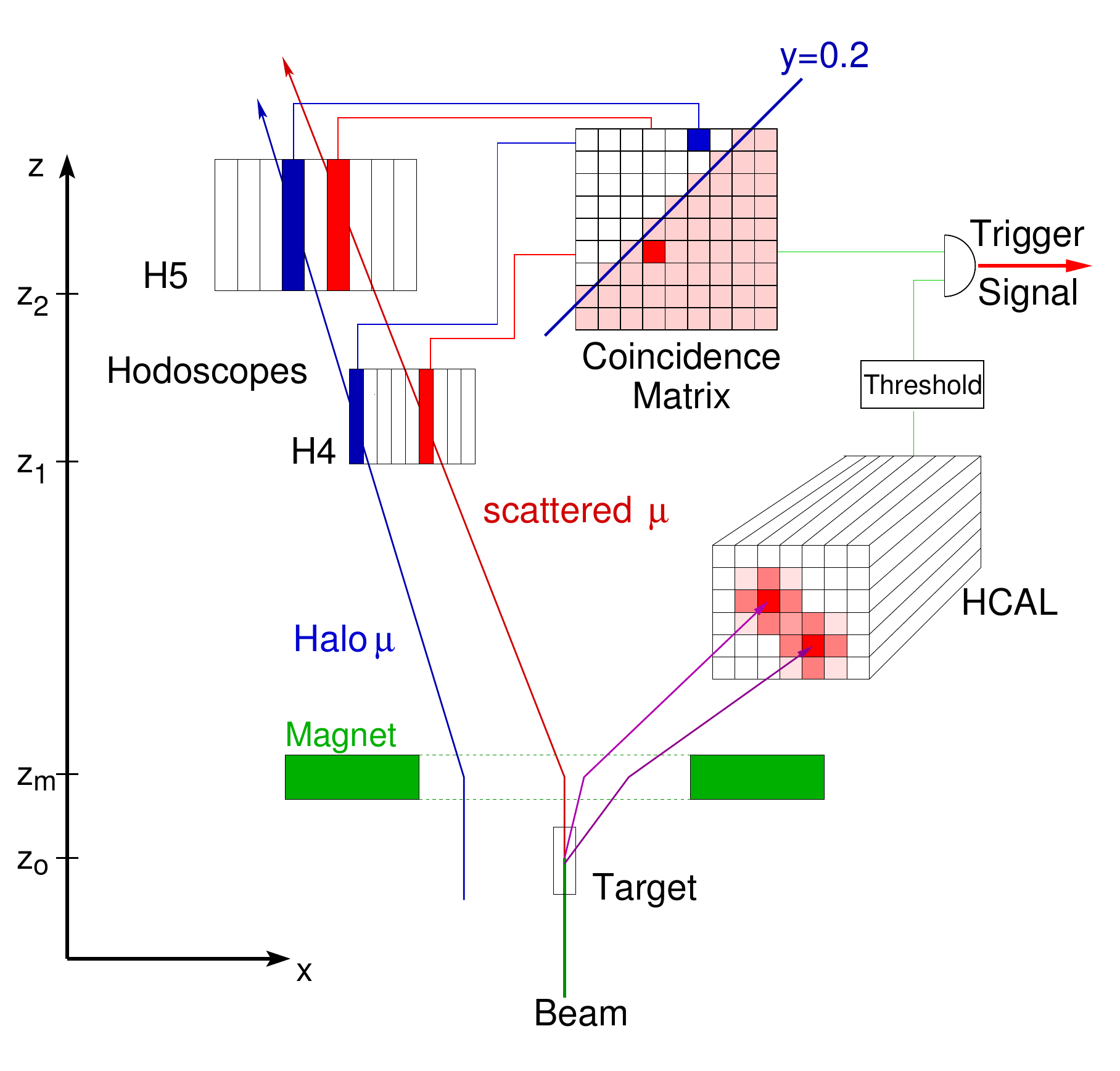}
\end{center}
\caption{\small Concept of the trigger for quasi-real photoproduction with
high energy loss. The scattered muon leads to a coincidence in the
activated
area of the coincidence matrix while the halo muon fails to do so. In addition,
a minimum hadron energy can be required in the calorimeter.}
\label{fig:trigger.muon.concept.schema}
\end{figure}

\begin{figure}[tbp]
  \begin{center}
    \includegraphics[width=\columnwidth]{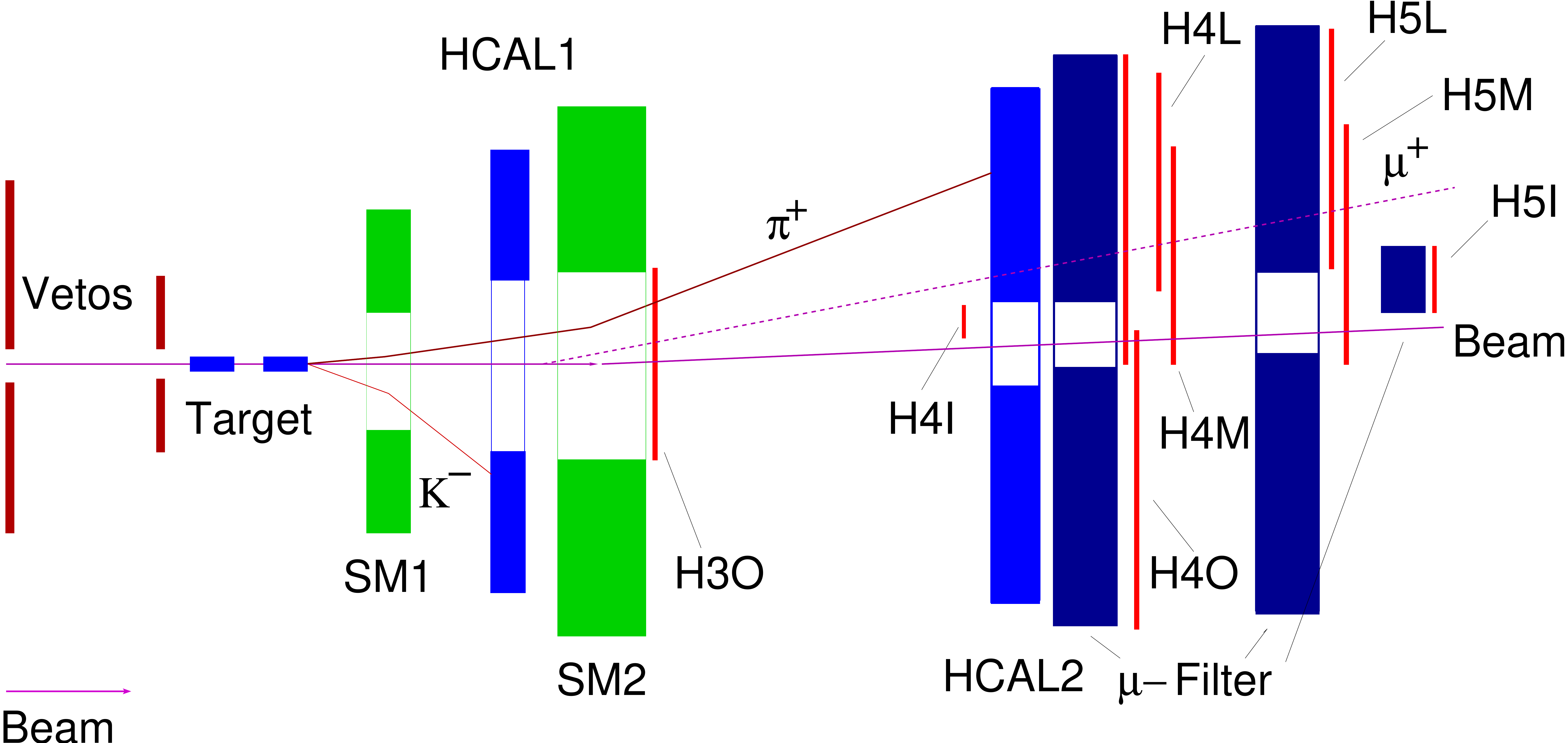}
  \end{center}
    \caption{\small Location of the components relevant for the trigger 
(schematically). For the true scale refer to
    Fig.~\ref{fig:layout.setup}.}
    \label{fig:trigger.muon.setup}
\end{figure}

At low $Q^2$, in the quasi-real photon regime, 
the muon scattering
angles are close to zero so that target pointing does not work any longer.
These events are selected by measuring the energy loss with two
vertical scintillator hodoscopes using the
bending of the muon track in the spectrometer magnets.
At these small angles there are  several background
processes such as elastic scattering off target electrons, elastic and
quasi-elastic radiative scattering off target nuclei and beam halo
contributing to the scattered muon signal. 
The trigger system requires energy clusters in the hadronic
calorimeter, which are absent in the background processes.
Thus, the quasi-real photon trigger consists of two parts, a trigger on the
energy loss by measuring the deflection of the scattered muon in the
two spectrometer magnets and a calorimetric trigger selecting hadron
energy clusters above a threshold (see
Fig.~\ref{fig:trigger.muon.concept.schema}). 
A detailed description of the trigger
system
is given in \cite{Bernet:05a}.
The location of the components of the trigger system in the COMPASS
experiment is shown schematically in Fig.~\ref{fig:trigger.muon.setup}.

\subsubsection{Hodoscope triggers}
\label{sec:trigger.muon.hodo}

In view of the high rates in the central region the hodoscopes of
the trigger system are subdivided into four subsystems consisting of two
hodoscope stations each, the inner (H4I, H5I), the ladder (H4L, H5L) , 
the middle (H4M, H5M) and the outer system (H3O, H4O).

With the inner and the ladder system the deflection of the scattered muon in
the two spectrometer magnets is estimated by requiring spatial
coincidences between the vertical elements in the two hodoscope
planes, 
where the precision is given by the element width of the hodoscopes.
The width of the beam at the target, the momentum spread
and the multiple scattering introduce an uncertainty into the spatial
correlation which has to be taken into account when choosing the
minimal relative energy loss $y_{\rm min}$.
In addition, muon selection requires an absorber in front of one of the
two hodoscopes to reject hadrons and electrons. The effect of the
additional
multiple scattering can be minimised by placing the 
absorber ($1.6\,\m$ Fe) directly in front of the second inner hodoscope. 
The 
detectors between the two inner trigger hodoscopes have a hole that
matches the size of these hodoscopes.
A fine grained structure of the two hodoscopes and a muon filter close
to the second one is only needed for the $y$ region where the cut is
applied. The inner system measuring the $y$ range from 0.2 to 0.5 is
therefore fine
grained with element widths of 6 and $12\,\mm$, while the ladder system
for $y$ between 0.5 and 0.9 
uses 20 to $87\,\mm$ strips. 

The middle system
selects quasi-real photon events as well as DIS events. It uses
horizontal planes (21.5 to $30\,\mm$ strips) to detect muons with
scattering angles between 4 and
$12\,\mrad$. To allow a coarse energy cut, vertical planes (62 and
$77\,\mm$ strips) are used.
The outer system selects muons up to $Q^2\approx 20\,(\GeV/c)^2$ by
vertical target pointing using 70 and $150\,\mm$ wide elements. 
The upper limit in $Q^2$ is fixed by the
size of the gap in the second spectrometer magnet.

The hodoscopes are read out using lightguides and photomultipliers 
on one or both sides of the elements. The signals are discriminated 
in customised constant fraction discriminators (CFD). 
A custom CFD module contains two CFDs and one optional meantimer used  for
the scintillators with two-sided readout.
 The signals are fed into coincidence matrices which select only 
those hit combinations
 which correspond to muon tracks that
 either just point back to the target 
 (for hodoscopes with horizontal strips)
 or that have suffered a minimum energy loss
 (for hodoscopes with vertical strips).

The discriminator boards holding 32~CFD modules and the coincidence matrices 
are custom-made VME boards.
Special care has gone into the design of the coincidence matrices,
 which use full custom CMOS chips to achieve excellent timing over
 the set of 1024 individual coincidences in each $32 \times 32$ matrix.
To be able to fully exploit the time resolution of the hodoscopes
 and the coincidence chips, a channel by channel timing compensation
 stage
 is included in each of the 64 input channels of each matrix board,
 using custom-designed CMOS chips with 16 delay and pulse shaping channels 
 with 32 timing steps in a $9\,\ns$ range.

\subsubsection{Calorimetric trigger}
\label{sec:trigger.muon.calo}

For hadron detection
 the signals produced in the two hadronic calorimeters are used
to require a cluster
with an energy deposition well beyond the value expected for a single muon.
In order to correlate an energy deposition with a scattering event on
the trigger level a
time resolution of order of $1\,$ns of the calorimeter is necessary to reduce
the number of accidental coincidences.
Also, the trigger has to deal with the very high rate of halo
muons of $2\cdot 10^7$/s passing through the calorimeter depositing
an energy of typically $1.8\,\GeV$.
The pulse length of an
sandwiched iron-scintillator with wavelength shifter readout is of order
50~ns. Therefore the  probability for a two muon pile-up is of order one.
The problem can be alleviated by requiring at least one local cluster 
with an energy exceeding
that of a single muon. The  size of an energy cluster corresponding to a
hadron is of order of one interaction length squared i.e. typically
$20 \times 20\,\cm^2$. With a suitable cluster forming and processing hardware 
the pile-up
problem can be reduced by more than a factor 10. 

The signals of $4 \times 4$ calorimeter cells are   summed in 
overlapping regions 
 in order to contain complete hadronic showers in one electronic
 channel
 while keeping the number of summed channels low in order to
 limit noise and pile-up of halo muon signals.
This is done in two stages:
At the back of the calorimeter itself
 the PMT signals are split into a larger fraction,
 which goes to the ADCs that measure the individual cells,
 and a smaller fraction that is summed up in a $2 \times 2$ pattern.
%%%%%%%%%%%rephrased:
At the second stage the $2 \times 2$ sums are summed up
into $4 \times 4$ sums.  There are four different possibilities to do the
summation to cover the calorimeter surface, thus forming four
different layers.
At the end, every shower is contained in at least one of
the sum-of-16 signals of one layer.

Each sum-of-16 signal of the four layers of both calorimeters is
discriminated with two CFDs and the discriminated signals are fed into
multiplicity units. For the calorimetric trigger in coincidence with a
hodoscope trigger, typically a $5.4\,\GeV$  threshold together with a
multiplicity 
$\ge 1$ is used.
This threshold suppresses the single muon response by more
than $90\,$\%.  The standalone calorimetric trigger uses a much higher
threshold ($16.2\,\GeV$ in 2003, $8\,\GeV$ in 2004) together with a
multiplicity $\ge 1$ on two layers. It extends the kinematic range to
large $Q^2$ values not covered by the hodoscopes and allows an
evaluation of the hodoscope trigger efficiencies.

\subsubsection{Veto system}
\label{sec:trigger.muon.veto}

\begin{figure}[tbp]
 \begin{center}
     \includegraphics[width=\columnwidth]{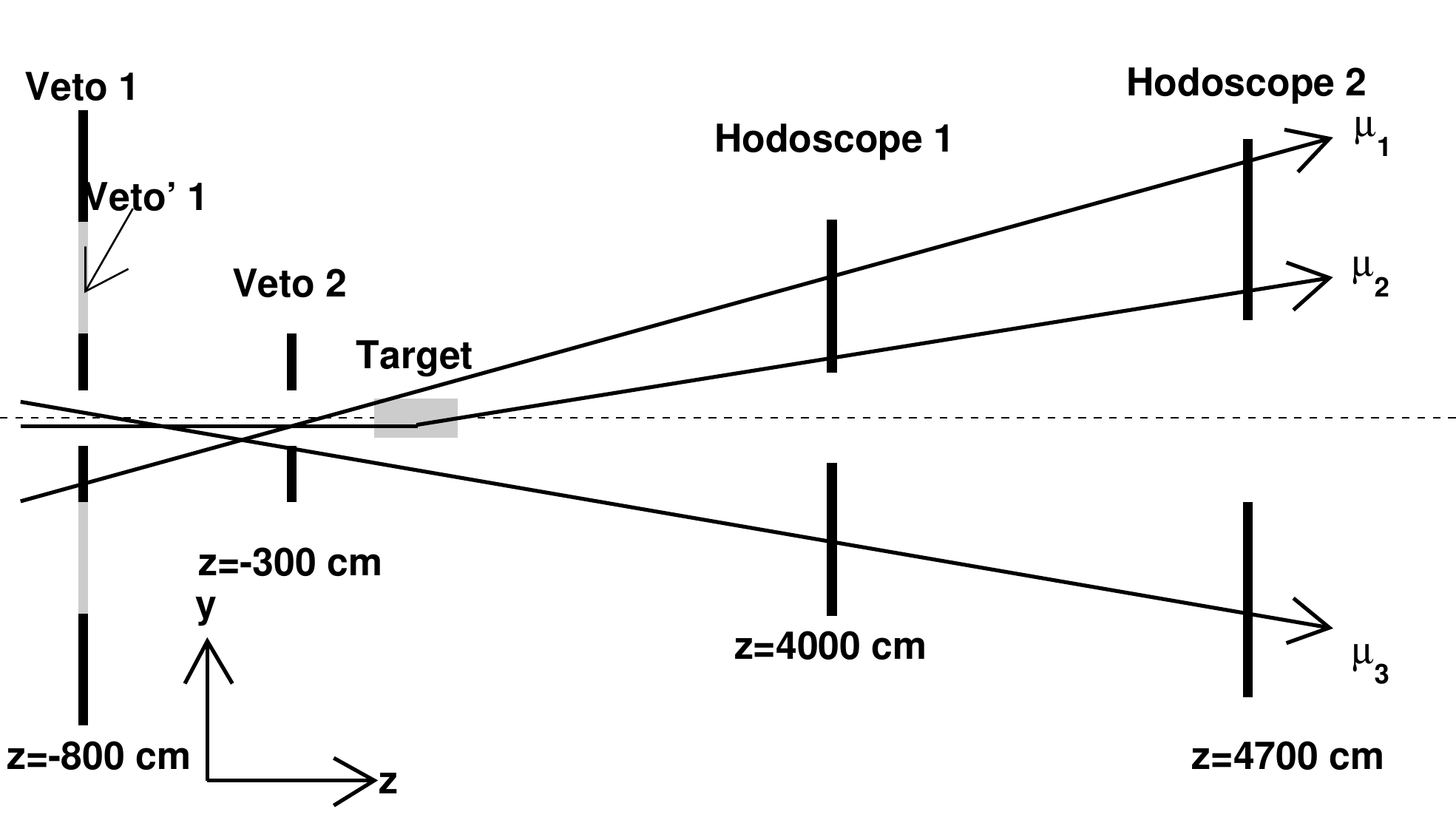}
 \end{center}
     \caption{\small Schematical layout of the veto system. 
The tracks $\mu_1$ and $\mu_3$ are vetoed, whereas the track $\mu_2$
fulfils the inclusive trigger condition.}
     \label{fig:trigger.muon.veto.schema}
 \end{figure}

Due to the sizeable muon beam emittance and the halo of about $25\,$\%
(near and far halo see Sec.~\ref{sec:beam})
many triggers in the hodoscope systems are caused by muons not
interacting in the target. They can be rejected by adding a veto
system to the trigger. It consists of two scintillator counters upstream
the target leaving the central region around the beam uncovered. The
first large detector (Veto 1, $250\,\cm \times 320\,\cm$) is located at
$-800\,\cm$
and the second smaller one (Veto 2, $30\,\cm \times 30\,\cm$)
at $-300\,\cm$. Two detectors  are needed to veto divergent beam
particles passing through the $4\,\cm$ diameter holes in one of them (for
illustration
see Fig.~\ref{fig:trigger.muon.veto.schema}).
This veto system is only fully efficient for tracks with a slope
greater than $8\,\mrad$ due to the limited space between the target
and the first beam line element. Therefore, a third counter (Veto BL, $50\,\cm
\times 50\,\cm$) with a $10\,\cm$ diameter hole was
placed further upstream at $-2000\,\cm$ (not shown in
Fig.~\ref{fig:trigger.muon.veto.schema}). All veto counters are
segmented, smaller elements are used close to the beam axis and larger
elements in the outside region.

A drawback of the veto system is the dead time associated to it. It is
given by the product of the rate of the system and the duration of the
time
gate during which the veto prohibits a trigger signal. The dead time
of the full veto system is about $20\,$\% at nominal beam intensity. This
veto
is only applied to inclusive triggers (middle and outer) which do
not require the calorimetric trigger. A subsample of the veto
elements was used for the ladder trigger due to the overlap of the
trigger hodoscopes with the calorimeters resulting in a dead time of
$6\,$\%.
No veto is needed for the inner trigger.
Using the full veto system the middle trigger rate is reduced from
1.4$\cdot 10^6$/spill to $18\,000$/spill.

\subsubsection{Performance}
\label{sec:trigger.muon.performance}

In 2004 the trigger rates were $14\,000$ and $7\,000$ per spill  
for the inner
and the ladder trigger in coincidence with the calorimeter trigger,
$18\,000$ and
$9\,500$ for the middle and outer trigger without calorimeter
condition and $22\,000$
for the standalone calorimetric trigger.

The efficiencies for all four hodoscope trigger systems
were obtained using the standalone calorimetric trigger  and yield 
values of above $99\,$\% for the inner and ladder
system, $>96\,\%$ for the middle and $>97\,\%$ for the outer system. These
lower numbers are due to the suppression of matrix elements which are
affected by large backgrounds.
The efficiency of the calorimetric trigger was studied in a similar way
using events from the middle trigger (without calorimeter condition)
with a vertex and more than 3 outgoing tracks. At $\nu\approx 40\,\GeV$
an efficiency of about $90\,$\% is reached.

After adjusting the delays at the input of the coincidence matrices,
a time resolution of approximately $500\,$ps between the trigger
signal and the beam momentum station is achieved 
for the inner middle and ladder system.
For the outer system, due to its larger elements, the resolution is $1\,$ns.
The standalone calorimetric trigger has a resolution of $2\,$ns.

The complete veto system has a time jitter of about $1.4\,$ns
with respect to the hodoscope systems.
Here the time resolution is limited by the use 
of large scintillator elements of up to $50\,\cm$ $\times$ $100\,\cm$
and elements with small light output due to their operation 
in the magnetic field of the target magnet. 

The efficiency of the veto system was estimated to be about $99\,$\%.
The resulting trigger purity is about $35\,$\% for the triggers using the
calorimetric trigger and $15\,$\% for the others.

\begin{figure}[tBP]
  \begin{center}
    \includegraphics[width=\columnwidth]{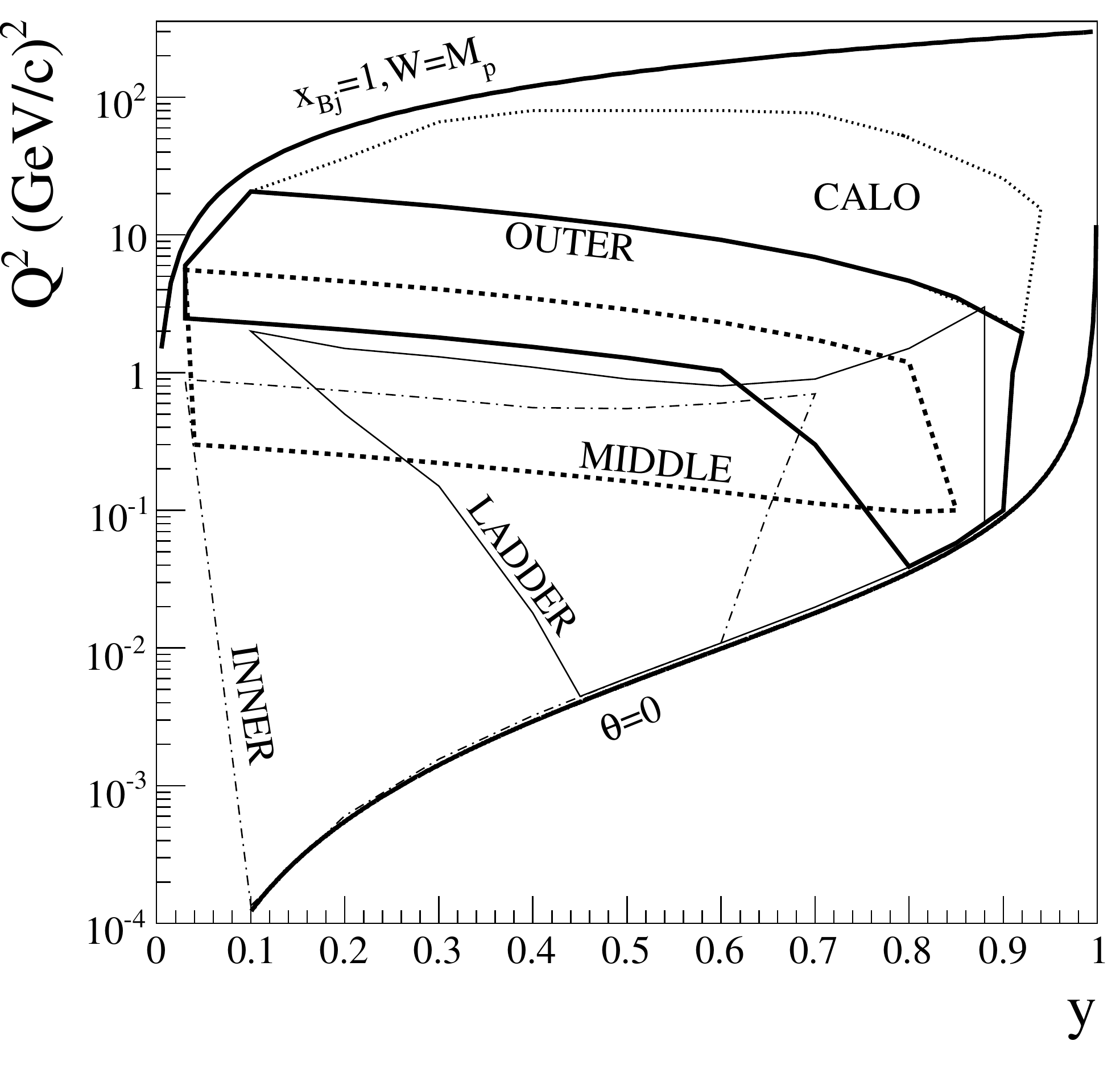}
  \end{center}
    \caption{The kinematical coverage in $y$ and $Q^2$ for the four 
    hodoscope trigger
    subsystems and the standalone calorimetric trigger.
    The two lines, $x_{\rm Bj}=1, W=M_{\rm p}$ and $\theta =0$ show 
    the kinematic limits of elastic scattering and forward scattering,
    respectively.}
\label{fig:trigger.muon.kinema}
\end{figure}

The kinematic range covered by the trigger system is
  illustrated in Fig.~\ref{fig:trigger.muon.kinema}. It shows the
  range in $y$ and $Q^2$ for the four
  hodoscope trigger subsystems and the standalone calorimeter trigger.

%%% Local Variables: 
%%% mode: latex
%%% TeX-master: t
%%% End: 

\subsection{Hadron beams}
\label{sec:trigger.hadron}

During the 2004 hadron run two different trigger systems have been
implemented, one designed to select Primakoff scattering events, and
the second to select diffractive $\eta$-meson production. The two
systems share a common set of devices that select beam
particles pointing to the target and reject events with particles 
emitted outside the
apparatus acceptance. In addition, some specific trigger component are
used to select the typical event topologies of Primakoff and
diffractive scattering. 

Beam particles are detected by the
  coincidence of two scintillator counters, with a diameter of about
$5 \,\cm$, centred on the beam trajectory. The Veto 1 system (see
  Sec.~\ref{sec:trigger.muon.veto}) with a central 
  hole of about $4\,\cm$ in diameter, placed upstream of the recoil veto
  (see Sec.~\ref{sec:targets.hadron}), is
  used to reject beam particles not crossing the target material.

Two additional veto counters made of lead-scintillator sandwiches and located
  downstream of the recoil veto, are used to reject events with charged
  particles or photons emitted at large angles and falling outside the
  acceptance of the electromagnetic and hadronic calorimeters. This
  condition is needed to precisely define the exclusivity of the
  reaction in the offline analysis. 

Non-interacting beam particles are rejected
  by means of a system of three small plastic scintillators (beam
  killers), with a diameter of $5\,\cm$ and a thickness of $5\,\mm$,
  centred on the 
  beam trajectory and placed between the SM2 magnet and the ECAL2
  calorimeter. 

The signal of the beam counters is used in anti-coincidence with the
sandwich vetos and the beam killers, to form the common part of the
trigger logic (common trigger). This information is combined with that
of the dedicated trigger devices for Primakoff and diffractive
scattering to produce the first level trigger for the COMPASS DAQ. 

%The trigger that has been implemented during the 2004 hadron run
%aimed at selecting Primakoff and Diffractive scattering events.  

\subsubsection{Primakoff triggers}
The Primakoff reaction is characterised by the emission of one single
high energy photon in coincidence with the scattered pion. Typical
photon energies are above $80\,\GeV$, corresponding to scattered pion
momenta below $110\,\GeV/c$. Events with smaller photon energies do not
significantly contribute to the measurement of the pion
polarisabilities. In order to cover the full energy range of the
emitted photons two different Primakoff triggers have been
implemented, called Primakoff~1 and Primakoff~2.  

\begin{figure*}
   \includegraphics[width=\textwidth]{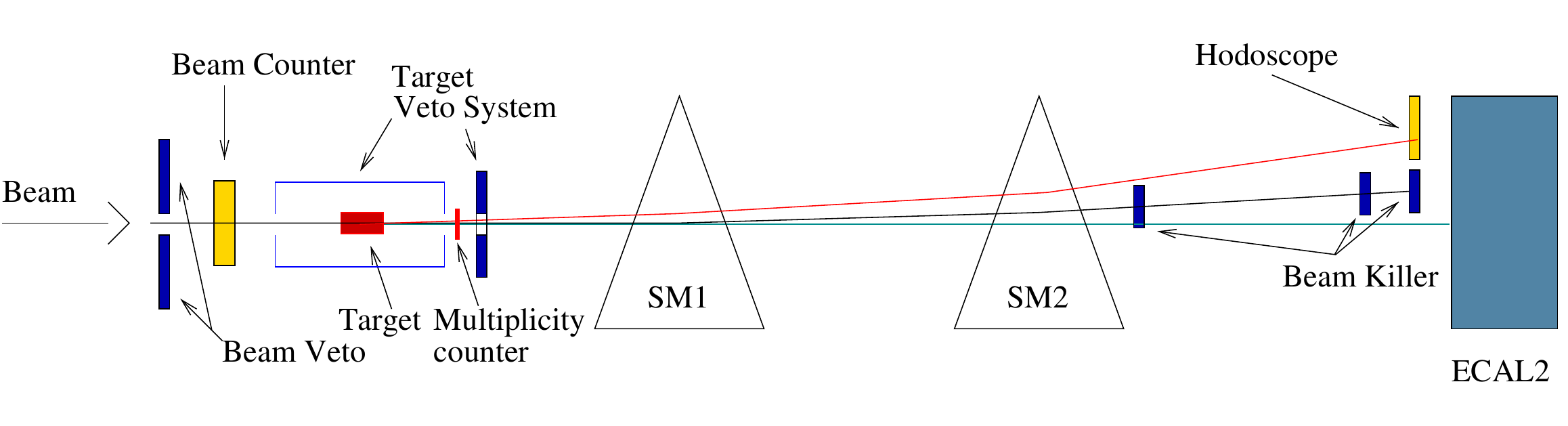}
   \caption{Schematic view of the Primakoff trigger.}
   \label{fig:primakoff_trigger}
 \end{figure*}

In the Primakoff~1 trigger the scattered pion is detected by means of a
 scintillator hodoscope located in front of the ECAL2 electromagnetic
 calorimeter (see Fig.~\ref{fig:primakoff_trigger}). The hodoscope is 
 composed of 20 slabs of $6 \times 90\,\cm^2$ in size and is able to
 detect pions in the energy range $20 - 110\,\GeV$. 
 Additionally, an energy deposit larger than $40\,\GeV$ in ECAL2 is
 required. The calorimeter threshold has been chosen half of the
 minimum photon energy of $80\,\GeV$ to avoid any bias in the event
 sample. The Primakoff~1 signal is then made by the coincidence of the
 common trigger, the scattered pion hodoscope, the ECAL2
 signal, and in addition an energy deposit above $18\,\GeV$ in HCAL2.

Events with very high photon energies are selected by the Primakoff~2
trigger. In this second case, the scattered pion hodoscope is excluded
from the logic and an higher energy threshold of $90\,\GeV$ is applied
to the summed ECAL2 signal. The overall Primakoff trigger is then
produced by the logical or of the Primakoff~1 and Primakoff~2 signals. 

Typical trigger rates of about $5.8\cdot 10^4$/spill for Primakoff~1
and $3.8\cdot 10^4$/spill for Primakoff~2 have been obtained during the
2004 hadron run. Due to the overlap in the kinematical region covered
by the two triggers, about 20\% of the Primakoff events fulfilled both
trigger conditions, leading to an overall Primakoff trigger rate of
about $7\cdot 10^4$/spill. 

  \subsubsection{Diffractive triggers}
 In addition to the Primakoff measurement, a test of the diffractive
meson production trigger was performed. The trigger was designed to select
the $(\eta\pi^-)$-system, with the $\eta$-meson decaying into
$\pi^+\pi^-\pi^0$. These events are characterised by the emission of three
charged particles and two photons from the interaction vertex. They
are selected by requiring that at least two charged particles hit a
scintillation counter read out by a photomultiplier tube and located
between the veto box and the downstream silicon telescope. The
multiplicity condition is obtained by setting a minimum threshold on
the photomultiplier signal. The multiplicity signal is then put in
coincidence with the common trigger to form the diffractive
trigger. Typical rates of about $2\cdot 10^4$/spill have been reached
during the 2004 hadron data taking. 

%%% Local Variables: 
%%% mode: latex
%%% TeX-master: t
%%% End: 

%%% Local Variables: 
%%% mode: latex
%%% TeX-master: t
%%% End: 

% Readout and DAQ
\section{Readout electronics and data acquisition}
\label{sec:daq}

\subsection{General ideas}
\label{sec:daq.overview}
% DAQ General ideas written by F.-H. Heinsius

The large number of 250~000 detector channels and the total amount of
up to $580\,\mathrm{TB}$ data recorded per year 
demanded to follow new directions in the
design of the data acquisition scheme. 
In order to cope with the high
particle fluxes of $2\cdot 10^8\,\mu$ per spill of $4.8\,\s$, a typical event
size of $35\,\mathrm{kB}$, trigger rates of about $10\,\kHz$ for the
muon beam, and a design value of $100\,\kHz$ triggers for the hadron
beam, a pipelined and nearly dead-time free readout scheme has
been adopted. 

\begin{figure}[tbp]
  \begin{center}
    \includegraphics[width=\columnwidth]{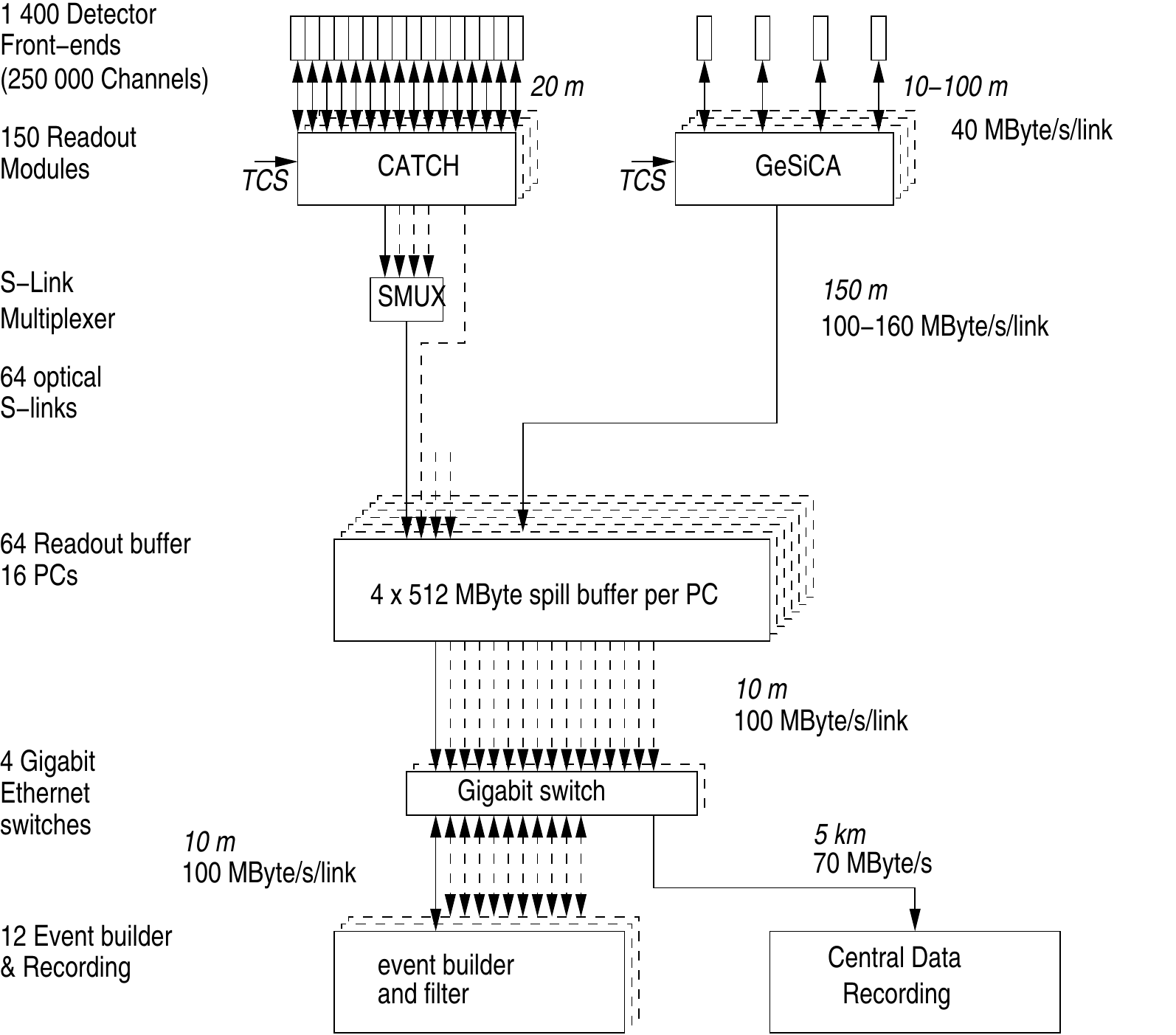}
  \end{center}
  \caption{\small General architecture of the DAQ system. Digitised
  data from the detector front-ends are combined on the readout modules
  named CATCH and GeSiCA close to the detectors. The storage of the
  data during the spill and the event building is performed locally. 
  The data are recorded at the CERN computer centre.}
  \label{fig:compass.daq.overview}
\end{figure}

An overview of the data flow is given in Fig.~\ref{fig:compass.daq.overview}.
The preamplifiers and discriminators are located close to the detectors.
The connection of the detector channels depends on the detector
type and is described in the corresponding detector chapters.
The data are constantly digitised and buffered, where possible directly at the
detector front-end electronics, in custom-designed TDC or ADC modules. 
The synchronisation of the digitising
and readout units is performed by  the trigger control system (TCS).
Upon arrival of the trigger signal the data are transferred via fast links to
readout-driver modules named CATCH and GeSiCA. 
These modules also distribute the trigger signals to the front-ends and
initialise them during the system startup. 
The readout-driver modules combine the data from
up to 16~front-end cards and transmit these sub events 
via optical S-LINK~\cite{Bij:96} at a maximum throughput of
$160\,\mathrm{MB}/\s$ to readout 
buffers. The data arriving from each link are stored in
$512\,\mathrm{MB}$ spill buffer cards. Data from readout modules serving low
occupancy detectors are combined by S-LINK multiplexer modules (SMUX)
before transmission through the S-LINK. 
In 2004 the total data transmitted during the spill to the
readout buffers corresponds to $230\,\mathrm{MB}/\s$.

The electronics components discussed above, apart from the S-LINK,
have been developed 
specially for COMPASS, while
the final event building system is based on high performance
PCs and standard Gigabit Ethernet components.
The event building takes place during the on- and off-spill time,
resulting in an average data rate of $70\,\mathrm{MB}/\s$.
These data are recorded on tape remotely at the CERN
central data recording facility located in the computer centre.

The architecture of the data acquisition system is very flexible and
expandable to  handle the modifications and upgrades during the lifetime of 
the experiment. New detectors can simply be added by including the
COMPASS standardised readout-driver modules and readout-buffer PCs,
while higher rates can be processed by adding more event builders and
utilising online filter capabilities.

\subsubsection{Ground connections}
Most of the readout components in COMPASS measure small and sensitive
electronics signals. Specific care was therefore taken for reducing
the overall electronics noise of the COMPASS environment. Beyond the
conventional safety and protection requirements, the grounding of the
COMPASS detectors was designed to minimise ground loops, unwanted
couplings and electrical interferences. A star topology of the
grounding cables was implemented, all protective grounds being
referenced to a unique point of the standard earth of the
experimental hall. As a result all electronics and detectors power
outlets (high and low voltages, electronics modules, data acquisition
racks) were separated from the standard ``structure'' system and
became a part of the ``electronics'' ground system in the star
topology of the COMPASS experimental setup. This topology is
complemented by electrical isolation of all electronics racks and
detectors. A galvanic isolation 
from the spill buffers and the subsequent DAQ elements
was achieved using optical links. The same
requirements were applied to the COMPASS slow-control system.  

%%% Local Variables: 
%%% mode: latex
%%% TeX-master: "compass_spec"
%%% End: 

\subsection{Trigger control system}
\label{sec:daq.tcs}
The main function of the Trigger Control System (TCS) is to distribute
trigger, time 
reference and event identification information to the readout-driver
modules (GeSiCA and CATCH) and to generate the strobes for gating some
of the analogue-to-digital converters.

The TCS employs an optical distribution system which broadcasts
information from a single source to a few hundred destinations using a
passive optical fibre network. The information is encoded on a serial
line with a speed of $155.52\,$MBaud. The clock with which the
information is encoded provides a time reference to all frond-end electronics.
The network has a star like architecture, where at the top of the
system the encoded signal is fanout to ten powerful laser
transmitters, then the optical signals are distributed via fibres to
different locations in the experimental hall. At every location the
light is split to 32~fibres by a passive optical coupler.
In 2004 the TCS distributed the information to 150 readout modules.

\begin{figure}[tbp]
  \begin{center}
    \includegraphics[width=\columnwidth]{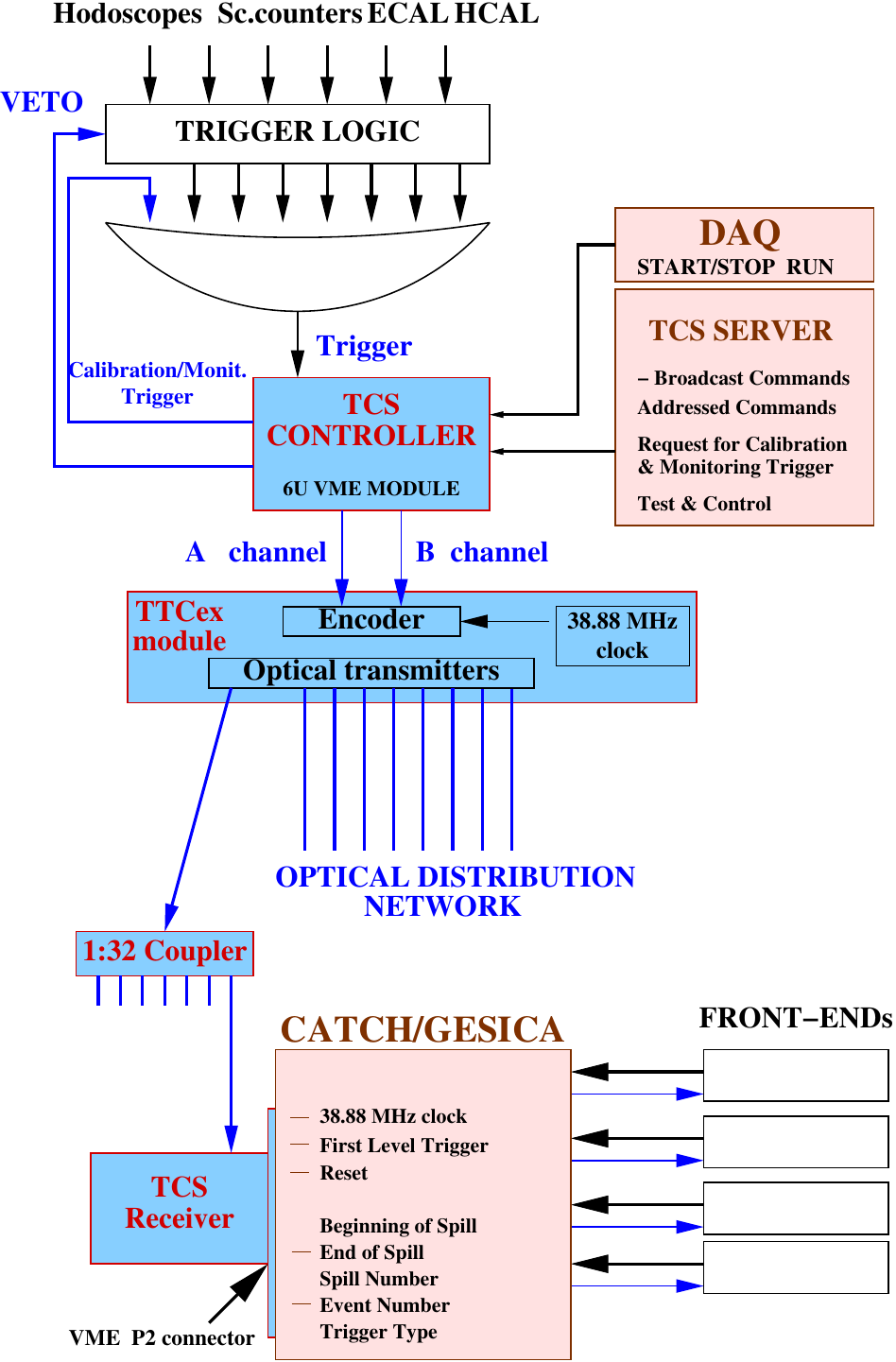}
    \caption{The architecture of the trigger control system.}
    \label{fig:daq.tcs.arch}
  \end{center}
\end{figure}

The trigger control system is based on the encoding method
and distribution principles of the Time and Trigger Control (TTC)
system developed for the LHC experiments~\cite{TTC:97}, namely the
TTCvi encoder and laser module and the passive optical splitters.
We have added the TCS controller, the TCS server and the TCS receivers 
(Fig.~\ref{fig:daq.tcs.arch}).

% ### bounding box, add SPS signal

The controller synchronises
data taking with the accelerator duty cycle, it encodes reset and
accepted trigger signals, counts the triggers, generates
the dead time, generates the event identification information,
retransmits the configuration commands from the TCS server and 
supports the MultiDAQ mode.
The TTCvi encoder and laser module encodes the signals from the TCS
controller and transmits them optically through the fibre network.
The TCS receiver is plugged at the back side of each slot
where GeSiCA or CATCH modules are located in a VME crate. The
receiver recovers the $155.52\,\MHz$ clock from the incoming data, decodes
the data, the trigger and provides this information to the readout-driver
module.

The TCS server program communicates with the TCS controller. It provides the
command interface between the TCS hardware and the DAQ and monitoring
software by means of a client-server protocol via TCP/IP.  It ensures
protected access to the TCS system and execution of all functions at
the right time.

Most of COMPASS front-end electronics are fully pipelined.  For some
detectors the pipeline principles could not be implemented at the time
when the readout electronics were built and a big effort was put
towards reduction of the dead time intervals between two consecutive
triggers.  Due to a mixture of two principles of the readout
electronics the TCS generates the dead time with two different
algorithms: 
the  minimum allowed time interval between two consecutive triggers and
the maximum number of triggers within a certain time interval.
The first dead time is needed for the not fully pipelined architecture of
the front-ends. The second one is needed for the pipeline readout
systems which have a limited depth of derandomised buffers and a limited
bandwidth of the data links.  During the 2004 data taking the dead time
conditions were set to:
\begin{itemize}
\item[-] $5\,\mu$s minimum time interval between triggers
\item[-] 3 events within $75\,\mu$s
\item[-] 6 events within $225\,\mu$s
\end{itemize}
This setting leads to $5$\% dead time at the nominal trigger rate of
$10\,\kHz$.

An extensive study of the clock jitter, as one of the most important
parameters of the system, was performed. The study demonstrated that the clock
jitter between two pairs of the TCS-receivers is about $44\,$ps RMS,
which is within the requirements of the COMPASS experiment to allow
time measurements with a resolution of $65\,$ps.

\subsection{Digitisation}
%\label{sec:daq.analog}
The analogue signals from the various detectors are preamplified and ---
for time measurements --- discriminated close to the detectors, as
described in the respective sections of the detectors. The
digitisation is also performed in most cases close to the detectors.
The gain is twofold: There is no loss of signal quality as no or only
short cables are required and the cost for cables is considerably reduced.
The specialised RICH readout 
system based on ADCs and DSPs has been detailed in
Sec.~\ref{sec:pid.rich.RO}. Here follows the description of the ADCs for the
silicon and GEM detectors and the two different ADC systems for the
calorimeters. All other detectors were read out by a common TDC
chip (see Sec.~\ref{sec:daq.digital.f1}).
\label{sec:daq.digitisation}

\subsubsection{ADC for GEM and silicon detectors}
\label{sec:daq.analog.sgadc}
The analogue differential output signals of the APV25~chips --- which are
used for the readout of the GEM and silicon detectors --- go via short
flat-cables through a repeater card to the ADC~card named SGADC. 
In the case of
the GEM detector one ADC~card handles all twelve APV~chips of a
chamber. Each silicon detector is read by two ADC~cards: one for the
ten APVs on the n-side and another for the eight APVs on the p-side.
%#### number of ADCs on one card?
The analogue signals are digitised by $10\,\mathrm{bit}$ differential ADCs at a
sample rate of $20~\MHz$. The digital data of each channel are
corrected for the pedestal and are reduced by a zero suppression
logic. They are then formatted to $32\,\mathrm{bit}$ words and are sent via
optical fibres to a GeSiCA readout-driver module (see
Sec.~\ref{sec:daq.gesica}). Apart from offering a high data
bandwidth of $40\,\mathrm{MB}/\s$, the optical fibres also insulate the
ADC~cards from the readout module. This simplifies the electrical
grounding and is of particular importance for the silicon front-end,
where, due to the applied depletion voltage, the two ADC~cards, that
read out one silicon detector, lie on different potentials.

The zero suppression in the ADC~cards is based on a threshold cut on
the strip amplitudes. Since fluctuations of the baseline of the APVs
are observed, the data have to be corrected for this so-called
``common mode'' noise, before a threshold cut can be applied. The
correction is performed at the hardware level and uses a median
algorithm based on the data of all channels of one APV chip~\cite{Grube:01}.
% new text by Boris:
A good approximation of the median channel amplitude is obtained from the
cumulative distribution of the amplitudes. Due to limited hardware
resources this histogram is generated only in a narrow amplitude window
around the baseline. In order to position the histogram window the mean
amplitude of the 128~channels is used as a first guess for the
APV~baseline. Hits will obviously pull the average amplitude away from the
baseline, but for sufficiently low occupancies the mean lies still close
enough to the median.
%If the frame contains no hits the average pulse
%height lies very close to the median amplitude, but signals obviously
%pull the average away from the median. For sufficiently low
%occupancies, however, the average pulse height lies close enough to
%the median amplitude. 
The algorithm is implemented in a FPGA as a pipeline, which is able
to process the data of six ADCs in parallel. The pipeline has several
stages, which perform the various steps of the zero suppression.

\subsubsection{ADC for calorimeters}
\label{sec:daq.analog.caladc}
All COMPASS calorimeters have a fast light
signal response, high light yield and a photomultiplier light
collection system. They have similar output signal properties, therefore 
the requirements for the ADC modules are compatible. 
For the first period of the COMPASS experiment the Fast
Integration ADC module (FIADC) was developed. About 3000~calorimeter
channels are read out by the FIADC modules. 
The development of a new Sampling ADC (SADC) started in 2003 when the
technological progress in the design of the commercial fast flash ADC chips
offered affordable components from the price and power
consumption point of view. 
In 2004 during the hadron pilot run 1000 ADC~channels of a newly developed
SADC modules were installed and operated. In 2006 the calorimeter setup
will be completed and additional 1500~SADC channels will be
installed.

{\bf Fast Integration ADC:}
\label{sec:daq.analog.caladc.fiadc}
The FIADC module is a 9U VME module, which houses 64~ADC channels.
Every channel consists of a gated charge sensitive integrator, which
converts the charge to a voltage, and an ADC chip.  Upon arrival of the
gate signal, which is derived from the trigger signal, the integrator
feedback switch opens and the charge is integrated. The integration
time is programmable in the range of 100--250$\,\ns$. At the end of the
integration time the voltage level at the output of the integrator is
converted by the ADC chip to $12\,\mathrm{bit}$ digital information.  The
digitised information is processed inside a FPGA chip, where the
pedestal value is subtracted and the result is compared with the
threshold. The values above threshold are combined in one data
block. The block header,  which includes the module identification
 and a local event
number, is attached to the data block and the complete event
information is sent to the CATCH module via a
serial HotLink interface with a maximum speed of $40\,\mathrm{MB}/\s$. 

The FIADC module has an intrinsic dead time after every trigger. The
dead time is the sum of the integration time and the conversion time of
the ADC chip. The maximum FIADC dead time in the COMPASS experiment is 
$400\,\ns$.
After digitisation the data are processed in a pipeline mode with
negligible dead time. 
%### 3 event depth?
The ADC modules have an integral non linearity of $10\,\mathrm{bit}$ %???????
and a typical noise level of $1.2\,\mathrm{ADC channels}$.

{\bf Sampling ADC system:}
\label{sec:daq.analog.caladc.sadc}
The principle of the SADC is a direct measurement of the signal shape
by converting an analogue signal into a sequence of digital values.
The digital value represents the amplitude of the analogue signal at a
particular time. One of the advantages of the SADC module is a simple
implementation of a digital delay line instead of using a coaxial cable for
the compensation of the trigger latency. Another advantage is the possibility
to use advanced digital signal
processing techniques for noise reduction and feature extraction. 
The SADC read out system consists of two modules, the shaper and the SADC
itself. 

For the electromagnetic calorimeter the signal behind the  photomultiplier
tube has a triangular shape 
with $15\,\ns$ rise time and about $70\,\ns$ fall time. 
An anti-aliasing filter is required to digitise the fast signals with a
$77.76\,$MSPS (mega samples per second) ADC.
The shaper employs a very simple circuit of an
integrator with a decay time constant of $40\,\ns$. The advantage of
the circuit is that it 
preserves the linear relation between input and output signal
integrals and stretches the signal to avoid aliasing problems. One
shaper module houses 32~channels and has the same format as the SADC.

One 6U SADC module includes 32~ADC chips which run at $77.76\,\MHz$ with
$10\,\mathrm{bit}$ resolution.  The ADC chips
are preceded by differential amplifiers,  which match the signal
amplitude to the ADC range and match the input impedance to the cable
impedance. The ADC continuously samples the analogue signals and
writes data to a ring buffer. The depth of the ring buffer is
programmable from 0 to $5\,\mu$s and should match the trigger
latency. Upon arrival of the trigger 30~samples are copied to a multi
event buffer. Then the data are processed by the zero suppression
logic, which accepts only channels with signals above threshold.
These data
are combined in one data block and transfered to the GeSiCA module via
a serial HotLink interface.
 
The SADC module has a pipelined architecture where every stage of data
processing works synchronously with the previous one. The system
architecture is intrinsically dead time less. The event rate
capability of the system is limited by the bandwidth of the serial
interface and the depth of of the derandomisation buffer.

The sampling clock of $77.76\,\MHz$ is derived from the $38.88\,\MHz$ TCS
clock. All SADC modules synchronously digitise the analogue signals
and this feature allows to measure the signal time
with a precision better than $1\,\ns$.
%### in reality??

\subsubsection{F1-TDC}
\label{sec:daq:digital}
\label{sec:daq.digital.f1}
The requirements for the time measurements widely vary between the
different detectors in the COMPASS experiment.
The scintillating fibres and trigger hodoscopes produce high rates of
up to $10\,\MHz$ and require a high resolution of about $100\,\ps$.
On the contrary for the multiwire proportional chambers and the mini
drift 
tubes a time information of $10\,\ns$ is sufficient, but here the large
number of channels calls for a cheap solution.
The Micromegas, drift chambers, straw tubes and drift tubes needs are
in between. 

Since no TDC with the high rate requirements and dead-time free
readout was available, COMPASS  developed its own. The F1-TDC
%jointly developed by the University of Freiburg and acam-messelectronic, 
was designed
with the flexibility to fulfil all requirements for the different
%%detectors~\cite{Fischer:02a}. %% wrong author list, BK, 22.12.2006
detectors~\cite{Fischer:00a}. 
It provides a cost effective solution, because it is used for all time
measurements. 
The main parameters of the F1-TDC are summarised in
Table~\ref{tab:daq.f1tdc}. 

The F1-TDC performs digitisation and readout asynchronously and 
without any dead-time. 
The phased-locked-loop in each of the F1-TDC 
is precisely synchronised to the $38.88\,\MHz$ 
reference clock provided by the trigger control system of the COMPASS
experiment. Synchronisation is performed 
by a signal at the beginning of each SPS accelerator cycle.
For each detector channel the arrivals time of the signals, as
determined by the combination of the internal phase-looked loop (PLL)
and a scaler, are stored
for about $1-2\,\mu$s in a 16 or 32 word deep hit buffer. 
Depending on the number of configured channels per chip, 4,
8 or 32, the digitisation widths are $65\,\ps$, $130\,\ps$ or
$4.2\,\ns$, respectively. In case of the 32 input channel operation
four channels are combined on one timing unit. Four bits of the time
information are reserved to indicate the channels which fired in a
programmable time interval. In this so called latch mode operation
the time is given for the signal arriving first out of the
group of four signals.

The trigger signal is
distributed via the trigger control system and the CATCH readout-driver
modules to the TDC. Here the hits are selected according to a
programmable gate. 
The resulting time information from the 8~time-measure\-ments units in
the TDC are stored in a common output buffer. The data from
up to 8~TDC-chips are directly sent via a 8~bit HotLink serialiser chip to
the readout-driver modules. Where higher rates are required the 24-bit
wide data of four TDC chips are written into a FIFO.

%For all gaseous detectors the TDC-chips are located on the front-end
%cards directly mounted to the detector and thus provide a
%data reduction and concentration at an early stage and reducing
%the cost for cables considerably.
The TDC with its integrated phased-locked-loop and three-wire
interface to a digital-to-analogue converter requires only a minimal amount
of external components. Thus only small amounts of space is required
for the front-end cards located at the detectors.

\begin{table}[htb]
\caption{Parameters of the F1-TDC.}

\label{tab:daq.f1tdc}
\begin{center}
\begin{tabular*}{\columnwidth}{@{\extracolsep{\fill}}ll}
 & \\
\hline
Parameter & value \\\hline
Number of channels & 4, 8 or 32\\
Digitisation width & 65 ps, 130 ps, or 4.2 ns\\
Hit buffer size & 32 or 16 hits per channel \\
Trigger buffer size & 4 triggers \\
Double pulse resolution & typical 22 ns \\
Hit input level & LVDS, LVPECL or TTL \\
Edge sensitivity & leading, trailing or both \\\hline
\end{tabular*}
\end{center}
\end{table}

%\clearpage

\subsection{CATCH readout-driver modules}
\label{sec:daq.catch}
% CATCH Modules

\begin{figure}[tbp]
  \begin{center}
    \includegraphics[width=\columnwidth]{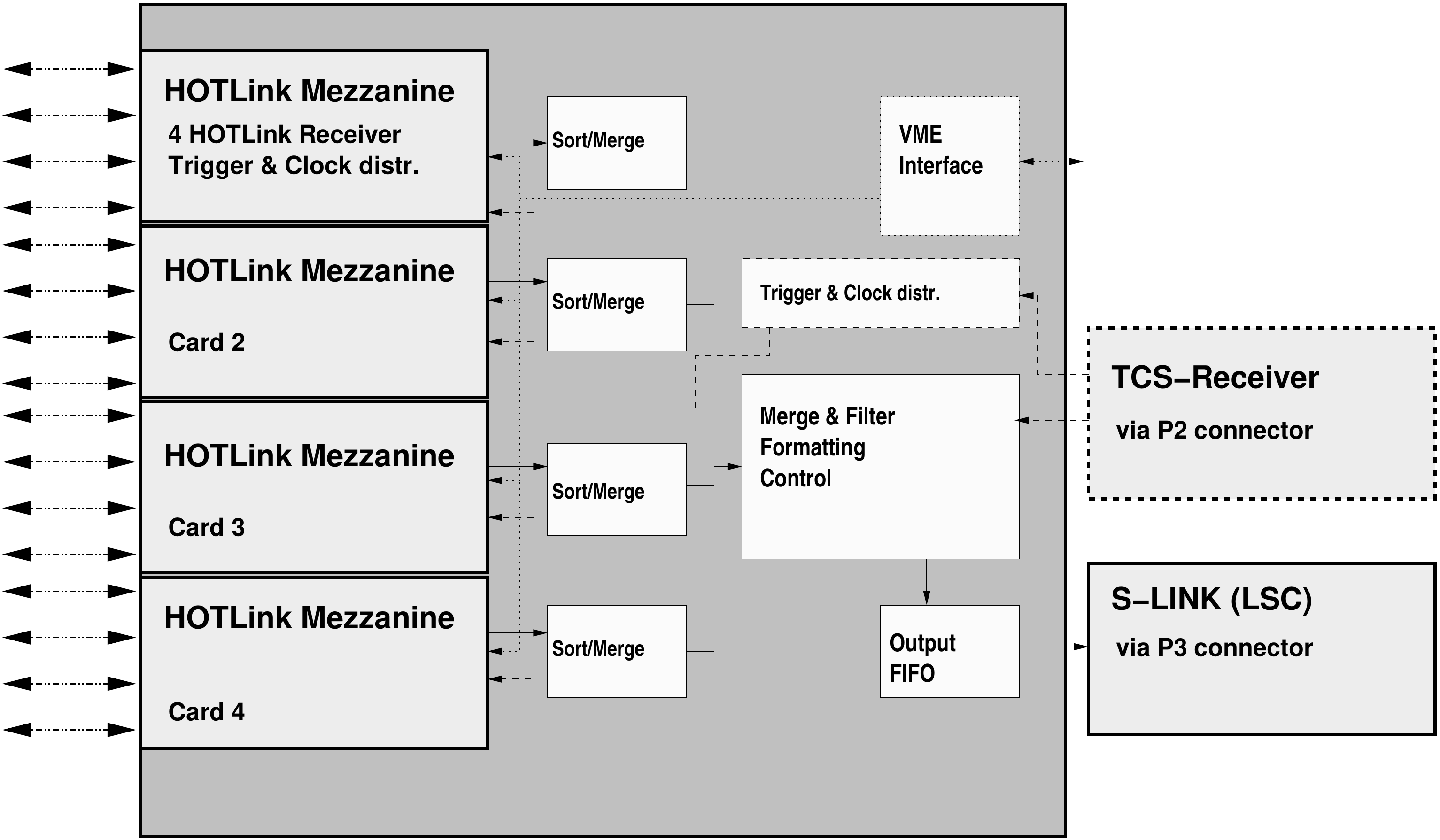}
  \end{center}
  \caption{\small CATCH readout-driver module mounted with for
    mezzanine cards. Shown are the path of the data (lines), the
    trigger and synchronisation signals (dashed) and the initialisation
    (dotted). On the left 16 cables connect to the front-ends.} 
  \label{fig:compass.daq.catch}
\end{figure}

To minimise development and maintenance costs we have chosen to unify
the readout of all TDC based front-ends as well as the FIADCs for the
calorimeter, the ADCs for the RICH and scalers for the trigger.
The CATCH readout-driver modules (COMPASS Accumulate, Transfer and
Control Hardware) 
distribute trigger and
timing signals to the front-ends, perform the initialisation of the
front-ends 
and serve as a concentrator for the data received from the digitising units 
of the front-ends.
For highest flexibility the inputs are implemented on mezzanine cards,
of which four are mounted on each CATCH module
(see Fig.~\ref{fig:compass.daq.catch}). 

The  hot-pluggable CATCH modules are housed in 9U
VME-crates, which provide power and an interface to the CATCH modules
via Linux VME CPUs. 
The configuration of the CATCH and the connected front-ends is
performed through the VME interface. 

The reference clock, synchronisation signals as well as spill and event 
numbers are transmitted 
optically on fibres from the trigger control system to the
TCS-receivers (see Sec. \ref{sec:daq.tcs}).
The TCS receivers are plugged on the backside of the
VME backplane at the P2 connector and transmit the decoded information
to each CATCH.
The CATCH acts as a fan-out and transmits the reference clock and
synchronisation signals to the front-ends.

%event building
The data arriving from the different front-ends are merged together
according to the event numbers. In total 7 FPGAs perform the event
sorting, header suppression and framing. The data are stored in
intermediate 16k$\,\times\,36\,$bit FIFOs at a maximum throughput of
$160\,\mathrm{MB}/\s$.  

%s-link
The complete events are transmitted with a S-LINK
transmitter to the readout buffers via optical fibres. Depending on
the S-LINK type the maximum transmission rate varies between 100 and
160$\,\mathrm{MB}/\s$. 
The S-LINK transmitters are plugged on the back of the VME P3 connector.
In addition, events are stored on the CATCH in a FIFO, 
which can be read via VME for
test setups with smaller rates and for debugging purposes.

% CMCs
\subsubsection{CATCH mezzanine cards}
\label{sec:daq.catch.cmc}
The input side of the CATCH modules is implemented on four
exchangeable CATCH Mezzanine Cards (CMC)
(Fig.~\ref{fig:compass.daq.catch}).  
To accommodate all requirements different types of mezzanine
cards have been developed~\cite{Fischer:00a}. %Pisa meeting
The standard operation mode is to digitise directly at the detectors
and transmit the serialised data from the front-ends via 20~m long
cables to the Hotlink-CMCs type of inputs. The Hotlink-CMC connects to
the front-end via electrical connections, while for the Hotfibre-CMC
and HOT-CMC optical connections are used. 
For detectors which produce high
amount of data like the scintillating fibre stations, which are located in the
beam, the time measurements are performed on the TDC mezzanine cards
(TDC-CMC) located on the CATCH. In the same category falls the
Scaler-CMC which counts signals on an event by event basis for 
normalisation purposes.

{\bf Hotlink-CMC:}
Most front-ends are connected via $600\,\MHz$
Ethernet type cables. Four 
front-end boards connect to each Hotlink-CMC. Thus to one CATCH 16
front-ends with 64 to 192 channels each are connected, resulting in a
concentration of up to 3072 detector channels per CATCH module.

\begin{figure*}
  \centering
  \includegraphics[width=\textwidth]{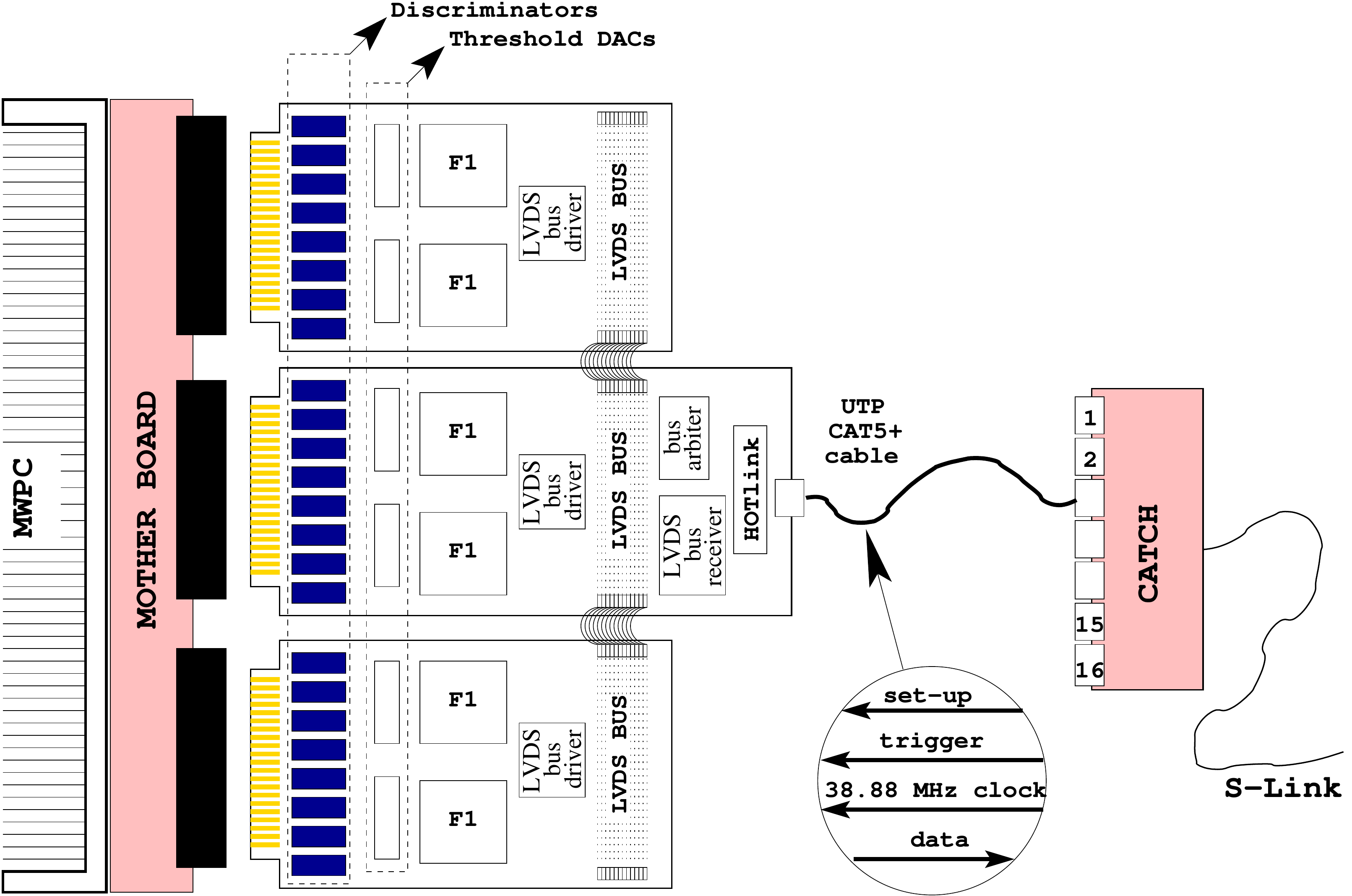}
  \caption{Scheme of the front-end card housing the
    preamplifier/discriminators, digitising electronics, and
    the interface to the CATCH via a cable with four twisted pairs of wires.}
  \label{fig:compass.daq.mwpc.readout}
\end{figure*}

As an example the scheme of the MWPC front-end cards and its interface
to the Hotlink-CMC is depicted in Fig.~\ref{fig:compass.daq.mwpc.readout}.
The signals of the four wire pairs on each cable have been standardised within
COMPASS: The CATCH readout-driver transmits to the 
to the front-end boards the 38.88~MHz
reference clock (in differential PECL), the coded trigger and
synchronisation signals (LVDS) and the $10\,$Mbaud (TTL) serial initialisation
string for the TDCs and discriminators.
The front-end sends its 24 or $32\,\mathrm{bit}$ wide data words through a $40\,\mathrm{MB}/\s$
serial link based on a 8b/10b DC balanced standard.

{\bf Hotfibre-CMC:}
For the RICH detector with its special requirements on noise decoupling
the Hotfibre-CMC was developed. Here the data are transmitted through optical
fibre pairs by making use of transceivers. 
Since the ADCs do not require the clock signal only the trigger
signal is transmitted through the fibre to the front-end. The
receiving fibre transfers the data from the front-end in the same
coding scheme as the Hotlink-CMC.

{\bf HOT-CMC:}
A new version of the Hotlink-CMC was developed in 2005. It is
also based on optical data transmission like the Hotfibre-CMC. 
But here we implemented the full functionality of
the Hotlink-CMC by coding the clock, trigger and initialisation string
on a single fibre. Starting in 2006 the readout of the central part of the
RICH utilises the HOT-CMC (see Sec.~\ref{sec:upgrade.rich}).

{\bf TDC-CMC:}
The TDC-CMCs measure the signal times of the high rate
scintillation fibres and trigger hodoscopes as well as the beam
momentum station. 
On the TDC-CMC four F1-TDC chips are mounted, providing
32~channels  
of $130\,\ps$ bin width or 16~channels of $65\,\ps$ digitisation width.
%(Fig.~\ref{fig:compass.daq.tdc-cmc}).  
The LVDS input signals are received from the amplifier and
discriminator boards mounted close to the detectors via flat, $20\,\m$
long twisted pair cables. 

{\bf Scaler-CMC:}
 The scalers measure the rates for the trigger
    hodoscopes and various counters needed to determine the luminosity.
Each Scaler-CMC provides 32 channels of dead-time-free $250\,\MHz$ scalers 
implemented on a Xilinx Virtex-E FPGA. Both the trigger and the gate
input can be adjusted for the timing delays on the cables
The Scaler-CMC accepts the same input signals and connection as the
 TDC-CMC. 

In 2004 a total number of 139 CATCH modules mounted in 15 VME crates
and equipped with 200 Hotlink-CMCs, 48 Hotfibre-CMCs, 288 TDC-CMCs and
16 Scaler-CMCs were in operation.

%%% Local Variables: 
%%% mode: latex
%%% TeX-master: "compass_spec"
%%% End: 

\subsection{GeSiCA readout-driver modules}
\label{sec:daq.gesica}
The GeSiCA readout-driver is an acronym for GEM and Silicon Control
and Acquisition module. 
Initially the module was designed as a readout module of the GEM and
silicon detectors which have a common front-end electronics based on
the APV25 chip \cite{French:01a}. These detectors feature high channel
density, high occupancy and high data rate. The readout chain of the
GEM and silicon detectors is very similar and includes the APV25
front-end card followed by the SGADC card and then the GeSiCA module.
Four SGADC cards are attached to one GeSiCA module via high speed
serial optical links.  One GeSiCA module reads up to 6000
detector channels.  
The second generation of the GeSiCA, called HotGeSiCA, 
extends the read out capability. In addition to
the SGADC card, it is able to read the SADC front-end modules of the
electromagnetic calorimeter.

The tasks of the GeSiCA module are to distribute the TCS clock,
the reset and the trigger signals to the front-end cards, to 
collect the data from the front-end cards and the transfer of
these data to the the readout buffer PCs via a S-LINK interface.
In addition it provides a slow bidirectional interface for the configuration of
the front-end cards.

The GeSiCA is a 9U VME module which has a set of interfaces: the VME
interface, four front-end interfaces, a
S-LINK interface and the TCS-receiver interface.
The front-end interface is a serial optical link with a speed of
$40\,\mathrm{Mbit}/\s$ from the GeSiCA to the front-end card and 
$400\,\mathrm{Mbit}/\s$ backward. The custom protocol provides two independent
data channels over this optical link. Via the first channel the
front-end card receives the trigger signal and sends back the
corresponding data from the detector. The second channel is used for
programming the front-end card, for loading firmware,
setting thresholds and verifying the status. The second channel is an
optical implementation of the I2C protocol. 
The I2C master is implemented in the GeSiCA FPGA and it is controlled
by a user program via the VME interface. 
The TCS receiver is attached to the P2 connector and provides the TCS
clock, the trigger signal and the event
identification information.  

During the data acquisition the GeSiCA receives data from four
front-end cards in parallel. The serial data are parallelised, checked for
consistency and put into the FIFO memories. Every event data block is
marked by a header and a trailer. The outputs of the four FIFO memories are
connected to the merger via a common bus. For every event the merger
creates an S-LINK header, which consists of the TCS event identifier,
the GeSiCA ID and the block size.  The merger sequentially
sends the S-LINK header and then four data blocks via the S-LINK card to
the readout buffer machine. 
The total bandwidth of the four front-end interfaces is equal to the
bandwidth of the S-LINK  
interface. Such balanced design minimises the internal buffer sizes and
excludes any data losses inside the GeSiCA. 

The HotGeSiCA has the same functionality as the GeSiCA module, however
it has new features.  First of all it is a 6U VME module.
The TCS receiver is integrated inside the  
module and does not occupy the P2 connector. Instead the P2 connector
is used for the S-LINK card. The number of the front-end interfaces was
increased to eight together with increasing the internal FIFO memory to
$512\,\mathrm{MB}$. During production the HotGeSiCA module can be equipped
either with optical front-end interfaces or copper connectors depending
on the type of the front-end electronics to be attached to the module.

In the 2004 run 15 GeSiCA and HotGeSiCA modules collected data from
more than 55\,000 channels. For the 2006 spectrometer upgrade of the
RICH-1 and ECAL1 the number of modules increased to 24 and the
number of attached channels exceeded 120\,000 (see
Sec.~\ref{sec:upgrade}). 

\subsection{S-LINK multiplexer}

The S-LINK multiplexer module (SMUX) was developed to minimise the
number of readout buffer cards.
It allows the data from up to four CATCHes to be transmitted via
one S-LINK interface.
The number of detector channels read out by one CATCH module varies between 
128 and 6912. Many of the CATCH modules read detectors with either low
occupancy or low channel
density and thus the amount of data transfered by a single
CATCH module is relatively small. Therefore it is possible to
multiplex the output of several CATCH modules.

The SMUX is a simple 3U card which houses a S-LINK source card, one
Xilinx FPGA chip and four parallel S-LINK interfaces. 
The SMUX is plugged directly to P3 connector of the CATCH module
instead of the S-LINK card. Up to three other CATCH modules can be
connected with flat cables. 
During data taking the SMUX expects one data block from every CATCH module
for every event. The data are combined in a bigger block and sent to the
readout buffer. 
The module does not need any special configuration, non-active CATCH
modules are automatically excluded.
In the 2004 beam time 34 SMUX cards were connected to 116 CATCH
modules by combining two, three or four CATCHes, depending on the data
rate.

\subsection{Data acquisition system}
\label{sec:daq.concept}
In order to match the high rate capabilities of the trigger system and readout
electronics, the data acquisition system has to rely on buffering and
parallelism. The inputs of the system are the optical fibres coming
from the readout-driver modules, carrying fragments of the data
pertaining to each trigger.  
The output are streams of data blocks which correspond to the
triggers as issued by the TCS; these are called events. The regrouping
of the data streams is called event building and is supported by the spill
buffer cards and the DAQ computers which form the event building network, as
described in the following three sections.

Due to impurities in the trigger system, the output streams contain
events which are useless for analysis. To save bandwidth, storage space and
reconstruction time, the output streams are filtered by our Online Filter,
which is described in Sec. \ref{sec:daq.filter}. The last step in the data
acquisition is the transfer of the filtered data to permanent storage and is
discussed in Sec. \ref{sec:daq.cdr}.

The main software used for the COMPASS data acquisition is the Alice DATE
package \cite{DATE}, which provides components for event building, run
control, information logging and event sampling. It had to be extended
by a program including a PCI driver to perform the spill buffer
readout. 
Modifications have been applied to the event building to optimise the
input/output and implement an interface to the online filter.
The run control is complemented by
an electronic logbook developed for COMPASS
(Sect.~\ref{sec:det_ctrl_mon.logbook}). 

\subsubsection{Spill buffer}
\label{sec:daq.spillbuffer}
The spill buffer is an S-LINK to PCI card which is
equipped with $512\,\mathrm{MB}$ of memory.  The card allows making use of
the SPS duty cycle: data are written into the memory during
the $4.8\,\s$ beam and are read out via a PCI interface during
the full  
cycle of $16.8\,\s$. This way, the required bandwidth is reduced by a
factor 
of three. In the COMPASS
conditions one spill buffer can store the data of 2-3 spills.  The DAQ
continuously monitors the amount of free memory in the spill buffers and
makes a forecast for the next spill.  If the amount of free memory
does not allow to store the data of the next spill the DAQ either
truncates the spill or completely blocks the triggers for the next spill. 

The SDRAM memory mounted on the spill buffer is organised in a FIFO
like way. The sustained memory 
access speed is $620\,\mathrm{MB}/\s$ which exceeds the bandwidth of
the S-LINK and the PCI by at least a factor of two. The card is fully
compliant to the 
PCI specification.

\subsubsection{DAQ computers} 

In 2004 we used 19 computers as readout buffers, each with $1\,\mathrm{GB}$
ECC SDRAM as 
main memory, two $1\,\GHz$ PIII CPUs and a 3COM Gigabit Ethernet
interface.  They have two PCI buses to read in the data on the first 32-bit
bus and then transfer it through the second 64-bit bus via Gigabit Ethernet to
the event builder computers. This configuration allows a simultaneous reading
from the detectors and writing to the event building network without overhead
or bandwidth losses. Four spill buffer cards are mounted per readout buffer.

Thirteen computers with two Athlon MP 1900+ processors, $1\,\mathrm{GB}$ ECC
DDR-SDRAM and an IDE-RAID with net $640\,$GB per machine were used as
event builders.  In total
$7.68\,\mathrm{TB}$ of disk space are available as buffer in case of
problems with 
tape recording giving a safety margin of about one day.  These computers also
host the online filter processes.

The greater number of detector channels and the higher trigger rates expected
for the data taking from 2006 on require an upgrade of the data
acquisition system. Ten new readout buffers with two $3.6\,$GHz
Pentium 4 XEON CPUs, and $4\,\mathrm{GB}$ main memory each are being installed.
Together with eleven new event builder computers in a similar configuration
with 1~TB of disk space each they will supplement the hardware for the
event building network. 
Also dedicated file and database servers, each mirrored and with identical
hardware configuration as the new event builder computers will be installed.

\subsubsection{Event building network}
\label{sec:daq.eventbuilding}

The data from the individual spill buffers are streams of S-LINK packets,
which are multiplexed into one stream of sub-events. The sub-events are then
distributed in a round robin fashion to the event builder computers, where the
sub-event streams from all readout buffers are multiplexed into streams of
complete events, which were on average $35\,\mathrm{kB}$ in size during the
muon run in 2004.

All DAQ computers are connected to a set of three 3COM 4900 switches with
twelve 1000 BaseT ports and four 1000 BaseSX up-links. These up-links connect
the front-end switches to a backbone 3COM 4900SX switch which provides the
necessary crosswise connectivity for all DAQ machines.  To achieve a balanced
configuration, the event builder computers are distributed evenly across the
three front-end switches, while the readout buffer computers are connected
according to their average output.  The stack of switches provides a total
number of 36 1000 BaseT ports to connect DAQ computers. The total theoretical
bandwidth available for event building amounts to $12\,\mathrm{Gbit}/\s$.

\subsubsection{Online filter}
\label{sec:daq.filter}

The online filter increases the purity of the
triggers and allows for a cost effective reduction of the amount of
tapes needed for recording.  For the physics programme with hadron beam
the online filter is required to reduce both the bandwidth needed for
transferring the data to the computer centre as well as the bandwidth 
needed to record the data on the tape drives. In addition one profits from a
reduced CPU time for the reconstruction of the data.

The basic unit for the online filter is the complete event as it is produced
by the event builder process. To facilitate the relocation of the filtering
task to its own dedicated computing farm at a later stage, the event stream
produced by each event builder is sent via socket I/O to the online filter
process, which up to the beam time of 2004 ran on the same computers as the
event builder processes. The filtered event stream is written to the
local hard disk arrays to await the asynchronous transfer to the central data
recording services.

At a trigger rate of $10\,$kHz, a particle extraction duty cycle of
$30$\% and with 
13~event builders sharing the load, the allowed average decision making time
is $4\,\ms$ per event. This constrains on the present hardware the
event analysis in two ways. First, 
only partial decoding of the data is possible. Second, tracking of charged
particles is impossible in regions where a magnetic field is present. This
also excludes the use of information from the RICH detector in the filter
decision. 

For the two physics programs, two filter algorithms have been developed. In
the muon programme the presence of a reconstructed beam track is required. For
this the silicon microstrip and scintillating fibre detectors upstream of the
target together with the beam momentum station must have recorded the a
sufficient number of hits from the beam particle. The algorithm is based on
the coincidence between the trigger time and the times measured by the
aforementioned detectors, respecting their different time resolutions and
allowing for the redundancy of the detector systems. Inevitable changes in the
timing of the detector signals due to changes in the environmental conditions
are compensated by an online re-calibration process. In 2004 we achieved a rate
reduction by $23$\% with an associated inefficiency of $0.4$\%.

For the hadron pilot run of 2004, the Primakoff trigger has not been filtered,
while the diffractive trigger was subject to a multiplicity cut. The hits
recorded by the silicon microstrip detectors downstream of the target are
filtered by the same coincidence algorithm as in the muon program's case.  The
truncated mean of the hits per detector plane then is a good approximation of
the track multiplicity after the target. Requiring this quantity to be greater
than one lead to a rate reduction of $45$\% for the diffractive trigger.

Presently these performance was enough, for the
future physics programme larger reduction rates will be achieved by the
implementation of a dedicated filter farm.

\subsubsection{Central data recording}
\label{sec:daq.cdr}

During data taking, the data flow is split in files of
$1\,\mathrm{GB}$ maximum each, which are written on the event builder disks. 
In parallel, headers of each event are read and written to a metafile
for later storage in an Oracle database. The files are copied in
parallel TCP/IP streams via a dedicated Gigabit link to disk servers
located at the CERN central computer centre. These disk servers are
part of the CASTOR hierarchical storage system~\cite{castor}.
A specific
configuration of CASTOR has been created for COMPASS and 6~tape writers are
reserved for the copy of the data files to tape. The configuration
has been optimised over the COMPASS data taking years to finally reach
performances of more than $8\,\mathrm{TB}/\mathrm{d}$, a rate close to the one
needed for the LHC experiments ATLAS and CMS.
When the
copy to tape of a given file is confirmed, its deletion is then authorised and
takes place when the space in the event builder disk is running low. In
parallel, when the data file is stored on tape, the corresponding metafile is
used to fill the Oracle database with information on the run and on each
event. The database, for performance reasons spread over 9~computers,
allows the reconstruction software to have a direct access to each of
these events, using selection criteria like trigger type or event number.

\subsubsection{Performance}
The maximum sustained data rate which has been recorded was
$8\,\mathrm{TB}/\mathrm{d}$ during 
the hadron pilot run, which corresponds to $740\,\mathrm{Mbit}/\s$. The limiting
factor was 
the network link to the computing centre, a single $1\,\mathrm{Gbit}/\s$ connection. Data
rates of up to $1.5\,\mathrm{Gbit}/\s$ have been achieved during tests without
transferring 
the data to permanent storage.

%

%%% Local Variables: 
%%% mode: latex
%%% TeX-master: "compass_spec"
%%% End: 

% Control and Monitoring
\section{Detector control and monitoring}
\label{sec:det_ctrl_mon}
The detector control and monitoring systems provide a user interface
to control the majority of the hardware parameters of the COMPASS
apparatus and ensure that the quality of COMPASS data stays 
%A complex experiment like Compass needs to insure that the quality of its data stays 
at a high level during data taking. Different aspects of the experiment are
constantly monitored: the operation of the front-end electronics and the
read-out chain, the stability of the
beam characteristics and the counting rate level of the different triggers.
The consistency and correctness of the data flow produced by the
front-ends is not sufficient to guarantee a reliable operation of the
detectors. This information has to be complemented by more
detector-specific parameters, like supplied voltages and currents, hit
profiles, time or amplitude 
information or noise spectra. The online monitoring is
performed on data provided on the fly by the DAQ system. 

The various tools used to control and monitor the COMPASS spectrometer
are described below.

\subsection{The detector control system}
\label{sec:det_ctrl_mon.dcs}
The main task of the Detector Control System (DCS) is to provide a complete and user friendly interface for setting and reading back all the relevant parameters for the operation of the various detectors and data acquisition elements, like high voltages, low voltages, VME crate status, gas pressures and mixtures, temperatures and magnetic fields. A software package allows to check and modify all parameters remotely, thus minimising the need to access the experimental hall.

\begin{figure*}[tbp]
  \begin{center}
    \includegraphics[width=\textwidth]{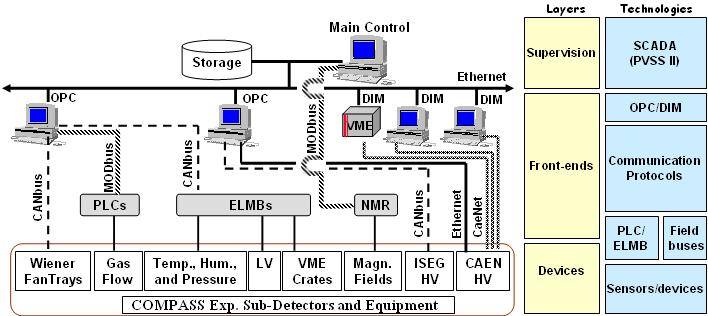}
  \end{center}
  \caption{The architecture of the Detector Control System.}
  \label{fig:dcs}
\end{figure*}

The DCS architecture (see also Fig.~\ref{fig:dcs}) is composed of three layers. 
The {\bf supervisory layer}, based on the commercial package PVSS-II~\cite{PVSS:05}, provides the graphical user interface for accessing and monitoring the hardware parameters. 
The {\bf front-ends layer} include the various software drivers
specific for each hardware element. It provides the supervisory layer
with a common communication interface to access the hardware. Whenever available, existing commercial drivers have been used, while custom drivers have been developed for unsupported hardware modules.
The hardware elements (crates, sensors, gas systems, etc.) form the {\bf devices layer}.

Such a modular architecture allows to control a large variety of devices in a coherent and transparent way:

\begin{itemize}
\item control of crates and power supplies; 
\item monitoring of voltages and currents in crates and power supplies; 
\item control of high voltage channels; 
\item monitoring of temperature, humidity, pressure and magnetic field 
in specific points of the experimental hall, sub-detectors and magnets; 
\item monitoring of gas fluxes and mixtures in gaseous chambers. 
\end{itemize}

%The Detector Control System (DCS) aims at guaranteeing a reliable and stable operation of the COMPASS detector, from one single remote point. All the control and monitoring tasks are performed from a central room, avoiding losses of beam time to access equipments placed in the experimental area. The DCS architecture consists of the supervisory layer, offering monitoring and control capabilities, with a graphical user interface, alert handling, data trending, processing and archiving of historical values; the front-ends layer, comprising the specific software and hardware for controls and interfacing; and the devices layer, including all the hardware to control, as well as sensors for the monitored parameters. For the supervisory layer, a commercial SCADA, the package PVSS-II~\cite{PVSS:05}, is used. Non-standard equipments are integrated in the DCS by using specifically developed front-end software.

%From the point of view of their operation, the COMPASS sub-detectors are independent from each other. There is no uniformity in the choices made by each responsible group for their sub-systems, and this variety of different equipments leads to a complex architecture at both the front-ends and the supervisory level of the DCS.

\subsubsection{The devices layer}
The DCS controls a large number of devices that are spread over about $200\,\m$ along the spectrometer and in the beam tunnel.
%The devices to control are distributed along $50~\,m$ in the experimental area, and also spread at the end of the beam tunnel and in the electronics rooms. The requirements for slow control include: 
A total of about 2000 high voltage (HV) and low voltage (LV) channels are constantly monitored by the system.
The HV power supplies are mainly different types of CAEN~\cite{CAEN}
%\footnote{CAEN, Via Vetraia, 11 - 55049 Viareggio (LU) - Italy, Tel. +39 0584 388398, Fax +39 0584 388959} 
modules housed in various main frame models. These main frames also contain 
part of the LV power supplies.  
Most of the CAEN main frames  
%are controlled via the proprietary CAENET fieldbus, or via ethernet for the newest. A total of 22 crates ($\sim 1200$ high voltage channels) 
are connected in daisy chains to 5 CAENET buses, while Ethernet is used for a few very recent models.
Several ISEG~\cite{ISEG} HV modules and the power supplies of the WIENER~\cite{WIENER} VME crates are controlled via standard CANbus lines.

ADC readings for analogue measurements are performed using the multi-purpose I/O system Embedded Local Monitor Board (ELMB)~\cite{Hallgren:01} developed at CERN for the ATLAS experiment. A total of 28 ELMBs are used, providing the readout of 800 analogue sensors.
 
%The high voltage systems use mainly CAEN modules, with the exception of two sub-detectors that use also ISEG modules. For CAEN high voltage power supplies, the proprietary fieldbus CAENET is used; and in the case of one crate of new type, offering for the first time this possibility, Ethernet is also used. 5 CAENET buses connect a total of 22 crates in daisy chains ($\approx$ 1200 high voltage channels). For the other power supplies, the connection to slow control is done using the industry standard CAN. Several CANbuses are used to connect ISEG high voltage crates ($\approx$ 340 channels), and WIENER fan-trays (for 15 VME 9U crates) to the slow control.

%COMPASS uses extensively the multi-purpose I/O system ELMB (Embedded Local Monitor Board)~\cite{Hallgren:01} developed at CERN for the ATLAS experiment. The ELMB solution was selected by COMPASS since it provides multiplexed ADC readings for analog measurements, at low cost per channel. A total of 28 ELMBs are used, providing 800 analog measurements.

The various gas systems are controlled by Programmable 
Logic Controllers (PLCs), that act on mass flow meters, pressure sensors and valves to ensure constant and 
correct gas flows and mixtures. The parameters relevant to monitoring are sent 
via serial connection to one of the front-end DCS computers.

\subsubsection{The front-ends layers}
OLE for Process Control (OPC)~\cite{OPC} servers exist for the majority of modules used in COMPASS.
%Several vendors of  high voltage power supplies commonly used in Physics experiments develop dedicated Ole for Process Control (OPC) servers, providing an abstraction layer between the real hardware and the user. 
%This is the case of CAEN, WIENER and ISEG, which provide OPC servers for their various types of power supplies. 
Wherever a commercial solution was not available, custom drivers have
been developed in collaboration with CERN, making use of the SLiC~\cite{Beharrel:05} framework developed at CERN and using OPC or Distributed Information Management (DIM)~\cite{Gaspar:01} protocols. For example, the CAENET-based high voltage system is managed by a SLiC DIM server with tunable read cycle speed for each of the controlled parameters. A fast cycle is used for the monitored voltages, currents and channel status with speeds of $10 - 20\,\ms$/channel. This allows a fast detection of high voltage trips and failures. A slow cycle, with a readout period of 3 minutes, is used for the read-back of setting values.

The OPC and DIM servers are running on Windows and Linux computers. The monitored values are exported to clients in the form of lists of items. Clients subscribed to a certain list are notified whenever an item changes its value or status. This event-driven protocol minimises the network traffic in the DCS.

\subsubsection{The supervisory layer}

For the supervisory layer the software package PVSS-II
%The highest level of abstraction is represented by the supervisory layer, implemented using the commercial package PVSS-II. This software framework 
has been extended at CERN with the JCOP-Framework~\cite{Myers:99}, which implements the functionalities specific to high energy physics experiments.

The supervisory layer consists of a core application and several OPC and DIM clients for the communication between the application and the  front-ends layer.
%connected to the corresponding servers. The communication between the core application and the front-ends layer is mediated by the OPC and DIM clients. 
The application provides a graphical user interface for  visualisation and control of the parameters, and an interface to the dedicated database were the monitored values are stored together with their time stamps.

An alarm system provides acoustic and visual alarms when critical parameters fall outside predefined boundary levels, or when hardware failures are detected. The alarm system reacts to a wide variety of events: high voltage trips, gas system failures, temperature and magnetic field changes and crate failures.
In addition, the trend of stored parameters as a function of time can be displayed via the main user interface, thus making it easy to recognise slow drifts or locate in time changes in the parameter values.

%PVSS-II is a commercial product originally designed for industrial controls, and it lacks some functionalities needed in high energy physics experiments. To overcome this, a software tool, the JCOP-Framework~\cite{Myers:99}, was developed at CERN, including guidelines, tools and libraries that simplify the development work needed to integrate standard equipments in PVSS. This and an additional small COMPASS framework are used in the DCS. 

%PVSS-II is device-oriented and its functioning is event-driven (as opposed to cyclic), thus avoiding unnecessary data traffic and offering good time performances. PVSS clients connect to the OPC and DIM servers, subscribing to lists of items. The DIM servers are distributed in Windows and Linux computers, in general placed close to the front-end electronics. The OPC servers and PVSS OPC clients run in several Windows computers. All these connect via TCP/IP to the main Linux DCS computer, where the PVSS core runs. Figure~\ref{figure:dcs} shows a schematic view of the DCS architecture.

%\subsubsection{Performance issues}

%During the data taking periods from 2002 to 2004, the DCS system has been extended and improved, becoming more robust and running stable. The reliability of the system is guaranteed by a regular backup policy of the whole DCS project; by the export of the historical data into external storage systems; and by connecting the main DCS computers to safety UPS units. The remaining problems that have been identified at the level of the software were solved with new release versions of the packages used. The recent PVSS-II version 3.0 lead to a new version of the JCOP-Framework, where a new architecture philosophy was applied, more adapted to high-energy physics needs. The 2005 SPS shutdown provided the opportunity to redesign the DCS system of COMPASS, making use of the new packages and structuring the project in a more coherent manner. 

\subsection{Beam and trigger stability}

The stability of the beam parameters, like the intensity and the position at the COMPASS target, is an essential requirement for the measurement of small physics asymmetries. 
Slow drifts of the beam characteristics can occur, as well as significant variations from one spill to the other. Therefore, a continuous monitoring of the beam parameters is needed.

%Beam characteristics as well as trigger rates are then monitored spill by spill. 

The beam and trigger monitoring software tool checks the beam current and position, and the rate for each trigger type on a spill-by-spill basis, and stores the read values into a local MySQL database~\cite{MySQL:05a}.
%At each spill, the beam current and position, and the rate for each trigger type is read and stored in a local MySQL database~\cite{MySQL:05a}.
Trigger rates normalised to the beam current are also computed to detect beam
changes and failures of the trigger system. The trend of those parameters over the last few hours is constantly displayed.

%The history of these characteristics over the last few hours are shown to the shift peoples and permanently updated, allowing them to detect easily any changes in the beam quality. 

%Comments coming from the SPS control room are also shown with this history. History over any periods can also be browsed via the graphics user interface. Automatic procedures can use this database to reject automatically data from spills where the beam quality was bad.

\subsection{Data monitoring}
\label{sec:det_ctrl_data_mon}

The read-out chain and the front-end electronics of the detectors are
monitored by a software tool named MurphyTV.
Samples of events are read on the fly during the data taking to monitor the data quality.
Failures of the read-out
chain are detected by error words generated by the electronic boards themselves, or by
inconsistent formats of the data structure (missing headers or trailers,
missing part of the data, incoherent event numbers in different portions
of the same event, etc.). MurphyTV reports these errors in
a graphical user interface.

%An important part of the task to monitor data quality consists to check if the data delivered by the different detectors stay stable and do not present any anomaly. 
A software package named COOOL \cite{Bernet:04a} has been designed to read COMPASS raw data streams and to produce different sets of ROOT histograms
\cite{Brun:05a} for each detector plane or group of detector planes. 
The COOOL program provides a graphical user interface that allows the user to interactively browse through the histograms and print them for future reference. The histograms are saved into a ROOT file upon program termination; the raw detector information can also be stored in the form of a ROOT tree, to allow the quick re-processing of the data. 
%Thanks to the graphic user interface included with COOOL (see Fig.\ref{fig:det_ctrl_mon.gui_coool}), these histograms can be browsed interactively, zoomed, printed, and also compared to reference histograms read from a reference run.
The generated histograms, like counting rate profiles, time spectra, signal
amplitude distributions, allow the detection of several types of failures,
like high voltage failures, low voltage
failures, large electronics noise, or failures of
front-end electronics cards. 
For each histogram, a reference distribution is plotted on top of the
current one, thus providing a quick comparison with the nominal detector behaviour. 
%Pre-configured sets of histograms can be plotted into separate windows, allowing the operators to quickly check the most relevant histograms for each detector type.
%This software includes a mechanism to show a preconfigured set of histograms for each detector type, allowing to operators to quickly check the most relevant histograms without having to browse all the existing ones. 
%A pool of "standard" pre-configured sets of histograms has been made by the detector experts, and is routinely used by the shift peoples to monitor the stability of the spectrometer. All these sets are regularly printed in books, allowing to keep an history of the detectors behaviours along the time.
The COOOL package is also extensively used for detector
commissioning and running-in. 
% \begin{figure}[tbp]
%   \begin{center}
%     \includegraphics[width=\columnwidth]{./figures/Online_gui_Coool}
%   \end{center}
%   \caption{\small Graphic user interface of the COOOL software. The main window is at
%   the bottom left, a 2D time vs channel histogram is browsed channel by channel on the
%   upper left window (channel selected by the slider),
%   profile histograms are compared to the reference (in red) on the
%   large upper right window in background.}
%   \label{fig:det_ctrl_mon.gui_coool}
% \end{figure}

% COOOL can produce on demand a ROOT file containing all the histograms. It can
% also produce a ROOT tree files allowing expert peoples to do fast analysis of
% their detectors. Treatments of given histograms are also implemented, to compute
% automatically some characteristics of the detectors (gains, missing channels,
% efficiencies, etc...). These characteristics can be then written in a text file.

% A batch version of COOOL, without GUI, is automatically launched for each data
% taking run. The corresponding histogram file are stored in the electronic run
% logbook (see section \ref{sec:det_ctrl_mon.logbook}), as well as the detectors
% characteristics which are automatically computed.

\subsection{Run logbook}
\label{sec:det_ctrl_mon.logbook}

To keep track of the data taking and of the conditions of the experimental setup a web-based logbook, connected to a MySQL database for the permanent storage and easy searching of the comments and annotations, was developed for COMPASS.

%Such an electronic logbook has been developed specifically for the COMPASS experiment. It is based on a MySQL database \cite{MySQL:05a} server which is filled using a set of Tcl/Tk scripts. 
A large variety of information is stored automatically at the
beginning of each data run (equivalent to 100 or 200 SPS spills), thus
providing a complete overview of the data taking conditions of each
single data run. In addition, some data quality checking and online monitoring tools (see Sec.~\ref{sec:det_ctrl_data_mon}) process a sub-sample of the collected data, to produce detector and DAQ performance histograms.

The list of automatically stored pieces of information is the following:
\begin{itemize}
\item information from the DAQ system
\item information provided by the shift crew
\item trigger information 
\item beam line information 
\item status of the COMPASS target
\item currents of spectrometer magnet SM1 and SM2
\item a ROOT file containing the COOOL histograms
\item detector specific information extracted from the COOOL histograms
\item text output from the MurphyTV software which monitors errors coming from
the front-end electronics
\end{itemize}

%Additional comments, either general or associated to a specific run, can be added by the users. They are tagged with a time stamp, the author name, a general topic name selected from a list and a title. In addition to the text, it is also possible to grab the entire screen or a portion of it, and insert the corresponding image into the comment. This is an extremely useful complement to the description of many problems, like detector misbehaviour or slow variations of parameters.

All stored information can be retrieved, searched and displayed through a web server.

%A text-based version of the COOOL software is used in batch mode by the electronic logbook software; the program is launched at each start of run and terminated at the run end, and the resulting ROOT file is stored in the database together with the other run-specific informations.

%%% Local Variables: 
%%% mode: latex
%%% TeX-master: "compass_spec"
%%% End: 

% Performance
\section{Event reconstruction and spectrometer performances}
\label{sec:performance}

\begin{figure}[tbp]
  \begin{center}
    \includegraphics[width=\columnwidth]{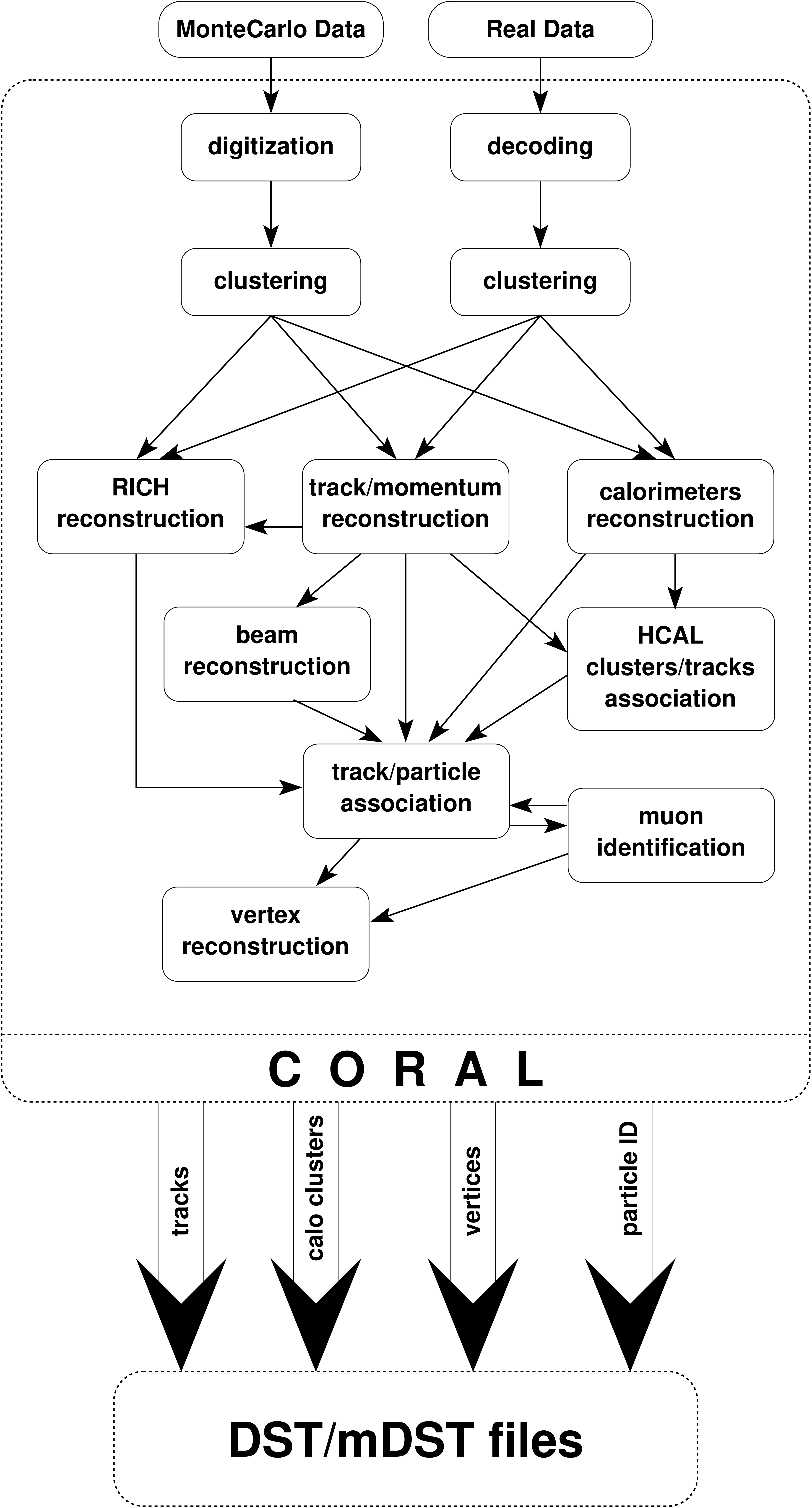}
  \end{center}
  \caption{Schematic representation of the COMPASS reconstruction
    software.}
  \label{fig:performance.flow_chart}
\end{figure}

The huge amount of data (about~$350\,\mathrm{TB}/\Year$) collected by the
experiment requires the availability of sufficient computing power to
reconstruct the events at a rate comparable to the data acquisition
rate. The required CPU power is estimated 
%at
to be 200k SPECint2000 units,
which are provided currently by 200 Linux Dual-CPU PCs out of the
CERN shared batch system. Event reconstruction is performed by a fully
object oriented program (CORAL) with a modular architecture and
written in C++. The schematic representation of the reconstruction
software, describing the various steps performed and their mutual
connections, is shown in Fig.~\ref{fig:performance.flow_chart}.

The input of the reconstruction software is represented either by the
raw data collected by the experiment, or by the output of the Monte
Carlo simulation software (see Sec.~\ref{sec:performance.MonteCarlo}).
The data files produced by the COMPASS acquisition software contain
the raw information from the detectors, digitised by the front-end
electronics. Two initial processing phases are needed to prepare the
input to the track finding algorithm. In the first phase, called {\bf
  decoding}, the information on the fired detector channel (either
wire, pad, or cell, depending on the detector type) is extracted from
the raw data. In the second phase, called {\bf clustering}, detector
channels that are fired by the same particle are grouped together. For
some detectors 
weighted centre-of-gravity algorithms are applied to 
get a better determination of the particle 
%penetration point.
parameters.
%improve the spatial resolution of some detectors. 
During the clusterisation phase, the information
on the geometrical position of each detector in space is used to calculate
the coordinate of the cluster in the main reference system of the
apparatus. The geometrical position of each detector plane is
retrieved from files that are generated either by the alignment
procedure (see Sec.~\ref{sec:performance.alignment}) or by the Monte
Carlo package.  The clusters are then pre-selected on the basis of the
time information. 
%The left/right ambiguity of clusters belonging to drift-like detectors is resolved by assuming either a target or an infinity pointing.

When Monte Carlo data files are processed, the decoding phase is
replaced by a {\bf digitisation} phase, in which the response of the
detector is simulated and hits are produced on the basis of the
particle trajectory and detector resolution.
% on the basis of the particle trajectory described in the Monte Carlo
% data file. %The clusterisation is then applied.

% The Monte Carlo data files only contain the exact coordinates of the
% interaction point between particles and detector active areas.  In
% the case of Monte Carlo data files, the exact coordinates of the
% interaction point between particles and detector planes is stored in
% the data files.  The simulation of the detector response to a
% crossing particle (called {\bf digitisation}) is performed inside
% CORAL, and is equivalent to the decoding for the real data. After
% that, clustering is applied to the generated hits, as for the real
% data.

After clusterisation, charged and neutral particles are reconstructed
and particle identification is performed. The information from
tracking detectors is used to reconstruct the trajectories of the charged
particles through the spectrometer and to determine their momenta, as described in
Sec.~\ref{sec:performance.reco}. Hadron calorimeter clusters
are used to separate muons and hadrons; electromagnetic
calorimeter clusters measure the energy and impact coordinate of
photons and electrons (see Sec.~\ref{sec:performance.calo}).
Hadron identification is performed by the RICH-1; a dedicated software package which combines information from the RICH photon detectors and from
reconstructed tracks and momenta, is used to calculate the most probable Cherenkov angle
and to assign probabilities to all possible particle hypotheses (see
Sec.~\ref{sec:performance.rich}).

A vertex finding procedure is applied to all reconstructed tracks, in order
to identify the primary interaction point and subsequent decays of
neutral particles, as described in Sec.~\ref{sec:performance.vertex}.

The result of the reconstruction phase (track parameters, vertices,
calorimeter clusters, PID probabilities, detector hit patterns,
etc.) is stored into output ROOT~\cite{Brun:05a} trees, called mini Data Summary Tapes
(mDST), that are distributed to home computing centres and serve as
input for all the physics data analyses. The data reduction factor
between the input raw data and the output mDSTs is about 100. Large
DST files, storing the detector digits and clusters in addition to the
tracking, vertex, and PID information are also created and kept at
CERN on tape.

The following subsections describe the main steps in the event
reconstruction and give a summary of the corresponding performances.

\subsection{Track and momentum reconstruction}
\label{sec:performance.reco}

The track reconstruction algorithm (TRAFFIC/TRAFDIC) is divided into three phases
corresponding to pattern recognition (i.e. finding track segments in
the various zones of the spectrometer), bridging (i.e. connecting
track segments from several distinct zones to build full tracks), and
fitting (i.e. computing the best estimators for the parameters of the
reconstructed tracks).

The {\bf pattern recognition} selects sets of clusters
% which pattern conforms with that of a track segment.
consistent with track segments.  To that end, the spectrometer is
divided into 5
% intervals of longitudinal abscissae, called "zones"
zones along the beam, where track segments are expected to follow
approximately straight lines.  The five zones comprise the regions
upstream of the target, from target to SM1, from SM1 to SM2, from SM2
to the second muon filter, and downstream of the second muon filter.
The first one is only used for beam tracks: tracks reconstructed in
the beam telescope are extrapolated upstream and associated,
on the basis of time and position, to momentum measurements performed
in the BMS (see Sec.~\ref{sec:beam.muon.bms}).

Reconstruction is first performed in {\bf projections}. For this
purpose the detector planes in each zone are divided into groups
having the same orientation, thus measuring the same projection of a
track. A pivot-plane algorithm is used to search for track
segments in these projections. In this approach, each pair of detector
planes is used successively as a pair of pivots. Each pair of clusters
from the pivot planes is used to define a straight road, thus associating
 clusters from all other planes.  At each successively
encountered detector, the road width is adjusted to take into account 
the detector resolution.
This projection search is performed for each projection.
A pre-selection of candidate track projections is then performed,
based on the number of clusters, and taking into account its variation as a function of the track angle.
% All successful combinations are then sorted according to a Quality
% Function, of which the basic ingredients are the number of clusters
% and the $\chi^2$.  The best combinations are selected by scanning
% the list in order of decreasing quality and retaining those which
% are encountered for the first time or which do not have more than a
% specified number of clusters in common with their already selected
% counterparts.

In the next step, all projections are combined to produce {\bf
  space track segments}. The search combines track segments taken from
pairs of projections to open a road in space; clusters
from all detectors in all projections are then collected within this road.
% All projection planes enter the combinations, including the stereo
% ones.
For each detection plane, the position corresponding to the measured coordinate is
assigned to the cluster, and a check whether the track impact point is
within the sensitive area of the corresponding detector is performed.
All track candidates are then compared to a dictionary of possible
tracks through the COMPASS spectrometer. This dictionary is organised
as a look-up table, thus increasing the speed of the fitting
procedure.

%{\it to be re-written\dots}\\
The building of full tracks out of the track segments found in the
preceding phase proceeds sequentially, first via straight line fit, and then via {\bf
bridging}
(connecting track segments) of two adjacent zones.  The successfully
combined track segments are then ordered according to a quality
function (QF) based on the $\chi^2$ and the number of hits associated 
to the track. 
The combination with the best QF is retained,
while combinations re-using one of the previously accepted segments
are discarded.
The process is iterated until the list is exhausted.
%\dots

In the final phase of the tracking algorithm, magnetic fields and material maps are used to get the best estimates of 
the track parameters ($x$ and $y$ positions, $\mathrm{d}x/\mathrm{d}z$ and $\mathrm{d}y/\mathrm{d}z$ slopes, inverse momentum $1/p$) 
%($x_0$, $y_0$, $\mathrm{d}x/\mathrm{d}z_0$, $\mathrm{d}y/\mathrm{d}z_0$ and $1/p$)
and their error matrix.
For this purpose,  the  {\bf Kalman fit} method is used 
\cite{Fruhwirth:87,WolinHo:93}. The fit starts from the first cluster of 
the found track. The track parameters are updated by this measurement 
and propagated to the next detector surface. This process is repeated
for all clusters belonging to the track.
The Kalman fit is performed twice, first in the downstream and then in the upstream direction,
in order to provide the track parameters at the first and last measured point
of the track.

The momentum of low energy particles emitted at large angles, and therefore not entering the SM1 aperture, is calculated from their trajectory in the fringe field of the magnet, although with smaller accuracy. These are called fringe field tracks~\cite{Kurek:02}.%NIM A 485 (2002) 720

Apart from spatial measurements (cluster coordinates), a number of 
COMPASS detectors also provide time information ({\bf cluster time}) with respect to the time of the trigger signal.
As those measurements are completely independent from coordinate measurements,
the time component is not included in the track parameterization but
a weighted mean time is calculated separately.
This track time information is important for rejection of pile-up tracks, 
beam momentum determination, trigger system performance studies, etc.

The list of tracks obtained at the end of the track reconstruction procedure is then scanned to search for the muon outgoing from the primary interaction (scattered muon). A track is identified as a scattered muon if it corresponds to a positively charged particle, and if its trajectory is compatible with the hodoscope hits as given in the trigger matrix. For the standalone calorimetric trigger (see Sec.~\ref{sec:trigger.muon}), a minimum number of hits is required downstream of either the first or the second hadron absorber. In addition, the muon track candidate must cross the entrance and the exit of the polarized target at a distance smaller than $5\,\cm$ from the beam axis. The beam track and the fringe field tracks are excluded from the search.

\subsection{Alignment procedure}
\label{sec:performance.alignment}

The alignment procedure determines the detector positions in space by
minimising the $\chi^2$ of all tracks simultaneously.  Initially
tracks are reconstructed using detector positions determined by a 
geometrical survey. The
minimisation of the $\chi^2$ is based on the analytical inversion of a
huge sparse matrix, using the Millipede program described
in~\cite{Blobel:02}.  The optimised parameters are: $i)$ the position
of the detector centre, $ii)$ the rotation between the detector
coordinate system and the global coordinate system, $iii)$ the
effective detector pitch, i.e. the distance between adjoining wires or
pads. The procedure is iterated until the changes in the detector
positions become negligible compared to their resolution.

The adjustment of detector parameters is first done using 
data samples, collected   
with the spectrometer magnets switched off, so that straight
trajectories through the spectrometer can be assumed, and then refined using standard data.
The alignment procedure is repeated after each long interruption of
the data taking and each time some detector has been moved.

\subsection {Vertex reconstruction}
\label{sec:performance.vertex}
                                                                               
The last phase in the event reconstruction is the location in space of
the primary interaction or of the two-body decays of neutral particles
(so-called $V^0$~vertices).  Here the aim is to get the best estimate
of the three coordinates of the vertex position from which each track
is assumed to originate, of the three components of the momentum
vector of each track at this vertex, and the corresponding covariance
matrices.

A first approximation of the primary vertex is obtained by computing the average Point Of Closest Approach (POCA) between one beam track and all possible outgoing tracks. All tracks having a POCA too far from the approximated vertex position are discarded; an exception is represented by the scattered muon, which, if present, cannot be removed from the vertex. %{\it define the scattered muon and how it is identified}

A vertex is called primary when it contains a beam track. Since more than one beam track can exist for one event, as many primary vertices as the number of existing beam tracks can be reconstructed in this phase; the selection of the best vertex is performed later at the level of physics data analysis. 

%All pairs of tracks with opposite charges (excluding the beam track) are also checked to identify two-body decays of neutral particles, like $\Lambda^0 \to p \pi^-$.

The tracks surviving the initial selection are used to perform the fit of the vertex position by an inverse Kalman filter algorithm. During the first iteration, all tracks are used to estimate the parameters of the vertex, and the relative $\chi^2$ contribution of each track to the fit is computed. If the largest $\chi^2$ contribution exceeds 
a threshold value, the corresponding track is removed from the list and the procedure is iterated once more. However, neither the reconstructed beam track nor the scattered muon track can be removed from the vertex. The algorithm stops when all remaining tracks survive the $\chi^2$ selection. 

The procedure described above may fail if the initial list of tracks contains a large number of fake tracks. In such a case the initial vertex position may be estimated to be too far from the real one, resulting in a rejection of the good tracks. In order to avoid this difficulty, a recovery
%A known limitation of the described procedure arises when the initial list contains a large fraction of fake tracks. In that case, the estimated vertex position during the first iterations can be far from the real one, and good tracks can be rejected due to that. Therefore, a recovery 
procedure is applied at the end of the vertex fit phase.
%At the end of the vertex fit, a recovery procedure is applied to unassigned tracks. In fact, if the initial list contains a large fraction of fake tracks, the estimated vertex position during the first iterations can be far from the real one, and good tracks can be rejected due to that. In the recovery phase, 
The unassigned tracks are re-inserted in the list one by one, and the vertex fit is re-calculated each time. A track surviving the $\chi^2$ selection is then finally re-assigned to the vertex. The recovery phase ends when all unassigned tracks have been checked.

%In an initial phase, all track candidates are extrapolated to a reference plane located just downstream the target and a first approximation of the vertex is obtained by searching the points of closest distance between the beam and each outgoing track.  The vertex fit is performed by a Kalman filter method where new tracks are added in turn to the already existing vertex. The parameters of the vertex and of the newly added track are updated while those of the previously used tracks remain unchanged.  At the end of the process, a filter procedure, based on the relative $\chi^2$ contribution of each track, is applied to decide wether every particular track is compatible with the fitted primary vertex. However, neither the reconstructed beam track nor the scattered muon track can be removed from the vertex. 

\begin{figure}[tbp]
  \begin{center}
    \includegraphics[width=\columnwidth]{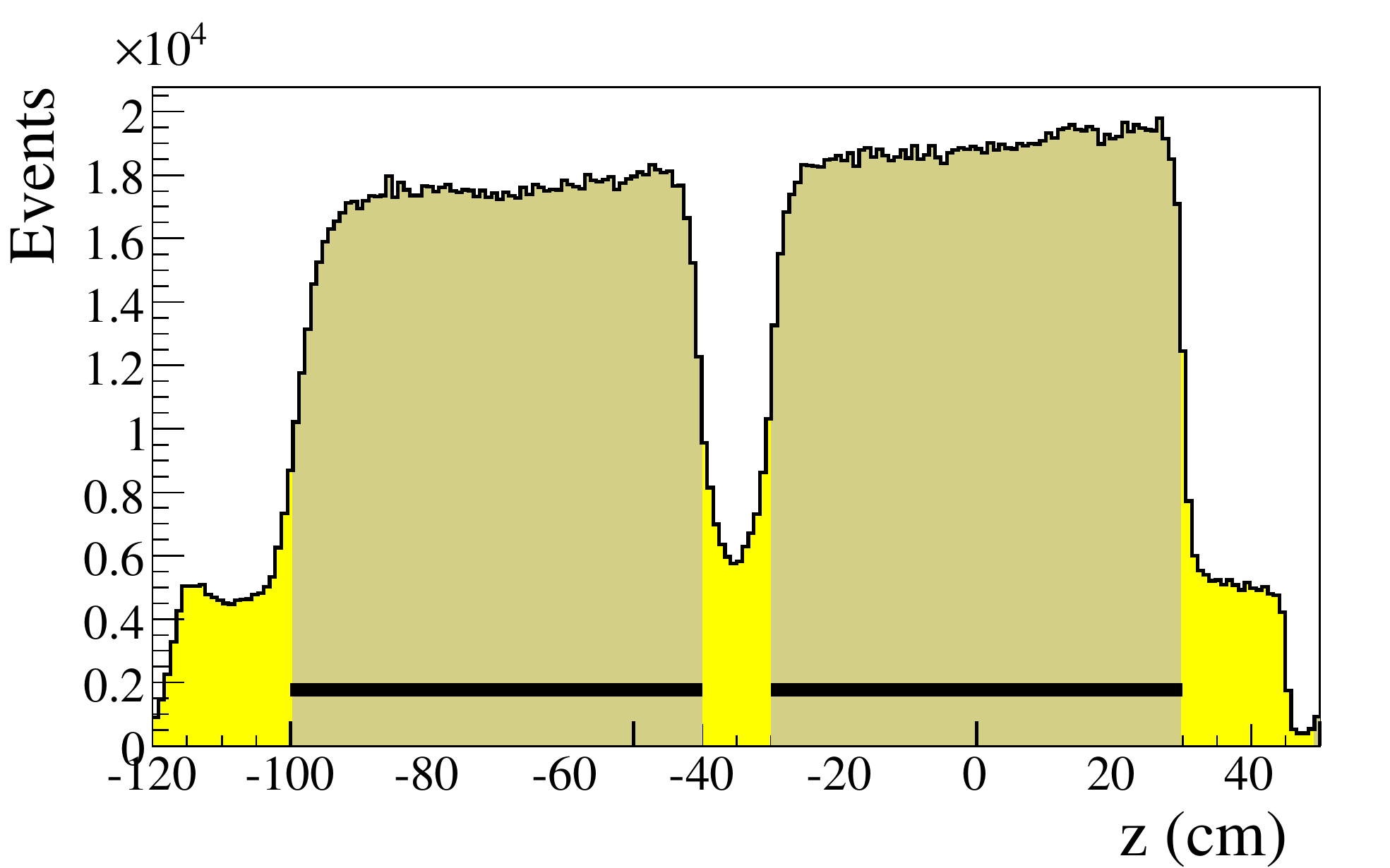}
  \end{center}
  \begin{center}
    \includegraphics[width=\columnwidth]{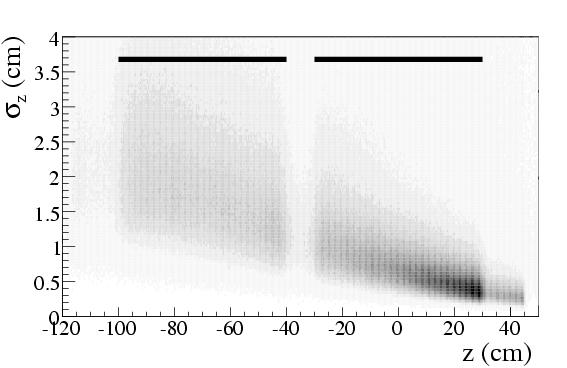}
  \end{center}
  \caption{\small{Distribution of the reconstructed vertex position
      $z$ along the beam axis (top); distribution of the error $\sigma_{z}$~vs.~$z$ (bottom). The solid lines show the position of the
      target cells.  }}
  \label{fig:z_dz.prec}
\end{figure}

The distribution of reconstructed primary vertices along the beam axis, for events with a $Q^2 > 1\,\GeV^2/c^2$ (Fig.~\ref{fig:z_dz.prec}~(top)), shows clearly the structure of the target cells, in good agreement with their real position. Vertices reconstructed outside the cells correspond to interactions with liquid helium or some other target component.

In the direction perpendicular
to the beam axis, the primary vertex coordinates are determined by the Kalman filter with an accuracy of about $0.1\,\mm$.  
%The r.m.s. value of the vertex coordinates, as determined by the
%Kalman filter, is found to be 0.1 mm  in the direction perpendicular
%to the beam axis.  
Along the beam axis the resolution depends on the
production angle of secondary particles, which is limited by the
opening angle of the target solenoid, and increases from about $5\,\mm$ at
the downstream end of the target to $25\,\mm$ in average at the upstream
end (Fig.~\ref{fig:z_dz.prec}~(bottom)).  The vertex resolution, estimated
using Monte Carlo simulations that include beam halo and pile-up
background, is compatible with the 
resolution determined with
the Kalman filter method.  The number of tracks not connected to any
physical vertex is smaller than 3\%.
% The vertex resolution, estimated from the error matrix computed by
% the Kalman filter method, was checked using Monte Carlo simulations
% that included beam halo and pile-up background. The vertex
% resolution has been evaluated in Monte-Carlo simulations including
% beam halo and pile-up backgrounds.

% Possible $V^0$~vertices are searched by combining all pairs of
% tracks with opposite charge, regardless of their association to the
% primary vertex. The most probable decay position, the tracks
% parameters at the vertex and the corresponding error matrices are
% calculated using the Kalman filter technique.

\begin{figure}[tbp]
  \begin{center}
    \includegraphics[width=\columnwidth,clip]{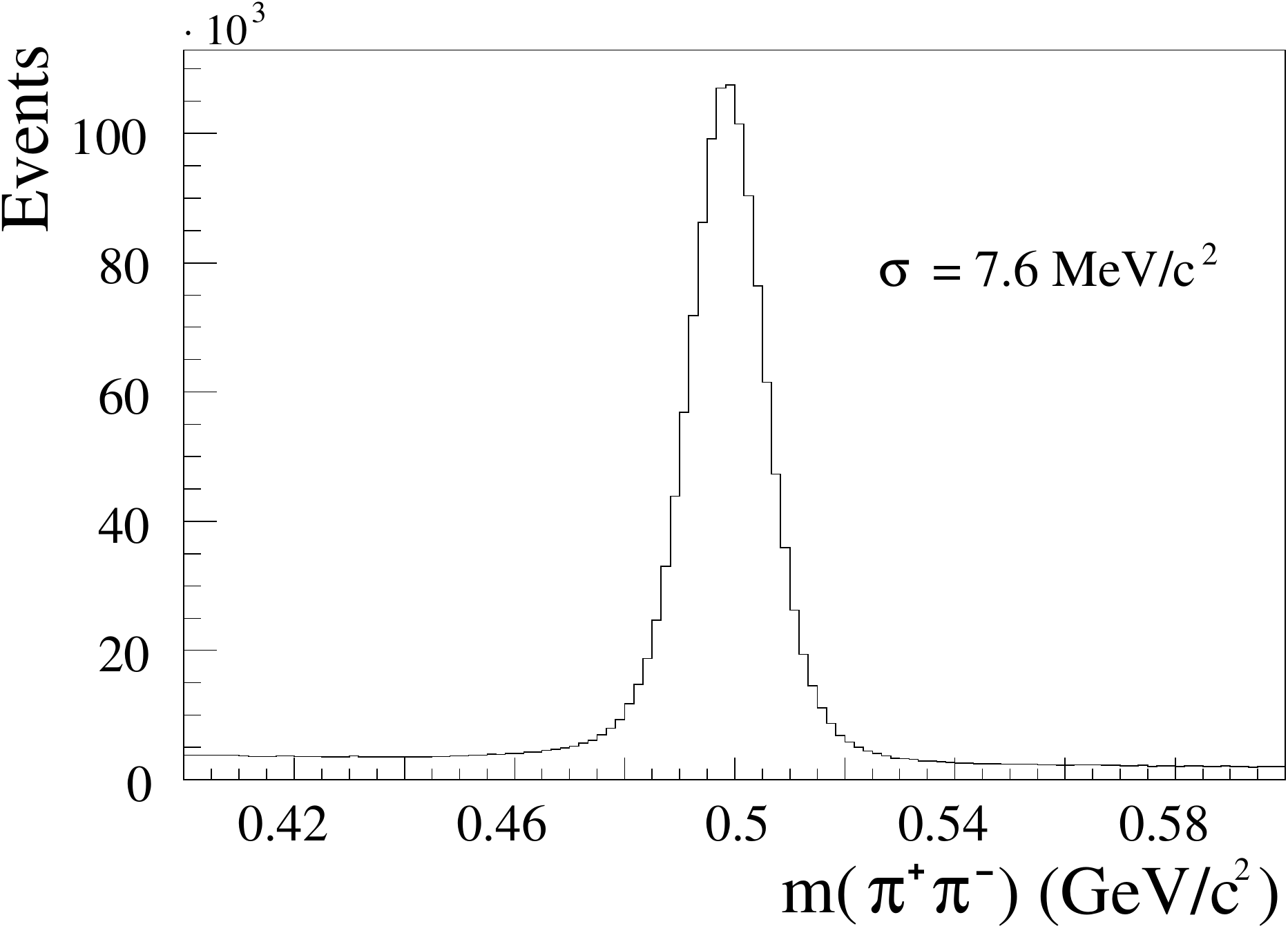}
  \end{center}
  \begin{center}
    \includegraphics[width=\columnwidth,clip]{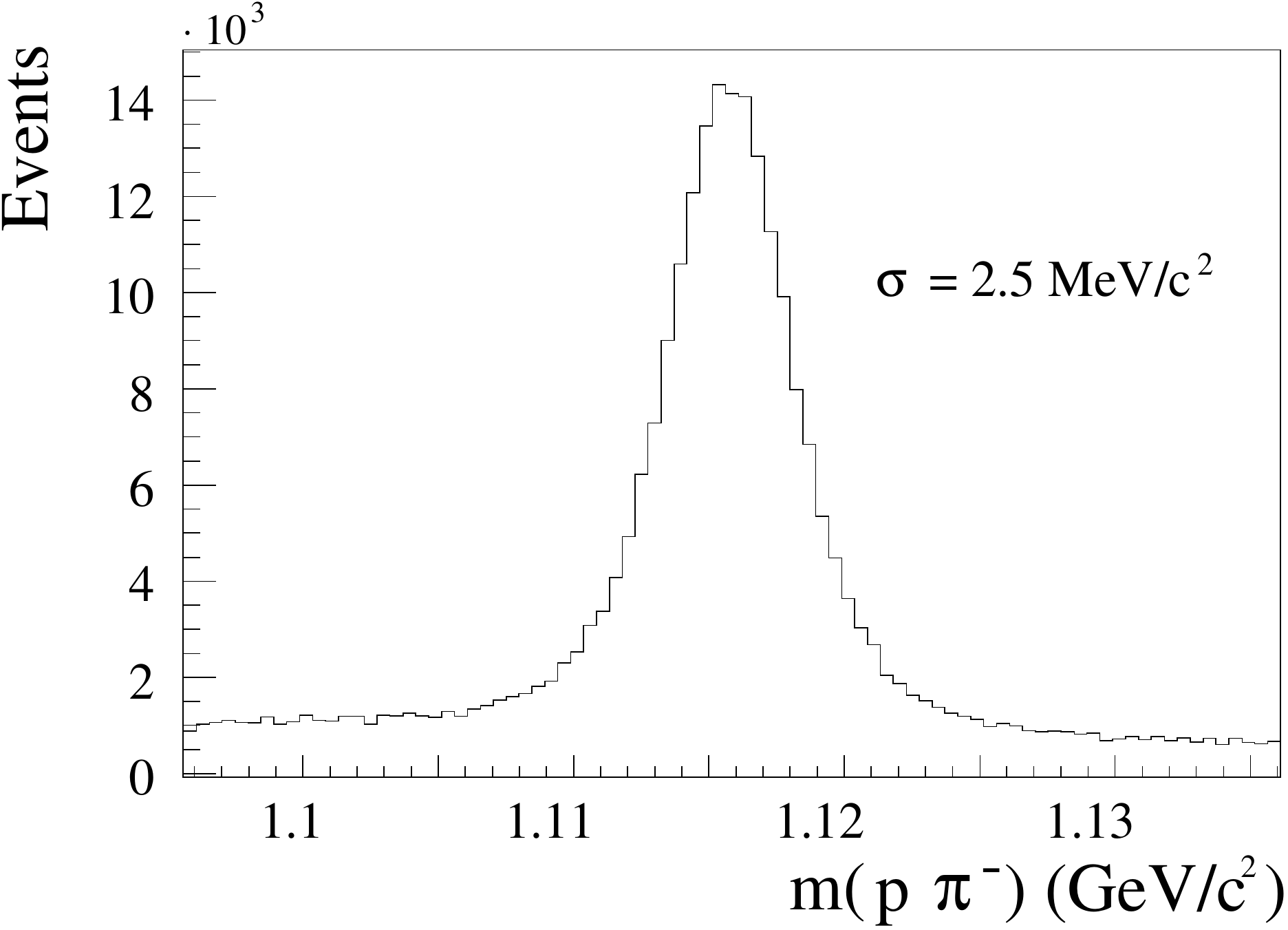}
  \end{center}
  \caption{Reconstructed invariant mass of $V^0$ decays, in the
    hypothesis of $\pi^+\pi^-$ decay (top) or $p\pi^-$ decay (bottom),
    respectively.}
  \label{fig:performance.v0_reco}
\end{figure}

Secondary vertices are reconstructed inside the target and several
meters downstream, so that nearly all $K^0_S$, $\Lambda$ and
${\overline \Lambda}$ decays into charged particles are observable.
The $V^0$~vertices are searched by combining all pairs of tracks with
opposite charge, regardless of their association to the primary
vertex. The most probable decay position, the track parameters at the
vertex and the corresponding error matrices are calculated using the
Kalman filter technique.  For decays occurring downstream of the
target, the mass resolution is found to be $7.6\,\MeV/c^2$ for the $K^0_S$
and $2.5\,\MeV/c^2$ for the $\Lambda$ and ${\overline \Lambda}$ (see
Fig.~\ref{fig:performance.v0_reco}).

% A search for a possible loss of tracks at the interaction vertex has
% been performed on a real data sample. From the study of tracks not
% connected to any physical vertex it was concluded that the loss of
% tracks at the interaction vertex does not exceed 3\%.

\subsection{The software package for the analysis of RICH-1 data}
\label{sec:performance.rich}

The pattern recognition and Particle Identification (PID) with RICH-1 is performed inside CORAL by a package called
RICHONE \cite{Baum:03b}.
The raw RICH-1 data provide the coordinates of the Photon Detectors (PD) pads with
signal above threshold and the signal amplitude; this information is combined with the particle
trajectory at the RICH entrance known from track
reconstruction.  At first a clustering procedure is applied to the
hit pads. The photon Cherenkov angle is reconstructed according to a
recipe from the literature \cite{Ypsilantis:94}: knowing the
reconstructed particle trajectory and the coordinates of the detected
photon at the photon detectors, assuming that the photon is emitted at
the mid point of the particle path in the radiator, an estimation of
the Cherenkov angle is obtained for each detected photons.  The
uncertainty in the photon emission point corresponds to the spherical
aberration of the image. The best estimate of the measured Cherenkov
angle is the mean value of all the Cherenkov angle estimations
obtained. This algorithm is at the base of both RICHONE particle ring
recognition and PID.  The ring recognition is
mostly used to monitor all the properties and the performances of the
RICH, like 
PD and mirror alignments, photon and ring angular
resolution, PID error calibration, ring reconstruction, pseudo-efficiency,
and also for a first approximation ${\chi}^2$-based PID.  The PID in
use is based on a likelihood method which includes several ingredients.
All the photons, within a large
fiducial area, are taken into account and compared to the mass
hypothesis. An accurate cluster background description is
obtained from the distribution measured on the PDs themselves.
A Poissonian
probability, taking into account the expected number of detected
photons, as a function of the particle mass and momentum and of the
detector acceptance (dead zones), is also included.  This likelihood method is
independent of the ring pattern recognition.

The RICHONE code has been developed and optimised using Monte Carlo
simulated events; physical data from the experiment have been largely
used, as soon as available, for tuning all geometrical and
physical quantities necessary for ring reconstruction and PID.
%on which the reconstruction and the PID are based.

\subsection{Monte Carlo simulation of the experimental apparatus}
\label{sec:performance.MonteCarlo}

A general interface to Geant~3.21~\cite{geant_man}, called COMGEANT, has been developed
in order to simulate the performance of the COMPASS
spectrometer. COMGEANT can 
be linked to any generator of lepton, photon or hadron interactions
such as 
Lepto~\cite{lepto_ref}, Aroma~\cite{aroma_ref} or Pythia~\cite{pythia_ref}. Radiative corrections can also be introduced in the generated events~\cite{radgen_ref}.
%For muon interactions, the beam track parameters used in COMGEANT also include the experimentally observed halo component. 

For the muon beam, the track parameters used to generate beam and halo particles
%The parameters relevant for the muon beam simulation 
are extracted from the real data events recorded with randomly generated triggers. In the case of beam particles, the parameters are taken
%The typical momentum and angular distributions of the beam particles are taken 
from the tracks reconstructed in the BMS, SciFis and 
Silicon trackers.
The parameters of the near halo component are extracted from halo tracks reconstructed along the spectrometer. For the far halo particles, falling outside of the spectrometer acceptance, the momentum is assigned using a parameterization taken from the  Monte Carlo simulation of the beam line.
A similar approach is used for the simulation of the hadron beams.

%The Monte Carlo simulation package creates also a reference geometry file, containing the nominal positions of each detector plane, that is used by the reconstruction software when simulated data is processed. 
 Interactions are generated randomly inside the target volume and
 secondary tracks propagated 
through the spectrometer. 
Special care was devoted to the realistic description of the regions of the spectrometer with high material
densities, through the use of material maps that describe the type and amount of material of each of the spectrometer elements. The maps consist of three dimensional grids
with variable cell size, depending on the homogeneity of the material being considered.
The relevant properties (radiation length, density) are stored for
every cell. Material maps are defined around the target, between 
the target and SM1, between
 SM1 and the RICH, for RICH itself and for both muon filter regions. These 
maps
 take into account all materials introduced into the spectrometer acceptance
 including detector frames, support structures and hadron absorbers.

For detector response simulation two basic quantities are used, efficiency
and resolution. 
In case of specific types of detectors like GEMs, Calorimeters and
RICH photon detectors, space and amplitude distributions of the signals
 are also simulated. All detector properties introduced in the simulation have been tuned using real data samples.
%The simulated detector hits are written out and subsequently processed as for real events. 

\subsection{Track reconstruction efficiency}
\label{sec:recon.eff}

\begin{figure}[tbp]
  \begin{center}
    \includegraphics[width=\columnwidth,clip]{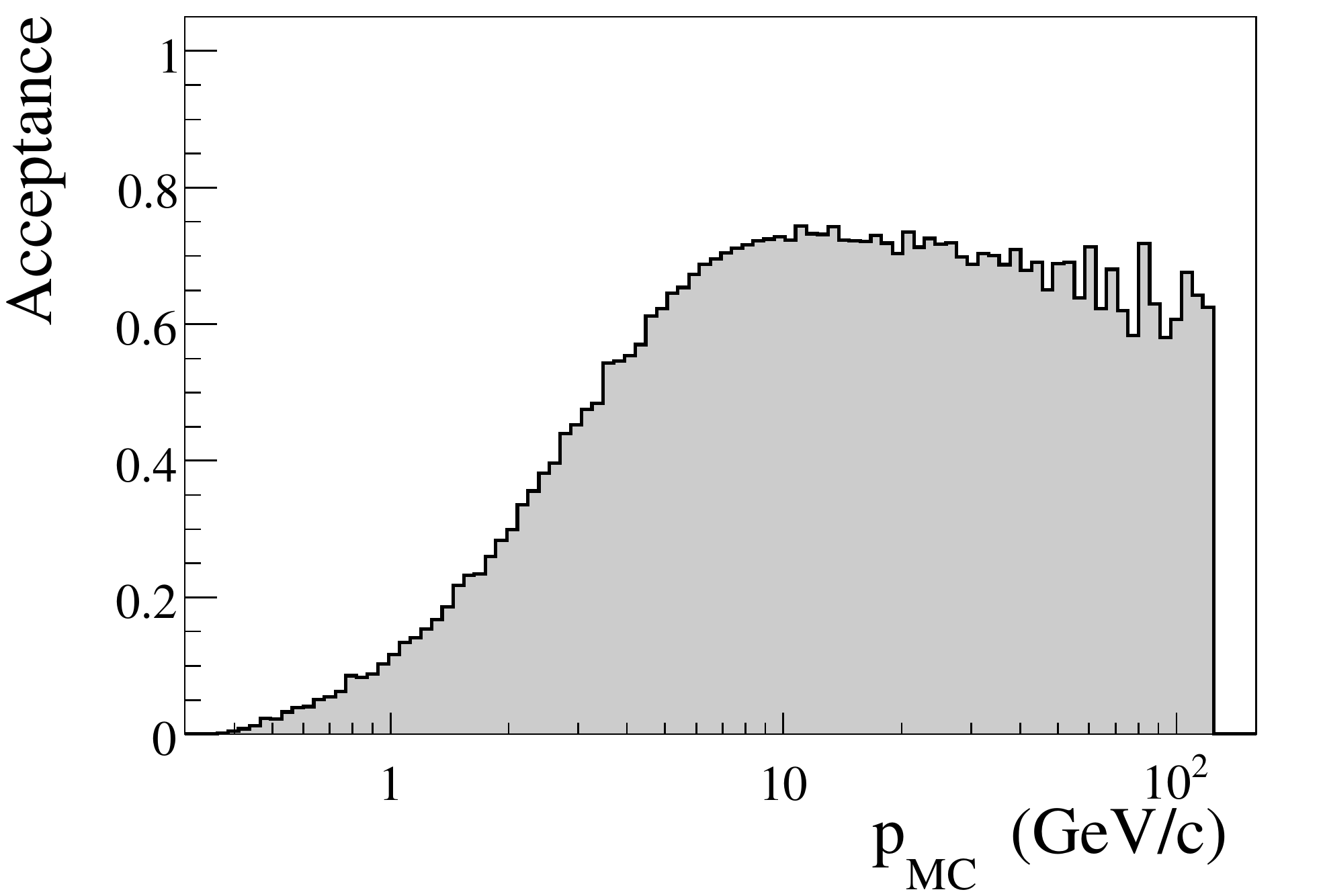}
  \end{center}
  \caption{Tracking acceptance for hadrons as a function of the generated particle momentum.}
  \label{fig:performance.prec1}
\end{figure}

The spectrometer acceptance is defined by the distribution of
reconstructible tracks.  To be reconstructible, tracks are required to
have segments formed by at least two clusters in two projections.  
%The fraction of reconstructible muon and hadron tracks is shown in Fig.~\ref{fig:performance.prec1} as a function of their momentum. For muons, the distribution starts around $30\,\GeV/c$ and the acceptance is close to 100\%. For hadrons, the distribution starts below $1\,\GeV/c$ but the acceptance remains much lower because about 25\% of them are lost in the target material mostly due to secondary interactions.
The fraction of reconstructible hadron tracks is shown in Fig.~\ref{fig:performance.prec1} as a function of their momentum. The distribution starts below $1\,\GeV/c$ and reaches a plateau of about 80\% around 10~GeV/c, the losses being mostly due to the absorption and 
re-interaction of hadrons in the polarized target material.
In the case of muons, the acceptance in the reconstruction is close to 100\% for $y > 0.1$.
%The values at low momenta but the acceptance at low momenta remains much lower because about 25\% of them are lost in the target material mostly due to secondary interactions.

The reconstruction efficiency is defined as the fraction of
Monte-Carlo generated tracks within the spectrometer acceptance which
are reconstructed by the tracking package. Here we consider only
charged tracks produced at a primary interaction point located inside
one of the two target cells.

\begin{figure}[tbp]
  \begin{center}
    \includegraphics[width=\columnwidth,clip]{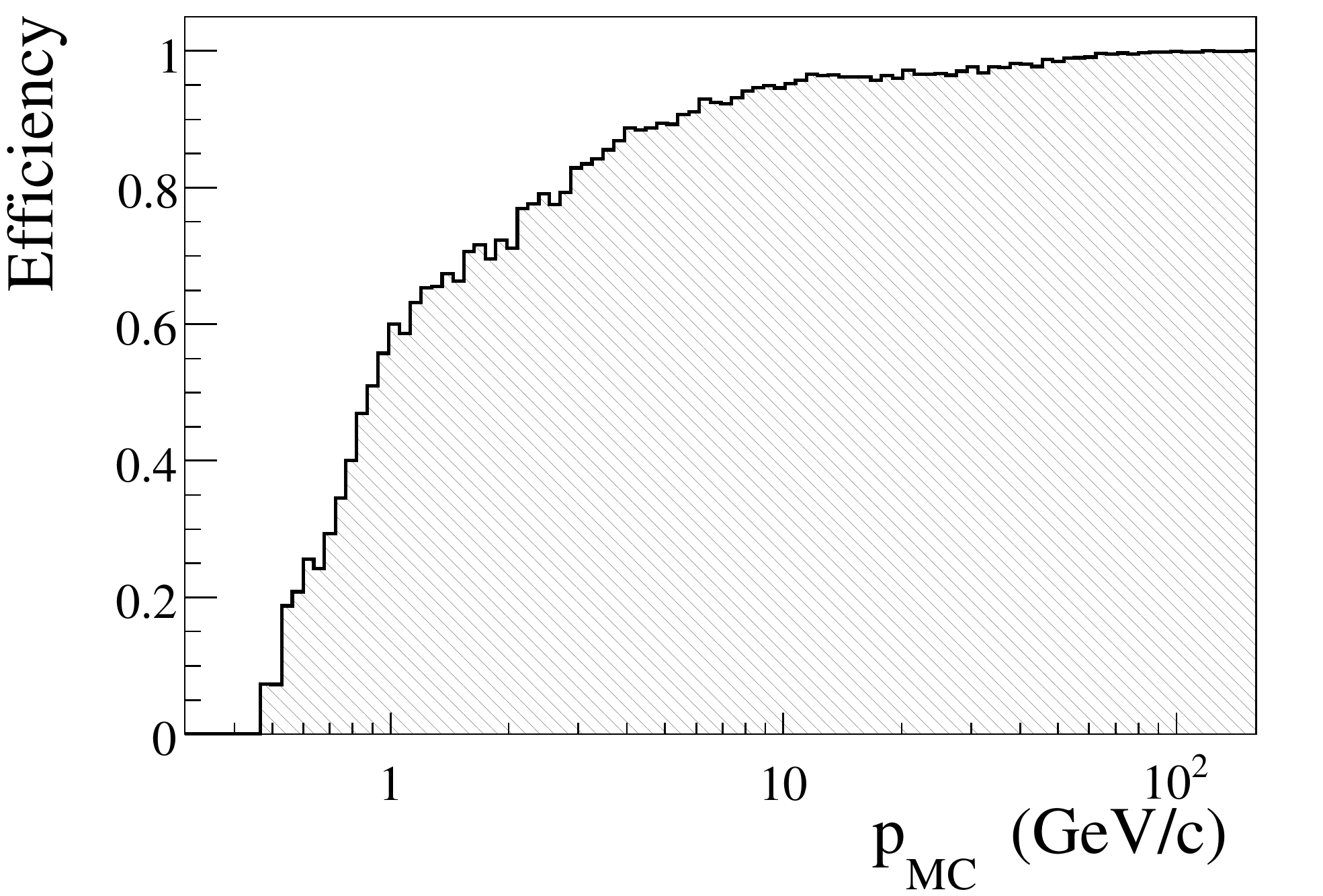}
  \end{center}
  \caption{Reconstruction efficiency as a function of the particle momentum.}
  \label{fig:performance.prec2}
\end{figure}

In view of the large uncertainty on the momentum of tracks
reconstructed in the fringe field of SM1, only tracks with segments on
both sides of one of the dipole magnets are considered here as
reconstructed. The reconstruction efficiency, which is independent of
the particle type, is found to be larger than 90\% for $p>5\,\GeV/c$
but drops rapidly for lower momenta (see
Fig.~\ref{fig:performance.prec2}).  

The inefficiency in track reconstruction may result either from a
failure in pattern recognition, due to an insufficient number of
clusters, or from a failure in the track fit. In general, low momentum
tracks cross fewer detectors, which makes pattern recognition more
difficult. The track fit may also fail, even for correctly associated
space segments, if the track has a break-point due to a re-scattering
at large angle because this process is not taken into account in the
reconstruction program. This mainly accounts for the reconstruction
inefficiency of high momentum tracks. Monte-Carlo studies show that at
lower momenta ($\sim 2-3\,\GeV/c$) pattern recognition and track fit
contribute about equally to the reconstruction inefficiency.  It may
be noted that the condition of only two clusters in two projections tends to everestimate the number of
reconstructible tracks. 
In order to limit the number of ghost tracks in the
reconstructed sample, the actual conditions on the
numbers of clusters, applied at different stages in pattern recognition,
are more rigorous; therefore they tend to reduce the efficiency shown in
Fig~\ref{fig:performance.prec2}.
   
\subsection {Track reconstruction accuracy}

The accuracy of the tracking is evaluated by comparing the track
parameters resulting from the Kalman fit on the reconstructed Monte Carlo
tracks with the corresponding values from 
the generated tracks.  The
r.m.s. widths of the distributions of residuals are found to agree
% The observed differences between generated and reconstructed values
% agree
with the errors estimated in the track fit, within factors of
$1.2-1.5$.

\begin{figure}[tbp]
  \begin{center}
    \includegraphics[width=\columnwidth,clip]{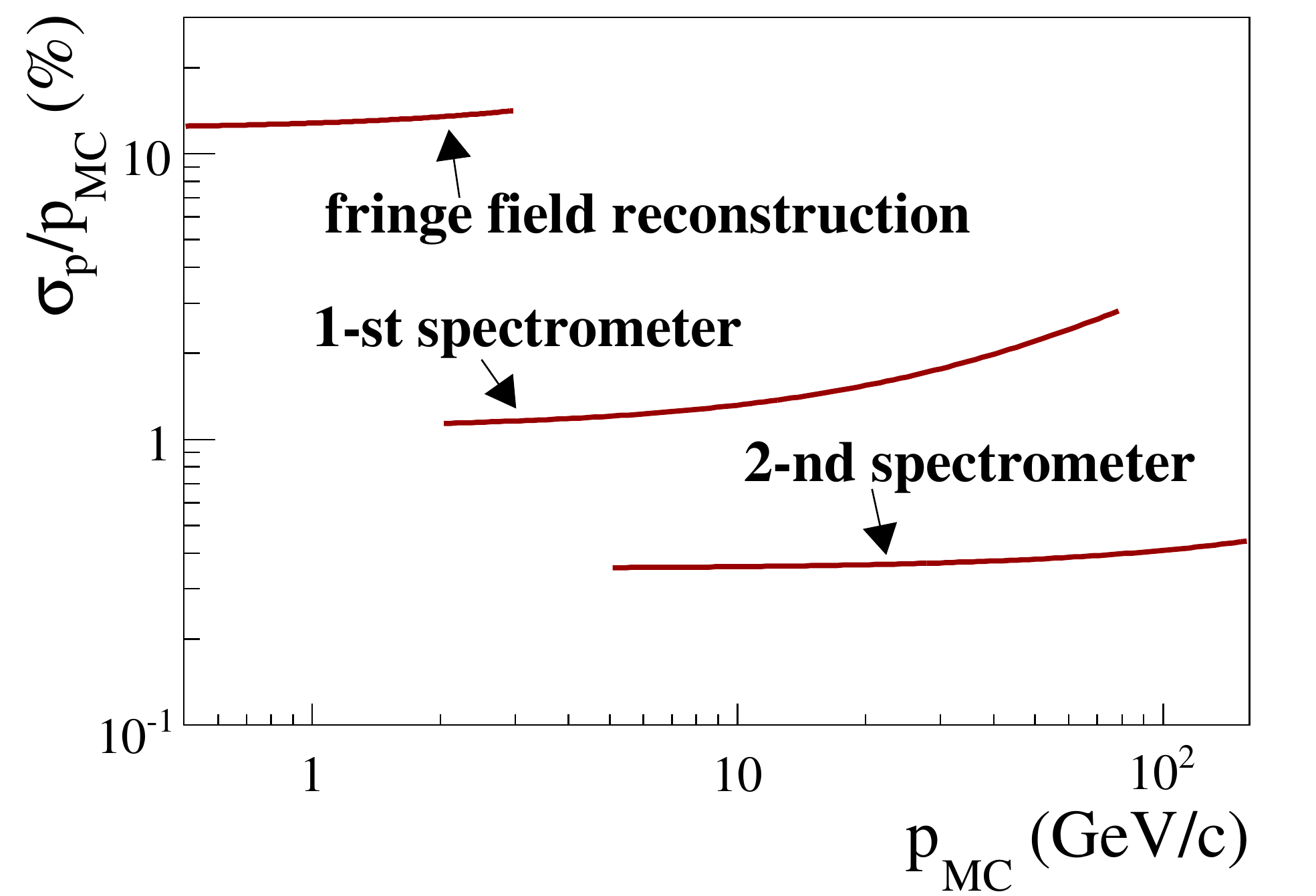}
  \end{center}
  \caption{Relative track momentum resolution $\sigma_p/p$ versus $p$.}
  \label{fig:performance.prec3}
\end{figure}

\begin{figure}[tbp]
  \begin{center}
    \includegraphics[width=\columnwidth,clip]{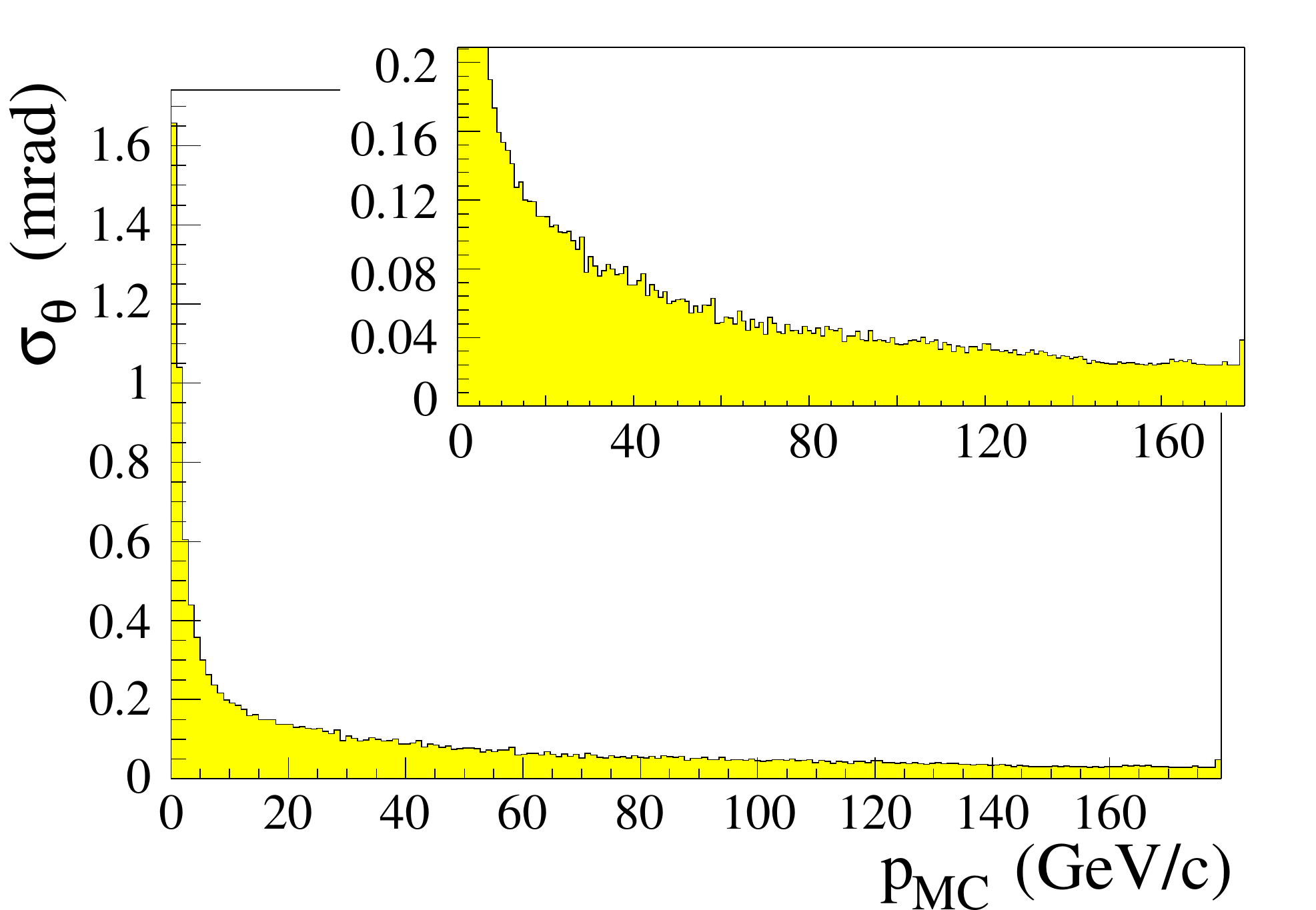}
  \end{center}
  \caption{Resolution of the reconstructed particle emission angle at the interaction vertex as a function of the particle momentum.}
  \label{fig:performance.prec4}
\end{figure}

The average errors on the track position in the transverse plane are
of the order of $80\,\mum$ at the 
$z$-coordinate of the first detector, in
agreement with the nominal resolution.  The relative momentum
resolution and the track angle resolution are shown in
Figs.~\ref{fig:performance.prec3}-\ref{fig:performance.prec4}
 as a function of the momentum.  The 
relative error $\sigma_p/p$ is about $0.5\%$ 
%for high momentum tracks reconstructed in both spectrometers ($p\ge 30\,\GeV/c$) 
for tracks reconstructed in both spectrometers ($p\ge5\,\GeV/c$)
and about
$1.2\%$ for low momentum tracks reconstructed in the LAS only. The
error on 
the track polar angle at the interaction vertex ($\sigma_{\theta}$) is
of the order of $0.1\,\mrad$ for $p \approx 30\,\GeV$ and increases
for lower momenta.

\subsection{RICH-1 performances}
\label{sec:recon.rich}

\begin{figure}[tbp]
  \begin{center}
    \includegraphics[width=\columnwidth]
    {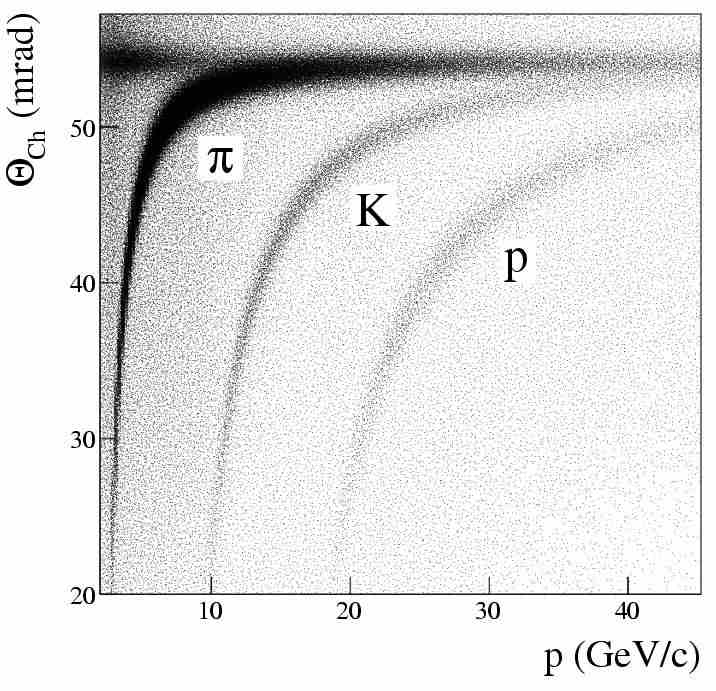} %%\vspace{-2ex}
  \end{center}
  \caption{A two-dimensional histogram of the Cherenkov angles as measured
    with RICH-1 versus the particle momenta.  The entries referring to particles identified as pions have been depressed by a factor of 3; those referring to particles identified as protons have been multiplied by a factor of 4.}
  \label{fig:performance.rich.ang-mom}
\end{figure}

Figure~\ref{fig:performance.rich.ang-mom} shows a two-dimensional
histogram of the Cherenkov angles as measured with RICH-1, versus the particle
momenta. The corresponding loci for
pions, kaons and protons are clearly visible.

\begin{figure*}
        \begin{minipage}{0.49\textwidth}
  \begin{center}
    \includegraphics[width=\textwidth]{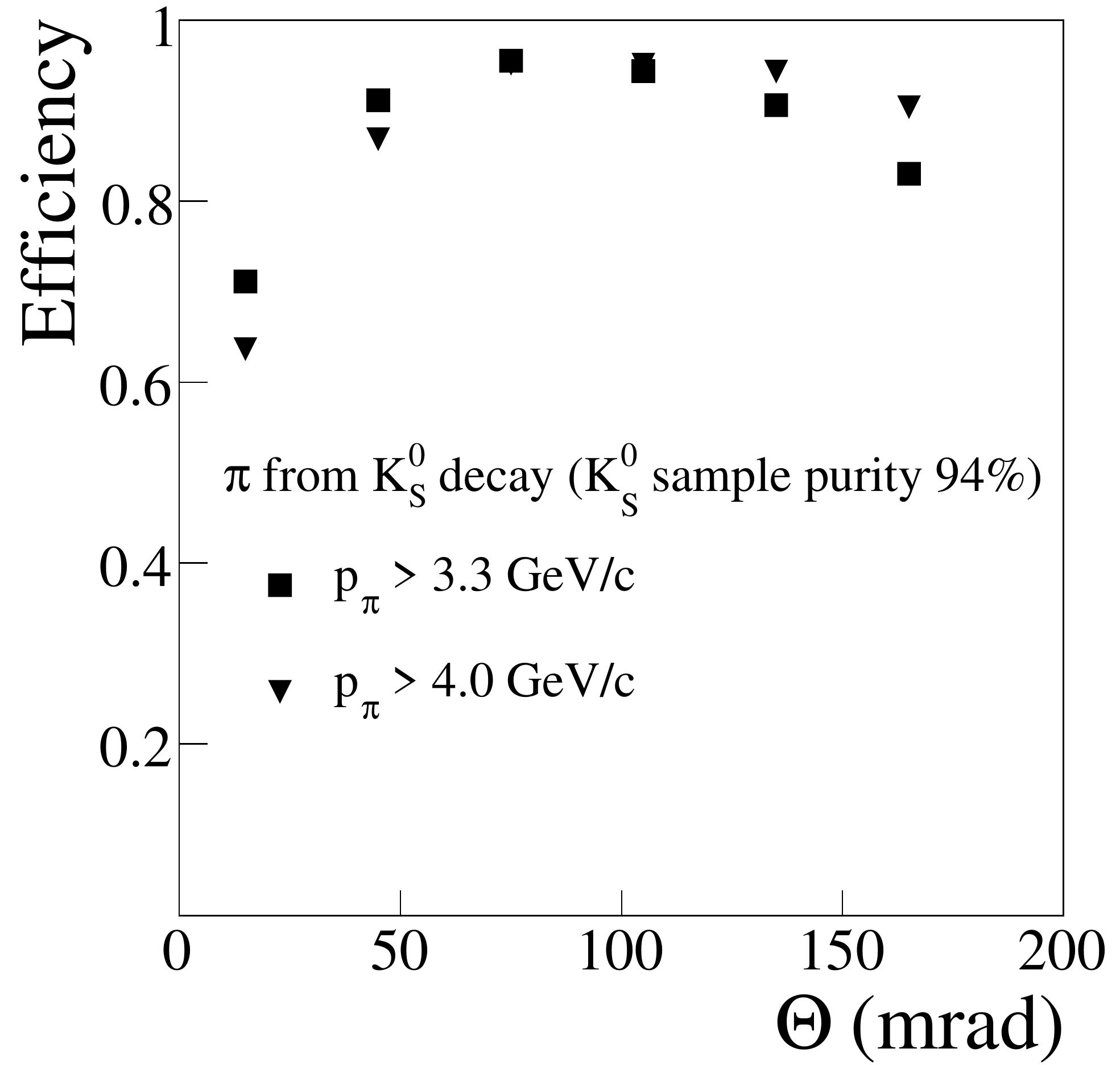}
        \end{center}
        \end{minipage}
        \hfill
        \begin{minipage}{0.49\textwidth}
  \begin{center}
    \includegraphics[width=\textwidth]{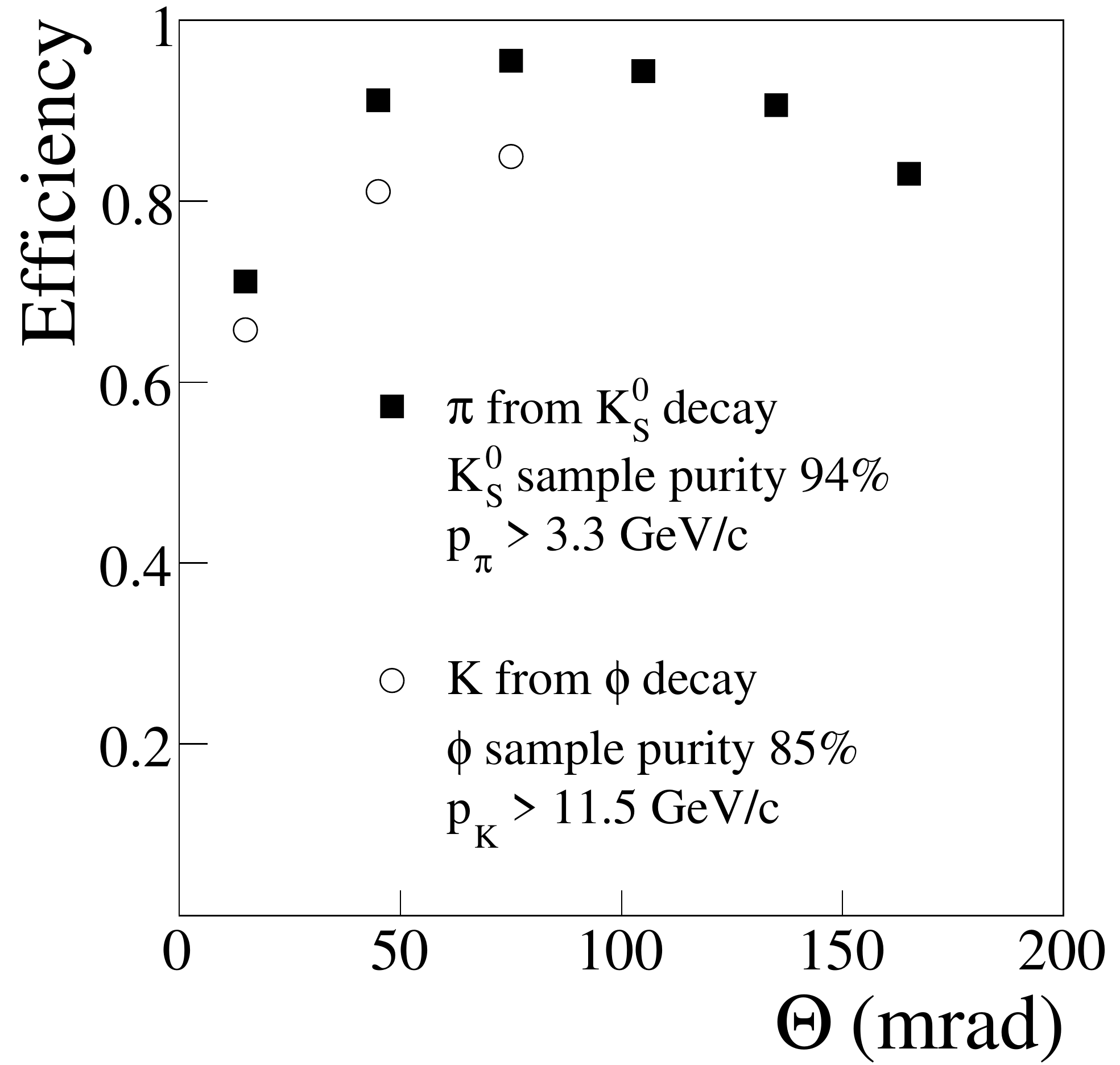}
        \end{center}
        \end{minipage}
    %\vspace{-2ex}
\caption{2003 data, PID efficiency for pions from $K_S^0$ 
decay 
%(K$_S$ sample purity: 94\%) 
versus the particle polar angle, pion momentum $p>3.3\,\GeV/c$ (squares);
left: comparison with PID efficiency for pions with momentum $p>4\,\GeV/c$ (triangles);
right: comparison with PID efficiency for kaons from $\phi$ decays, kaon momentum $p>11.5\,\GeV/c$ (open circles).}
\label{fig:performance.rich.rich_efficiency}
\end{figure*}

Figure \ref{fig:performance.rich.rich_efficiency}~(left) shows the PID
efficiency for pions from $K_S^0$ decay versus the particle polar angle
applying two different cuts to the particle momentum: $3.3\,\GeV/c$,
corresponding to a mean value of 5 detected photoelectrons per
particle and $4\,\GeV/c$, corresponding to a mean number of 8 detected
photons.  The $K_S^0$ sample has a purity of 94\%.  The efficiency is
plotted versus the particle polar angle because different angles
correspond to different regions of the photon detectors. 
For particles scattered at small polar angles, the signal
to noise ratio is 
less favourable due to the overlap with the images
generated by the huge beam halo of the SPS muon beam.  Moreover, the
presence of the beam pipe inside the RICH vessel and a corresponding
hole in the mirror wall reduces the number of detectable photons for
these particles. Particles scattered at large polar angle have, on
average, low momentum.

Comparing PID efficiency for particles selected with different
momentum cuts, it is shown that the slightly lower efficiency at large
polar angles is related to samples with lower mean momentum value,
thus with lower number of detected Cherenkov photons. For example, for
the pions contributing to the point at largest polar angles, the mean
value of the momentum distribution is as low as $4.3\,\GeV/c$ .

Figure \ref{fig:performance.rich.rich_efficiency}~(right) presents the PID
efficiency for kaons from $\phi$ decay versus the particle polar
angle. Momenta larger than $11.5\,\GeV/c$, corresponding to a mean value of 5 detectable photoelectrons, were selected.  The $\phi$ sample has a purity of 85\%.  The kinematic range
explored is limited due to the phase space spanned by the diffractive
$\phi$ production. The apparent reduced efficiency for K PID is
compatible with the poorer purity of the $\phi$ sample.

\begin{figure}[tbp]
  \begin{center}
    \includegraphics[width=\columnwidth]{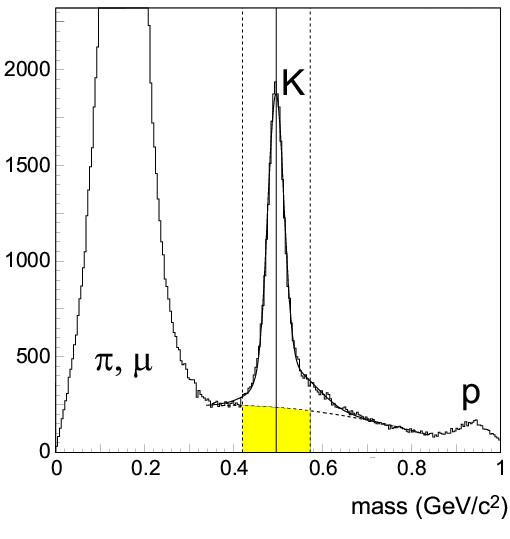}
%%    \vspace{-2ex}
  \end{center}
  \caption{Reconstructed mass spectrum for particles with momenta
    $11.5\,\GeV/c$ and polar angle $>$~25~mrad.}
  \label{fig:performance.rich.purity}
\end{figure}

Figure \ref{fig:performance.rich.purity} illustrates the achieved PID
purity: the reconstructed mass spectrum, K region, for particles with
momenta larger than $11.5\,\GeV/c$ and polar angle $>$~25~mrad is shown. The
signal S over background B ratio is 2.5 and the purity, defined as
S~/~(S~+~B), is $>$~0.7.  
The corresponding phase space
region includes $\sim$75\% of the kaons from $D^0$ and
$\overline{D}^0$ decay, as selected applying standard COMPASS
analysis cuts.

\subsection{Energy reconstruction in the COMPASS calorimeters}
\label{sec:performance.calo}

In COMPASS, the energies of charged and neutral particles are
reconstructed by means of two hadron calorimeters (HCAL1 and HCAL2,
see Sec.~\ref{sec:pid.calo.hcal1} and \ref{sec:pid.calo.hcal2}) and
one electromagnetic calorimeter (ECAL2, see
Sec.~\ref{sec:pid.calo.ecal}). The energies and positions of the
calorimeter showers are reconstructed by a dedicated software package,
common to both types of calorimeters. The shower parameters
are estimated from the energy deposited in clusters of neighbouring
cells, assuming a known parameterization of the shower profile.

% The particle reconstruction in the COMPASS calorimeters is based on
% the reconstruction of the parameters (energy and position) of the
% shower (electromagnetic or hadronic, depending on the calorimeter
% type) that was initiated by the particle.

% The first step of the shower reconstruction is the selection of the
% cells with an energy deposit above a predefined threshold, in order
% to approximately identify the local maxima of the induced showers.
During the first step of the shower reconstruction, the local maxima
are approximately identified by selecting the cells with an energy
deposit above a predefined threshold.  At this stage, two showers are
not separated if the local maxima are located at nearby cells.

The energy and position of the cell corresponding to the local maximum
are used as initial parameters for the shower reconstruction. The
remaining part of the shower is searched in a region of $3 \times 3$
or $5 \times 5$ cells surrounding the local maxima. The region size
depends on the estimated shower energy; for higher energies a larger
region is selected. Although a local maximum can be attributed to one
shower only, the energy deposited in the surrounding cells can be
shared between two or more showers. Therefore, the initial values of
energy and position for the shower are used to estimate the amount of
energy deposited in the cells surrounding the local maximum. This procedure is
repeated for all local maxima; the relative
contribution of each shower to the shared cells is thus computed.

% Once cells are attributed to expected showers, the fraction of cell
% energy attributed to each shower is determined in the following
% manner. The initial values of energy and position for the shower are
% used to estimate the amount of energy deposited in the cells
% surrounding the local maxima. This is repeated for each and every
% expected shower, and the relative contribution to the common cells
% is computed.

Once the amount of energy deposited in each cell by the expected
showers is computed, a fitting procedure is applied to calculate the
exact parameters (energy and position). The energy is simply
calculated from the sum of cell energies, corrected for the shower
leaks outside the calorimeter and for the shower overlaps. For the
determination of the coordinate parameters, the inverse
one-dimensional cumulative shower profile function~\cite{Lednev:95} is
used. In this approach, the theoretical two-dimensional shower surface
is projected onto two planes, perpendicular to the calorimeter
surface, one directed horizontally and one vertically. The projection
of the shower surface is symmetric with respect to the shower axis,
and the two boundaries are parameterized with analytical functions, 
the actual parameters of which are 
optimised for the hadron and electromagnetic cases. The position of
the shower axis which gives the best fit to the energy distribution in the cells
attributed to the shower determines the X and Y coordinates, depending
on the projection surface.

\begin{figure}
  \begin{center}
    \includegraphics[width=\columnwidth]{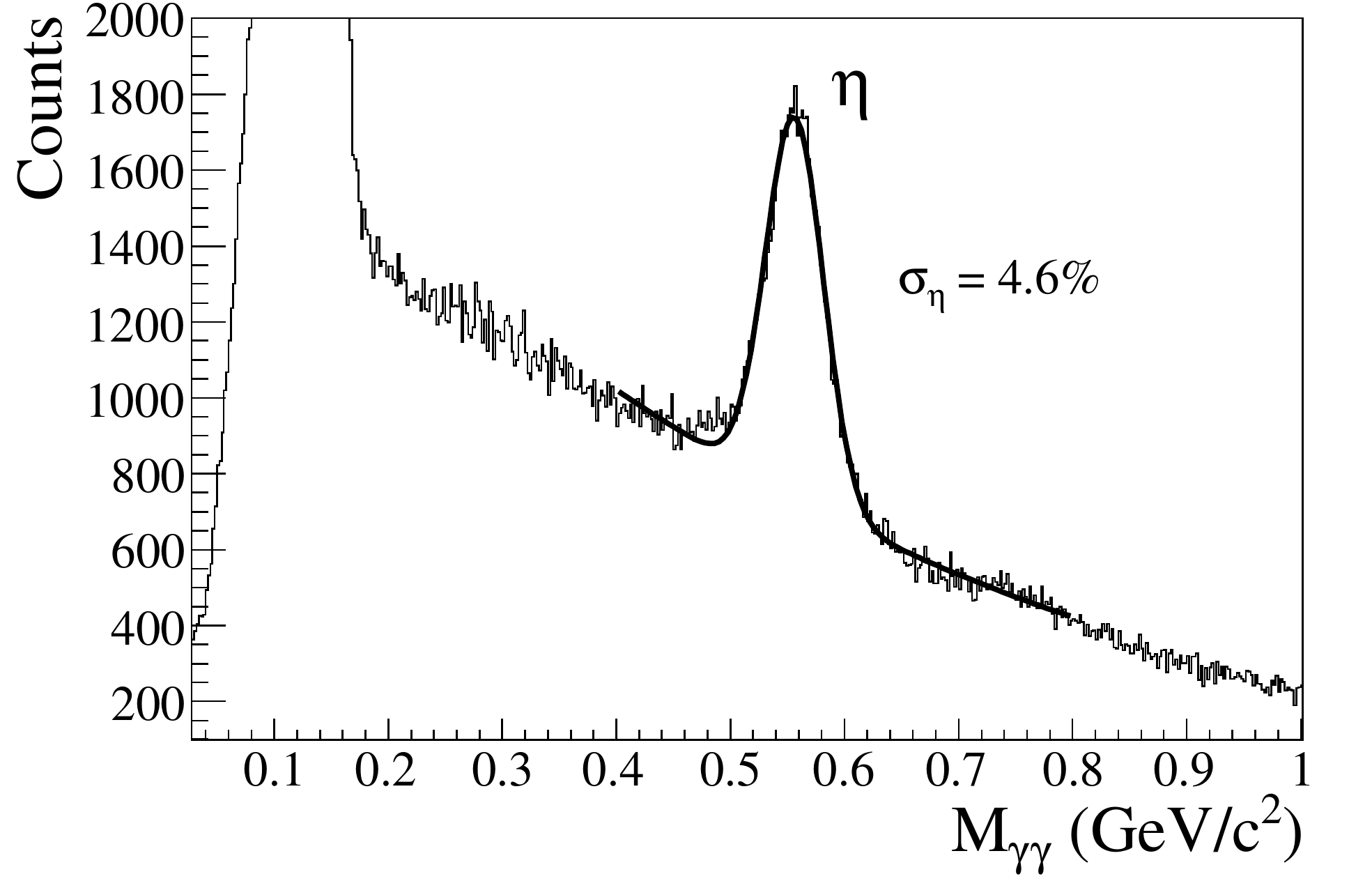}
  \end{center}
  \caption{Invariant mass distribution in the hypothesis of $2\gamma$ neutral decays. The $\eta$ mass is
    determined from the gamma energy reconstructed in the
    electromagnetic calorimeter ECAL2, in the hypothesis that the
    photons originated from the primary interaction vertex.}
  \label{fig:performance.eta_mass}
\end{figure}

A good estimation
of the overall ECAL2 performances is given by the mass resolution for
reconstructed $\eta \to \gamma\gamma$ decays. To estimate the width of
the $\eta$ mass peak and minimise the combinatorial background, only
events with exactly two clusters originating from neutral particles have been selected. The
invariant mass is then calculated under the assumption that the two
clusters were produced by two photons emitted from the reconstructed
primary interaction vertex. The resulting mass distribution is shown
in Fig.~\ref{fig:performance.eta_mass}. The fit of the mass peak with
a Gaussian function and a polynomial background gives a relative
resolution of 4.6\%, quite close to the theoretical limit of about 4\%
given by the intrinsic energy resolution of the calorimeter elements.

%%% Local Variables: 
%%% mode: latex
%%% TeX-master: "compass_spec"
%%% End: 

% Upgrade
\section{Spectrometer upgrade}
\label{sec:upgrade}
%%%%%%%%%%%%%%%%%%%%%%%%%%%%%%%%%%%%%%%
%%%%%%%To be written by S. Dalla Torre.
%%%%%%%%%%%%%%%%%%%%%%%%%%%%%%%%%%%%%%%
The COMPASS setup employed during the data 
taking periods in 2002, 2003 and 2004,
and described in the previous sections, 
is not as complete as the one foreseen
in the original design of the
COMPASS spectrometer. Part of the components have been staged.
For the data taking from 2006 on, an
important upgrade of the COMPASS apparatus is implemented.    
It concerns the polarised target (see Sec.~\ref{sec:upgrade.pt}), the 
large area trackers of the LAS (see Sec.~\ref{sec:upgrade.LAT}), the
RICH-1 detector (see Sec.~\ref{sec:upgrade.rich}), the electromagnetic
calorimeter ECAL1 (see Sec.~\ref{sec:upgrade.ecal1}), and the
new tracker and preshower station located downstream of RICH-1
(see Sec.~\ref{sec:upgrade.rw}). 

%%%%%%%%%%%%%%%%%%%%%%%%%%%%%%%%%%%%%%%
\subsection{Polarised target}
\label{sec:upgrade.pt}
The original design of the COMPASS spectrometer included a large-aperture 
superconducting solenoid magnet, which 
%establishes the magnetic field  and 
maintains the polarisation of the target. The solenoid has been designed 
to match the required $\pm180\,\mrad$ acceptance for the most downstream edge 
of the polarised target (see Sec.~\ref{sec:targets.pt} for a description of the polarised 
target), while providing 
a homogeneity 
at least as good as the  
magnet used until 2004 in COMPASS, despite a volume five times larger.

The solenoid was delivered to the collaboration by Oxford Danfysik \cite{Oxford} 
at the end of 2004. Extensive tests were performed in order to fully 
characterise the intrinsic performances of the magnet. The relative intrinsic 
field uniformity was tuned to within $\pm3\EE{-5}$ over the useful 
volume of the target cells ($130\,\cm$ total length, $3\,\cm$ diameter). A
fully automated control system  
for the handling of the cryogenics and a fast 
safety system designed to trigger the heaters in case of a quench were 
implemented. A specific procedure involving the heaters on the shim coils 
had to be worked out in order to accommodate for a short circuit in one 
of the shims. After installing the magnet in its final position in the 
COMPASS hall, new optimisations of the shim coil currents were performed 
in order to account for the presence of the large SM1 dipole magnet yoke. 
A final relative field uniformity of $\pm4\EE{-5}$ was achieved, valid for
operation at both field  
orientations. Field profiles have also been measured in the region between 
the solenoid and SM1, providing additional input for the tracking software. 
As for the magnet used until 2004, a transverse field is required for 
field rotations and for measurements with transverse orientation of the target 
polarisation. A dipole coil provides such a transverse field of $0.6\,\T$. 

The polarised target configuration used until 2004 consisted of two identical 
cells with opposite polarisation (see Sec. \ref{sec:targets.pt}). The variation of the spectrometer 
acceptance for reaction products originating from the two cells is a source of 
false asymmetry, largely removed by reversing the cell polarisation every eight hours. 
A new target system was built for the new solenoid. It consists of three target cells 
and a redesigned microwave cavity adapted to the enlarged acceptance. In this 
new configuration, the central cell is twice longer than the two external cells 
and its target material is polarised in opposite direction to the external
ones. Simulations show that  
such a configuration leads to a further reduction - by an order of magnitude - of 
the false asymmetry.

%%%%%%%%%%%%%%%%%%%%%%%%%%%%%%%%%%%%%%%
\subsection{Large drift chamber and straw tubes}
\label{sec:upgrade.LAT}
The two tracker telescopes of the LAS, located upstream and 
downstream of the SM1 analysing magnet, have to match the
enlarged acceptance made possible by the availability of the COMPASS solenoid:
$\pm 180 \, \mrad$ in the horizontal and vertical projections between the
polarised target and SM1 and $\pm 250\,\mrad$ in the horizontal projection
between SM1 and RICH-1. Upstream of SM1, the existing tracker telescope is
completed with an additional medium size drift chamber positioned immediately
behind the first Micromegas station. Downstream of SM1 only one out of the
three stations of the corresponding tracking telescope (see
Sec.~\ref{sec:tracking.lat}) has a transverse surface big enough to cover the
enlarged angular acceptance. The two drift chambers are replaced by a new
large drift chamber and by a second straw tube station, identical to the ones 
described in Sec.~\ref{sec:tracking.lat.straw}. 

%%%%%%%%%%%%%%%%%%%%%%%%%%%%%%%%%%%%%%%
%%\subsubsection{Large drift chamber}
%%\label{sec:upgrade.LAT.DC4}
The new drift chamber has a design derived from 
trackers DC1, DC2 and DC3, already in operation in the 
COMPASS spectrometer (see Sec.~\ref{sec:tracking.lat.dc}).
The active surface is enlarged 
to $248\times 208\,\cm^2$, compared to $180\times 127\,\cm^2$ for DC1--3. 
The chamber has eight active wire planes  
for the measurement of four coordinates (horizontal wires, vertical 
wires and wires inclined at
$\pm 10$~degrees with respect to the vertical axis). The drift cells have a pitch
of $8\,\mm$ and a gap of $8\,\mm$. The central dead zone has a diameter of
$30\,\cm$. 
The total material budget of the chamber is $0.32\%$ of a radiation length.
The readout system is identical to the one in use for DC1-3 chambers, namely
ASD8 front-end chips coupled to F1-TDCs. Performances (efficiency, resolution)
are expected to be identical to the ones obtained with stations DC1-3.   

%%%%%%%%%%%%%%%%%%%%%%%%%%%%%%%%%%%%%%%
\subsection{RICH-1}
\label{sec:upgrade.rich}

The overall RICH-1 upgrade project aims at a significant increase of the 
RICH performances in the years to come. It was designed to improve two
important  limitations of the RICH-1 detector. The first limitation
comes from the large dispersed halo that generates high pad
occupancies and deteriorates the signal-to-noise ratio (see  \ref{sec:pid.rich.characterization}), 
particularly in the central region. The halo also increases the probability for electrical 
instabilities of the photon detectors. The second limitation is due to the large dead time 
imposed by the architecture of the GASSIPLEX front-end chip (see Sec. \ref{sec:pid.rich.RO}).  

In order to minimise the overall cost of the project, two complementary technologies have 
been chosen. In the most sensitive central region both the photon detectors and the readout 
system are replaced with a new photon detection system based on Multi-Anode Photo-Multiplier 
Tubes (MAPMT). In the peripheral region the existing photon detectors are kept, but their  
associated electronics is replaced with a new, much faster readout system based on the APV25 chip.

%%%%%%%%%%%%%%%%%%%%%%%%%%%%%%%%%%%%%%%
\subsubsection{RICH-1 upgrade in the central region}
\label{sec:upgrade.rich.MAPMT}
The central part of the RICH-1  detector is instrumented with a 
system based on MAPMTs for fast photon detection. The time resolution, of a few ns, 
allows an efficient rejection of the high background due to uncorrelated events. 
Each MAPMT is  
coupled to a telescope formed by a field lens and a concentrator lens. 
These new detectors replace the four central photocathodes of the CsI MWPCs, corresponding 
to $25\%$ of the total active surface. 

The approach to detect photons in RICH counters with MAPMTs and lenses is not new (HERA-B \cite{HeraB:00},  
studies for LHCb \cite{LHCB:00,Albrecht:02}). In our design, two new elements are introduced: the detection of visible and UV photons and a largely increased ratio of the collection surface to the photocathode one. We extend the detected 
range of the Cherenkov light spectrum to the UV domain 
(down to $\sim200\,\nm$) by using UV extended MAPMTs and fused silica lenses. 
We have selected the R7600-03-M16 MAPMT developed by Hamamatsu \cite{Hamamatsu}, 
characterised by a bialkali photocathode with $18 \times 18\,\mm^2$ active surface, 
16 pixels, 70\% collection efficiency of photoelectrons
 and a UV extended glass entrance window extending the sensitive
range down to $200\,\nm$.
The effective pixel size of a MAPMT coupled to the lens telescope is 12$\times$12~mm$^2$. 
The resulting ratio of the entrance lens surface to the MAPMT photocathode surface has a value 
of about 7, thus reducing the number of MAPMTs and keeping the system at affordable cost.   

The lens telescope has been designed in order to satisfy several important requirements. 
The image distortion has to be minimised, keeping the telescope length around $10\,\cm$, in 
spite of the large image demagnification and the large angular acceptance. Simulations show that 
only about $10\%$ of the photons are not detected in the corresponding MAPMT pixel, but in a nearby 
one.  The telescope must have a large angular acceptance; the value achieved is about $\pm165 \,\mrad$, resulting in an estimated ring loss below $5\%$. The dead zone should be as small as possible; an accurate 
mechanical design assures only $2\%$ dead zone between field lenses.

The MAPMTs are read with a MAD4 amplifier-discriminator \cite{Gonella:01}, 
housed on small front-end boards mounted directly onto the resistive divider boards. 
Power and thresholds are distributed via deck boards, while digital 
boards housing F1-TDC chips are directly coupled to the deck boards; 
the arrangement is without cables, thus very compact and reliable. 
The readout architecture 
is similar to that already successfully 
implemented in COMPASS for the Micromegas and DC detectors (see
Sec. \ref{sec:daq:digital}), and also 
used for the Rich wall detector . 
The photon detection principle has been validated in 
two test beam measurements at 
the CERN PS beam line T11 in summer 2003 and 2004 \cite{Alekseev:05}. 
The MAPMTs have been studied and characterised for single photoelectron detection
in the laboratory \cite{Abbon:05a,Abbon:05c}.
Comparing Monte Carlo studies and test results, it has been shown that 
single photoelectrons collected by the MAPMT dynode system 
are detected with full efficiency. The long plateau of the threshold curves clearly indicates that 
noise and crosstalk can be efficiently rejected 
without photoelectron losses. 
At high rate, the MAPMT efficiency shows no losses up to
single photoelectron rates of $5 \,\MHz$ per pixel, while the front-end
chip is limited to about $1\,\MHz$. A new version of the chip overcoming
this limitation, C-MAD, is being developed and will be used in 2008 and beyond.

The new detection system collects a number 
of detected photons per ring that is a factor of 4 larger 
than the previous one; at saturation, and far from the central 
region affected by dead zones, this results in about 60 photons 
per ring. Due to 
the enlarged pixel size, the resolution on the measured Cherenkov angle from 
single photons is worsened by about a factor 1.5. 
This reduction is largely compensated by the increased number 
of detected photons. The angular resolution also benefits 
from the negligible uncorrelated background level. The 
expected resolution for a whole ring at saturation, and far 
from the central region is about $0.3 \,\mrad$ 
(presently $0.45\,\mrad$). The improved   
resolution makes the $\pi$/K separation at $2.5 \,\sigma$ 
possible up to at least $50\,\GeV/c$, to be compared with the present limit of $43\,\GeV/c$. 
Finally, the increased number of detected photons will lower the effective 
RICH thresholds; the new thresholds nearly coincide with the physical 
thresholds of the Cherenkov effect. 

%%%%%%%%%%%%%%%%%%%%%%%%%%%%%%%%%%%%%%%
\subsubsection{RICH-1 upgrade in the peripheral region}
\label{sec:upgrade.rich.APV}

The existing system of MWPC with CsI photocathodes continues to 
equip the peripheral 75\% of the active area of the RICH-1 detector. 
The readout system will be replaced by a new analogue sampling readout based on the
APV25 chip \cite{Abbon:05b}, already employed to read COMPASS silicon
(see Sec.~\ref{sec:tracking.vsat.silicon}) and GEM detectors
(see Sec.~\ref{sec:tracking.sat.gem}).  
The time information obtained from the new system allows us to reduce the effective time gate 
of the RICH readout from presently $\approx 3.5\,\mus$ to less than $400\,\ns$, leading to 
a large suppression of the  
uncorrelated background. In addition, the system significantly reduces
the readout  dead time, making it similar to the rest of the experiment, i.e. 5\% at $50\,\kHz$ 
trigger rate. Multiplexing 128 detector channels onto a single output channel also 
makes this solution very cost-effective. 

The APV25 chip, although originally designed for silicon microstrip
detectors, is used here for the first time to read ``slow'' signals
from a MWPC, maintaining a signal-to-noise ratio of about 9. This is made
possible by the fact that 
the time constants of the preamplifier and shaper stages of the chip
can be increased from the standard value of $50\,\ns$ up to $500\,\ns$  
by varying bias currents through feedback transistors. 
The sampling of the input signals is still performed at $40\,\MHz$, as
for the silicon and GEM detectors; the three samples transferred
per incoming trigger, however, are separated in time by $150\,\ns$ instead of
$25\,\ns$ in order to cover the rising edge of the MWPC signal. 
The multiplexed information from each APV25 chip 
is digitised using a pipelined ADC, similar to the one described for
the silicon and GEM detectors in Sec.~\ref{sec:daq.analog.sgadc}. To
allow a precise tuning of thresholds and efficient zero suppression,
necessary due to the typical 
Polya amplitude distribution for single photon events, which has an
approximately 
exponential shape at low gain, 12-bit ADCs are used for the RICH,
in contrast to 10-bit ADCs for silicon and GEM detectors. 
The sampling rate of the ADCs is increased from 
$20\,\MHz$ to $40\,\MHz$ in order 
to keep the dead time at the expected value. 

To match the configuration of the RICH-1 CsI photon 
detector, four APV25 chips are mounted 
on one front-end card, each APV25 reading 108 RICH pads. 
For optimum noise performance and reliability, four front-end cards
are then directly connected to one ADC card,  
digitising 16 APV25 chips and distributing the trigger and control signals
to the front-end cards. Two Virtex-4 FPGAs perform online common mode
noise correction and zero suppression. 

A full test of the 
proposed readout system was performed in nominal muon beam 
conditions in the central region of RICH-1 in 2004, with  
half the area of one photon 
detector equipped with APV25 chips and 5184 channels read out by 10-bit
ADCs. For the
optimal APV25 settings a signal-to-noise ratio very similar to the
original GASSIPLEX-based readout was obtained, the higher ballistic
deficit of the APV25 readout being fully compensated by a lower average
noise figure. As a consequence, the detection efficiency for single
photoelectrons  
was found to be 
very close to that obtained with the GASSIPLEX. 
The signal-to-background ratio, however, is strongly increased by 
the new readout due to its increased  time resolution. 
A gain by a factor of 5-6 was measured in the central region, where
pile-up due to the muon beam halo is most significant. 

%%%%%%%%%%%%%%%%%%%%%%%%%%%%%%%%%%%%%%%
\subsection{Electromagnetic calorimeter}
\label{sec:upgrade.ecal1}
An electromagnetic calorimeter (ECAL1), with overall dimensions  
of $4.00 \times 2.91\,\m^2$ is assembled and positioned upstream of
the HCAL1 hadronic calorimeter. The ECAL1 calorimeter covers the
angular acceptance of $\pm 180\,\mrad$ made possible by the use of the
new solenoid magnet. It allows measurements of reaction channels
with the production of low energy prompt photons and/or neutral pions.   

The ECAL1 calorimeter is formed by blocks of three different
sizes. The most central region is equipped with 576 blocks of
$38.2\times 38.2\,\mm^2$, originating from the GAMS
calorimeter \cite{Binon:86}. In the intermediate region 580 blocks of
$75\times 75\,\mm^2$ from a calorimeter built for the WA89 experiment
at CERN \cite{Brueckner:92} are used. The most external region is
filled with 320 blocks from the OLGA
calorimeter \cite{Astbury:85} with dimensions of $143 \times
143\,\mm^2$. The signal amplitude from all the calorimeter blocks are 
read  by fast sampling SADCs, identical to the ones described in
Sec. \ref{sec:daq.analog.caladc.sadc}.

%%%%%%%%%%%%%%%%%%%%%%%%%%%%%%%%%%%%%%%
\subsection{Rich wall}
\label{sec:upgrade.rw}

A new large-size (4.86 $\times$ 4.22 m$^2$) tracking station (Rich wall),
is positioned downstream of RICH-1, directly in front of 
the ECAL1 electromagnetic calorimeter. 
It consists of eight layers of MDT modules (see Sec. \ref{sec:pid.muon.mw1},
Fig.~\ref{fig:compass.mw1_mdt}) that have the same design as MW1,
apart from a thinner Al profile 
($0.45\,\mm$ instead of $0.6\,\mm$). Mechanically, the Rich wall detector is similar to 
one station of MW1 (see Fig.~\ref{fig:compass.mw1_mw1}), with a converter inserted
in front of each of the four pairs of layers. Each converter is a sandwich of three plates 
(steel/lead/steel), resulting in a total converter thickness of about 3
radiation lengths. The Rich wall detector provides up to four coordinate
points per track in two projections, 
$X$ and $Y$. The required coordinate accuracy (of about $0.8\,\mm$ r.m.s. per 
plane) is obtained by reading out the MDT modules in drift mode. Ageing tests
performed with an Ar/CO$_2$ (70/30) gas mixture have shown no degradation effects
up to $1\,\C/\cm$. The beam-induced MDT
charge, integrated over the lifetime of the COMPASS experiment, is comparable
to that value; therefore no performance degradation is expected for the
duration of the experiment.  

The Rich wall tracker station performs a twofold function in the 
COMPASS spectrometer. First, it measures the particle trajectories
downstream of the RICH, thus allowing a better reconstruction of the
particle trajectories in the RICH volume. Second, it acts as
preshower for the ECAL1 electromagnetic calorimeter, thus improving
the spatial resolution of the calorimeter itself.

%References
%Oxford Danfysik, Ferry Mills, Osney Mead, Oxford, Oxfordshire, OX2 0ES, UK, http://www.applegate.co.uk/
%Hamamatsu Photonics,325-6,Sunayama-cho,Hamamatsu City,Shizuoka Pref.,430-8587,Japan,http://sales.hamamatsu.com/

%%% Local Variables: 
%%% mode: latex
%%% TeX-master: t
%%% End: 

% Summary
\section{Summary and outlook}
\label{sec:summary}
An overview of the COMPASS experimental setup, as used in the 2004 run, was given
in this article. 
The setup was designed with the purpose of detecting one or more outgoing hadrons 
in coincidence with the scattered beam particle, in a large momentum range, for large 
angular acceptances and at high incident flows.  In order to fulfil these requirements 
the outgoing particles are detected in a two-stage spectrometer, built around 
two large dipole magnets. The setup was built using various detector technologies, 
standard and novel ones.
Charged particle tracks are measured using scintillation fibre hodoscopes, planes of
high-precision silicon microstrips,
large-size, high-resolution, minimal 
material budget GEM and Micromegas micropattern gas detectors, MWPCs,
various sets of large-size drift chambers, straw drift tubes, planes of mini drift tubes
and of stainless steel drift tubes.
Particle identification is  an important requirement 
for most of the COMPASS physics programme.
It is performed making use of a large dimensions RICH 
detector, based on MWPCs with CsI photocathodes
as photon detection technology, 
and tuned for momenta between $5$ and $43\,\GeV/c$, of an 
electromagnetic calorimeter, two hadron calorimeters
and two muon filters. 
The trigger system makes use of scintillating counter hodoscopes 
and of the hadron calorimeter information.
Several custom-made electronics devices such as the SFE16 front-end, F1-TDC and trigger 
matrix chips, the CATCH, GeSiCA and SADC readout modules and
several specific front-end boards have been developed in order 
to match the COMPASS specifications, in particular the capability to stand 
high beam and trigger rates and the request of minimising the 
dead time generated by the data acquisition system. 

Most of the equipments, designed for use in both muon and hadron experimental 
programmes, were commissioned between 2001 and 2002; some additional detector 
planes have been added for the 2003 and for the 2004 data taking. Up to 2004, 
the apparatus has been optimised 
and successfully 
operated mostly with a muon beam. A first pilot run with a hadron 
beam was also carried out. 

All the COMPASS detectors, targets, 
control and monitoring packages, reconstruction and analysis software 
were described.   Emphasis was put on their general features and 
on the characteristics 
obtained in nominal muon beam conditions. Performances of the whole setup were 
summarised.

The upgrade, as put in operation for the data taking phase 
starting in 2006, was also described. With this setup fully
operational, COMPASS can complete the study of hadron structure using
leptonic probes.
The physics programme with hadron beams has
been only marginally addressed until now, with the measurement of the pion
polarisabilities during the 2004 pilot hadron run, and is scheduled from 2007 on. 
It comprises a variety of
measurements, including central production of light mesons and production of
heavy flavours, which require further upgrades of the apparatus in addition to 
the ones completed for the 2006 run.
Thin targets surrounded by radiation-hard high-precision silicon
trackers are needed to achieve the required mass resolutions and to separate
the production and decay vertices of charmed hadrons.
Precise tracking close to the beam in the spectrometer requires the use
of light detectors with little material budget in order to minimise
secondary interactions outside the target.
Radiation-hard electromagnetic
calorimeter modules are foreseen to 
cope with the required beam intensities.
Additional trigger devices will be necessary to select the
desired final states and kinematical regions.

In summary, the COMPASS setup is fully operational 
and is being used for a broad nucleon structure and spectroscopy 
programme. The upgrade completed in 2006
insures data taking with further improved performances 
in the forthcoming years.

%%% Local Variables: 
%%% mode: latex
%%% TeX-master: "compass_spec"
%%% End: 

% Acknowledgement

\begin{ack}

We gratefully acknowledge the CERN 
laboratory and the CERN AB, AT, PH and TS departments for providing
constant and efficient support during the construction 
phase of our experimental setup and later during data taking. Special
thanks are due to the CEA/Saclay DSM/DAPNIA technical services SACM,
SEDI and SIS and to the CERN AT/ECR group for their help for the
preparation and during the installation of the new superconducting
solenoid magnet. We are also grateful to F.~Kircher, T.~Taylor, and
A.~Yamamoto for their important reviewing work. We express our
gratitude to the  numerous engineers and technicians from our home
institutions, who have contributed to the  
construction and later to the maintenance of our detectors and
equipments. Special thanks go to J.M.~Demolis and V.~Pesaro for their
technical assistance during the installation and the running of the
experiment.  

We acknowledge support from MEYS (Czech Republic), CEA (France), BMBF and Maier-Leibnitz-Laboratorium (Germany), ISF (Israel), INFN and MIUR(Italy), MECSST, Daiko Foundation and Yamada Science Foundation (Japan), DST-FIST and SBET (India), MNII research funds for 2005-2007 and KBN (Poland), FCT (Portugal), Presidential grant NSh 5911.2006.2 (Russian Federation) and from the European Community-Research Infrastructure Activity under the FP6 "Structuring the European Research Area" programme Hadron Physics (contract number RII3-CT-2004-506078). 
\end{ack}

% Bibliography
%\bibliographystyle{elsart-num}
\bibliographystyle{nima_short}
\bibliography{compass_spec}

\end{document}